\documentclass[a4paper,12pt,titlepage,oneside]{kuphdthesis}

\usepackage{url}
\usepackage[bottom]{footmisc}
\usepackage[caption=false,font=footnotesize]{subfig}
\usepackage{graphicx, epstopdf, fancyhdr, setspace, color, pgf, tikz, pdfpages}
\usepackage{amsmath,float}
\usepackage{breqn, bbm}
\usepackage[inline]{enumitem}   
\usepackage{amssymb}
\usepackage{amsfonts}
\usepackage{breakcites}
\usepackage{wrapfig}
\usepackage{lscape}
\usepackage{rotating}
\usepackage{pdflscape}
\usepackage{afterpage}
\usepackage{multirow}
\usepackage{listings}
\usepackage{pgfplots}
\usepackage{booktabs}

\usepackage[linesnumbered, ruled,vlined, algo2e]{algorithm2e}
\SetKwInput{KwInput}{Input}                
\SetKwInput{KwOutput}{Output}              

\usepackage[english]{babel}
\usepackage[style=chicago-authordate, natbib=true, backend=biber, giveninits=false, uniquename=false, autocite=footnote, doi=false,isbn=false,url=false, eprint=false, abbreviate=false, maxbibnames=20]{biblatex}
\renewbibmacro{in:}{}
\AtEveryBibitem{\clearfield{month}}
\AtEveryCitekey{\clearfield{month}}

\AtEveryBibitem{\clearfield{series}}
\AtEveryCitekey{\clearfield{series}}

\graphicspath{{./figures/}}

\newtheorem{theorem}{Theorem}

\usepackage{xspace}
\usepackage{epstopdf}

\newcommand \COMMENT  [1] {}       

\begin{document}

\Title{Modelling and Control of Production Systems based on Observed Inter-event Times: An Analytical and Empirical Investigation} \Author{Nima Manafzadeh Dizbin} \Year{September 22, 2020}
\Program{Operations and Information Systems}

\TTitle{Üretim Sistemlerinin Gözlemlenen Olaylar Arası Sürelere Dayalı Modellenmesi ve Kontrolü: Analitik ve Ampirik Bir Araştırma} \TYear{September 22, 2020}
\TProgram{Operasyon ve Bilgi Sistemleri}

\Signature{Prof. Bar\i\c{s} Tan (Advisor)}
\Signature{Prof. Fikri Ahmet Karaesmen}
\Signature{Prof. Zeynep Ak\c{s}in Karaesmen}
\Signature{Prof. Ahmet Bar\i\c{s} Balc\i o\u{g}lu}
\Signature{Assoc. Prof. Aybek Korugan}

\prelimpages
\titlepage
\abstract{Technological advances allow manufacturers to collect and access data from a production system effectively. The objective of data collection is deploying the collected data in developing decision support systems for performance evaluation, problem identification, and production control.  The goal of this dissertation is to investigate how the collected data can be used to evaluate performance and optimize manufacturing systems, analytically and empirically.
 
In the first part of the thesis, I investigate the question: \textit{How can the collected data from the shop-floor be used in efficient control and design of manufacturing systems?} In order to investigate the impact of possible dependency in the inter-event times on the optimal control and performance measures of the system, first, a manufacturing system that is controlled by using a single-threshold production control policy is analyzed. It is shown that ignoring autocorrelation in interarrival or service times can lead to overestimation of the optimal threshold level for negatively correlated processes, and underestimation of the optimal threshold level for the positively correlated processes. Then the optimal control problem of a production/inventory control problem system with correlated inter-arrival and service times modeled as Markovian Arrival Processes is considered.  It is shown that the optimal control policy that minimizes the expected average cost of the system in the steady-state is a state-dependent threshold policy.
 
In the second part of the thesis, an exploratory data analysis is conducted by using a large industrial data set that includes 17 million rows of data related to flow of 17000 unique products that are processed in 500 different machines at a semiconductor manufacturing plant. The product flow dynamics that include inter-arrival, service, and inter-departure distributions and autocorrelations, and work-in-process and cycle time dynamics are investigated at different levels of detail. Then, the data-driven cycle time prediction problem is considered. In order to develop effective prediction methods, a methodology is developed to determine the most important product and system-state related features. The performance of different prediction algorithms is compared by using the selected features.
 
The analytical and empirical results presented in this dissertation show that the effective use of the collected data from a manufacturing system enables controlling the manufacturing system effectively and predicting its main performance measures accurately. 
}

\acknowledgments{
First and foremost, I would like to thank my academic advisor, for his phenomenal support and guidance over the course of my master's and doctoral work at Ko\c{c} University. Throughout the past six years, his optimism and positive energy inspired me to build a better version of myself every and each day. I would like to thank the jury members of my thesis committee for accepting to evaluate my thesis. I would also like to thank the faculty members and PhD students at the Graduate School of Business in Ko\c{c} University for their constant support and Didem G\"urses for her constant help in the administrative matters. Apart from academics, at Bosch Center for Artificial Intelligence (BCAI), I have met some of the most inspiring people. I would like to thank my colleges in BCAI whom I have worked with, academically and personally. Finally, I would like to thank my family.

Research leading to this thesis has received funding from the EU ECSEL Joint Undertaking under Grant Agreement No. 737459 (Project Productive 4.0) and from TUBITAK (217M145).
}
\tableofcontents
\listoftables
\listoffigures
\abbreviations{
\begin{tabular}{lp{1cm}l}
	MAP &  &Markovian Arrival Process\\
	Ph & & Phase Type Distribution\\
	MDP & & Markov Decision Process\\
	WF & & Wafer Fabrication \\	
\end{tabular}
}
\textpages

\chapter{Introduction}\label{chapter:introduction}

Technological advances allow manufacturers to collect and access data from a production system more easily and effectively. The objective of data collection is deploying the collected data in developing decision support systems for performance evaluation, problem identification, and production control. As a result, data-driven modeling and control methods are now considered as enabling technologies to address the technology challenges for implementing factory of the future \citep{IEC}. 

Over the years, manufacturers have become more successful in efficient control of their supply chain and deploying new methodologies that match supply with demand by adopting data-driven methodologies. The collected data is used for different purposes from predictive maintenance to performance evaluation, production control, to supply chain optimization. The supply chain efficiency can be further improved by using the recent development in data analytics, optimization, and machine learning methodologies. Manufacturing systems are affected by various types of uncertainty in their supply chain. Effective control of the manufacturing system and supply chain requires a thorough understanding of the effects of these uncertainties in strategic decision making related to demand fulfillment and production planning. Strategic decisions are based on the demand forecasting that needs to be coordinated with the production unit to account for long lead times of the products. 

There is a lack of documented, comprehensive, empirical research on manufacturing systems that uses detailed data from shop-floor to evaluate performance and optimize the manufacturing systems efficiency. One of the main research questions that arises in the study of a digital manufacturing systems is: \textit{How can the collected data from manufacturing systems be characterized and modeled for analytical methods and simulation in the most possible accurate way?} The answer to the first question gives rise to the following questions: \textit{How can the collected data from the shop-floor be used in efficient control and design of manufacturing systems?} and \textit{How can the collected data be used directly in identifying the most important features, evaluating the performance, and predicting the performance measures?} The goal of this dissertation is to investigate the answers to these questions. Each question is explained in further detail in the following subsections. I use stochastic models and the product flow data of the Reutlingen semiconductor manufacturing system of the Robert Bosch Company in investigating these questions, empirically and analytically. The inter-event data used in the empirical analysis consists of the product movement of every product processed in the 200 mm wafer fabrication. The raw dataset consists of 17\,223\,658 rows of inter-event data which contains the data related to almost 17000 unique products, categorized into 216 different parts. The products are being processed in 160 different equipment-groups which contain 500 different machines and can process 2159 different recipes.

\section{How can the collected data be used from the shop-floor in efficient control and design of manufacturing systems?}
\begin{figure}
	\centering
	\caption{Empirical distribution and dependence of the processing, inter-arrival, and inter-departure times of a specific equipment at the Reutlingen semiconductor manufacturing plant}
	\includegraphics[width=1.08\linewidth]{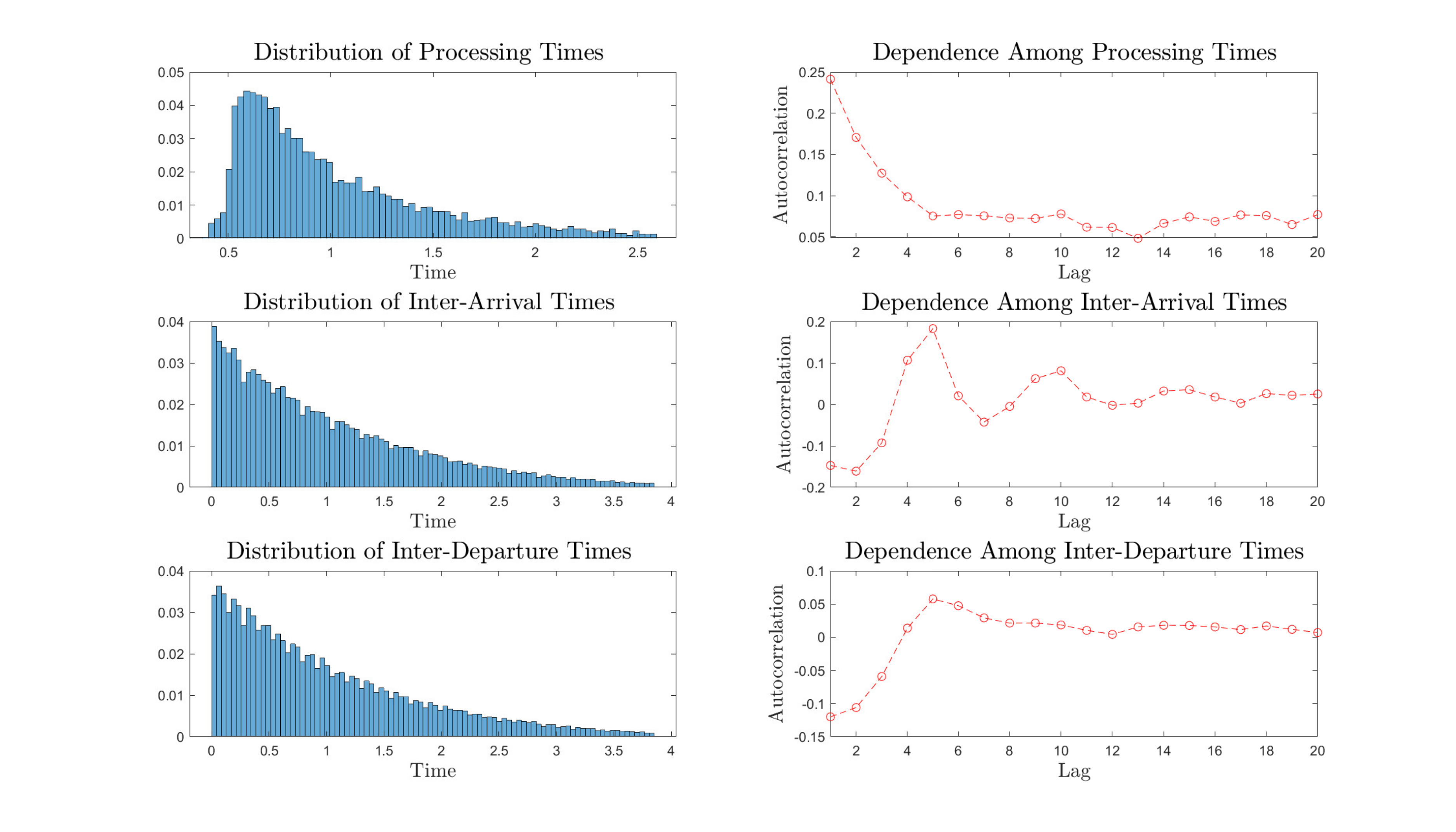}
	\label{empfig_intro}%
\end{figure}

The usual approach in the study of the manufacturing systems has been modelling the complex reality with a simple queuing model that represents the complex reality. The parameters of the adopted models, such as the rate of exponential distribution for modeling the processing times, are estimated by using the collected inter-event data. \citet{Shanthikumar2007QueueingProblems} state that the current queuing models for manufacturing systems ignore some important characteristics of the inter-event times data such as the correlation between the inter-event times. My empirical analysis of the statistical properties inter-event times of the dataset analyzed in this thesis shows that the inter-event times of a production system such as the inter-arrival and processing times may demonstrate a significant dependency between themselves. Figure \ref{empfig_intro} demonstrates the empirical distribution and autocorrelation of the inter-event times of a particular equipment at the Reutlingen plant. Such a dependency has been ignored in most of the analytical studies in control and design of production systems. Hence, the questions that arise are: \textit{How can we model the dependency between the inter-event times in the manufacturing systems?}, and \textit{How does the dependency between the inter-event times impact the optimal control of the system?}

I investigated the answer to the question "\textit{How can we model the dependency between the inter-event times in the manufacturing systems?}" in my master's thesis \citep{Reference98}. There, I showed how the inter-event data collected from a manufacturing system can be used to build a Markovian Arrival Processes (MAP) model that captures correlation in the inter-event times. The obtained MAP model can then be used to control the production system in an effective way. I also presented a comprehensive review on MAP modeling and MAP fitting methods applicable to manufacturing systems. Then, I presented results on the effectiveness of these fitting methods and discussed how the collected inter-event data can be used to represent the flow dynamics of a production system accurately \citep{Reference98}. 

In this thesis, I answer the question "\textit{How does the dependency between the inter-event times impact the optimal control of the system?}" by investigating the following questions:
\begin{itemize}
	\item How does ignoring the dependency between the inter-event times impact the control of a production system controlled by single-threshold policy?
	\item What is the optimal control policy of a production system with correlated inter-arrival and processing times?
	\item How does the single-threshold policy perform in comparison to the optimal policy? 
\end{itemize} 

\subsubsection{Impact of Correlation in Inter-Event Times on the Performance Measures of Production Systems Controlled by Single-Threshold Policy}
Chapter \ref{correlation_impact_single_threshold} analyzes a manufacturing system that is controlled by using a single-threshold policy to investigate the impact of capturing the flow dynamics accurately on the performance of a production control system. I study the impact of correlation in inter-event times on the optimal single-threshold level of the system numerically by employing the structural properties of the MAP. I show that ignoring correlated arrival or service process can lead to overestimation of the optimal threshold level for negatively correlated processes, and underestimation of the optimal threshold level for the positively correlated processes. In other words, I show that ignoring autocorrelation of a correlated inter-event times results in setting the base-stock level at a higher or lower level in comparison to the optimal threshold level. I conclude that MAPs can be used to develop data-driven models and control manufacturing systems more effectively by using the shop-floor inter-event data. This study contributes to the literature by presenting results that show the impact of correlation in the inter-event times on the production control for the first time in the literature.

\subsubsection{Optimal Control of Production-Inventory Systems with Correlated Demand Inter-Arrival and Processing Time}
In order to answer the question "\textit{How can we optimally control a production system with correlated inter-event times?}", I investigate the optimal control policy of a production system with correlated inter-arrival and service times in Chapter \ref{OptimalMAPControl}. I consider a production/inventory control problem with correlated demand arrival and service process modeled as Markovian arrival processes. The objective of the control problem is minimizing the expected average cost of the system in steady-state by controlling when to produce a new product. I show that the optimal control policy of a fully observable system is a state-dependent threshold policy. I compare the performance of the optimal policy with that of the optimal single-threshold or the base-stock policy where the threshold level is set independent of the state of the system. I evaluate the performance measures of the system controlled by the optimal state-dependent base-stock policy by using a Matrix Geometric method. I investigate how the autocorrelation of the arrival and service process impacts the optimal threshold levels of the system. The numerical analysis demonstrates that the state-independent policy performs near-optimal for the negatively correlated processes. However, when the inter-event times are positively correlated, using a state-dependent threshold policy improves the performance. I consider showing the optimality of the state-dependent base-stock policies for production/inventory systems with inter-arrival and service times that are modeled as MAPs, evaluating the performance of the single-threshold policy in controlling the system, and impact of correlation of the inter-event times on the optimal control levels as the main contributions of this study.

\section{How can the collected data be used directly in identifying the most important features, evaluating the performance, and predicting the performance measures?}
I use the product movement data from the semiconductor manufacturing system of the Robert Bosch Company in Reutlingen, Germany to investigate the answers to this question. This dataset allows us to perform exploratory data analysis (EDA) on the main features that are impacting the main performance metrics of the semiconductor wafer fabrication in different levels of detail in Chapter \ref{eda_chapter}. 

Semiconductor manufacturing systems are the most complex manufacturing systems in existence. The most important step of the semiconductor manufacturing is the wafer fabrication which accounts for more than 75\% of the total production time of the products. The capital intensive machines in the wafer fabrication necessitates the same machines to be used for similar processing steps resulting in a production network rather than assembly line type of a system. Hence, a queuing-network view of product flow is necessary for studying the performance of the wafer fabrication. The aim of this study is to identify the important features that are impacting the main performance metrics of the wafer fabrication such as the total cycle times of the products in different levels of detail.

The EDA reveals that half of the products spend more than 70\% of their time inside wafer fabrication waiting to be processed. The waiting times of the products is impacted by a different variety of factors from different product types to the dispatching rules used for assigning products into machines. The analysis reveals that some of the layers (subset of different equipment) in the product route may contribute significantly more than others to the waiting and eventually the total cycle times of the products. The layer level analysis suggests that the production manager may need pay extra attention to the bottleneck recipes inside the layers.  

The machine level EDA on the statistical properties of the inter-event times shows that processing times demonstrate a significant amount of positive dependency between themselves. Inter-arrival and inter-departure times may also demonstrate a significant correlation between themselves as well.  

This chapter contributes to the literature by empirically investigating the most important features that are impacting the performance metrics and control of the system. My empirical findings based on the product flow data of 500 machines reveal that significant amount of correlation may exist between the inter-event times in the machine level which has usually been ignored in the literature. In addition, EDA shows that major part of the products cycle times consists of the waiting times of the products. Based on the EDA results, I propose possible research directions that needs to be investigated to decrease the waiting and eventually total cycle times of the products. 

In Chapter \ref{ct_estimation_chapter}, I use the lot-trace data from the wafer fabrication to prepare a dataset for predicting the total cycle times of the products. The dataset consists of two sets of features that can explain the total cycle time of the products in the wafer fabrication, namely the product-related and system-state related features. The product-related features capture different product attributes such as the product type, the production route of the product and the distribution of the processing times on the product route. On the other hand, the system-state related features capture the state of the system upon arrival of the lot. I use this dataset to find prediction methods that are able to predict the cycle times of the lots with acceptable accuracy. In particular, rolling average based and learning based methods are adopted. The rolling average based methods predict the cycle times of the products by using the cycle times of the previous products that have completed all or some part of their processing. On the other hand, learning-based methods identify the parameters of a certain model by minimizing a specified objective function such as sum of squared errors. They predict the cycle times of the products by using the product and system state related features as independent variable and cycle times of the products as dependent variable. The contribution of this chapter is two-folds. First, it is shown how the most important product and system related features can be determined to predict the cycle time in a large-scale manufacturing system. Second, the compiled large dataset in this study can be used for evaluating the performance of different prediction algorithms that will be developed by other researchers. 


The findings of this thesis has been made possible through the participation of Ko\c{c} University in the \textit{Productive 4.0}\footnote{\url{https://productive40.eu/}} project and the corresponding \textit{T\"{U}BITAK} project. My one year PhD visit at Bosch Center for Artificial Intelligence has also been another contributing factor. The visiting position in Stuttgart, Germany enabled me to visit the wafer fabrication more frequently and discuss the related issues with the experts in understanding and cleansing the data and integrating different datasets. 

The remainder of this thesis is organized as follows. Chapter \ref{correlation_impact_single_threshold} presents the main findings of the impact of the dependence between inter-arrival times on the base-stock policy. Chapter \ref{OptimalMAPControl} identifies the optimal control of a manufacturing system with correlated demand-arrival and service times. Chapter \ref{eda_chapter} performs EDA on the semiconductor manufacturing system of the Robert Bosch Company to identify the most important features that impact the performance of system in different levels of detail. Chapter \ref{ct_estimation_chapter} develops prediction methods for predicting the cycle times of the products in the wafer fabrication.  Finally, Chapter \ref{conclusion_chapter} concludes the thesis. 

\chapter[Impact of Correlation on the Optimal Base-Stock Policy]{Impact of Correlation in Inter-Event Times on the Performance Measures of Production Systems Controlled by Single-Threshold Policy\footnotemark} \footnotetext{The results in this chapter are published in: Manafzadeh Dizbin, N. and Tan, B. (2019). "Modelling and analysis of the impact of correlated inter-event data on production control using Markovian arrival processes". Flexible Services and Manufacturing Journal 31.4, pp. 1042–1076 }\label{correlation_impact_single_threshold}

\section*{Abstract}
Empirical studies show that the inter-event times of a production system are correlated. However, most of the analytical studies for the analysis and control of production systems ignore correlation.  In this chapter, I analyze the impact of correlation on a manufacturing system that is controlled by using a base-stock policy. I model the correlation in inter-arrival and service times using Markovian Arrival Process (MAP). I study the impact of correlation in inter-event times on the optimal base-stock level of the system numerically by employing the structural properties of the MAP. I show that ignoring correlated arrival or service process can lead to overestimation of the optimal base-stock level for negatively correlated processes, and underestimation for the positively correlated processes. I conclude this chapter by stating that MAPs can be used to develop data-driven models and control manufacturing systems more effectively.
\section{Introduction}
\begin{figure}
	\caption{The observed and fitted distribution and autocorrelations ($\rho_k$) of inter-arrival time of a specific equipment at the Robert Bosch Reutlingen semiconductor manufacturing plant}
	\includegraphics[scale=0.8]{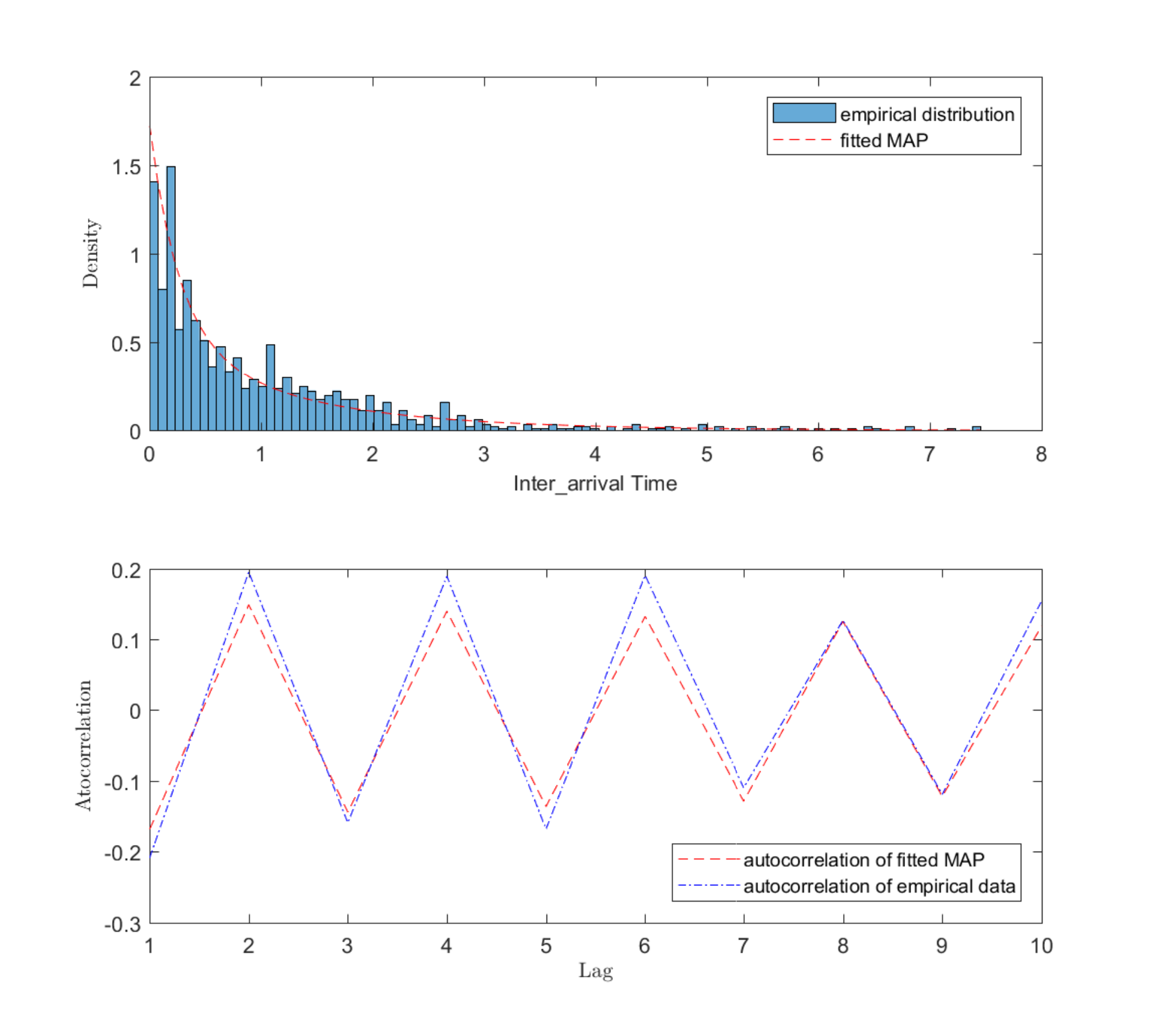}
	\label{Bosch_ia}
\end{figure}

In this chapter, I investigate the impact correlation in the inter-arrival and service times of the production systems. Empirical studies show that the inter-event times of a production system may demonstrate a significant correlation \citep{Reference6, Inman1999EmpiricalSystems}. For instance, Figure \ref{Bosch_ia} shows the empirical distribution of the inter-arrival times and their autocorrelations by using the data collected from a specific equipment at the Robert Bosch Reutlingen semiconductor manufacturing plant. The inter-arrival times demonstrate significant autocorrelations between themselves. 
In addition to empirical studies, analytical studies of the output dynamics of production system with i.i.d. service time distribution demonstrate that the output process from a production system can be correlated \citep{Hendricks1993TheBuffers, Tan2017OnBlocking}. Since the output from a production system is an input to another one, the inter-arrival times of work centers may also exhibit autocorrelation.  However, most of the studies in the literature make simplifying assumptions for the inter-event distributions.  More specifically, they assume independence of the inter-event times in order to achieve analytical tractability.  However, these simplifying assumptions may increase the risk of model misspecification that leads to errors in setting control parameters. 
\begin{table}
	\centering
	\caption{Statistical descriptors of the inter-arrival time of an equipment and its fitted MAP}
\scalebox{.8}{
	\begin{tabular}{lcccccccc}
		& \multicolumn{3}{c}{Moments} & \multicolumn{5}{c}{Autocorrelation (lag)}       \\ \hline
		& 1       & 2       & 3       & 1       & 2       & 3       & 4       & 5       \\ \hline
		Real & 1.00    & 2.37    & 8.87    & -0.2088 & 0.1950 & -0.1593 & 0.1902 & -0.1690 \\
		$\hat{\rm{MAP}}$ \citep{butools}  & 1.00    & 2.65 & 11.78  & -0.1684 & 0.1488 & -0.1439 & 0.1398 & -0.1360 
	\end{tabular}
		}	\label{9TEPQI44}
\end{table}
In this chapter, I focus on the inter-event data such as the inter-arrival, inter-departure, and service completion times that can be collected from the shop floor. My objective is to answer the following question: \emph{what is the benefit of capturing correlation in inter-event times on the performance of a production system controlled by single-threshold or base-stock policy?}  
I adopt Markovian Arrival Processes (MAP) to model the inter-event times. MAPs are point processes that can approximate any inter-event process arbitrarily close enough \citep{Reference12}. There exist numerous MAP fitting methods in the literature of telecommunications systems that can be exploited to fit a MAP for the collected inter-event data. This approach yields accurate models that can generate flow dynamics that is statistically very close to the collected data. These models can be used in process visualization as building blocks for analytical and numerical simulation.  Since the inter-event data can be collected for a single machine, for a group of machines, or for a production area, this approach can also be used to develop aggregate models for planning.  The MAP representation of a production system can also be used to determine the parameters of the control policies for a given production system. 

In order to answer the second research question on the \emph{impact} of capturing the flow dynamics accurately on the performance of a production control system, I then focus my analysis to the control of a manufacturing system that uses a single-level base-stock policy. This policy is fully specified with a single threshold, referred as the base-stock level where production stops when inventory hits the threshold. This policy (or its alternative representations) is commonly implemented in practice.

\section{Related Literature}\label{literature}

\subsection{Presence of Correlation in Inter-Event Times of Queuing and Manufacturing Systems}
\citet{Reference6} analyze the cycle time of semiconductor manufacturing systems and show that the cycle times are highly correlated. \citet{Inman1999EmpiricalSystems} demonstrates the presence of autocorrelation in the output data of some manufacturing stations in automotive industry. \citet{Reference19} present data depicting the presence of significant empirical first lag autocorrelation in times between machine alignment (registration) failures, aviation equipment times to failure, and inter-arrival times of packaged food items. \citet{Tan2017OnBlocking} present evidence of correlated inter-departure times of cars leaving an assembly line of an automotive manufacturer. \citet{Hendricks1993TheBuffers} demonstrate that the output process of a production line with i.i.d. arrival and service processes can be correlated. \citet{Tan2017OnBlocking} propose an analytical method to calculate the inter-departure time correlation analytically and discuss how buffer capacity and process variability affect autocorrelation structure.

\subsection{Impact of Correlation in Inter-Event Times on Performance Measures}
The impact of correlation on the performance of queuing and production systems has been investigated by using simulation and analytical methods in the literature. Simulation studies introduce correlation to queues by methods like Transform Expand Sample and Minification \citep{Reference20}, vector-auto-regressive-to-anything \citep{Reference64} and Markovian Arrival Processes. On the other hand, analytical studies use Markov Renewal Processes, Markovian Arrival Processes, and Supplementary Variables to take dependence into account \citep{Reference62}.    

\subsection{Simulation Studies}
\citet{Reference20} shows that correlation in inter-arrival and service times may severely affect the distribution of the queue length and consequently waiting times of the products inside the system. \citet{Reference25} analyze a production system controlled by Kanban cards with concurrent and sequential ordering processes. They demonstrate that correlated demand arrivals can have significant impact on the WIP levels and the expected waiting time of the products. \citet{Reference19} study the effects of correlated job arrivals in an M/M/1 queue, the effects of correlated production process in an M/G/1 queue, and in a pull-type production system. \citet{Reference32} study the impact of arrival dependence on the performance measures of G/M/1 queues. They demonstrate that ignoring long range dependence can significantly alter the behavior of system. \citet{Reference33} analyze the impact of long-memory arrivals on the performance measures of queuing systems. \citet{Reference32} implement a similar study, stating positive autocorrelation among inter-arrival times increases the waiting time and queue size of the process. \citet{Reference21} study the impact of dependence in a single server queue and conclude that positively correlated arrivals increase the expected waiting time of a customer in the system. \citet{Reference51} conduct a simulation study with MAP arrival, PH service with infinitely many servers and finite buffer capacities. They show that positively correlated arrivals spent more expected time in the system. \citet{Reference22} evaluate the impact of correlation on assembly processes, serial lines and disassembly processes. They demonstrate that considering autocorrelation in modeling the process enhances the accuracy of estimated performance measures of the real system.

\subsection{Analytical Approaches}
\citet{Reference35, Reference36} study the effect of dependence on the expected waiting time of a system with integer Markov dependent arrival intervals and independent exponential service. \citet{Reference38} studies the impact of positive correlation on the waiting time distribution of a M/M/1 queue. He shows that by increasing the correlation, the pace of convergence of waiting time distribution to its liming value increases. \citet{Reference76} studies a correlated arrival process which is represented by a two-state Markov Renewal process. \citet{Reference65} investigate the impact of correlation on mean queue length of queues with correlated Markov renewal arrival process, and exponential service time (MR/M/1). They show that mean queue length of a system with correlated arrival, independent of the arrival distribution, can be over 30 times greater than renewal processes' queue length. \citet{Reference27, Reference74} study the  the impact of positively correlated arrival times on the performance measures of MR/GI/1 queues. They demonstrate that the higher correlation in arrival streams may result in more variability of the waiting times and higher mean queue lengths. \citet{Bauerle1997MonotonicityQueues} study the effect of the transition matrix of the environmental process on the waiting time of the $n^{th}$ customer and on the stationary waiting times. They generalize the results of \citet{Reference74} and state that the greater the dependency in the arrival process, the larger the stationary waiting times with respect to the increasing convex order. \citet{Susto2016DealingProblems} studies the correlation among the jobs at separate facilities and evaluate the effect of correlation on a variety of system performance measures such as queue length. \citet{Reference40} study a single server queue with MAP inter-arrival and generally distributed service times with a cross-correlation between arrival and service process. They evaluate the impact of autocorrelation and cross-correlation on the mean waiting times. \citet{Reference41} consider a discrete time queuing system with two-state discrete time Markov modulated batch arrival with autoregressive input. They demonstrate that the mean queue length of the processes is quite different in correlated arrival processes. 

My study in this chapter differs from the existing literature in the sense that I evaluate the impact of autocorrelation on the optimal control of a production system. In particular, I evaluate the impact of correlation in the inter-arrival and service times in a production system controlled by the single-threshold or base-stock policy. The influence of demand variability on the performance of a make-to-stock systems has already been investigated \citep{JEMAI}. However, the impact of correlation in demand inter-arrival times and service times on the optimal control of the system has not been  investigated.   

\section{Model and Evaluation Methodology} \label{model}
I now present the basic model and the evaluation methodology used to answer the main research question: \emph{what is the benefit of capturing correlation in inter-event times on the performance of a production control system?}  

\subsection{Markovian Arrival Processes}
I model the correlated demand inter-arrival and service times as MAPs \citep{Neuts1979AProcess}. MAPs are generalization of PH distributions \citep{Neuts1975ProbabilityType} that can capture correlated inter-event times. They contain most of the commonly used arrival processes such as Erlang processes, Coxian distributions and Markov-modulated Poisson processes (MMPP) as subclasses. A MAP consists of two different sub-processes each of which has a discrete state space called phases. One sub-process represents the dynamics of the phase process denoted by $ D_0 $, i.e transition between phases without an event, while the other corresponds to the occurrence of an event denoted by $ D_1$. A MAP can be interpreted as a continuous-time Markov-chain with the generator matrix $D = D_0 + D_1$, and $|D|$ states. Let $D$ be an irreducible generator matrix with initial probability vector $\beta_0$. The Markov chain starts at state $i$ with probability $\beta_0(i)$, spends an exponential time with rate $\lambda_i = -D_0(i ,i)$ there, and moves to state $j$ with probability $p_{ij}$ defined as:

\begin{equation}
p_{ij}= 
\begin{cases}
\dfrac{D_1(i ,i)}{\lambda_i},& \text{ } j=i\\
\dfrac{D_0(i ,j)+D_1(i ,j)}{\lambda_i},  & \text{} j\ne i
\end{cases}
\end{equation}  
When the Markov chain experiences a state transition from the state $i$ to $j$, an arrival occurs with probability $\dfrac{D_1(i ,j)}{\lambda_i}$. Let $P = -(D_0)^{-1}D_1$ and $ \beta P= \beta $ and $ \beta \mathbbm{1}=1 $ where $\mathbbm{1}$ is vector of ones with appropriate size. The phase distribution at arrival instants form a discrete time Markov chain with transition probability matrix $P$. A MAP can be fully specified with $D_0$ and $D_1$ if I let $\beta_0= \beta$. I denote a MAP with these sub-processes with MAP($ D_0,D_1$). For further detailed information regarding, distribution, moments, autocorrelation structure, and other features of MAPs the reader is referred to \citet{Reference1}, \citet{Reference2} and \citet{Reference14}. 

\subsubsection{MAP Representation of the Output Process of a Production System}\label{2M0B}
The output process from a certain production system with limited number of states can be represented a MAP. To do so, one needs to differentiate between the transitions that lead to departure of a product and the rest of transitions of the system. By capturing the transitions that do not lead to departure in matrix $D_0$, and transitions that lead to departure in matrix $D_1$, I represent the output process as MAP($D_0,D_1$). 

For instance, the MAP representation of a two-station production line with exponential service times with rates $\mu_1$ and $\mu_2$, respectively, and no inter-station buffer is given below. The state of this system consists of a tuple $(s_1,s_2)$ where $s_i$ demonstrates the state of machine $i$, $s_i \in \{0,1\}$ where $s_i=1$ represent that machine $i$ is working, $s_1=0$ represents machine 1 is blocked and $s_2=0$ represents machine 2 is idle. The $D_0$, and $D_1$ matrices of this line can be written as:

\[
D_0=\begin{bmatrix}
-\mu_1 &\mu_1 & 0\\
0& -(\mu_1+\mu_2) & \mu_1 \\
0& 0&-\mu_2 
\end{bmatrix}   
\qquad
D_1=\begin{bmatrix}
0 &0  & 0\\
\mu_2& 0 & 0 \\
0&\mu_2&0 
\end{bmatrix}     
\]
\\
\noindent where the states are ordered as $(1,0), (1,1), (0,1)$.  The $D_1$ matrix captures transitions that are related to process completions on machine 2 that generates an output from the production line. \citet{Tan2017OnBlocking} use similar approach to determine the autocorrelation structure of the output process from open and closed queuing networks subject to blocking.  The output process of the two-station production line with no inter-station buffers with the MAP($D_0,D_1$) representation given above demonstrates a negative first-lag autocorrelation resulting from the blocking phenomena in the system.
\subsection{Base-Stock Model}
I consider a production/inventory system with correlated demand inter-arrival and service times that are modeled as MAPs. An arriving demand will be satisfied from the inventory according to the first-come-first-served (FIFO) rule if a product is available in the inventory. Otherwise, if the inventory is empty, it will be backlogged until it is satisfied. I will assume that the production is controlled by the single-threshold or base-stock policy. Under this policy, the producer produces when there is an outstanding production order i.e., inventory is under a given threshold. I employ this control policy since it (or its equivalent representations) is commonly implemented control policy in practice. Note that single-threshold policy is not necessarily the optimal control policy in this setting. I demonstrate in \citet{ManafzadehDizbin2020OptimalTimes} that the optimal policy to control a production/inventory system with correlated inter-arrival and service times is state-dependent threshold policy.

I assume that raw materials are supplied from an unlimited stock with zero lead-time and the system is continuously reviewed. The cost structure of the system consists of inventory holding and backlog costs. The inventory holding cost is $h$ per unit per unit of time and the backlog cost is $ b$ per unit per unit of time. The objective of the problem is minimizing the long run average cost of the system. 

\subsubsection{Optimal Base-Stock Level of a System with MAP Arrival and Service Times}
In this section, I demonstrate how to calculate the optimal base-stock level of a system with correlated arrival and service processes modeled as MAP. Correlated arrival processes can be an output process from an earlier stage of the production line. Correlated service can be the result of production time variations of the machines. Let $S$ be the base-stock level of the system, $O(t)$ be the number of outstanding production orders at time $t$ to reach base-stock level $S$, $X(t)=S-O(t)$ be net inventory level at time $t$, and $G_X(S)$ be the stationary distribution of $X$ given base-stock level $S$. 

It is known that the optimal base stock level that minimizes the expected cost under the single threshold policy is given by:
\begin{equation} \label{s*}
S^\ast=\operatornamewithlimits{argmin}_S\left\{G_X(S)\geq\dfrac{b}{b+h}\right\}.
\end{equation}
The $G_X(S)$ of a system with MAP arrival and service processes is equivalent to the queue length distribution of the MAP/MAP/1 queue given in Equation \ref{GxS} which is a subclass of Quasi-Birth-Death (QBD) processes. 

The generator matrix of a QBD process consists of three types of matrices. Forward matrices $ (F) $ capture the transitions that lead to arrival of new demand. Local matrices $ (L) $ capture the transition between phases of the process without any arrivals or departure from the system. Backward matrices $ (B) $ capture transitions leading to a service completion. These matrices for a system with MAP ($ D_0 $, $ D_1 $) arrival, and MAP ($ A_0 $, $ A_1 $) service processes can be calculated as: 
\begin{eqnarray} \label{matdef}
F=D_1 \otimes I_A \nonumber \\ B=I_D \otimes A_1 \nonumber \\ L=D_0\oplus A_0 \nonumber \\ L_0= D_0 \oplus I_A 
\end{eqnarray}
where $\otimes$, $\oplus$, $I_D$, and $I_A$  are Kronecker product, Kronecker sum, and square identity matrices with size $|D_0|$ and $|A_0|$, respectively. The generator matrix of a process with MAP arrival and service can be written as follows:

\begin{equation}
Q=\begin{pmatrix}
L_0  &  F  &        &        &        \\ 
B   &  L  &  F       &        &        \\ 
&  B  &  L       &  F       &        \\ 
&   &   \ddots   &   \ddots   &   \ddots   \\ 
&   &        &        &        \\ 
\end{pmatrix}.
\label{MAP/MAP/1 Generator Matrix}
\end{equation}
\\
The limiting probability distribution of QBD processes with block matrices $ F,L,B $ and level zero local matrix $L_0$ can be calculated as 
\begin{equation} \label{probdist} \pi_n=\pi_0R^n \end{equation} 
where $ R $ is the solution of quadratic matrix equation 
\begin{equation} \label{MGR}
F+RL+R^2B=0
\end{equation}
and $ \pi_0 $ is the solution of the following set of equations:
\begin{eqnarray} \label{MGRpi}
\pi_0(L_0+RB)=0, \nonumber\\
\pi_0(I-R)^{-1}\mathbbm{1}=1.
\end{eqnarray}
There exist a number of computational algorithms for computing the geometric matrix $R$. I refer the reader to \citet{Reference57} for a review of the methods and their algorithmic implementations. The stationary distribution of the $X$ as a function of $\pi_0$ and $R$ can be written as: 
\begin{equation} \label{GxS}
G_X(S)=\sum_{i=0}^{S}\pi_0R^i\mathbbm{1}.
\end{equation}

In my study, I first generate the matrices $D_0$ and $D_1$ for the given arrival process and the matrices $ A_0 $, $A_1 $ for the given service process with their autocorrelation structures and inter-event time distributions. Then, I generate the block matrices $ F,L,B, L_0$ from $D_0, D_1,A_0,A_1$ by using Equation (\ref{matdef}). Then, I determine the steady-state distribution for a given base-stock level $S$ by using Equations (\ref{probdist}), (\ref{MGR}), (\ref{MGRpi}), and (\ref{GxS}).  Finally, I determine the optimal base-stock level by using Equation (\ref{s*}). Once the optimal base-stock level, $S^\ast$ is determined, the performance measures can be evaluated from the distribution of $G_X(S^\ast)$. The expected inventory level ($E[X^+]$), the expected backlog level ($E[X^-]$), and the probability of not having inventory in the system ($Pr[X<0]$) can be calculated using the geometric matrix $R$ as:  

\begin{equation}
Pr[X<0] =\sum\limits_{s=S^\ast+1}^{\infty}{\pi_s}\mathbbm{1} =  \pi_0R^{S^\ast+1}(I-R)^{-1}\mathbbm{1},
\end{equation}

\begin{equation}
E[X^-]=\sum\limits_{s=S^\ast+1}^{\infty}{\pi_s(s-S^\ast)}\mathbbm{1} = \pi_0R^{S^\ast+1}(I-R)^{-2}\mathbbm{1},
\end{equation}

\begin{equation}
E[X^+]=\sum\limits_{s=0}^{S^\ast-1}{(S^\ast-s)\pi_s\mathbbm{1}} = S^\ast\pi_0(I-R)^{-1}\mathbbm{1} -\pi_0(I-R)^{-2}(I-R^{S^\ast})R\mathbbm{1}.
\end{equation}

\subsection{Evaluation Methodology} \label{impact}
\subsubsection{Impact of Autocorrelation on the Optimal Base-Stock Level}
In order to measure the impact of autocorrelation on the optimal base-stock level, I employ processes with the same marginal distribution and different magnitude of autocorrelation. In order to generate processes with the same marginal distribution and different magnitudes of autocorrelation, I use the fact that first-lag autocorrelation of a MAP is a linear function of the elements of $D_1$ matrix \citep{Reference3}. I use processes with the same $D_0$ (which is associated with the marginal distribution of the process) and different $D_1$ matrices. I generate $D_1^{new}$ matrices in the following form :
\begin{equation}
D_1^{new}=\theta D_1+(1-\theta)D_1^{ren}
\end{equation}
where $\theta\in[0,1]$, and  $D_1^{ren}$ is the $D_1$ matrix of a MAP with the same distribution and zero autocorrelations (renewal MAP). The $D_1^{ren}$ of the renewal MAP can be calculated as:  
\begin{equation}
\label{renewal MAP}
D_1^{ren}=D_1\mathbbm{1} \beta
\end{equation}  
where $\beta$ is the phase distribution immediately after an arrival of the MAP. I use MAP($D_0,D_1^{new}$) as an arrival process to measure the effect of autocorrelation on the optimal control policy. The first-lag autocorrelation of the MAP($D_0,D_1^{new}$) is 
\begin{equation}
\rho_{1}^{new}(\theta)=\theta\rho_1
\end{equation}
where $\rho_1$ is the first-lag autocorrelation of the MAP($D_0,D_1$).

As an example, consider the system with two machine and no buffer in between, presented in Section \ref{2M0B}, with rates $\mu_1=\mu_2 = 1.5$. The $D_0$, $D_1$, and $D_1^{ren}$ matrices of this system can be written as follows 

\[
D_0=\begin{bmatrix}
-1.5 &1.5  & 0\\
0& -3 & 1.5 \\
0& 0&-1.5 
\end{bmatrix}   
\qquad
D_1=\begin{bmatrix}
0 &0  & 0\\
1.5& 0 & 0 \\
0& 1.5&0 
\end{bmatrix}     
\qquad
D_1^{ren}=\begin{bmatrix}
0 &0  & 0\\
0.75& 0.75 & 0 \\
0.75& 0.75&0 
\end{bmatrix}     
\]

\subsubsection{Performance of Renewal Approximations in Estimating Correlated Processes}
In the numerical experiments, I evaluate the performance of commonly adopted distributions in the literature in modeling a correlated arrival process of a system controlled by the single threshold base-stock policy. In particular, I consider the performance of modeling the correlated arrival process with processes that capture the exact marginal distribution, the first-two moments, and the first lag autocorrelation of the correlated process. I evaluate the performance of the system controlled by the threshold calculated by using the approximated processes by comparing their performance measures to the values obtained by using the threshold set by using the original correlated process. The performance measures that I consider are the total cost, the expected inventory and backlog, and the probability of not having an inventory in the system. For brevity, I present the results for systems with correlated inter-arrival and exponentially distributed service times. However, the results are similar for the correlated service processes \citep{Reference98}.  I calculate the performance measures of a system with the correlated arrival and exponentially distributed service times, by employing the queue length distribution of a MAP/M/1 queue.

I employ PH distribution with the $D_0$ equal to that of the correlated process and $t=D_1\mathbbm{1}$ to model a correlated process by means of its marginal distribution. I employ the queue length distribution of the PH/M/1 queue in order to calculate the optimal base-stock level. Then, I calculate the performance measures of the original system controlled by this base-stock level. I employ a PH distribution with the same first-two moments ($\rm{PH}_2$) to model the arrival process by means of it's first-two moments. I employ the method presented in \citet{butools} to generate a PH distribution with the first-two moments of the correlated arrival process. I employ the queue length distribution of the $\rm{PH}_2$/M/1 to calculate the base-stock level of the system. Finally, I model the correlated arrival process by means of its first moment by using exponential distribution. I employ the queue-length distribution of the M/M/1 queue to calculate the optimal base-stock level.

\section{Impact of Correlation in Inter-Arrival Times on the Control of a Production/Inventory System}\label{4}
In this section, I analyze the impact of first-lag autocorrelation on optimal base-stock level. I employ two types of processes with positive and negative first-lag autocorrelations. I consider autocorrelation structures with positive and negative first-lag and decaying magnitude of higher lags. I analyze the impact of autocorrelation with different values of coefficient of variation of the arrival and service processes, and the traffic intensity of the system. 

\begin{table}[] 
	\centering
	\caption{The parameters used in the numerical experiments}
	\begin{tabular}{|ll|l|}
		\cline{1-3} 
		&$cv_a$ & \{0.2, 0.5, 0.8, 1, 1.3, 1.8\} \\\cline{1-3} 
		& $cv_s$& \{0.2, 0.5, 0.8, 1, 1.3, 1.8\}\\\cline{1-3} 
		& $\nu$ & \{0.5, 0.65, 0.8\}\\\cline{1-3} 
		& $\rho^+_1$ & \{0, 0.02, 0.04, \dots, 0.68, 0.70\}\\\cline{1-3} 
		& $\rho^-_1$ & \{0, -0.02, -0.04, \dots, -0.48, -0.50\}\\\cline{1-3} 
	\end{tabular} \label{poscorrange}
\end{table}

\subsection{Impact of Positive Correlation in Inter-Arrival Times on the Control of a Production/Inventory System }
In this part, I consider a production system with positively correlated arrival and PH distributed service processes. In order to analyze the effect of autocorrelation on the optimal base-stock level, I design an experiment with different first-lag autocorrelations, coefficient of variation of arrival and service processes, and traffic intensities. Table \ref{poscorrange} gives the range of the coefficient of variation of the arrival ($cv_a$) and service ($cv_s$) times, first-lag autocorrelation of negatively ($\rho^-_1$) and positively ($\rho^+_1$) correlated processes, and traffic intensity ($\nu$) used in the numerical experiments.  This experiment setup includes  $3\times6\times36\times6 = 3888$ different cases analyzed.
\begin{table}[] 
	\centering
	\caption{ Accuracy of the renewal estimation in estimating the optimal base-stock level of the correlated system}
	\begin{tabular}{|lc|c|}
		\hline
		& & {\% cases: $S^{cor.} = S^{renewal}$ }   \\\hline 
		&$\nu = 0.50$ & 41.98   \\ \hline 
		& $\nu = 0.65$& 23.07   \\ \hline 
		& $\nu = 0.80$ & 12.96   \\ \hline 
	\end{tabular} \label{bstcaccuracy}
\end{table}
Table \ref{bstcaccuracy} demonstrates the percentage of the optimal base-stock levels calculated for the correlated processes ($S^{cor.}$) that are equal to the base stock level that is calculated by using the renewal approximation that ignores the autocorrelation ($S^{renewal}$) for different traffic intensities. When the utilization is 0.5, in 42\% of the cases, the optimal base-stock levels calculated by using the autocorrelation structure and by using the renewal approximation are the same.  However,  as the traffic intensity of the system increases, the percentage of the base-stock levels of correlated processes equal to that of the renewal process decreases. When the utilization is 0.8, only in 13\% of the cases, the optimal base-stock levels calculated by using the autocorrelation structure and by using the renewal approximation are the same.   

\begin{table}[htb]
	\centering
	\caption{Range of the base-stock levels of systems with different coefficient of variation of the arrival ($cv_a$) and service ($cv_s$) times ($\nu =0.8$, $\rho_1^+ \in [0, 0.7]$)}
	
	\begin{tabular}{c|c|cccccc|}
		\multicolumn{1}{l}{}   & \multicolumn{7}{c}{$cv_s$}                          \\ \cline{2-8} 
		\multirow{7}{*}{$cv_a$} &  & 0.2  & 0.5 & 0.8   & 1.0 & 1.3& 1.8   \\ \cline{2-8} 
		& 0.2 & 2-10 & 3-11  & 4-12  & 5-12  & 6-13  & 7-15  \\ \cline{2-8} 
		& 0.5 & 4-23 & 4-24  & 5-25  & 6-25  & 7-26  & 8-28  \\ \cline{2-8} 
		& 0.8 & 5-35 & 6-36  & 7-37  & 7-38  & 8-39  & 10-40 \\ \cline{2-8} 
		& 1.0 & 6-43 & 6-44  & 7-45  & 8-46  & 9-47  & 10-48 \\ \cline{2-8} 
		& 1.3 & 7-55 & 8-56  & 9-57  & 9-58  & 10-59 & 12-60 \\ \cline{2-8} 
		& 1.8 & 9-75 & 10-76 & 11-77 & 11-78 & 12-79 & 14-80 \\ \cline{2-8}  
	\end{tabular} 	\label{bs range}
\end{table}

\afterpage{%
	\clearpage
	\thispagestyle{empty}
	\begin{landscape}
		\begin{figure}
			\centering
			\includegraphics[scale=.7]{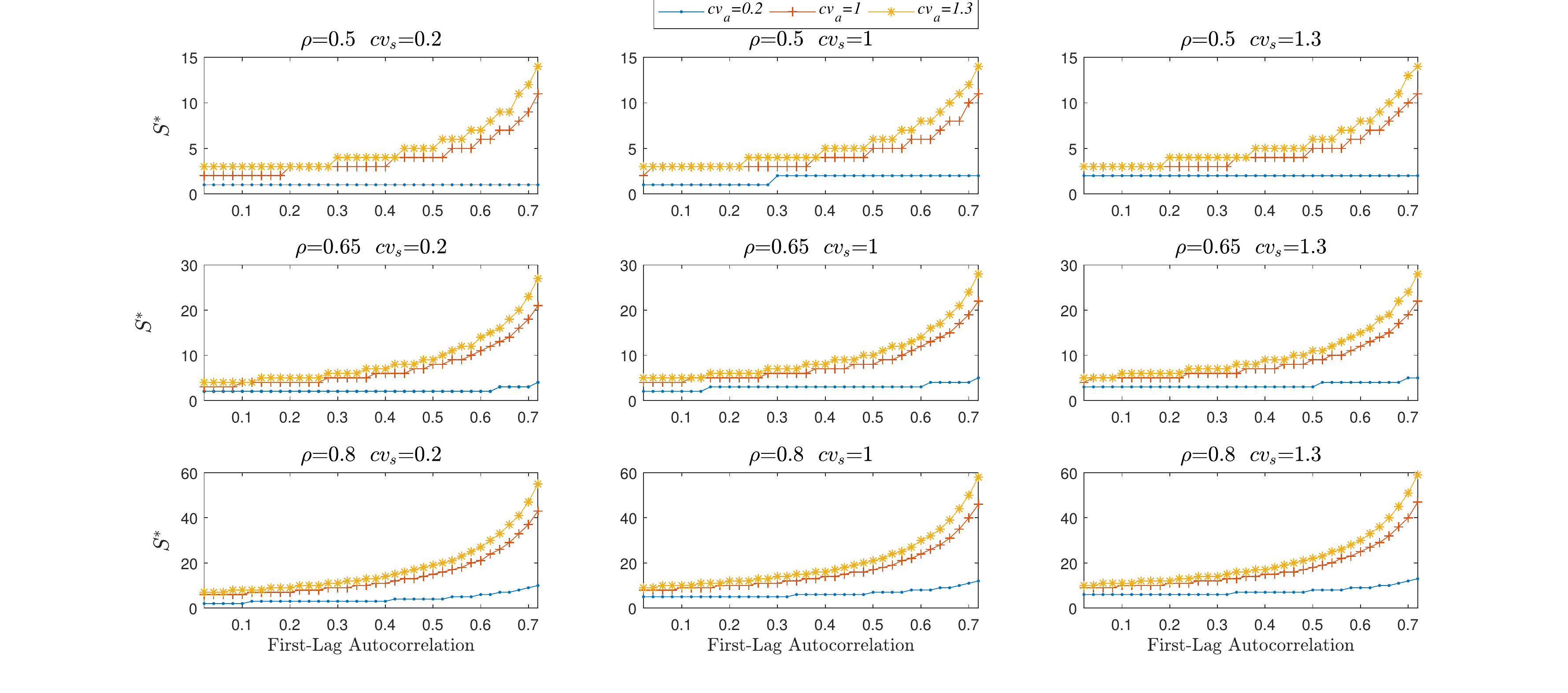}
			\caption{Impact of the first-lag autocorrelation of a positively correlated arrival process on the optimal base stock levels of a manufacturing system}
			\label{ArrImpactrho1p}
		\end{figure}
	\end{landscape}
	\clearpage
}
Table \ref{bs range} demonstrates the range of the optimal base-stock levels for correlated processes with their first-lag autocorrelation in the range [0, 0.7] for traffic intensity $\nu =0.8$. The base-stock level of a system with $cv_a$ = $cv_s$= 0.2 and zero first-lag autocorrelation is equal to 2. The optimal base-stock level for the same processes and first-lag autocorrelation 0.7 is equal to 10. The optimal base-stock level increases as a function of coefficient of variation of the arrival and service process. This result is in line with the findings of \citet{JEMAI} who study the effect demand variability on performance measures of make-to-stock systems. 

The range of the optimal base-stock level for processes with no autocorrelation is between 2 and 14. For the correlated processes, the range of the optimal base-stock level is from 2 to 80. This is an indication of significant impact of positive autocorrelation in inter-arrival times on the optimal base-stock levels. 

My numerical analysis shows that the impact of ignoring autocorrelation increases as a function of the first-lag autocorrelation.  Figure \ref{ArrImpactrho1p} demonstrates the optimal base-stock level of the system as a function of the first-lag autocorrelation, for some of the cases. In all of the figures, the optimal base-stock level increases as a function of the first-lag autocorrelation. This behavior is due to the fact that positive autocorrelation increases the probability of having higher queue lengths in the output processes of the system since a short inter-arrival time is expected to be followed by a short  inter-arrival time, and similarly a long inter-arrival time is expected to be followed by a long  inter-arrival time in positively correlated processes. In other words, the process creates clusters of short and long inter-arrival times. This leads to an increase in the probability of higher queue lengths in comparison with independent inter-arrival times.

\subsection{Impact of Negative Correlation in Inter-Arrival Times on the Control of a Production/Inventory System}
In this part, I analyze a production system with negatively correlated arrival and PH distributed service processes. Table \ref{poscorrange} gives the values of the parameters used in the analysis. Figure \ref{ArrImpactrho1n} demonstrates the impact of negative autocorrelation on the optimal base-stock level for some of the cases. In all of the cases, the optimal base-stock level of the system decreases as a function of the first-lag autocorrelation. In other words, ignoring correlation results in overestimation of the optimal base-stock level of a system with negatively correlated arrival process. This is due to the fact that in negatively correlated processes a short inter-arrival time is expected to be followed by a long inter-arrival time. Therefore, the probability of having higher queue lengths in the system decreases, which results in having lower optimal base-stock level in comparison with the same process with lower magnitude of correlation. 

\afterpage{%
	\clearpage
	\thispagestyle{empty}
	\begin{landscape}
		\begin{figure}[ht]
			\includegraphics[scale=.7]{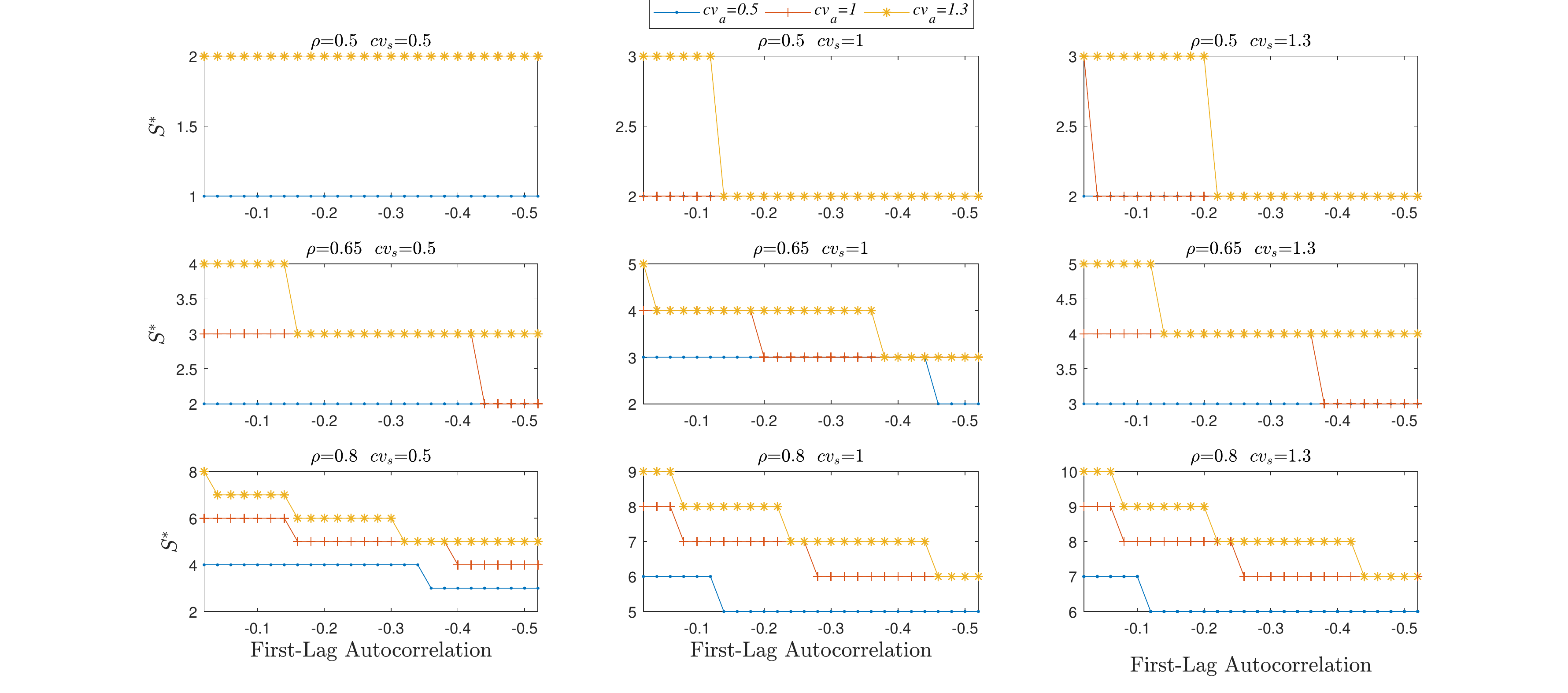}
			\caption{Impact of the first-lag autocorrelation of negatively correlated arrival process on the optimal base stock levels of a manufacturing System}
			\label{ArrImpactrho1n}
		\end{figure}
	\end{landscape}
	\clearpage
}

I conclude that main drivers of the optimal base-stock level of a system with correlated arrival (or service) are the autocorrelation structure, coefficient of variation of arrival and service processes, and traffic intensity of the system. My numerical analysis demonstrates that autocorrelation structure, significantly impacts the optimal base-stock level. The optimal base-stock level increase as the first-lag autocorrelation of the arrival (or service) process becomes more positive.

\section{Effects of Using Different Renewal Processes to Approximate Correlated Processes on the Performance}\label{sec:renewal}
In this section, I evaluate the performance of modeling a correlated process by using different distributions ignoring autocorrelation. As we saw earlier, ignoring autocorrelation in positively (negatively) correlated processes, underestimates (overestimates) the optimal base stock level of a production system controlled by base-stock policy. In this section, I analyze the performance of modeling a correlated process by means of its marginal distribution (PH), first-two moments ($\rm{PH}_2$), and first moment (M) in estimating the optimal base-stock level of the system. I consider processes with negative and positive first-lag autocorrelation, and coefficient of variation ($cv_a$) greater and less than one for the arrival process. For simplicity, I assume that service times are exponentially distributed and traffic intensity is $\nu =0.8$. The performance measures that I consider are total cost of the system ($TC$), expected inventory ( $E[X^+]$), expected backlog ( $E[X^-]$), and probability of not having inventory in the system ($Pr[X<0]$). I let the inventory holding cost, and backlog cost to be $h=1$, and $b=5$, respectively.   
\subsection{Impact of Approximating Positively Correlated Arrival Process with Different Renewal Processes on the Performance}
\subsubsection{The low variability case: $cv_a <1$:}
\begin{figure} 
	\begin{center}
		\includegraphics[scale=0.4]{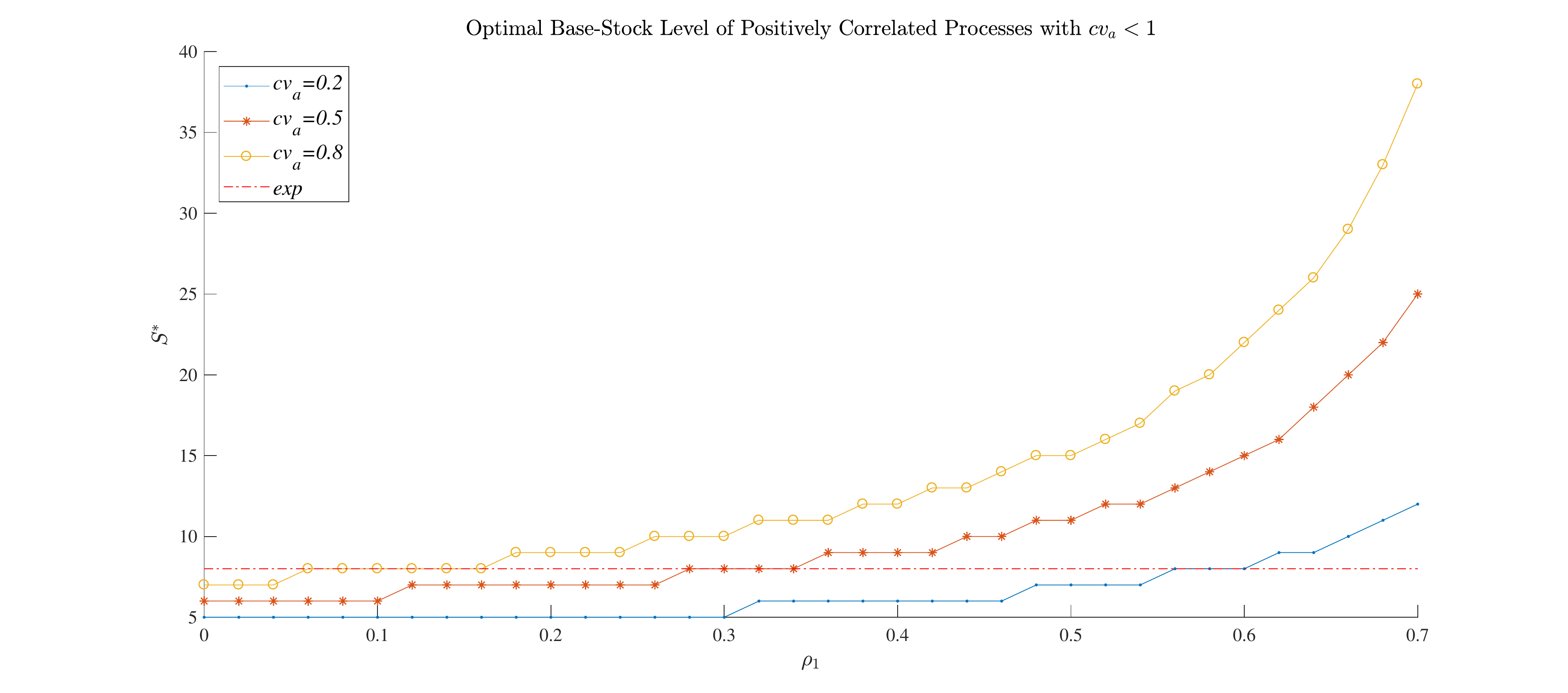} \end{center}
	\caption{Optimal base-stock level of positively correlated processes with $cv_a<1$ and their estimation with the exponential model ($exp$: exponential distribution)}
	\label{Pcv_a_l_1}
\end{figure}

I consider an arrival process with $cv_a=0.5$ and first-lag autocorrelation $\rho_{1}=0.3$. The performance measures of a system with such an arrival process are shown in Table \ref{pcval1}. Estimating the optimal base-stock level using the marginal distribution and first-two moments of the process results in underestimating the optimal base-stock level. Controlling the system by a base-stock level calculated using these models results in a higher cost, and lower service level. Interestingly, estimating the correlated arrival process by means of exponential distribution results in better estimation of the optimal base-stock level. This is due to the fact that coefficient of variation of the exponential distribution is greater than coefficient of variation of the correlated process. Higher coefficient of variation increases the optimal base-stock level which results in a better estimation of base-stock level than the exact marginal distribution. Hence, in positively correlated arrival processes with $cv_a<1$, employing the exact distribution of the process does not necessarily result in a better estimation of the optimal base-stock level than the exponential distribution. Figure \ref{Pcv_a_l_1} demonstrates the optimal base-stock level of positively correlated systems with $cv_a<1$ and traffic intensity $\nu = 0.8$. The optimal base-stock level associated with modeling the correlated arrival with exact marginal distribution is equal to that of the a process with zero first-lag-autocorrelation. Modeling the arrival process with exponential distribution gives a better approximation of the optimal base-stock level as the first-lag autocorrelation increases.  

\begin{table}
	\centering
	\caption{Impact of positively correlated arrival process with $cv_a<1$ on the performance measures of a production system controlled by base stock policy ($\nu =0.8$) }
	\scalebox{0.92}{
	\begin{tabular}{lccccccccc}
		& $S^\ast$ &TC & Error &  $E[X^-]$  & Error &  $E[X^+]$  & Error & $Pr[X<0]$ & Error  \\ \hline
		MAP/M/1 & 8     & 8.7344 &       & 0.8062 &       & 4.7035 &       & 0.1468 &  \\ \hline
		PH/M/1 & 6     & 9.1283 & 5\%   & 1.2052 & 49\%  & 3.1024 & -34\% & 0.2195 & 49\% \\
		$\rm{PH}_2$/M/1 & 6     & 9.1283 & 5\%   & 1.2052 & 49\%  & 3.1024 & -34\% & 0.2195 & 49\% \\
		M/M/1 & 8     & 8.7344 & 0\%   & 0.8062 & 0\%   & 4.7035 & 0\%   & 0.1468 & 0\% \\
	\end{tabular}
	}
\label{pcval1}	
\end{table}

\subsubsection{The high variability case $ cv_a >1$:}
I consider an arrival process with $cv_a=1.3$ and first-lag autocorrelation $\rho_{1}=0.3$. The performance measures of a system with such an arrival process is shown in Table \ref{pcvag1}. Modeling the arrival process by means of renewal processes result in underestimation of the optimal base-stock level of the system. In contrary to the case with $cv_a<1$, the marginal distribution and first-two moments gives a better approximation of the correlated process than exponential distribution. This result holds for all of the cases with $cv_a>1$ that I evaluated.    

\begin{table}
	\centering
	\caption{Impact of positively correlated arrival process with $cv_a>1$ on performance measures of a production system controlled by base stock policy ($\nu =0.8$) }
	\scalebox{0.92}{
	\begin{tabular}{lccccccccc}
		& $S^\ast$ &TC & Error &  $E[X^-]$  & Error &  $E[X^+]$  & Error & $Pr[X<0]$ & Error   \\ \hline
		MAP/M/1 & 14    & 15.8010 &       & 1.5021 &       & 8.2907 &       & 0.1586 &  \\ \hline
		PH/M/1 & 9     & 17.5352 & 11\%  & 2.6244 & 75\%  & 4.4130 & -47\% & 0.2771 & 75\% \\
		$\rm{PH}_2$/M/1 & 9     & 17.5352 & 11\%  & 2.6244 & 75\%  & 4.4130 & -47\% & 0.2771 & 75\% \\
		M/M/1 & 8     & 18.3945 & 16\%  & 2.9343 & 95\%  & 3.7229 & -55\% & 0.3099 & 95\% \\
	\end{tabular}%
	}
	\label{pcvag1}
\end{table}

\subsection{Impact of Approximating Negatively Correlated Arrival Process with Different Renewal Processes on the Performance}
\subsubsection{The low variability case $ cv_a <1$:}
I generate negatively correlated MAP from the output process of a production line with three stations and zero buffers. I let the distribution of each machine to be acyclic PH distribution. The statistical descriptors of this output process is shown in the first row of Table \ref{nfitcvl1}. Its coefficient of variation and first-lag autocorrelation are $cv_a=0.29$, $\rho_{1}=-0.30$, respectively. I adopt this process as the arrival process. The performance measures of a system with such an arrival process is given in Table \ref{ncval1}. Modeling the arrival process by means of renewal processes results in overestimation of the optimal base-stock level of the system. Estimating the arrival process by means of its marginal distribution and first-two moments results in better estimation of the base-stock level than the exponential distribution. This result holds for all of the cases with $cv_a<1$ that I evaluated.  

\begin{table}[htbp]
	\centering
	\caption{Impact of negatively correlated arrival process with $cv_a<1$ on performance measures of a production system controlled by base stock policy ($\nu =0.8$)}
	\scalebox{0.92}{
	\begin{tabular}{lccccccccc}
		&$  S^\ast $   & TC    & Error & $E[X^-]$ & Error & $E[X^+]$ & Error & $Pr[X<0]$ & Error \\ \hline
		MAP/M/1 & 4     & 4.2646 & 0\%   & 0.4417 & 0\%   & 2.0559 & 0\%   & 0.1535 & 0\% \\ \hline
		PH/M/1 & 5     & 4.3439 & 2\%   & 0.2883 & -35\% & 2.9024 & 41\%  & 0.1001 & -35\% \\
		$\rm{PH}_2$/M/1 & 5     & 4.3439 & 2\%   & 0.2883 & -35\% & 2.9024 & 41\%  & 0.1001 & -35\% \\
		M/M/1 & 8     & 6.0949 & 43\%  & 0.0801 & -82\% & 5.6942 & 177\% & 0.0278 & -82\% \\
	\end{tabular}%
		}
	\label{ncval1}%
\end{table}%

\subsubsection{The high variability case $ cv_a >1$}
I consider an arrival process with $cv_a=1.3$ and first-lag autocorrelation $\rho_{1}=-0.30$. Table \ref{ncavag1} shows the performance measures of a system  with such an arrival process and its renewal approximations. The results demonstrate that modeling negatively correlated arrival process by means of its first-two moments, and marginal distribution does not necessarily result in a better estimation of the optimal base-stock level than exponential distribution. 

\begin{figure}[ht] \begin{center}
		\includegraphics[scale=0.4]{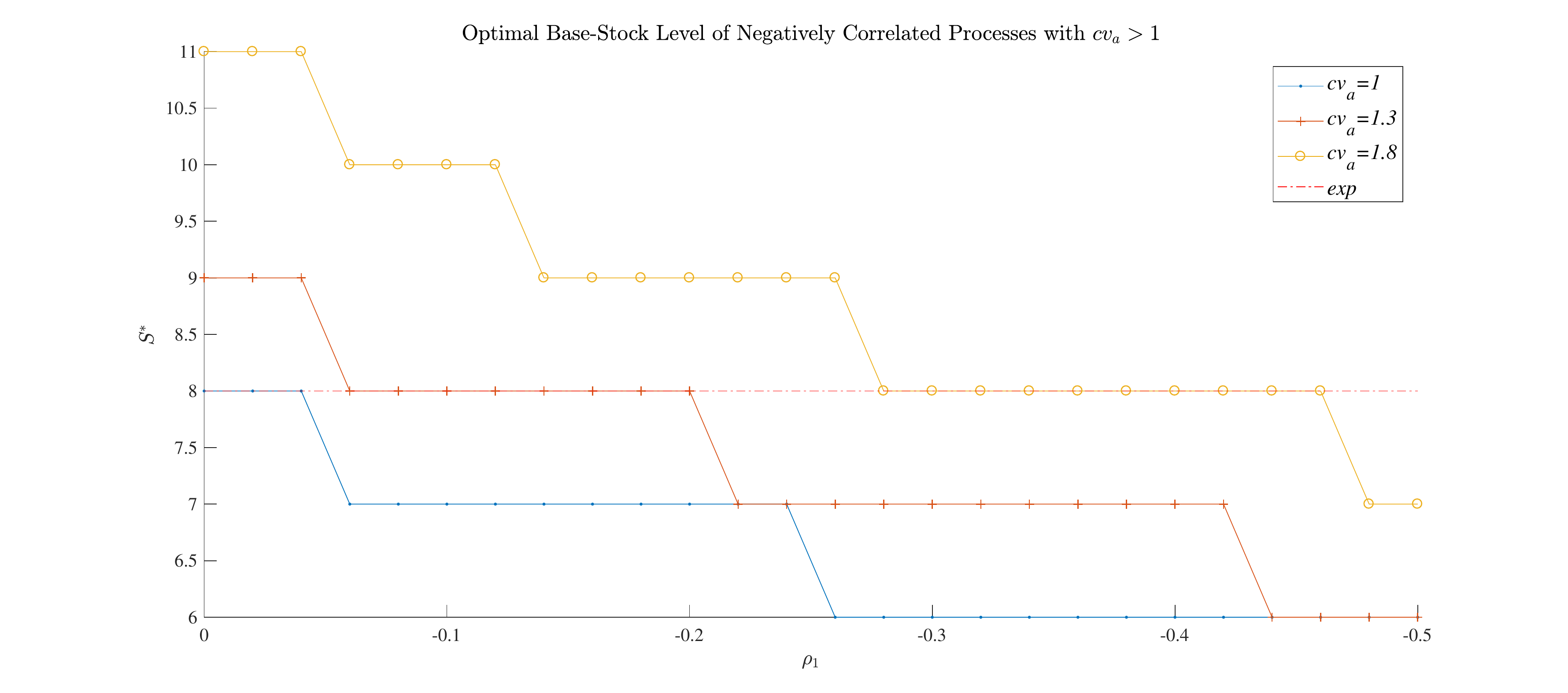} \end{center}
	\caption{Optimal base-stock level of negatively correlated processes with $cv_a\geq1$ and their estimation with exponential model}
	\label{Ncv_a_g_1}
\end{figure}

Estimating the arrival process by an exponential distribution gives a better approximation than using the first-two moments or marginal distribution of the process. This result is similar to the result of estimating the optimal base-stock level of a positively correlated arrival process with $cv_a<1$ by its mean. Figure \ref{Ncv_a_g_1} demonstrates the optimal base-stock level of positively correlated systems with $cv_a>1$ and traffic intensity $\nu = 0.8$. The optimal base-stock level associated with modeling the correlated arrival with exact marginal distribution is equal to that of the a process with zero first-lag-autocorrelation. Modeling the arrival process with exponential distribution gives a better approximation of the optimal base-stock level as the first-lag autocorrelation decreases. 

\begin{table}[htbp]
	\centering
	\caption{Impact of negatively correlated service process with $cv_a<1$ on performance measures of a production system controlled by base stock policy ($\nu =0.8$)}
	\scalebox{0.92}{
	\begin{tabular}{lccccccccc}
		& $ S^\ast  $  & TC    & Error & $E[X^-]$ & Error & $E[X^+]$ & Error & $Pr[X<0]$ & Error \\ \hline
		MAP/M/1 & 7     & 7.2548 & 0\%   & 0.6719 & 0\%   & 3.8956 & 0\%   & 0.1475 & 0\% \\ \hline
		PH/M/1 & 9     & 7.6795 & 6\%   & 0.4093 & -39\% & 5.6330 & 45\%  & 0.0898 & -39\% \\
		$\rm{PH}_2$/M/1 & 9     & 7.6795 & 6\%   & 0.4093 & -39\% & 5.6330 & 45\%  & 0.0898 & -39\% \\
		M/M/1 & 8     & 7.3701 & 2\%   & 0.5244 & -22\% & 4.7481 & 22\%  & 0.1151 & -22\% \\		
	\end{tabular}%
		}
	\label{ncavag1}%
\end{table}%

Finally, it worth mentioning that, estimating the arrival process by means of its first-two moments and marginal distribution results in the same optimal base-stock levels in my numerical experiments.

\section{Summary of Findings} \label{sec:summary}
In this section, I summarize my findings based on the numerical experiments reported in the preceding sections:

\begin{enumerate}
	\item In systems with negatively correlated arrival or service process, ignoring autocorrelation can lead to overestimation of the optimal base-stock level. In other words, renewal approximation of a negatively  correlated process overestimates the optimal base stock level, and consequently the total cost of the system. 
	\item  The overestimation of optimal base-stock level in negatively correlated process is due to the impact of negative autocorrelation on the stationary queue length probability distribution. 
	\item In systems with positively correlated arrival or service process, ignoring autocorrelation can lead to underestimation of the optimal base-stock level. In other words, renewal approximation of a positively correlated process underestimates the optimal base stock level, and consequently the total cost of the system. 
	\item The underestimation of optimal base-stock level in positively correlated process is due to impact of positive autocorrelation on the stationary queue length probability distribution. 
	\item The queue length distribution of a system with positively correlated arrivals stochastically dominates the same process with less magnitude of autocorrelation. In other words, positive correlation  increases the probability of having higher queue length in the system as opposed to negative autocorrelation which increases the probability of having lower queue lengths. 	
	\item  Capturing the inter-departure time distribution more accurately while ignoring the autocorrelations does not necessarily improve the accuracy of the results.  Modeling the correlated arrival process with a PH distribution does not necessarily result in more accurate estimation of the optimal control parameter. For instance, modeling a positively correlated arrival with a coefficient of variation less than one, with an exponential distribution may result in better estimates of the base-stock level than PH distribution.  
\end{enumerate}

\section{Conclusions} \label{ch1conclusion}
In this chapter, my objective is controlling a production system by using shop-floor inter-event data such as the inter-arrival, inter-departure, and service completion times. Although, empirical studies show that inter-event times of a production system are correlated, most of the analytical studies for the analysis and control of production systems ignore correlation.  I use Markovian Arrival Processes (MAP) to model shop-floor inter-event data because of their ability in estimating any inter-event process arbitrarily close enough \citep{Reference12}. 

I analyze the impact of autocorrelation on the optimal base-stock level by using the structural properties of the MAPs. My analysis show that ignoring autocorrelation in modeling inter-arrival and service times of production-inventory systems can lead to misleading results. I conclude by stating that MAP is suitable process to model shop-floor inter-event data. MAP can capture the correlation in inter-arrival and service times which affect the optimal base-stock level of the system, significantly. My findings on the effects of autocorrelation on the system can be summarized as follows: \begin{enumerate} [label=(\roman*)] 
	\item The autocorrelation structure should be taken into account while designing a production system.
	\item  Positively (Negatively) correlated systems need a larger (smaller) buffer size in comparison with the correlated processes with lower magnitude of first-lag autocorrelation. In other words, the expected inventory in front of a service station increases (decreases) as a function of the positive (negative) autocorrelation.  
	\item Positive autocorrelation may increase the average cycle time of the products in the system. 
	\item The arrival of products into the system can be modulated in order to control the correlation in inter-arrival times. For instance, if low buffer levels is needed in front of a line creating negative autocorrelation may decrease the optimal buffer level.   
\end{enumerate}

I developed an estimation-then-optimization framework for controlling a production line in which production decision can be taken by employing full statistical properties of the shop-floor inter-event data. A possible research direction is to adopt model free methods instead of estimation-then-optimization approach in setting the base-stock level. Model-free methods decrease the model misspecification errors at a cost of computational complexity. Integrating MAPs into a model-free method may reduce their computational costs and result in more efficient algorithms in estimating the optimal base-stock levels. More information from shop-floor such as the state of the machines, and quality of the products can be modeled by using model-free methods. 

My results are presented here for single-product problems. My framework can be extended to a multi-product problem by adopting marked MAP. \citet{Reference77}, \citet{Reference71} propose methods to fit a marked MAP into data. These methods can be adopted in controlling a production line with multiple products by using shop-floor inter-event data. Further research directions include performance of the base-stock policy in controlling a system with correlated arrival or service processes.

\chapter[Optimal Control of Production Systems with Correlated Inter-Event Times]{Optimal Control of Production-Inventory Systems with Correlated Demand Inter-Arrival and Processing Times\footnotemark} \footnotetext{The results in this chapter are published in: Manafzadeh Dizbin, N., and Tan, B. (2020). "Optimal  control  of  production-inventory  systems  with  correlated demand  inter-arrival  and  processing  times". International Journal of Production Economics 228.107692}\label{OptimalMAPControl}

\section{Abstract}
I consider the production control problem of a production-inventory system with correlated demand inter-arrival and processing times that are modeled as Markovian Arrival Processes. The control problem is minimizing the expected average cost of the system in the steady-state by controlling when to produce an available part. I prove that the optimal control policy is the state-dependent threshold policy. I evaluate the performance of the system controlled by the state-dependent threshold policy by using the Matrix Geometric method. I determine the optimal threshold levels of the system by using policy iteration. I then investigate how the autocorrelation of the arrival and service processes impact the performance of the system. Finally, I compare the performance of the optimal policy with 3 benchmark policies: a state-dependent policy that uses the distribution of the inter-event times but assumes {i.i.d. }inter-event times, a single-threshold policy that  uses both the distribution and also the autocorrelation, and a single-threshold policy that uses the distribution of the inter-event times but assumes they are not correlated. Our analysis demonstrates that ignoring autocorrelation in setting the parameters of the production policy causes significant errors in the expected inventory and backlog costs.  A single-threshold policy that sets the threshold based on the distribution and also the autocorrelation performs satisfactorily for systems with negative autocorrelation.  However, ignoring positive correlation yields high errors for the total cost.  Our study shows that an effective production control policy must take correlations in service and demand processes into account.  

\section{Introduction}
Controlling production systems to match supply and demand in an uncertain environment received considerable attention in the manufacturing systems literature. Control policies such as the Control-Point Policy, Generalized Kanban Policy, and Base-Stock Policy are suggested to control the material flow in a production system, e.g., \citep{Gershwin2000DesignPolicy, Duri2000ComparisonKanbanb, Liberopoulos2000ASystems} among others.

\begin{figure}
	\centering
	\caption{Empirical distribution and dependence of the processing, inter-arrival, and inter-departure times of a specific equipment at a semiconductor manufacturing plant}
	\includegraphics[width=1.08\linewidth]{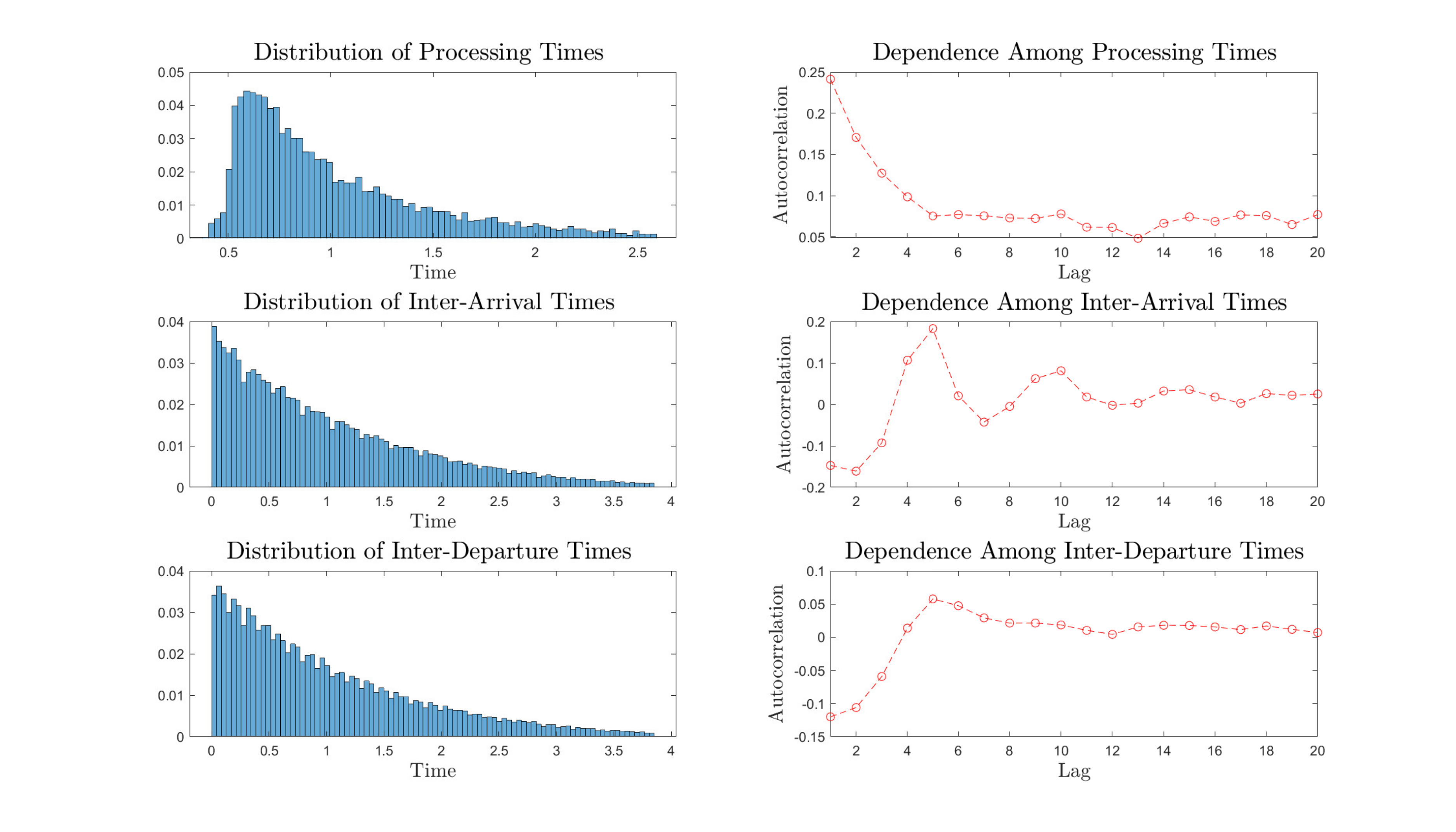}
	\label{empfig}%
\end{figure}

The analytical models that evaluate the performance of production systems controlled to match supply and demand usually model the demand inter-arrival and processing times as independent random variables. As a result, dependence among the inter-arrival and processing times is not often taken into account. However, autocorrelation can be observed in processing, inter-arrival and inter-departure times. Figure \ref{empfig} depicts the empirical distributions and autocorrelations of the  processing, inter-arrival, and inter-departure times of certain equipment at the Robert Bosch semiconductor manufacturing plant.  \citet{Reference6} and \citet{Inman1999EmpiricalSystems} also report dependence in observed cycle and inter-event times.  The simulation and analytical studies also show negative dependence among the inter-departure times of the products leaving a production line \citep{Hendricks1993TheBuffers, Tan2017OnBlocking, ManafzadehDizbin2019ModellingProcesses}.

Complicated processing tasks such as batch processing, parallel processing, and merging may create high dependence between the processing times of the products. Moreover, dispatching rules and the production network for different products yield dependence among the inter-arrival times observed at different stations. Correlated inter-arrival and processing times then result in a correlated output process. The correlated output process creates the arrival process at other stages of the production and causes further dependence among inter-departure times.  

Ignoring dependence among inter-event times has been one of the shortcoming of the classical queuing theory in analyzing manufacturing systems \citep{Shanthikumar2007QueueingProblems}. Although the optimal inventory control policies with independent and identically distributed ({i.i.d.}) inter-event times have been investigated thoroughly in the literature, the optimal production control policies of production systems with correlated inter-event times have not been studied. The objective of this chapter is to fill this gap by deriving the optimal control policy of a manufacturing system with correlated inter-arrival and processing times and analyzing the effects of correlation on the production control. 

The control problem studied in this chapter is minimizing the expected holding and backlog costs of a production-inventory system with correlated processing and demand inter-arrival times that are modeled as Markovian Arrival Processes (MAP) in the long run. The action space consists of whether or not to produce depending on the state of the system. I prove that a manufacturing system with MAP demand inter-arrivals and MAP processing times can be controlled optimally by using a state-dependent threshold policy. I use a matrix-geometric method to evaluate the performance of a production-inventory system controlled by the state-dependent threshold policy. I determine the optimal threshold levels by using a policy iteration method. I then evaluate the impact of positive and negative autocorrelations in inter-arrival and processing times. In addition, I compare the performance of the optimal policy in controlling a system with 3 benchmark production policies: a state-dependent policy that uses the distribution but assumes {i.i.d.} inter-event times, a single-threshold policy that  uses both the distribution and also the autocorrelation, and a single-threshold policy that uses the distribution but assumes i.i.d. inter-event times.

I consider proving the optimal control policy of a production-inventory system with correlated inter-arrival and processing times that are modeled as MAPs, proposing the Matrix Geometric method to evaluate the performance of the system when it is controlled optimally, and investigating the impact of autocorrelation among the inter-event times on the performance of a production-inventory system as the main contributions of this paper.

The structure of the chapter is as follows. The literature related to the optimal control of production and queuing-inventory systems is reviewed in Section \ref{lit}. In Section \ref{Problem}, the problem is defined and it is proven that the state-dependent threshold policy is the optimal control policy. The Matrix Geometric method to evaluate the performance of the system for given thresholds and the policy iteration approach used to determine the optimal thresholds are presented in Section \ref{Q-sdbs}. Section \ref{numerical} evaluates the impact of positive and  negative autocorrelation in demand inter-arrival and processing times on the performance of the state-dependent threshold policy and compares the performance of the benchmark policies with the optimal policy. Finally, Section \ref{conclusion} concludes the chapter and discusses future research directions. 

\section{Literature Review}\label{lit}
I discuss the pertinent literature in three related areas. The first area is related to the papers that investigate the optimal control policy of inventory systems with correlated demand-arrivals where the demand is modeled as a Markov-modulated process. The second area is related to the papers that study the optimal control of production systems.  Finally, the third area is related to the papers that evaluate the performance of queuing-inventory systems by using matrix-geometric methods.

\subsection{Optimal Control of Inventory Systems}
In this section, I cover the inventory control literature that uses models where the demand is modeled as Markov-modulated (MM) processes. The optimal inventory control policy  for these models is proven to be state-dependent base-stock or $(s, S)$ policies under different assumptions about the cost criterion, lead time, timing horizon, and production capacity. Accordingly, an order of an appropriate quantity is triggered when the inventory position is below a given threshold. 

\citet{Song1993InventoryEnvironment}  investigate the optimality of state-dependent
base-stock and  $(s, S)$ policies for an inventory system under continuous review with Markov-modulated Poisson Process (MMPP) demand process, and fixed ordering costs over finite and infinite horizon. They show that the state-dependent base-stock policy is optimal when there is no fixed cost and the state-dependent $(s, S)$ policy is optimal when there is a fixed cost. \citet{Song1996EvaluationPolicies} evaluate the steady-state performance measures of multi-echelon systems with MMPP demand and state-independent base-stock policies in multi-echelon inventory systems. \citet{Song1996EvaluationPoliciesb} evaluate the performance measures of a two-echelon inventory system with MMPP demand where the first (second) stage is controlled by a state-independent (dependent) base-stock policy. \citet{Bayraktar2010InventoryDemand} consider a continuous-time model for inventory management with Markov-modulated non-stationary demand where the state of the modulating processes is unobserved. They prove the optimality of a time-and-belief-dependent $(s, S)$ strategy and develop a numerical method to calculate the optimal policy. \citet{Nasr2015ContinuousDemand} consider a continuous inventory replenishment system with a MMPP demand process. 

\citet{Sethi1997Demand} consider a discrete-time model with non-stationary demand and no lead time and investigate the optimality of $(s, S)$ policies for finite and infinite horizon and non-stationary systems. \citet{Beyer1997AverageDemands} consider a system with convex surplus cost and prove that the $(s, S)$ model with MMPP demand and average cost criterion minimizes the inventory cost. \citet{Ozekici1999InventoryEnvironment} consider a model involving random variations in supply  and prove the  optimality of the state-dependent base-stock policies and state-dependent $(s, S)$ type policies with and without fixed setup costs. The optimality of the $(s, S)$ policy is generalized to systems involving general costs \citep{Beyer1998InventoryGrowth}, and lost sales \citep{Cheng1999OptimalitySales}. \citet{Chen2001OptimalDemand} show the optimality of the state dependent base-stock policies for serial systems with Markov modulated demand and deterministic lead time under a finite and infinite horizon and average cost criterion. \citet{Muharremoglu2008ASystems} study a multi-echelon inventory system with Markov-modulated demand under periodic review and propose a single item-single customer approach to prove the optimality of state-dependent base-stock policies. \citet{Janakiraman2009ASystems} utilize the method developed in \citet{Muharremoglu2008ASystems} to demonstrate the optimality of the state dependent base-stock policies in a two-echelon serial system with identical ordering/production capacities. \citet{Hu2016sDemands} study a class of periodic review $(s, S)$ inventory systems with a Markov-modulated demand process. They develop an algorithm to calculate the moments of the inventory level. 

The main objective in these studies is determining the amount of inventory to be shipped to the next level of supply chain where the inventory is supplied exogenously without any capacity restrictions. One of the main assumptions of the infinite-capacity problems is the independence between lead times of the products shipped between two consecutive echelons of the supply chain.  As a result, orders given at different times can pass each other.  This property is crucial in proving the optimal policy. Hence, the researchers that study the finite capacity problems in an inventory control setting usually model the lead time as deterministic or consider a finite time horizon to deal with this problem. The main difference in our study is considering the optimal production control problem of a finite-capacity producer that has correlated processing times and meets a demand stream with correlated inter-arrival times.

\subsection{Optimal Control of Production Systems}
The second stream of the literature is related to the studies that consider the optimal control of production systems.  These studies use models with discrete or continuous flow of materials. It is shown that threshold-type policies that are referred as the \emph{base-stock} policies for the models with discrete material flow and the \emph{hedging}-type policies for the models with continuous material flow are optimal.  Accordingly, production is allowed when the inventory level is below a threshold determined for the given state.

For a system with discrete material flow, \citet{Veatch1994OptimalSystem} study a make-to-stock manufacturing system with an exogenous Poisson demand and two stations. Each station is modeled as a queue with controllable production rate and exponential service times. The objective of the study is to control the production rates to minimize the inventory holding and backordering costs. \citet{Berman2001DynamicChains} consider the dynamic replenishment of parts in a supply chain with single class of customers where parts are procured by a supplier with an Erlang processing distribution. They assume Poisson customer arrivals and exponential processing times and model the problem as a Markov decision process. They show that the optimal ordering policy that minimizes the customer waiting, inventory holding, and order replenishment costs has a monotonic threshold structure.  \citet{He2002OptimalSystem} examine several inventory replenishment policies for a make-to-order production-inventory system with Poisson demand-arrival. They derive the optimal replenishment policy, which minimizes the average total cost per product of the warehouse.\citet{deVericourt2002OptimalSystem} consider a capacitated supply system with a single product and several classes of customers where each customer class has a different backorder cost. They study the optimal allocation policy of products and show the optimality of a threshold policy.  \citet{Karabag2019PurchasingPrices} analyze the purchasing, production, and sales policies for a continuous-review discrete material production-inventory system with exponentially distributed demand inter-arrival, and processing times. They show that the optimal purchasing, production, and sales strategies are state-dependent threshold policies.

For a system with continuous material flow and constant demand for a single product, \citet{Sharifnia1988ProductionStates} study the production control of a manufacturing system with arbitrary number of machine states. He shows that the optimal production policy that minimizes the average inventory and backlog costs of the system is the hedging-point policy.  \citet{Tan2002ProductionUncertainty} considers a manufacturing system with two-state Markov modulated demand, uncertain repair and failure times and continuous material. He shows that the optimal production flow control policy that minimizes the expected average inventory holding and backlog costs is a double-hedging policy. \citet{Gurkan2007OptimalOptimization} use simulation-based optimization to determine the threshold levels of a production-inventory system where stochasticity in the system is modeled using semi-Markov processes. \citet{Gershwin2009ProductionDemand} consider a manufacturing system with deterministic production time and stochastic Markov modulated demand. They show that the hedging point is the optimal control policy of the system.  \citet{Tan2018ProductionUncertainty} studies the  optimal production flow control problem of a make-to-stock manufacturing system with price, cost, and demand uncertainty. He models the stochastic dynamics of the system with a time-homogeneous Markov chains and shows that the optimal production policy is a state-dependent hedging policy. 

All of these models assume deterministic or i.i.d. demand inter-arrival and service times and do not consider correlation explicitly.  Most of the studies further assume that the inter-event times are exponentially distributed random variables.  Our study differs from the production control literature in that I analyze a discrete material-flow continuous-time production-inventory system with correlated demand inter-arrival and service times modeled as Markovian Arrival Processes. 

\subsection{Matrix-Geometric Methods}
I now focus on the literature that utilize the matrix-geometric methods for performance evaluation of the queuing-inventory system. \citet{He2000PerformanceSystem} use the matrix-analytic methods to evaluate the performance of a make-to-order production-inventory system with Poisson arrivals and exponential processing times. 
\citet{Manuel2007ATimes, Manuel2008ACustomers} study a perishable  $(s, S)$ inventory system under continuous review with a finite buffer and a single server. They consider two types of customers arrivals modeled as a MAP and service process with phase-type distribution. They evaluate the joint probability distributions of the number of customers in the system and the inventory level in the steady state. \citet{Zhao2011ACustomers} consider a queuing-inventory system with Poisson demand arrival, exponential processing times and  $(r,Q)$  replenishment policy. Their objective is  minimizing the long-run expected waiting cost. They formulate the problem as a level-dependent Quasi-Birth-and-Death process and investigate the control of such a process. \citet{Liu2014FlexibleClasses} investigate a Markovian inventory system with two classes of demands, replenishment policy, and a flexible service discipline. They derive the steady state probability distribution of the inventory levels by using Markovian processes, and adopt a mix integer optimization model to find the optimal inventory control levels. \citet{Jiang2015AnBalance} consider a two-echelon queuing-inventory system with demand that follows a compound Poisson process, a two-echelon inventory system consisting of a central warehouse and several sub-warehouses. They propose an algorithm for minimizing the mean total cost of the inventory system. \citet{Xia2017OptimalQueue} study the service rate control problem of the MAP/M/1 queue. They evaluate the impact of service rates on the long-run average total cost of the system and show the optimality of quasi-threshold-type policy under some conditions. \citet{ManafzadehDizbin2019ModellingProcesses} study a production-inventory control system with correlated service and processing times controlled with a single base-stock level that is not the optimal policy for this system.  They evaluate the performance of the models with the Matrix Geometric method and evaluate the effect of autocorrelation on the performance when the system is controlled with this sub-optimal policy.   

These studies that use the matrix-geometrics methods in the literature focus on the performance evaluation of queuing-inventory systems under a given production policy where arrival and service processes are modeled usually as independent distributions.  Our study differs from this stream of the literature in that I derive the optimal control policy and use the Matrix Geometric method to determine the performance measures of a system with correlated demand-arrival and processing times.  

\section{Model} \label{Problem}

I consider a single machine with an unlimited buffer where the raw material is supplied from an unlimited stock with zero lead-time. An arriving demand to the system is satisfied immediately, if there is enough inventory in the buffer to meet the demand. Otherwise, the demand is backlogged. Since backlog is allowed, all demand is satisfied eventually according to the first-come-first-served (FIFO) rule. The difference between the cumulative production and demand is referred as the inventory position and denoted by $X(t)$.  The on-hand inventory is $X^+(t)=\max\{X(t),0\}$ and the backlogged demand level is $X^-(t)=\max\{-X(t),0\}$.

The cost structure of the system consists of the holding and backlog costs. The holding cost is $h$ per unit per unit of time, and the backlog cost is $b$ per unit per unit of time. The cost function at any time $t$ is a function of  $X(t)$ and given as
\begin{equation}
C(X(t))= 
\begin{cases}
bX^-(t), & \text{if } X(t)<0, \\
hX^+(t), & \text{if } X(t)\geq 0,
\end{cases}
\end{equation}

The demand inter-arrival times and processing times are modeled as discrete state-space and continuous-time processes.  The state of the demand arrival process is denoted with $j_a \in J_a$.  The state of the production time process, also referred as the service process, is denoted with $j_s \in J_s$.  At any given time, the demand arrival process can be in one of $m_a=|J_a|$ discrete states and the service process can be in one of $m_s=|J_s|$ states. When the machine is available, it may or may not start producing a new part depending on the control policy. 

I assume that the system is continuously reviewed, and the state of the system is fully observed at any time $t$.  The state of the system is fully specified by the inventory position and the states of the demand arrival and service processes. 

\subsection{Production Control Problem}
The main goal of the production control problem is to determine the production policy $\Pi$ that minimizes the long run average cost of the system in the steady state by deciding on whether to produce or continue to producing $(u=1)$ or not $(u=0)$ depending on the state of the system $(X, j_a, j_s)$. The average cost of this system in steady state under policy $\Pi$ can be written as Equation (\ref{average_cost}).
\begin{equation}
\label{average_cost}
V^{\Pi}(X, j_a, j_s) = E^{\Pi}\left[\lim\limits_{{T\to\infty}}\frac{1}{T}\int_{0}^{T}C\left(X(t)|X(0)=0, J(0)= (j_a^0, j_s^0) \right)dt \right].
\end{equation}
Then the objective of the problem is to identify the optimal policy $\Pi^\ast$ defined as
\begin{equation}
\label{Piast}
\Pi^\ast = \sup\limits_{\Pi} V^{\Pi}(X, j_a, j_s).
\end{equation} 
\subsection{Demand Inter-Arrival and Processing Times}
In order to capture the autocorrelation in inter-event times, the demand inter-arrival and processing time processes are modeled as Markov Arrival Processes.
MAPs are generalization of phase-type (Ph) distributions \citep{Neuts1979AProcess}. MAPs contain most of the commonly used arrival processes such as Erlang processes, Coxian distributions and Markov-modulated Poisson processes as subclasses.  MAPs can approximate a given inter-event process arbitrarily close enough \citep{Asmussen1993MarkedStreams}. MAPs are used commonly in the telecommunications literature to model positively correlated inter-arrivals \citep{Reference2}. MAPs can also be used in modeling negatively correlated arrivals in manufacturing systems \citep{Hendricks1993TheBuffers, ManafzadehDizbin2019ModellingProcesses}. \citet{ManafzadehDizbin2019ModellingProcesses} review various algorithms to construct a MAP by using the observed inter-event time data.

A MAP consists of two different sub-processes each of which has a discrete state space referred as phases.  The MAP of the demand inter-arrival time is denoted as MAP($D_0,D_1$). The non-diagonal elements of matrix $ D_0 $ include the transition rates between the phases that do not generate a demand arrival. The diagonal elements of the $D_0$ correspond to the rates of the exponentially distributed sojourn times in corresponding states. The elements of $D_1$ capture the transition rates that generate a demand arrival. 

A MAP can be interpreted as a continuous-time Markov-chain with the generator matrix $D = D_0 + D_1$, and $|D|$ states. The joint probability density function of the consecutive inter-arrival times $T_i$, $i=0, 1, \ldots$, of the MAP($D_0,D_1$) is written as:
\begin{equation}\label{MAPjointdensity}
f(t_1, t_2, \dots, t_k)= 
\beta\textbf{exp}^{D_0t_1}D_1\textbf{exp}^{D_0t_2}D_1\dots\textbf{exp}^{D_0t_k}D_1\mathbbm{1},  \text{  for } t_i\geq 0, i\in\{1, \dots, k\}
\end{equation}   
where $\beta$ is the solution of $\beta (-D_0)^{-1}D_1= \beta $ and $ \beta \mathbbm{1}=1 $ and $\mathbbm{1}$ is a vector of ones with an appropriate size. $\beta$ can be interpreted as the probability distribution of the  phases immediately after an arrival. 

The $n^{th}$ moment of $T$, $E[T^n]$ is calculated from the matrix $D_0$  as:
\begin{equation}
E[T^n]=n!\beta\left(-D_0\right)^{-n}\mathbbm{1}. 
\end{equation}
Accordingly, the expected value and the variance of $T$, $E[T]$ and $Var[T]$ are 
\begin{align*}
E[T] &=-\beta\left(D_0\right)\mathbbm{1,} \\
Var(T) &= 2 \beta D_0^{ - 2}\mathbbm{1} - \left(\beta D_0^{ - 1}\mathbbm{1}\right)^2.
\end{align*}
The squared coefficient of variation of the inter-event times is $scv={Var(T)}/{E^2[T]}$. The covariance of $T_0$ and $T_k$, $\operatorname{Cov}\left(T_{0}, T_{k}\right)$ is also determined from the matrices $D_0$ and $D_1$ as
\begin{equation}
Cov\left(T_{0}, T_{k}\right)=\mathbf{E}\left(T_{0} T_{k}\right)-\mathbf{E}(T)^{2}=\mathbf{E}(T) \frac{\beta\left(-D_0\right)^{-1}}{\beta\left(-D_{0}\right)^{-1}\mathbbm{1}} \left((-D_0)^{-1}D_1\right)^{k}\left(-D_{0}\right)^{-1}\mathbbm{1}-\mathbf{E}(T)^{2}.
\end{equation}
The $k$th-lag autocorrelation coefficient $\rho_k$ is defined as 
\begin{equation}
\rho_{k} =   \frac{Cov(T_{0}, T_{k})}{Var(T)}.  
\end{equation}
\color{black}
The production time is denoted with $T_s$ with mean $E[T_s]$, variance $Var[T_s]$ and the squared coefficient of $scv_s$. The MAP of the production time process is denoted as MAP($A_0,A_1$) and the matrices $A_0$ and $A_1$ are defined similar to the definition of $D_0$ and $D_1$.  Let the infinitesimal generator of the underlying Markov chain matrix of the production process be $A=A_0+A_1$. Then, the state transition of the system is governed by the underlying Markov chain that has an infinitesimal generator $M= D\oplus A$ ($\oplus$ is the Kronecker sum) with a finite state space $J=J_a\times J_s$. The  underlying Markov chain consists of $m=|M|=m_a\times m_s$ states and the state of the underlying process is $(j_a, j_s)$.

\subsection{Structure of the Optimal Control Policy}
In order to identify the optimal control policy that determines whether to produce or continue to produce or not depending on the state of the system $(X, j_a, j_s)$, I discretize the continuous-time Markov process by using the uniformization technique and write the Bellman optimality equation of this system as  
\begin{multline}
\label{MAP/MAP/1-optimality equation}
V(X,j_a, j_s) + g = \frac{C(X)}{\alpha}\\+\sum\limits_{j\in J_a}P_0(j_a,j)V(X,j,j_s)\\+\sum\limits_{j\in J_a} P_1(j_a,j)V(X-1,j,j_s) \\ +
\min\begin{cases}
0 &  u=0 \\
\sum\limits_{j'\in J_s}R_0(j_s,j')V(X,j_a,j')\\+\sum\limits_{j'\in J_s}R_1(j_s,j')V(X+1,j_a,j')-V(X,j_a, j_s)&  u=1
\end{cases},
\end{multline}
where $g$ is the optimal average cost of the system over an infinite horizon and $V(X, j_a, j_s)$ is the differential cost of being in state $(X, j_a, j_s)$, the matrices $P_0$, $R_0$, $P_1$, and $R_1$ capture transitions related to events without and with the demand-arrival and service completion.  The probability transition matrices of the arrival and service processes are written as 
\begin{eqnarray} \label{transitionprobabilityblocks}
P_0&=&\frac{1}{\alpha}D_0+I_{m_a}, \nonumber\\
P_1&=&\frac{1}{\alpha}D_1, \nonumber\\
R_0&=&\frac{1}{\alpha}A_0+I_{m_s}, \nonumber\\
R_1&=&\frac{1}{\alpha}A_1
\end{eqnarray}
where  $\alpha \geq \max_i\left(-M(i,i)\right)$ is the uniformization constant, $I_{m_a}$ and $I_{m_s}$ are identity matrices of the size $m_a$ and $m_s$, respectively. 

Our main result given in Theorem \ref{sdbspolicy} proves that the state-dependent threshold policy is the optimal policy to control this system.

\begin{theorem}\label{sdbspolicy}
	A state-dependent threshold policy is the optimal control policy of a production system with correlated demand-arrival and processing times modeled as MAP. According to the state-dependent threshold policy, described with the binary variable $u(x, j_a, j_s)$, the state of the modulating Markov process ($j_a, j_s$) and the inventory level determine whether to start, stop, or continue the production. That is, when the optimal policy is used, the production starts or continues ($ u(x, j_a, j_s)=1 $) if the state of the modulating Markov process is $(j_a,j_s)$ and the inventory level ($X$) is less than  the threshold-level associated with the state $ (j_a, j_s)$ denoted as $Z(j_a, j_s)$. Similarly, the production stops ($ u(x, j_a, j_s)=0 $) if the inventory level is greater than or equal to $ Z(j_a, j_s)$. The policy can be stated as follows:
	\begin{equation}
	u(x, j_a, j_s)= 
	\begin{cases}
	1, & \text{if } x<Z(j_a, j_s), X=x \\
	0,              & \text{otherwise}.
	\end{cases} \label{th1eq}
	\end{equation}
	
	Proof. Given in \ref{proofTh1}.
\end{theorem}


Let the states of the Markov-modulating process be indexed from $1$ to $m$.  The state-dependent threshold policy is determined by the vector of the thresholds denoted by $\bar{Z}=(Z^{(m)},Z^{(m-1)},\dots,Z^{(2)},Z^{(1)})$ where $Z^{(i)}$ is the $(m-i+1)^{th}$ biggest threshold level. The states of the Markov-modulating process are ordered according to the ordering of $\bar{Z}$. Since the steady-state performance of the system depends on the selection of these thresholds, the optimal policy is determined by selecting the optimal thresholds. 

\begin{figure}
	\centering
	\caption{Production-Inventory System with MAP Inter-Arrivals with Two States and Exponential Service Times Controlled by the State-Dependent Threshold Policy ($(Z^{(2)},Z^{(1)})=(10,5)$, $E[T] = 1.1250$, $scv = 1.5$, $\rho_1 = 0.15$, $E[T_s]=1$) }
	\includegraphics[width=\linewidth]{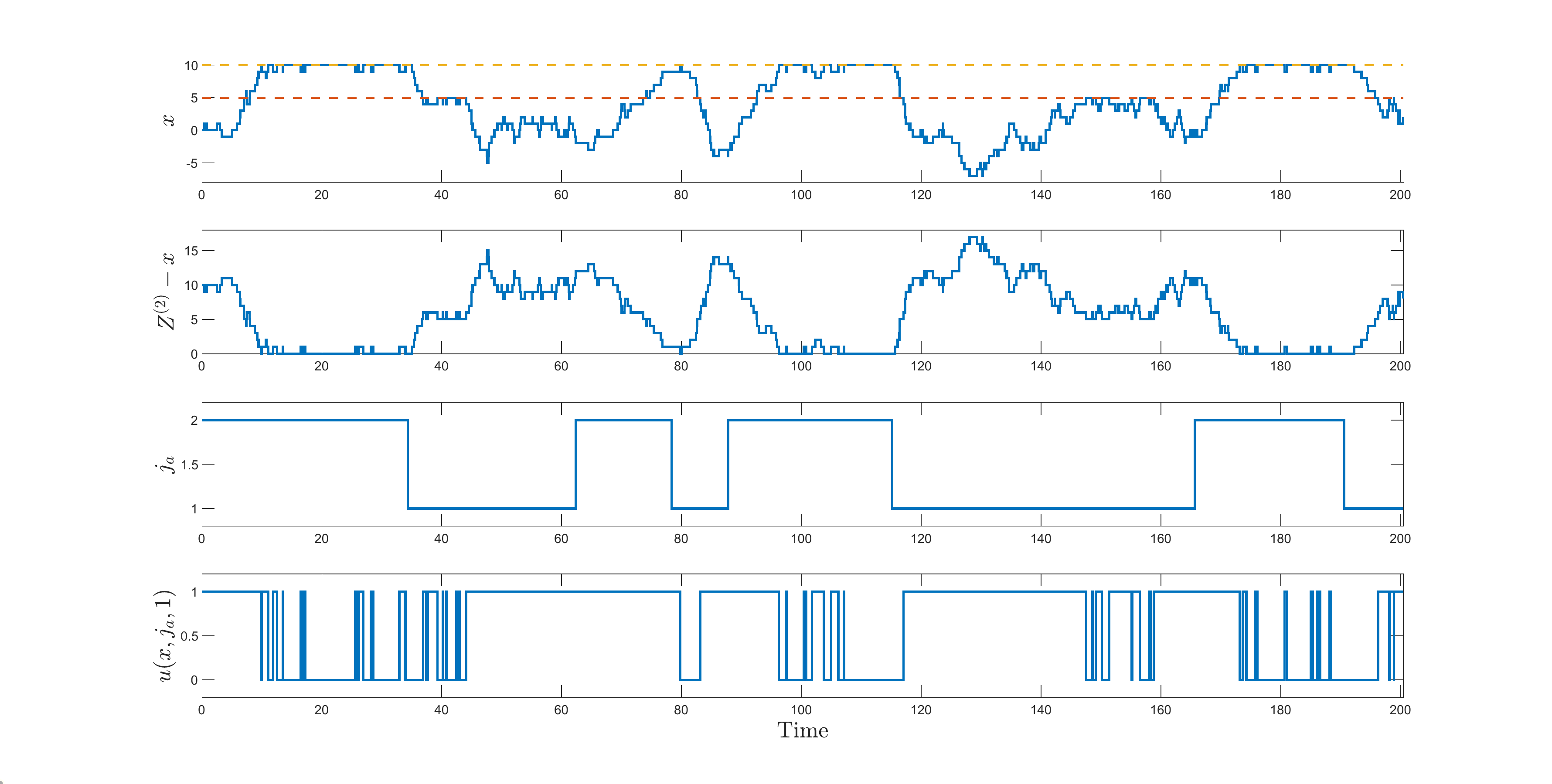}
	\label{sdbs}%
\end{figure}

Figure \ref{sdbs} shows the evolution of the inventory position, the shortfall, the state of the arrival process, and the control policy of a production-inventory system with a two-state MAP inter-arrival and exponential service times controlled by the state-dependent threshold policy. The threshold levels corresponding to state 2 and 1 are set to $(Z^{(2)}, Z^{(1)}) = (10, 5)$. Depending on the state of the arrival process, I stop the production if the  inventory position is equal to or above the corresponding threshold level.  

\section{A Matrix Geometric Approach to Evaluate the Steady-State Performance of the System}
\label{Q-sdbs}
In this section, I present the Matrix Geometric approach to evaluate the performance of a system controlled by the optimal state-dependent threshold policy given in Theorem \ref{sdbspolicy}. I first present a method to generate the infinitesimal generator matrix of a production system controlled with the given thresholds of the state-dependent threshold policy, denoted by $Q$.  I then discuss how to determine the steady-state probabilities by using the Matrix Geometric approach.

\subsection{The Steady-State Probability Distribution}

In order to capture the dynamics of the system, I focus on the shortfall from the largest threshold.  The shortfall level $k$ is equivalent to the inventory position $Z^{(m)}-k$ as shown in Figure \ref{sdbs} for the specific case.  The steady-state probability of being in state $(Z^{(m)}-k, j_a, j_s)$ is denoted as $\pi(k, j_a, j_s)$.  The steady-state probability vector $\pi_k = \{\pi(k, j_a, j_s)\}$, $j_a \in J_a$, $j_s \in J_s$ contains the probabilities at the shortfall level $k$.  The probabilities in $\pi_k $ are ordered according to the ordering of states in the underlying Markov chain of the system.  

The steady-state probability distribution is given by the vector $\pi = \{\pi_k\}$, $k=0, 1,\ldots$.  The steady-state probabilities satisfy  $\pi Q = 0$ and $\pi \mathbbm{1} = 1$.  Therefore, once the infinitesimal generator matrix of the system is constructed, the steady-state probabilities can be calculated, and the performance of the system can be evaluated from the steady-state probability distribution.

\subsection{Infinitesimal Generator Matrix}	
The generator matrix $Q$ is determined by the submatrices that are associated with three types of events: demand arrival, production completion, or an event that does not change the inventory position for each level of the inventory position.

The submatrix that captures transitions associated with demand arrivals is referred as, the forward matrix.  The state transition rates of the arrival and production process without a demand arrival or service completion are captured in  the local matrix. Finally, the backward matrix captures the transition rates related to service completion. The forward, local, and backward matrices at the  shortfall level $k$ are denoted by $F_k$, $L_k$, and $B_k$ respectively.  Accordingly, the infinitesimal generator matrix has the following block tri-diagonal form:

\begin{equation}
Q=\begin{pmatrix}
L_{0}    & F_0&         &          &          &         &       \\ 
B_1  & L_1& F_1 &          &          &         &       \\ 
& B_2& L_2& F_2 &          &          &              \\ 
&        & \ddots &\ddots   & \ddots   &          &               \\ 
&        &        &B_k  &L_k &F_k &                  \\ 
&        &        &         &\ddots        &\ddots    &\ddots        \\ 
\end{pmatrix}.
\label{Qmatrix} 
\end{equation}

\noindent The submatrices $F_k$, $L_k$, and $B_k$ are determined by the matrices that describe the demand arrival and service MAPs and also by the production policy.  

The transition rates related to demand arrivals are given in the $D_1$ matrix of the demand arrival MAP.  Therefore, the forward matrix $F_k$ that includes the transitions that increases the shortfall (decreases the inventory position) by one unit is completely determined by the $D_1$ matrix:

\begin{equation}
F_k = D_1\otimes I_{m_s}. \label{defFk}
\end{equation}
On the other hand, the shortfall decreases by one unit after a production completion. The production decision is authorized by the production policy at each decision epoch. The state-dependent threshold policy described in Equation (\ref{th1eq}) makes production authorization decisions based on the inventory position and the threshold levels. Let  $U_k$ be a diagonal matrix of size $m\times m$ at the shortfall level $k$ that authorizes the production for each state. Let $m, \ldots, 1$ be the states of the underlying Markov process corresponding to states with the threshold levels of $Z^{(m)}, \ldots, Z^{(1)}$. $U_k$ is defined as
\begin{equation}
U_k = 
\begin{cases}
U^{(i+1)}, & \text{if } Z^{(m)}-Z^{(i)}\geq k> Z^{(m)}-Z^{(i+1)},  i=1,\dots,m-1\\
U^{(1)}, & \text{if } k> Z^{(m)}-Z^{(1)} 
\end{cases}
\hspace{0.1cm} \label{defUk}
\end{equation}
where $U^{(i)}$ is a diagonal matrix of the size $U_k$.  The $(j,j)$ element of $U^{(i)}$ corresponds to the control decision of state $m-j+1$ as given below: 
\begin{equation}
U^{(i)}(j,j) = 
\begin{cases}
1, & \text{if } \ i \leq m-j+1 \leq m \\
0, & \text{if } 1 \leq m-j+1 \leq i-1
\end{cases},\hspace{0.1cm} j=1,\dots,m. \label{defUi}
\end{equation}

The backward matrix is determined by $A_1$ matrix of the production time matrix and also the production authorization matrix $U_k$ as
\begin{equation}  
B_k = U_k\left(I_{m_a}\otimes A_1\right).  \label{defBk}
\end{equation} 
Similarly, the local matrices for the transition rates of the events that do not change the inventory position are given as 
\begin{eqnarray}
L_k&=&D_0\otimes I_{m_s}+U_k\left(I_{m_a}\otimes A_0\right), \label{defLk} \\
L_{0} &=& D_0\otimes I_{m_s}. \label{defL0}
\end{eqnarray}
\noindent The production authorization matrix appears in the definition of the local matrix $L_k$ given in Equation (\ref{defLk}) in order to make the transition matrix a valid infinitesimal generator and ensure that the transition rates are zero in the states where production is not authorized.

\subsection{Determining the Performance Measures} \label{MatrixGeo}	
For given thresholds, the production policy imposed by these thresholds is implemented by defining the production authorization matrix $U_k$ in Equation (\ref{defUk}).  Then the generator matrix $Q$ for the given thresholds is constructed by determining the forward, backward, and local matrices as given in Equations (\ref{defFk}), (\ref{defBk}), (\ref{defLk}), and (\ref{defL0}).  

In order to compute the steady-state probability distribution, the inventory position that can take values between $Z^{(m)}$ and $-\infty$ can be truncated at a lower bound $K^\ast$.  In this case, the size of the generator matrix will be $(Z^{(m)} + K^\ast)m \times (Z^{(m)} + K^\ast)m$.   In principle, the steady-state probabilities $\pi$ can be determined by solving $\pi Q = 0$ and $\pi \mathbbm{1} = 1$. However, this approach can be computationally demanding if $(Z^{(m)} + K^\ast)m$ is large.

Alternatively, the tri-diagonal structure of the generator matrix can be exploited to determine $\pi$ more efficiently. Equations $\pi Q = 0$ and $\pi \mathbbm{1} = 1$ can be  rewritten by using the tri-diagonal structure.  Accordingly,   the steady-state probabilities $\pi_k$ satisfy the following equations:
\begin{eqnarray}
\pi_0 L_0 + \pi_1 B_1 = 0& \label{recur1}  \\
\pi_k F_k + \pi_{k+1} L_{k+1} + \pi_{k+2} B_{k+2} = 0&, \hspace{0.3cm}k=0,1,\ldots \label{recur2} \\
\sum\limits_{x=0}^{\infty} \pi_k = 1 
\end{eqnarray}

The equations for the forward, backward, and local matrices given in Equations (\ref{defFk}), (\ref{defBk}), (\ref{defLk}), and (\ref{defL0}) show that the forward matrix remains constant at each level.  However, the backward and local matrices differ depending on the inventory position and the state of the underlying Markov chain. This structure is equivalent to the structure of a level-dependent Quasi Birth and Death (LDQBD) process. 

There are efficient methods developed for the analysis of LDQBD processes.  	I determine the steady-state probabilities of this process by using the Matrix Geometric method developed for LDQBD processes \citep{Bright1995CalculatingProcesses, Baumann2010NumericalProcesses}.

Once the steady-state probabilities are obtained, the performance measures of the system are then calculated by using the steady-state probability distribution for the given threshold vector $\bar{Z}$ used in the threshold-based production policy.  As functions of $\bar{Z}$, the expected inventory level, ${\rm{X}}^+(\bar{Z})$, the expected backlog level, ${\rm{X}}^-(\bar{Z})$, the probability of not having inventory in the system, $P_0(\bar{Z})$, and the total cost, $TC(\bar{Z})$ are given as:  
\begin{eqnarray} \label{eq:perform}
{\rm{X}}^+(\bar{Z})&=&\sum\limits_{k=0}^{Z^{(m)}-1}{(Z^{(m)}-k)\pi_k\mathbbm{1}}, \nonumber \\
{\rm{X}}^-(\bar{Z})&=&\sum\limits_{k=Z^{(m)}+1}^{\infty}{(k-Z^{(m)})}\pi_k\mathbbm{1}, \nonumber \\
P_0(\bar{Z}) &=&\sum\limits_{x=Z^{(1)}+1}^{\infty}{\pi_k}\mathbbm{1}, \nonumber \\
TC(\bar{Z})&=& b {\rm{X}}^-(\bar{Z}) + h {\rm{X}}^+(\bar{Z}).
\end{eqnarray}

\subsection{Determining the Optimal State-Dependent Threshold Levels} \label{policy_iteration}

The performance measures given in Equation (\ref{eq:perform}) are obtained for \emph{given} thresholds.  In order to determine the \emph{optimal} thresholds that minimize the total cost, an effective search method is needed.  I use the policy iteration algorithm \citep{Bertsekas2005DynamicControl} to determine the optimal threshold levels of the state-dependent threshold policy.
The policy iteration works with a finite Markov Decision Process (MDP) and  a finite number of policies, and guarantees that an optimal policy and optimal value function are determined in a finite number of iterations.  

I generate the initial MDP by truncating the inventory level between a maximum threshold level of the state-dependent threshold policy and a lower level of the inventory position.  An appropriate level of the lower bound $K^{\ast}$ is set to accumulate the probabilities of being in levels lower than $K^{\ast}$ into level $K^{\ast}$ by using the approach given in \citep{Heindl2004ETAQAProcess}.  Then at each iteration, the transition probabilities of the system are determined by using the block matrices given in Equation (\ref{Qmatrix}). 
Once the finite-size generator matrix is determined in this way, at each iteration of the algorithm, the steady-state probabilities and the total cost are calculated by using the methodology given in Section \ref{MatrixGeo} for the given thresholds.  The algorithm is initialized by setting the state-dependent threshold levels equal to the optimal single-threshold level. The new policies are generated by implementing policy improvement. This results in increasing or decreasing the state-dependent threshold levels compared to the previous iteration.  The policy improvement iterations are continued until the cost cannot be improved further.  

The steady-state probabilities of the system controlled by the optimal policy obtained from the policy iteration are calculated and the performance measures for the system controlled with the optimal thresholds are obtained by using Equation (\ref{eq:perform}) with these steady-state probabilities.

\section{Numerical Experiments}\label{numerical}
In this section, I analyze the effect of autocorrelation in inter-arrival and processing times on the performance of a production system controlled by the state-dependent threshold policy. I consider processes with positive and negative first-lag autocorrelations. Our objectives in these experiments are two-fold: investigating the impact of positive and negative autocorrelation on the optimal performance and also comparing the optimal performance with the benchmark production policies.
\subsection{Experimental Setup}
\subsubsection{Benchmark Production Control Policies}
I use three sub-optimal production control policies as benchmarks to compare the performance of the optimal policy.  The first benchmark policy, referred as \underline{M}ultiple-\underline{T}hreshold \underline{N}o \underline{A}utocorrelation (MTNA), uses a state-dependent threshold policy where the thresholds are determined by considering the inter-event time distributions but assuming that the inter-event times are not correlated.  In order to determine the state-dependent threshold for this case, the correlated inter-event time is modeled with a MAP with zero autocorrelation ($\rm{MAP}^{ren}$).  The thresholds are determined by using the methodology in Section \ref{policy_iteration} and the steady-state probabilities are obtained for the system controlled with these thresholds.

The second approximation, referred as \underline{S}ingle-\underline{T}hreshold \underline{W}ith \underline{A}utocorrelation (STWA), uses a single-threshold policy where the single-threshold is set optimally considering the inter-event distribution and the autocorrelation to represent the inter-event time

Finally, the third approximation, referred as \underline{S}ingle-\underline{T}hreshold \underline{N}o \underline{A}utocorrelation (STNA), uses a single-threshold policy where the optimal value of the single-threshold is determined by considering the inter-event distribution but assuming that the inter-event times are not correlated.

These different benchmark policies yield different threshold vectors to be used by the production policy.  The total costs, the expected inventory levels, the expected backlog levels, and the probability of not having inventory in the system are determined for each benchmark case by evaluating the original system with the correlated inter-event times when the system is controlled with the thresholds given by the benchmark cases.  These performance measures are compared with the performance measures obtained by using the optimal state-dependent threshold policy (Multiple-Threshold With Autocorrelation).  The percentage deviation of the expected inventory, expected backlog, probability of having no on-hand inventory, and the total cost with a benchmark policy with respect to these measures obtained with the optimal policy are denoted with $\Delta_{X^+}$, $\Delta_{X^-}$, $\Delta_{P_0}$, and $\Delta_{TC}$ respectively.

\subsubsection{Demand Arrival and Production Time Processes}\label{MAPsinNumericalAnalysis}

I consider four different MAPs with high ($scv>1$) and low ($scv<1$) variability, and positive and negative first-lag autocorrelation in our analysis.  I consider systems with correlated inter-arrivals and exponential processing times (MAP/M/1), Poisson arrivals and correlated processing times (M/MAP/1), and correlated arrival and processing times (MAP/MAP/1). This setup allows us to capture the impact of autocorrelation in demand arrivals or production times separately and jointly.

The MAP representation, the squared coefficient of variation $scv$, and the first-lag autocorrelation of the base MAPs, $\rho_1$  used in the analysis are given in Table \ref{MAPs}.   The traffic intensity of the system is set to be $\rho = 0.8$ in all cases. The mean of the inter-event times that have these MAPs is set to 1. I rescale the mean of inter-event times of the MAPs \citep{butools} to generate systems with the given traffic intensity. This rescaling preserves coefficient of variation and autocorrelation structure of the original process. It is done by multiplying the elements of the $D_0$ and $D_1$ matrices with the ratio of the original mean and the new mean and normalizing the $D_0$ matrix to make $D_0+D_1$ an infinitesimal generator matrix. Normalization is done by replacing the diagonal element of the new $D_0$ matrix by the negative of the summation of the other elements in a given row.  
\begin{table}[htbp]
	\centering
	\caption{The MAP representation, squared coefficient of variation, and first-lag autocorrelation of the MAPs employed in numerical analysis}
	\scalebox{0.8}{\begin{tabular}{|p{3.35cm}|c|c|c|c|c|}
			\hline
			Process & $D_0$ & $D_1$ &$E[T]$ & $scv$ & $\rho_1$ \\
			\midrule		
			Positively correlated MAP with $scv<1$ & $\begin{bmatrix}-1.4968&0	&0.0426\\0.0033&-1.5339&	1.4213\\0&	0&	-1.5340\end{bmatrix}$  & $\begin{bmatrix}1.4213&	0.0329&	0\\	0&	0&0.1093\\0.0533&	1.4807&0
			\end{bmatrix}$ & 1&0.76 & 0.10\\
			\midrule
			Positively correlated MAP with $scv>1$ & $\begin{bmatrix}-0.4531&0.0395\\
			0  & -1.2612 \end{bmatrix}$  & $\begin{bmatrix}0.4135  & 0 \\
			0.0176 & 1.2436\end{bmatrix}$ & 1&1.50 & 0.15  \\
			\midrule		
			Negatively correlated MAP with $scv<1$ & $\begin{bmatrix}-1.5 &1.5  & 0\\0& -3 & 1.5 \\0& 0&-1.5 \end{bmatrix}$  & $\begin{bmatrix}0 &0  & 0\\1.5& 0 & 0 \\0& 1.5&0 \end{bmatrix}$ & 1& 0.78 & -0.14 \\
			\midrule
			Negatively correlated MAP with $scv>1$ & $\begin{bmatrix}-0.5214 &   0.5214  &  0\\
			0  &-21.1159  & 0\\	0   &   0&  -21.1159\end{bmatrix}$  & $\begin{bmatrix}0  & 0 & 0\\
			1.3035&   0 &  19.8124\\19.5518 & 0  & 1.5641\end{bmatrix}$ &1 &2.75& -0.29  \\
			\hline
	\end{tabular}
} %
	\label{MAPs}
\end{table}%

\subsubsection{Generating MAPs with Different First-Lag Autocorrelations} \label{ExpSetMAP}

I analyze the impact of autocorrelation on the performance of the system by generating different MAPs with the same inter-event time distribution and different magnitudes of autocorrelation and analyzing the system with these generated MAPs following a similar approach used in \citep{ManafzadehDizbin2019ModellingProcesses}. The $D_1$ matrix  of a MAP with the same distribution and zero autocorrelation denoted by $D_1^{ren}$ is calculated as 
\begin{equation}
\label{renewal MAP}
D_1^{ren}=D_1\mathbbm{1} \beta,
\end{equation}
where $\beta$ is defined in Equation (\ref{MAPjointdensity}).  MAP($D_0$, $D_1^{ren}$) is referred as $\rm{MAP}^{ren}$.  

The first-lag autocorrelation of a MAP is a linear function of the elements of $D_1$ matrix \citep{Reference3}. When the first-lag autocorrelation of  MAP($D_0$,$D_1$) is $\rho_1$, in order to generate MAP($D_0$,$D_1^{new}$) with the same inter-event time distribution but with the first-lag autocorrelation of $\theta\rho_1$, $D_1^{new}$ matrix is constructed as
\begin{equation}\label{D_1ren}
D_1^{new}=\theta D_1+(1-\theta)D_1^{ren}
\end{equation}
where $D_1^{ren}$ is given in Equation (\ref{renewal MAP}).

\subsection{Impact of Correlated Demand Arrival Process}
In this section, I investigate the impact of positive and negative autocorrelation in demand inter-arrival times on the performance of a MAP/M/1 system. In order to focus on the impact of autocorrelation in demand arrival process, the production times are set to be {i.i.d.} exponential random variables.  

Tables \ref{p1}, \ref{p2}, \ref{exp2-8}, and \ref{n2A} show the performance measures when a production system with positively or negatively correlated demand arrivals and exponential service times is controlled with the optimal policy and their comparisons with the results obtained with the benchmark policies for the cases when the squared coefficient of variation $scv$ is less than 1 and greater than 1 respectively.  The MAPs used for the demand arrival processes in these experiments are given in Table \ref{MAPs} and $\rm{MAP}^{\rm{ren}}$s are constructed by using the procedure given in Section \ref{ExpSetMAP}. 

Tables \ref{p1}, \ref{p2}, \ref{exp2-8}, and \ref{n2A} show that the single-threshold policy that sets the single threshold based on the demand inter-arrival time distribution and also autocorrelation (Benchmark policy STWA) performs satisfactorily for all cases.  The percentage error obtained for the total cost is 2-3\% for the positively correlated demand arrival cases and less than 0.1\% for the negatively correlated demand arrival cases.  However, when the autocorrelation is not incorporated in the production policy, as in benchmark policies MTNA and STNA, the percentage errors increase.  These errors are more significant for the positively correlated demand arrival cases.   The effect of ignoring autocorrelation on the expected inventory and backlog levels are more pronounced.  For example, when the autocorrelation is not incorporated in production policy, the resulting backlog level is 101\% higher than the optimal backlog level for the positively correlated demand arrival case with a high squared coefficient of variation (Table \ref{p2}).  Similarly, while a single-threshold policy that does not incorporate autocorrelation (STNA) yields 5\% error for the total cost for the negatively correlated demand arrival case with high variability (Table \ref{n2A}), the resulting inventory level is 40\% higher than the optimal inventory level.

\color{black}
\begin{table}[H]
	\centering
	\caption{Performance Measures of a Positively Correlated Demand Inter-arrival and Exponential Production Time System (MAP/M/1) Controlled with the State-dependent Threshold Policy and Benchmark Policies (Demand Inter-arrival Time $ scv<1$) }
	\scalebox{0.8}{
		\begin{tabular}{llccccccccc}
			Policy & System & $ \bar{Z}^\ast $ & $TC(\bar{Z}^\ast)$    & $\Delta_{TC}$ & ${\rm{X}}^-(\bar{Z}^\ast)$ & $\Delta_{X^-}$ & ${\rm{X}}^+(\bar{Z}^\ast)$ & $\Delta_{X^+}$ & $P_0(\bar{Z})$ & $\Delta_{P_0}$ \\ \hline
			Optimal   & MAP/M/1 & (16, 11, 10) & 13.4067 &       & 1.3201 &       & 6.8064 &       & 0.1490 &  \\ \hline
			MTNA  & $\rm{MAP}^{ren}$/M/1 & (7, 7, 6) & 14.9538 & 12\%  & 2.2120 & 68\%  & 3.8941 & -43\% & 0.2511 & 69\% \\
			STWA  & MAP/M/1 & 11    & 13.7295 & 2\%   & 1.3981 & 6\%   & 6.7392 & -1\%  & 0.1577 & 6\% \\
			STNA  & $\rm{MAP}^{ren}$/M/1 & 7     & 14.9017 & 11\%  & 2.2601 & 71\%  & 3.6013 & -47\% & 0.2566 & 72\% \\
	\end{tabular} }%
	\label{p1}%
\end{table}%
\begin{table}[H]
	\centering
	\caption{Performance Measures of a Positively Correlated Demand Inter-arrival and Exponential Production Time System (MAP/M/1) Controlled with the State-dependent Threshold Policy and Benchmark Policies (Demand Inter-arrival Time $ scv>1$)}
	\scalebox{0.8}{
		\begin{tabular}{llccccccccc}
			Policy & System & $ \bar{Z}^\ast $ & $TC(\bar{Z}^\ast)$    & $\Delta_{TC}$ & ${\rm{X}}^-(\bar{Z}^\ast)$ & $\Delta_{X^-}$ & ${\rm{X}}^+(\bar{Z}^\ast)$ & $\Delta_{X^+}$ & $P_0(\bar{Z})$ & $\Delta_{P_0}$ \\ \hline
			Optimal & MAP/M/1 & (19, 12) & 17.8130 &    & 1.7380 &   & 9.1231 &   & 0.1583 & \\ \hline
			MTNA & $\rm{MAP}^{ren}$/M/1 & (9, 8) & 21.4375 & 20\%   & 3.4962 & 101\%  & 3.9565 & -57\%  & 0.3184 & 101\%\\ 
			STWA & MAP/M/1 & 16    & 18.3744 & 3\%   & 1.7677 & 2\%   & 9.5361 & 5\%   & 0.1610 & 2\% \\
			STNA & $\rm{MAP}^{ren}$/M/1 & 9     & 21.4637 & 20\%  & 3.4492 & 98\%  & 4.2176 & -54\% & 0.3142 & 98\% \\
	\end{tabular} }%
	\label{p2}%
\end{table}%
\begin{table}[H]
	\centering
	\caption{Performance Measures of a Negatively Correlated Demand Inter-arrival and Exponential Production Time System (MAP/M/1) Controlled with the State-dependent Threshold Policy and Benchmark Policies (Demand Inter-arrival Time $ scv<1$)}
	\scalebox{0.8}{
		\begin{tabular}{llccccccccc}
			Policy & System & $ \bar{Z}^\ast $ & $TC(\bar{Z}^\ast)$    & $\Delta_{TC}$ & ${\rm{X}}^-(\bar{Z}^\ast)$ & $\Delta_{X^-}$ & ${\rm{X}}^+(\bar{Z}^\ast)$ & $\Delta_{X^+}$ & $P_0(\bar{Z})$ & $\Delta_{P_0}$ \\ \hline
			Optimal & MAP/M/1 & (6, 6, 5) & 6.1660 &       & 0.6114 &       & 3.1088 &       & 0.1543 &  \\ \hline
			MTNA & $\rm{MAP}^{ren}$/M/1 & (7, 7, 6) & 6.2399 & 1\%   & 0.4571 & -25\% & 3.9544 & 27\%  & 0.1154 & -25\% \\
			STWA & MAP/M/1 & 6      & 6.1775 & 0\%   & 0.5692 & -7\%  & 3.3316 & 7\%   & 0.1437 & -7\% \\
			STNA & $\rm{MAP}^{ren}$/M/1 & 7       & 6.3155 & 2\%   & 0.4255 & -30\% & 4.1879 & 35\%  & 0.1074 & -30\% 
		\end{tabular}%
	}
	\label{exp2-8}%
\end{table}%
\begin{table}[H]
	\centering
	\caption{Performance Measures of a Negatively Correlated Demand Inter-arrival and Exponential Production Time System (MAP/M/1) Controlled with the State-dependent Threshold Policy and Benchmark Policies (Demand Inter-arrival Time $ scv>1$)}
	\scalebox{0.8}{
		\begin{tabular}{llccccccccc}
			Policy & System & $ \bar{Z}^\ast $ & $TC(\bar{Z}^\ast)$    & $\Delta_{TC}$ & ${\rm{X}}^-(\bar{Z}^\ast)$ & $\Delta_{X^-}$ & ${\rm{X}}^+(\bar{Z}^\ast)$ & $\Delta_{X^+}$ & $P_0(\bar{Z})$ & $\Delta_{P_0}$ \\ \hline
			Optimal   & MAP/M/1 & (13, 12, 11) & 11.6780 &       & 1.0670 &       & 6.3431 &       & 0.1513 &  \\ \hline
			MTNA  & $\rm{MAP}^{ren}$/M/1 & (13, 16, 16) & 12.3225 & 6\%   & 0.6743 & -37\% & 8.9509 & 41\%  & 0.0956 & -37\% \\
			STWA  & MAP/M/1 & 11    & 11.6845 & 0\%   & 1.0762 & 1\%   & 6.3034 & -1\%  & 0.1526 & 1\% \\
			STNA  & $\rm{MAP}^{ren}$/M/1 & 14    & 12.3084 & 5\%   & 0.6802 & -36\% & 8.9074 & 40\%  & 0.0965 & -36\% \\
	\end{tabular}}
	\label{n2A}%
\end{table}

The deviations between the results obtained by using the optimal policy and the benchmark policies are caused by the differences in the threshold levels set by the optimal and benchmark policies.  For example, for the case with positively correlated demand arrivals with high variability (Table \ref{p2}), the optimal policy uses two thresholds and sets them to (19, 12).  A single-threshold policy incorporating the autocorrelation (STWA) sets the threshold to 16 that is in between the optimal thresholds.  As a result, the percentage errors for the total cost, the expected backlog, and inventory levels are low (3\%, 2\%, and 5\% respectively).   However, for the positively correlated demand arrival case, when the autocorrelation is ignored in the production policy, using a state-dependent policy depending on the distribution of the demand inter-arrival time (MTNA) or using a single-threshold policy (STNA) do not yield good results since the thresholds are set far away from the optimal levels.  For the case with negatively correlated demand arrivals (Tables \ref{exp2-8} and \ref{n2A}), the threshold levels set by the benchmark policies are close to the optimal levels and therefore the percentage errors are smaller.

\begin{figure}
	\centering
	\caption{Impact of the first-lag autocorrelation of a positively correlated demand inter-arrival process on the threshold levels and the percentage difference between the total costs of MAP/M/1 system controlled with the optimal state-dependent threshold policy and the single-threshold with autocorrelation (STWA) policy ($h=1$, $b=5$)}
	\includegraphics[width=\linewidth]{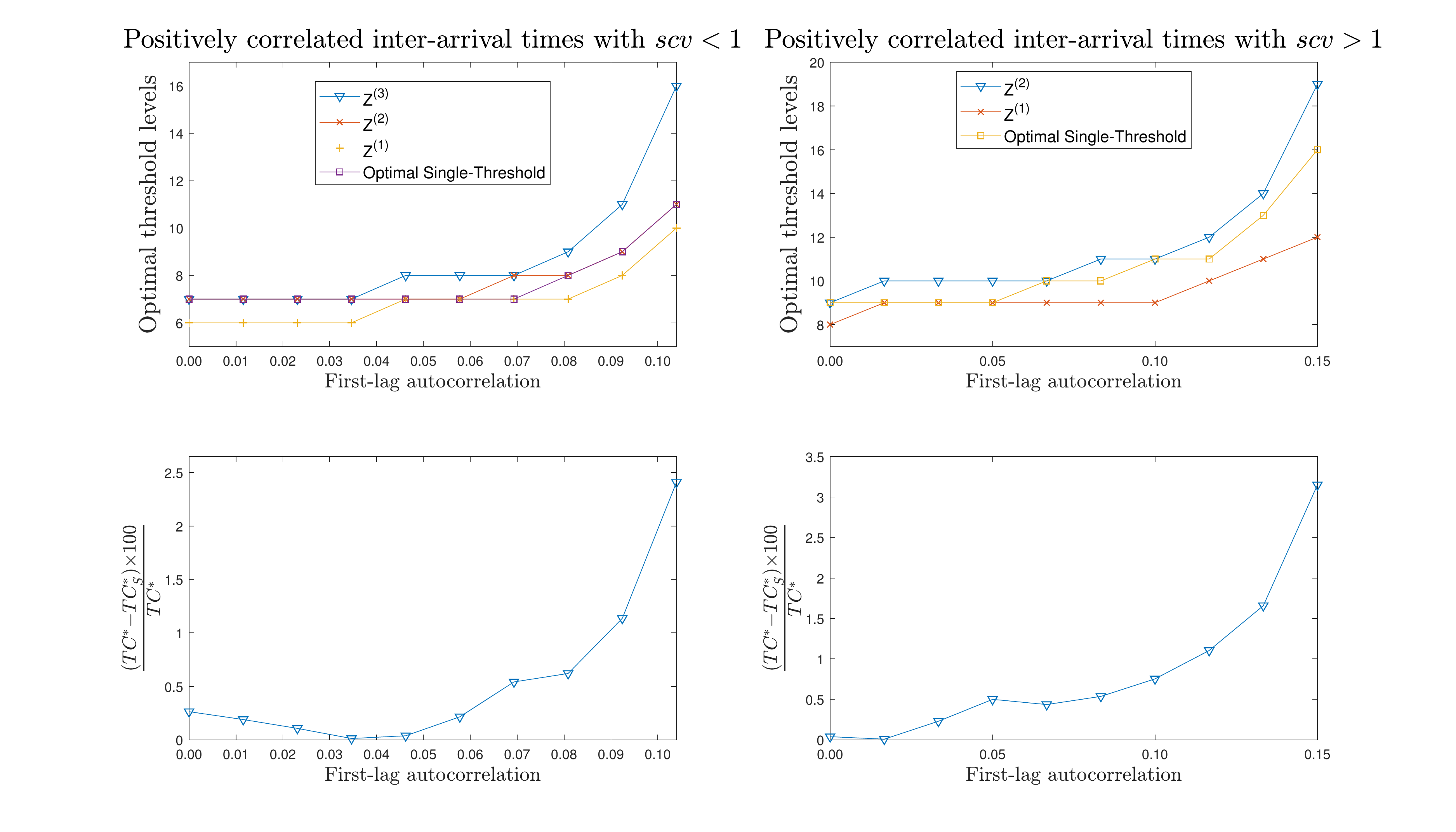}
	\label{sdbsap}%
\end{figure}

As shown in Figure \ref{sdbsap}, in a system with positively correlated arrival process, the biggest optimal state-dependent threshold level set by the state-dependent threshold policy increases more than the increase of the optimal state-independent threshold level.  This is due to the positive autocorrelation of demand arrivals that causes a short (long) inter-arrival time is to be followed by a short (long) inter-arrival time. Hence, the state-dependent threshold policy increases the biggest optimal state-dependent threshold level, greater than that of the optimal state-independent threshold level to cope with the variability in the inter-arrival times.    For the negatively correlated demand arrival case, demand accumulation due to variability is reduced by the arrival pattern and as a result the errors are lower, and the single-threshold policy performs well.

\begin{figure}
	\centering
	\caption{Impact of the first-lag autocorrelation of a negatively correlated demand inter-arrival process on the threshold levels and the percentage difference between the total costs of MAP/M/1 system controlled with the optimal state-dependent threshold policy and the single-threshold with autocorrelation (STWA) policy  ($h=1$, $b=5$)}
	\includegraphics[width=\linewidth]{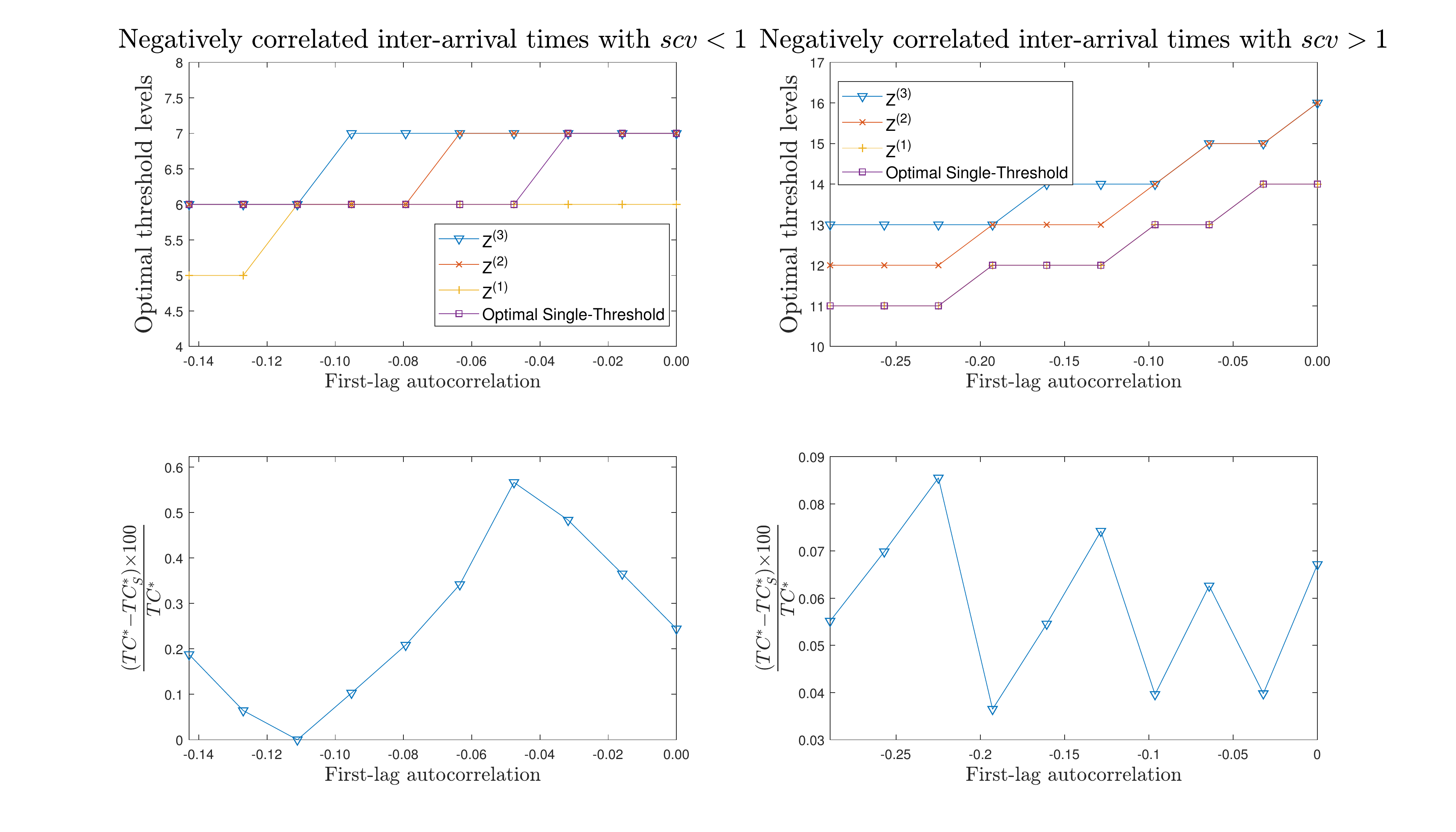}
	\label{sdbsan}%
\end{figure}

Figure \ref{sdbsap} and Figure \ref{sdbsan} show the impact of incremental increase in the first-lag autocorrelation of positively and negatively correlated demand arrival processes respectively. In order to conduct these experiments, the MAPs given in Table \ref{MAPs} are used and the new MAPs are generated by using the procedure given in Section \ref{ExpSetMAP}.  The distance between the zero and the first-lag autocorrelation of MAP is divided into 9 equal distances to generate MAPs with incrementally increasing the first-lag autocorrelations.  The figures show the effects of autocorrelation on the threshold levels and percentage difference between the optimal total cost of the system, $TC^\ast$ controlled with the optimal state-dependent policy and the cost of the state-independent threshold policy, $TC_S^\ast$, where the optimal threshold is set by incorporating the demand inter-arrival distribution and autocorrelation (STWA). The performance of STWA deteriorates in general as the positive correlation of the system increases. However, the policy does not show a monotone behavior for negatively correlated systems.  This erratic behavior maybe the result of the significant decrease of the cost for negative correlated systems. 


\subsection{Impact of Correlated Production Time Process}

I now focus on the impact of autocorrelation in production time process on the performance of a M/MAP/1 system controlled with  the optimal policy and the benchmark policies.  Similar to the analysis of the effect of correlated demand arrivals, in order to focus on the impact of autocorrelation in the service process, the demand inter-arrival times are set to be i.i.d. exponential random variables. 

Tables \ref{p1S}, \ref{p2S}, \ref{n1S}, and \ref{n2s} show the performance measures when a production system with positively or negatively correlated production times and exponential demand inter-arrival times is controlled with the optimal and the benchmark policies for the cases $scv<1$ and $scv>1$ respectively.  The MAPs used for the production time processes are given in Table \ref{MAPs} and $\rm{MAP}^{\rm{ren}}$s are constructed by using the procedure given in Section \ref{ExpSetMAP}.

\begin{table}[H]
	\centering
	\caption{Performance Measures of a Positively Correlated Production Time and Exponential Demand Inter-arrival Time System (M/MAP/1) Controlled with the State-dependent Threshold Policy and Benchmark Policies (Production Time $ scv<1$) }
	\scalebox{0.8}{
		\begin{tabular}{llccccccccc}
			Policy & System & $ \bar{Z}^\ast $ & $TC(\bar{Z}^\ast)$    & $\Delta_{TC}$ & ${\rm{X}}^-(\bar{Z}^\ast)$ & $\Delta_{X^-}$ & ${\rm{X}}^+(\bar{Z}^\ast)$ & $\Delta_{X^+}$ & $P_0(\bar{Z})$ & $\Delta_{P_0}$ \\ \hline
			Optimal   &  M/MAP/1 & (12, 11, 7) & 10.1854 &       & 1.0483 &       & 4.9438 &       & 0.1593 &  \\ \hline
			MTNA  & M/$\rm{MAP}^{ren}$/1 & (8, 7, 6) & 11.3075 & 11\%  & 1.6547 & 58\%  & 3.0339 & -39\% & 0.2516 & 58\% \\
			STWA  &  M/MAP/1 & 9     & 10.5654 & 4\%   & 1.0761 & 3\%   & 5.1849 & 5\%   & 0.1630 & 2\% \\
			STNA  & M/$\rm{MAP}^{ren}$/1 & 7     & 11.0796 & 9\%   & 1.4951 & 43\%  & 3.6039 & -27\% & 0.2268 & 42\% \\
	\end{tabular} }%
	\label{p1S}%
\end{table}%
\begin{table}[H]
	\centering
	\caption{Performance Measures of a Positively Correlated Production Time and Exponential Demand Inter-arrival Time System (M/MAP/1) Controlled with the State-dependent Threshold Policy and Benchmark Policies (Production Time $ scv>1$) }
	\scalebox{0.8}{
		\begin{tabular}{llccccccccc}
			Optimal   &  M/MAP/1 & (30,  21) & 29.4551 &       & 3.1322 &       & 13.7943 &       & 0.1626 &  \\ \hline
			MTNA  & M/{MAP}$^{\text{ren}}$/1 & (12, 9) & 34.3890 & 17\%  & 5.9574 & 90\%  & 4.6020 &-67\% & 0.3125 & 92\% \\
			STWA  &  M/MAP/1 & 22    & 29.9031 & 2\%   & 3.1155 &-1\%  & 14.3256 & 4\%   & 0.1617 &-1\% \\
			STNA  & M/{MAP}$^{\text{ren}}$/1 & 9     & 35.6392 & 21\%  & 6.2382 & 99\%  & 4.4482 &-68\% & 0.3269 & 101\% \\
	\end{tabular} }%
	\label{p2S}%
\end{table}%
\begin{table}[H]
	\centering
	\caption{Performance Measures of a Negatively Correlated Production Time and Exponential Demand Inter-arrival Time System (M/MAP/1) Controlled with the State-dependent Threshold Policy and Benchmark Policies (Production Time $ scv<1$) }
	\scalebox{0.8}{
		\begin{tabular}{llccccccccc}
			Policy & System & $ \bar{Z}^\ast $ & $TC(\bar{Z}^\ast)$    & $\Delta_{TC}$ & ${\rm{X}}^-(\bar{Z}^\ast)$ & $\Delta_{X^-}$ & ${\rm{X}}^+(\bar{Z}^\ast)$ & $\Delta_{X^+}$ & $P_0(\bar{Z})$ & $\Delta_{P_0}$ \\ \hline
			Optimal   &  M/MAP/1 & (7, 6, 5) & 6.2734 &       & 0.6233 &       & 3.1572 &       & 0.1537 &  \\ \hline
			MTNA  & M/$\rm{MAP}^{ren}$/1 & (7, 7, 6) & 6.2925 & 0\%   & 0.5525 & -11\% & 3.5302 & 12\%  & 0.1362 & -11\% \\
			STWA  &  M/MAP/1 & 6     & 6.3932 & 2\%   & 0.6230 & 0\%   & 3.2780 & 4\%   & 0.1536 & 0\% \\
			STNA  & M/$\rm{MAP}^{ren}$/1 & 7     & 6.4719 & 3\%   & 0.4695 & -25\% & 4.1245 & 31\%  & 0.1157 & -25\% \\	
	\end{tabular} }%
	\label{n1S}%
\end{table}%
\begin{table}[H]
	\centering
	\caption{Performance Measures of a Negatively Correlated Production Time and Exponential Demand Inter-arrival Time System (M/MAP/1) Controlled with the State-dependent Threshold Policy and Benchmark Policies (Production Time $ scv>1$) }
	\scalebox{0.75}{
		\begin{tabular}{llccccccccc}
			Policy & System & $ \bar{Z}^\ast $ & $TC(\bar{Z}^\ast)$    & $\Delta_{TC}$ & ${\rm{X}}^-(\bar{Z}^\ast)$ & $\Delta_{X^-}$ & ${\rm{X}}^+(\bar{Z}^\ast)$ & $\Delta_{X^+}$ & $P_0(\bar{Z})$ & $\Delta_{P_0}$ \\ \hline
			Optimal   & MAP/MAP/1 & (12, 11, 9) & 11.1589 &       & 1.0147 &       & 6.0855 &       & 0.1461 &  \\ \hline
			MTNA  & $\rm{MAP}^{ren}$/MAP/1 & (13, 12, 12) & 11.6535 & 4\%   & 0.7389 & -27\% & 7.9593 & 31\%  & 0.1064 & -27\% \\
			STWA  & MAP/MAP/1 & 11    & 11.4695 & 3\%   & 1.0019 & -1\%  & 6.4602 & 6\%   & 0.1442 & -1\% \\
			STNA  & $\rm{MAP}^{ren}$/MAP/1 & 13    & 11.8632 & 6\%   & 0.7341 & -28\% & 8.1925 & 35\%  & 0.1057 & -28\% \\
			
	\end{tabular} }%
	\label{n2s}%
\end{table}%

\begin{figure}
	\centering
	\caption{Impact of the first-lag autocorrelation of a positively correlated production time process on the threshold levels and the percentage difference between the total costs of M/MAP/1 system controlled with the optimal state-dependent threshold policy and the single-threshold policy with autocorrelation (STWA) policy ($h=1$, $b=5$)}
	\includegraphics[width=\linewidth]{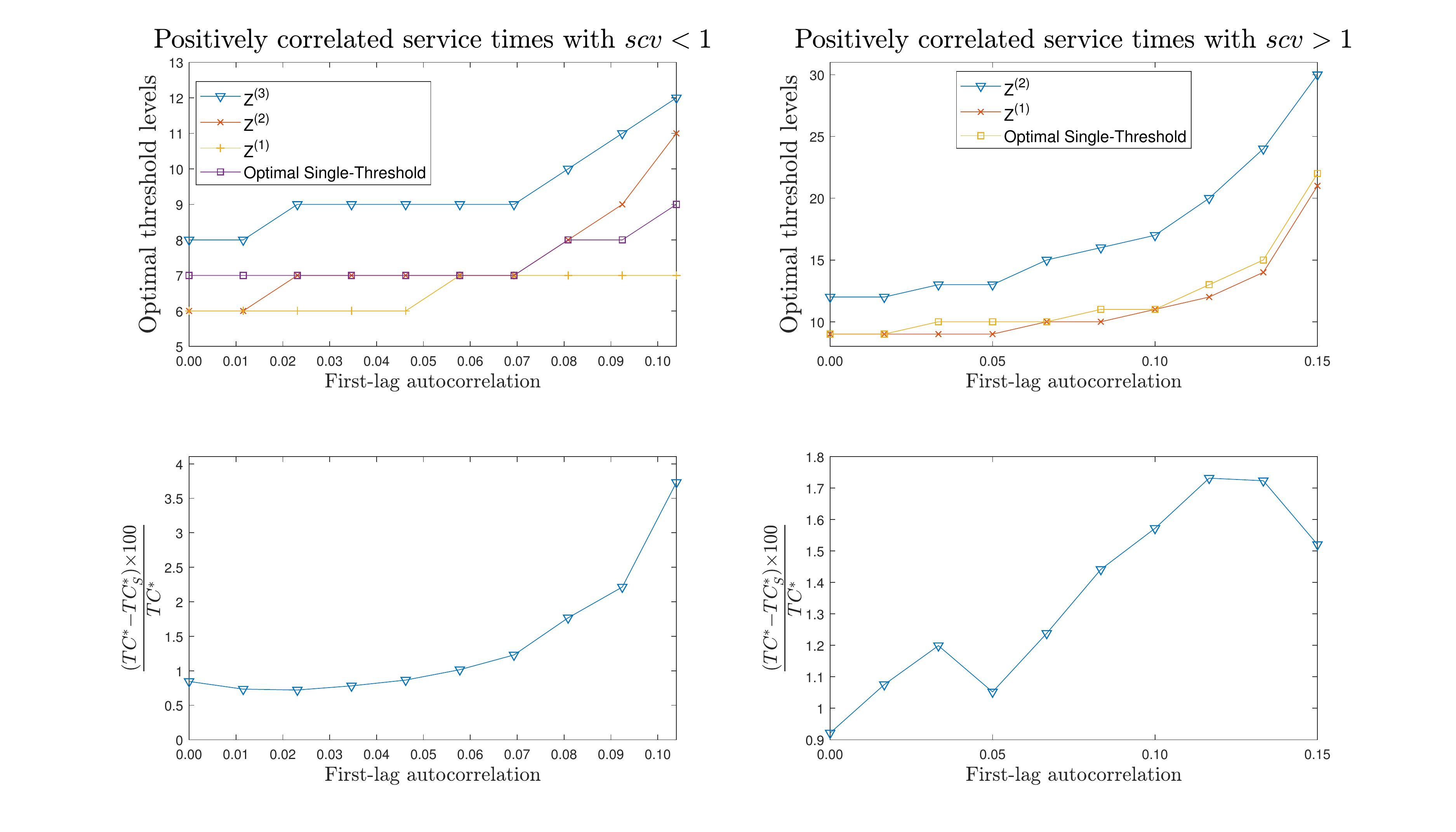}
	\label{sdbssp}%
\end{figure}%
\begin{figure}
	\centering
	\caption{Impact of the first-lag autocorrelation of a negatively correlated production time process on the threshold levels and the percentage difference between the total costs of M/MAP/1 system controlled with the optimal state-dependent threshold policy and the state-independent threshold policy with autocorrelation (STWA) policy ($h=1$, $b=5$)}
	\includegraphics[width=\linewidth]{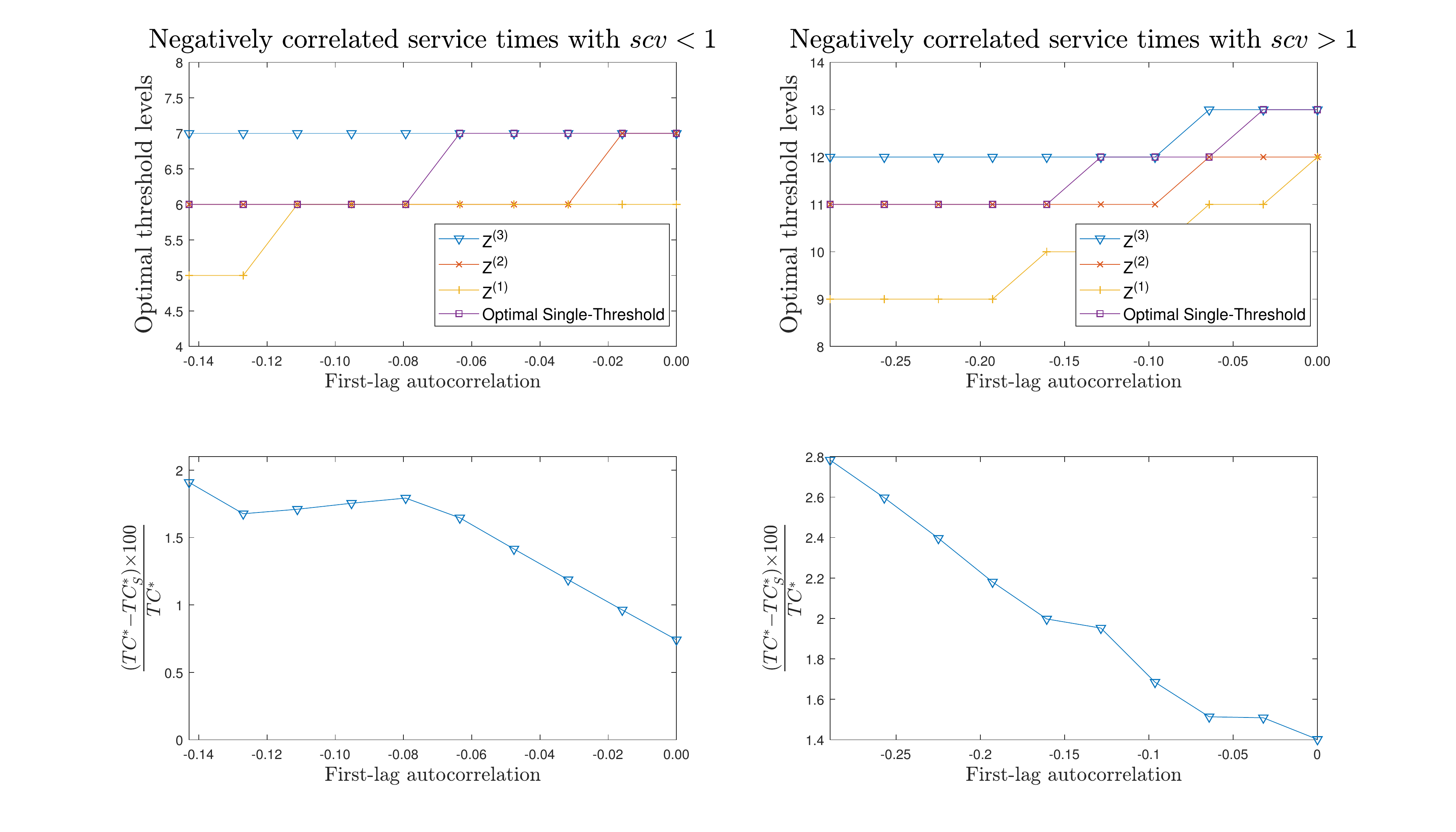}
	\label{sdbssn}%
\end{figure}%

Similar to the effect of autocorrelation in the demand arrivals, the effect of ignoring positive autocorrelation in production time process in the production policy is more significant compared to the effect of negative autocorrelation.  The single-threshold policy where the threshold is set based on the service time distribution and autocorrelation (STWA) yields a total cost that is 2-4\% higher than the optimal cost.  The resulting expected backlog and inventory levels are also within 6\% of the optimal levels.  However, if the autocorrelation information is not used in selecting the policy parameters, as in the benchmark policies MTNA and STNA, the threshold levels set by these policies are far away than the optimal levels.  Accordingly, the total cost obtained by using these policies is 9-21\% above the optimal cost for the positively correlated service times.  

When the service times are negatively correlated, the thresholds set by the benchmark policies are closer to the optimal levels and the total costs are within 6\% of the optimal cost.  For positively and negatively correlated service times, ignoring autocorrelation in production policies yield significant errors for the expected backlog and inventory levels.   

Figure \ref{sdbssp} and \ref{sdbssn} show the impact of incremental increase in the first-lag autocorrelation of positively and negatively correlated service processes on the threshold levels and on the percentage difference between the total costs of systems controlled by the optimal state-dependent and state-independent threshold policy with autocorrelation (STWA). Figure \ref{sdbssp} shows that the optimal threshold levels of the system increase as the first-lag autocorrelation increases. For the positively correlated service times, the performance of the single-threshold policy with autocorrelation (STWA) in controlling the system deteriorates as the first-lag autocorrelation increases as well.  For the negatively correlated service times, Figure \ref{sdbssn} shows that the optimal state-dependent threshold-levels also increases as a function of the first-lag autocorrelation and the performance of the single-threshold policy in controlling the system deteriorates as the first-lag autocorrelation increases. However, the relation is not monotone.

\subsection{Impact of Correlated Arrival and Service Processes on the Optimal Control of the System}
Finally, in this section, I focus on the combined effects of autocorrelation in both the demand arrival and also production processes on the performance of a MAP/MAP/1 production system controlled with the optimal and the benchmark policies.  I consider four different cases: positively correlated arrival and service, positively correlated arrival and negatively correlated service, negatively correlated arrival and positively correlated service, and negatively correlated arrival and service. The MAPS given in Table \ref{MAPs} are used for both demand arrival and service time processes in these experiments.

\begin{figure}
	\centering
	\caption{Percentage difference between the total cost of MAP/MAP/1 system with correlated inter-arrival and service times controlled with the optimal policy and the Multiple-Threshold NO Autocorrelation policy (MTNA)}
	\includegraphics[width=\linewidth]{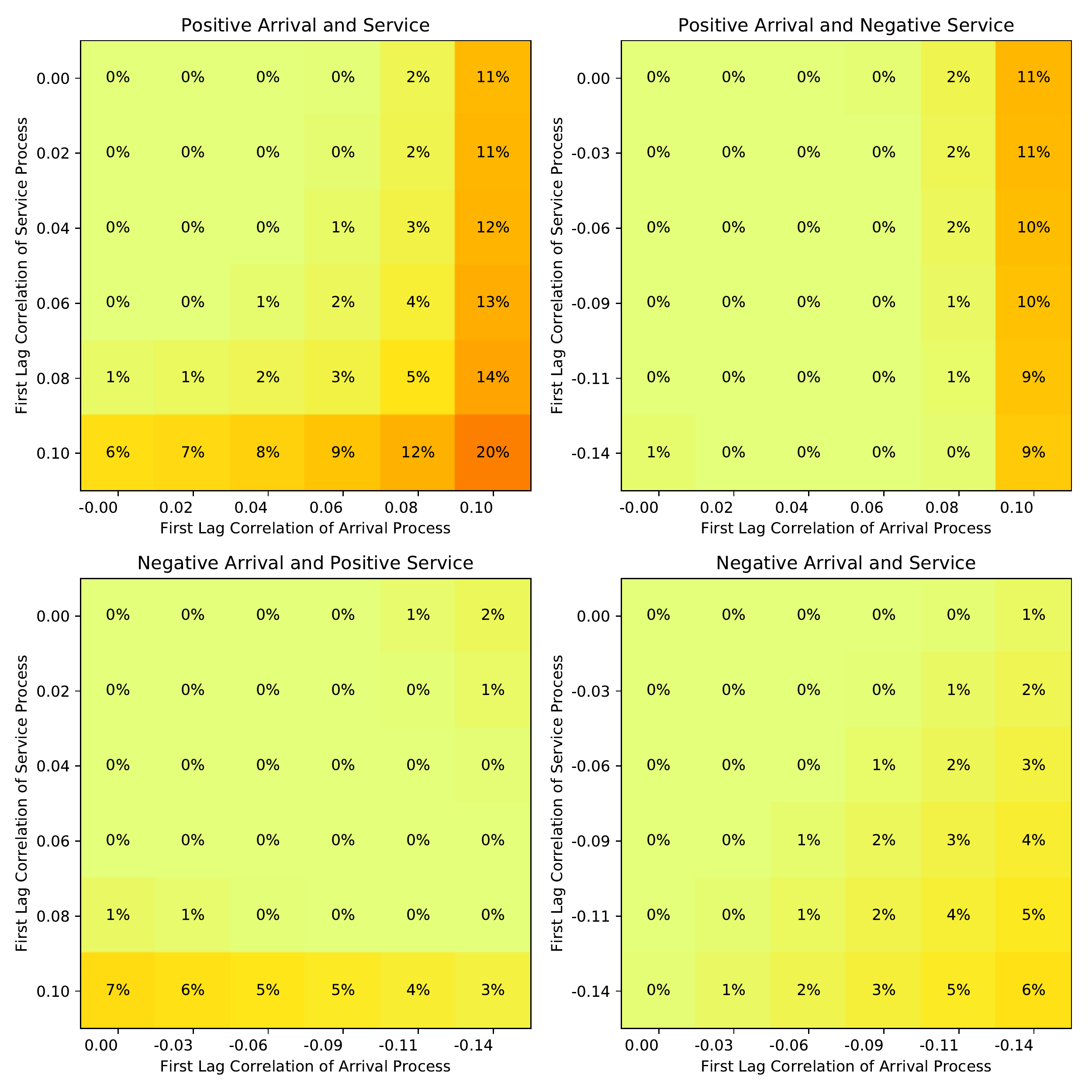}
	\label{corrAS}%
\end{figure}

Figure \ref{corrAS} shows the percentage increase in the total cost of a correlated system controlled by the threshold levels that are optimal for the uncorrelated system (Benchmark Policy MTNA) for the low variability case $scv<1$. Since the MAPs given in Table \ref{MAPs} for $scv<1$ have 3 states, the optimal state-dependent threshold policy that incorporates autocorrelation in service and demand arrival times sets 9 thresholds depending on the state of the demand and the production time process. 
The figure shows that positive correlation has more impact on the performance of the system. Controlling a system with positively correlated arrival and service times with first-lag autocorrelation of 0.1 results in a 20\% increase in the total cost. Results are less dramatic for a system with negatively correlated arrival and service times. Controlling a system with negatively correlated arrival and service times with the first-lag autocorrelation of -0.14 result in a 6\% increase in the total cost of the system. In addition to the impact of ignoring correlation, Figure \ref{corrAS} captures the interaction between negative and positive correlations as well. For the cases where the signs of the demand arrival and service process autocorrelations are different, i.e., positive/negative and negative/positive, positive and negative correlations neutralize the impact of each other, resulting in a lesser increase in the deviations. Similar results follow for processes with $scv>1$.

\begin{figure}
	\centering
	\caption{Percentage difference between the total cost of MAP/MAP/1 system with correlated inter-arrival and service times controlled with the optimal policy and the Single-Threshold With Autocorrelation policy (STWA)}
	\includegraphics[width=\linewidth]{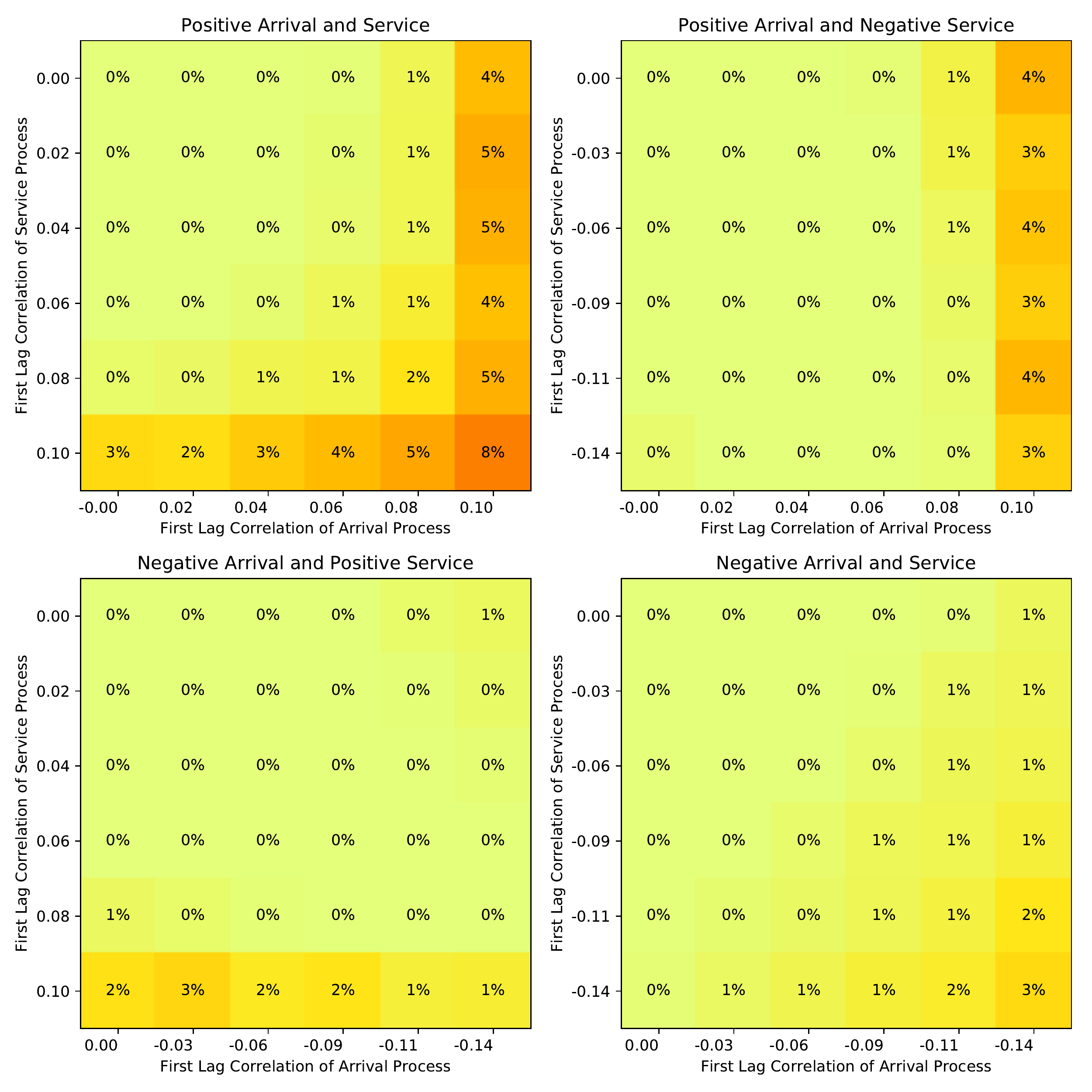}
	\label{corrASSTWA}%
\end{figure}

\begin{figure}
	\centering
	\caption{Percentage difference between the total cost of MAP/MAP/1 system with correlated inter-arrival and service times controlled with the optimal policy and the Single-Threshold No Autocorrelation policy (STNA)}
	\includegraphics[width=\linewidth]{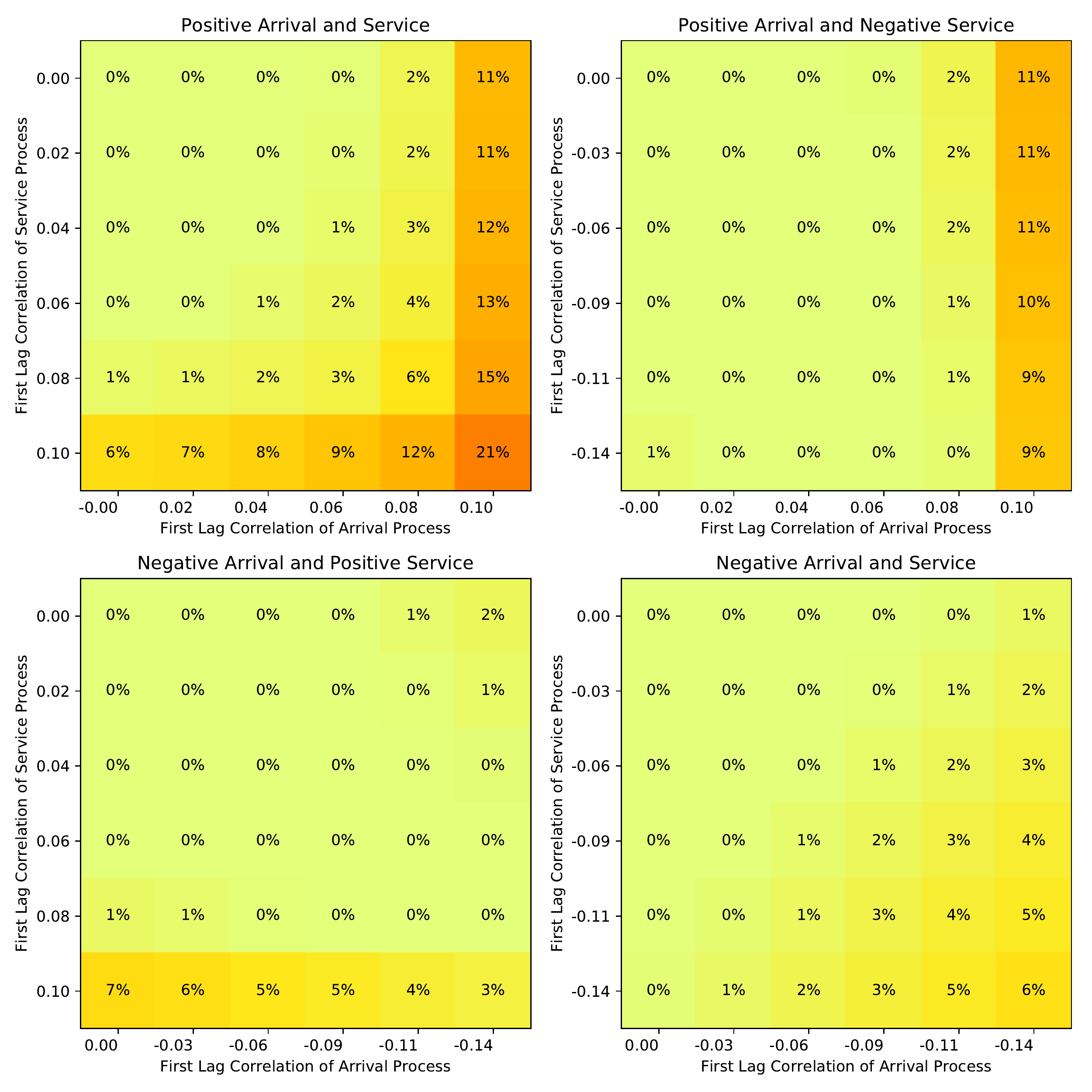}
	\label{corrASSTNA}%
\end{figure}

Figure \ref{corrASSTWA} demonstrates the percentage difference between the total cost of MAP/MAP/1 system with correlated inter-arrival and service times controlled with the optimal policy and the single threshold policy (STWA). Similar to the previous cases the performance of the single threshold policy deteriorates as the correlation increase. Controlling a system where both of the arrival and service processes are positively correlated with a first-lag of 0.1 with a single threshold policy results in an 8\% increase in the cost in comparison to the optimal policy. Figure \ref{corrASSTNA} demonstrates the percentage difference between the total cost of the correlated system with that of the system controlled by single threshold policy that assumes i.i.d. inter-event times (STNA). The policy performs well for cases with low magnitude of correlation and cases that negative and positive correlations interact with each other. Its performance gets worse as the correlation increases.

\section{Conclusions} \label{conclusion}
In this paper, I investigate the optimal control policy of a system with correlated demand inter-arrival and processing times modeled as Markovian Arrival Processes. I prove that the optimal policy that minimizes the expected average cost of the production-inventory system in the long run is a state-dependent threshold policy.  I then propose a matrix analytic method to evaluate the performance of the system controlled by the state-dependent threshold policy and use the policy iteration algorithm to determine the optimal thresholds. 

I use the structural properties of MAP to generate demand inter-arrival and production time processes with the same distribution and different magnitudes of autocorrelation. These MAPs are then used to evaluate the impact of autocorrelation in the inter-arrival and processing times on the optimal threshold levels and performance of the state-independent policy.

Our numerical analysis demonstrates that the optimal performance measures of a system with correlated inter-arrival times and service process are dependent on the autocorrelation structure in the system. Positive autocorrelations in the inter-arrival and processing times increase the optimal threshold levels of the  system. In contrast, negative autocorrelations in inter-arrival and processing times decrease the optimal threshold levels. In this study, I report the results for a system with moderate traffic intensity.  Our extensive numerical results showed that the impact of the correlation in this system increases significantly as I increase the traffic intensity of the system. 

I evaluate the performance of the  optimal single-threshold policy in controlling the system by comparing the total costs of systems controlled by 3 benchmark policies: a state-dependent policy that uses the distribution but assumes i.i.d. inter-event times, a single-threshold policy that  is set by using both the distribution and also the autocorrelation, and a single-threshold policy that uses the distribution but assume that the inter-event times are i.i.d.

Our analysis demonstrates that ignoring autocorrelation in setting the parameters of the production policy causes significant errors in the expected inventory and backlog costs.  A single-threshold policy that sets the threshold based on the distribution and also the autocorrelation performs satisfactorily.  However, ignoring positive correlation in setting the state-dependent thresholds levels and the single threshold based on the distribution yields high errors.  The difference between the total costs of the systems controlled by the state-dependent and the state-independent policies may increase as the magnitude of the first-lag autocorrelation increases.  Our study shows that an effective production control policy must take correlations in service and demand processes into account.   

This research can be extended in a number of directions. The approach used in deriving the optimal policy can be extended to determine the optimal threshold levels in a partially observable system, where the buffer level is observable while the state of the underlying Markovian process is unobservable.  The methodology used here for evaluating the optimal policy can also be adopted to evaluate the optimal control policy of a system with different classes of customers. In order to implement the state-dependent threshold policy or an approximate policy that uses the autocorrelation and distribution information, a methodology needs to be developed to determine the thresholds by using the collected data from the shop floor.  These are left for future research.

\section*{Proof of Theorem \ref{sdbspolicy}}\label{proofTh1}
I adopt the methodology presented in \citet{Koole2006MonotonicityApplications} to identify the optimal control policy of a system with correlated demand-arrival and service processes modeled as Markovian Arrival Process (MAP). Recall that the state of the production-inventory system presented in Section \ref{Problem} can be fully specified by the inventory level $(X)$ at the buffer and state of the underlying Markovian process corresponding to demand-arrival ($j_a$) and service ($j_s$) processes. I represent the state of the system with $(X, j_a, j_s)$.  
\begin{equation}
\label{MAP/M/1}
\resizebox{1 \textwidth}{!} 
{$ V(X,j_a)+g=\min\begin{cases}
	\frac{C(X)}{\alpha}+\sum\limits_{j\in J_a}P_0(j_a,j)V(X,j)+\sum\limits_{j\in J_a} P_1(j_a,j)V(X-1,j)&  u = 0, \\
	\frac{C(X)}{\alpha}+\sum\limits_{j\in J_a}P_0(j_a,j)V(X,j)+\sum\limits_{j\in J_a} P_1(j_a,j)V(X-1,j)+(\frac{\mu}{\alpha})V(X+1,j_a)-(\frac{\mu}{\alpha})V(X,j_a)&  u = 1.
	\end{cases}
	$}
\end{equation} 
For simplicity, I first analyze a system with $ MAP(D_0,D_1)$ arrival and exponential service process with rate $\mu$. The state of the system in this case can be represented by the inventory level at the buffer and the state of the arrival process $(X, j_a)$. The action space of the system consists of two actions denoted by $u=0$, and $u=1$. The action $u=0$ corresponds to stopping or not continuing the production, and the action $u=1$ corresponds to starting or  continuing the production. The optimality equation of this system can be written as in Equation (\ref{MAP/M/1}) using the uniformization technique. Note that the uniformization operator preserves the monotonicity and convexity characteristics of the value function of the continuous-time model \citep{Koole2006MonotonicityApplications}. Hence, the optimality results carry over to the continuous time problem. Starting from state $(X, j_a)$ the system spends an exponential time with rate $\alpha$ in this state, which results in $\frac{C(X)}{\alpha}$ cost, before moving to the next state. The transition probabilities to the next state is determined by the action $u$. If $u=0$ the transition probabilities of the system are related to the demand-arrival. In this case, the transition probabilities corresponding to state-change without new demand-arrivals and with demand-arrivals are captured with non-diagonal elements of the $P_0$ matrix, and $P_1$ matrix, respectively. The system stays at the same state with probability $1+\frac{D_0(j_a, j_a)}{\alpha} = P_0(j_a, j_a)$. The transition probabilities related to demand-arrival are the same when action $u=1$ is taken. In this case, the system moves to state $(X+1, j_a)$ with probability $\frac{\mu}{\alpha}$ and remain in the same state with probability $1-\frac{\mu-D_0(j_a, j_a)}{\alpha} = P_0(j_a, j_a)-\frac{\mu}{\alpha}$.

By rewriting the Equation (\ref{MAP/M/1}) and letting $V_{n+1}(X,i)$ be the minimal expected total cost if there are $n+1$ more events to go, $V_{n+1}(X, j_a)$ can be written in terms of $V_{n}(X, j_a)$ as in Equation (\ref{MAP/M/1-3}). 
\begin{multline}
\label{MAP/M/1-3}
V_{n+1}(X, j_a)+g=\frac{C(X)}{\alpha}+\sum\limits_{j\in J_a}P_0(i,j)V_n(X,j)+\sum\limits_{j\in J_a} P_1(j_a,j)V_n(X-1,j)\\+\frac{\mu}{\alpha}V_n(X,j_a)+(\frac{\mu}{\alpha})\min\{V_n(X,j_a),V_n(X+1,j_a)\}.
\end{multline}  
Equation (\ref{MAP/M/1-3}) can be expressed by using the operators given in Equation (\ref{operators}) which are similar to operators defined in \citet{Koole2006MonotonicityApplications}. These operators preserve the convexity and increasing property of a given function $f$. Let $x$ be the  inventory level and $j$ represent the state of the Markovian process.  
\begin{eqnarray}\label{operators}
T_{costs}f(x)&=&C(x)+f(x),  \nonumber\\
T_{unif}(f_1,\dots,f_l)(x)&=&\sum\limits_{j=1}^{l}p(j)f_j(x),  \ \ \ \ where   \ \  \sum\limits_{j=1}^{l}p(j)=1, \nonumber\\
T_{env}f(x, j) &=& f(x,j'),\nonumber\\
T_{A}f(x, j) &=& f(x-1, j), \nonumber\\
T_{P}f(x, j) &=& f(x+1, j), \nonumber\\
T_{CP}f(x, j)&=& \min\{f(x, j), T_{P}(T_{env})\} = \min\{f(x, j),f(x+1, j')\}, \nonumber\\
T_{menv}f(x, j)&=& \min\{f(x, j), T_{env}\} = \min\{f(x, j),f(x, j')\}, \nonumber\\
T_{DA}f(x, j) &=& T_{A}(T_{env}) = f(x-1, j').
\end{eqnarray}
Equation (\ref{MAP/M/1-3}) can be written in terms of the operator in Equation (\ref{operators}) as follows: 
\begin{equation}
V_{n+1}+g=T_{costs}\left(T_{unif}\left(T_{env}V_n ,\dots, T_{env}V_n,T_{DA}V_n,\dots, T_{DA}V_n\right) +\frac{\mu}{\alpha} \left(V_n +T_{CP}V_n\right)\right).
\end{equation}
I show by induction that $V_{n+1}(X, j_a)$ is convex in $X$ for a given $j_a\in J_a$. Let $V_0(X, j_a) = C(X)$. Convexity of the $V_1(X, j_a)$ is established trivially. Assume by induction that $V_{n'}(X, j_a)$ is convex in $X$ for $n'\in(2\dots n)$ and $j_a \in J_a$. Since all of the operators preserve the convexity property, convexity of $V_{n+1}$ is followed. For $n\rightarrow \infty$ this policy converges to the policy that minimizes the long-run average cost of the system. By Theorem 8.1 in \citet{Koole2006MonotonicityApplications}, convexity of $ V(X, j_a)$ in $X$ results in a threshold type optimal policy.  

The optimality equation for a system with $MAP(A_0,A_1)$ service process differs from the optimality equation of a system with exponential service process only in the second line of the Equation (\ref{MAP/M/1}). The state of the system in this case is $(X,j_a, j_s)$. The second line of the optimality equation of this system includes transition probabilities related to state-change (off-diagonal elements of $R_0$) and service-completion ($R_1$) of the service process and probability of the state remaining the same $(1-\frac{-D_0(j_a, j_a)-A_0(j_s, j_s)}{\alpha})$.
\begin{multline}
\label{MAP/MAP/1-}
V(X,j_a, j_s)+g=\frac{C(x)}{\alpha}+\sum\limits_{j\in J_a/\{j_a\}}P_0(j_a,j)V(X,j, j_s)+\sum\limits_{j\in J_a} P_1(j_a,j)V(X-1,j,j_s)\\+\sum\limits_{j'\in J_s/\{j_s\}}R_0(j_s,j')V(X,j_a, j')+\sum\limits_{j'\in J_s}R_1(j_s, j')V(X+1,j_a,j')\\+(1-\frac{-D_0(j_a,j_a)-A_0(j_s,j_s)}{\alpha})V(X,j_a, j_s),
\end{multline}
where $J_s$ is the set of Markovian states of the service process. Hence, the optimality equation is of the following form:
\begin{multline}
V(X,j_a, j_s)+g=\label{MAP/MAP/1-4}
\frac{C(x)}{\alpha}+\sum\limits_{j\in J_a}P_0(j_a,j)V(X,j,j_s)+\sum\limits_{j\in J_a} P_1(j_a,j)V(X-1,j,j_s)\\+
\min\Bigg\{0,\sum\limits_{j'\in J_s /\{j_s\}}R_0(j_s,j')V(X,j_a, j')+\sum\limits_{j'\in J_s} R_1(j_s, j')V(X+1,j_a,j')\\ -(\frac{-A_0(j_a,j_a)}{\alpha})V(X,j_a, j_s)\Bigg\}
\end{multline}
By using the following equation
\begin{equation}
\sum\limits_{j'\in J_s /\{j_s\}}R_0(j_s,j')+\sum\limits_{j'\in J_s}R_1(j_s,j')=\frac{-A_0(j_s, j_s)}{\alpha}, 
\end{equation}
I can re-write the Equation (\ref{MAP/MAP/1-4}) as 
\begin{multline}
\label{MAP/MAP/1-5}
V(X,j_a, j_s)+g=\frac{C(x)}{\alpha}+\sum\limits_{j\in J_a}P_0(j_a, j)V(X,j,j_s)\\+\sum\limits_{j\in J_a} P_1(j_a, j)V(X-1,j,j_s)\\+
\min\Bigg\{0,\sum\limits_{j'\in J_s/\{j_s\}}R_0(j_s,j')\left(V(X,j_a,j')-V(X,j_a, j_s)\right)\\+\sum\limits_{j'\in J_s}R_1(j_s,j')\left(V(X+1,j_a,j')-V(X,j_a, j_s)\right)\Bigg\}
\end{multline}
By taking the minimum function inside each function and adding and subtracting $(\frac{A_0(j_s,j_s)}{\alpha})V(X,j_a, j_s)$ I get the following equation.
\begin{multline}
V(X,j_a, j_s)+g=\label{MAP/MAP/1-7}
\frac{C(x)}{\alpha}+\sum\limits_{j\in J_a}P_0(j_a,j)V(X,j,j_s)\\+\sum\limits_{j\in J_a} P_1(j_a,j)V(X-1,j,j_s)-(\frac{A_0(j_s,j_s)}{\alpha})V(X,j_a, j_s)\\
\\+\sum\limits_{j'\in J_s/\{j_s\}}R_0(j_s,j')\min\Big\{V(X,j_a, j_s), V(X,j_a,j')\Big\}\\+\sum\limits_{j'\in J_s}R_1(j_s,j')\min\Big\{V(X,j_a, j_s), V(X+1,j_a,j')\Big\}
\end{multline}
which can be written in the following form:
\begin{multline}
V(X,j_a, j_s)+g=\label{MAP/MAP/1-7-2}
\frac{C(x)}{\alpha}\\+\sum\limits_{j\in J_a}P_0(j_a,j)V(X,j,j_s)+\sum\limits_{j\in J_a} P_1(j_a,j)V(X-1,j,j_s)\\
+\sum\limits_{j'\in J_s}R_0(j_s,j')\min\Big\{V(X,j_a, j_s), V(X,j_a,j')\Big\}\\+\sum\limits_{j'\in J_s}R_1(j_s,j')\min\Big\{V(X,j_a, j_s), V(X+1,j_a,j')\Big\}
\end{multline}
Equation (\ref{MAP/MAP/1-7-2}) can be written as function of the given operators in Equation (\ref{operators}) as given in Equation \ref{MAPMAP1opearators}. Convexity of the $V_{n+1}$ follows from the fact that the operators preserve convexity.  
\begin{multline}
\label{MAPMAP1opearators}
V_{n+1}+g=T_{costs}\big(T_{unif}\left(T_{env}V_n ,\dots, T_{env}V_n,T_{DA}V_n,\dots, T_{DA}V_n\right)\\
+T_{unif}\left(T_{menv}V_n ,\dots, T_{menv}V_n, T_{CP}V_n,\dots, T_{CP}V_n\right) \big).
\end{multline}
By Theorem 8.1 of \citet{Koole2006MonotonicityApplications} the optimal policy is of the threshold type for a given Markovian state, proving the Theorem \ref{sdbspolicy}.

\chapter[Empirical Analysis of Product Flows at a Production System]{Empirical Analysis of Product Flows at a Semiconductor Production System}\label{eda_chapter}
\section*{Abstract}
Semiconductor manufacturing systems are the most complex manufacturing systems in existence. The most important step of the semiconductor manufacturing is the wafer fabrication which accounts for more than 75\% of the total production time of the products. The capital intensive machines in the wafer fabrication necessitates the same machines to be used for similar processing steps resulting in a production network. Hence, a queueing-network view of product flows is necessary for analyzing the performance of the wafer fabrication. The aim of this study is to identify the important features that are impacting the main performance metrics of the wafer fabrication.  I focus on the total cycle times of the products and the inter-event time processes as the main performance metrics. 

This study is based on the extensive dataset obtained from the semiconductor manufacturing system of the Robert Bosch Company in Reutlingen, Germany. This dataset allows us to perform an Exploratory Data Analysis (EDA) on the main features that are impacting the main performance metrics of the wafer fabrication in different levels of detail. 

The EDA reveals that half of the products spend more than 70\% of their time inside wafer fabrication waiting to be processed. The waiting times of the products is impacted by a different variety of factors from different product types to the dispatching rules used for assigning products into machines. The analysis reveals that some of the layers in the product route contribute significantly more than others to the waiting and eventually the total cycle times of the products. The layer level analysis suggests the bottleneck recipes can be recognized using the waiting times of the layers.  

The machine level EDA on the statistical properties of the inter-event times show that processing times demonstrate a significant amount positive autocorrelation. Inter-arrival and inter-departure times may also demonstrate a significant autocorrelation as well.

This empirical study shows the need to develop modeling and analysis that capture the main dynamics observed in complex manufacturing systems.

\section{Introduction} 
Over the years, manufacturers have become more successful in efficient control of their supply chains and deploying new methodologies that match supply with demand. However, the supply chain efficiency can be further improved by using the recent developments in technology. Effective control of the manufacturing system requires a thorough understanding of the uncertainties that impact the performance of the system in strategic decision making related to demand fulfillment and production planning. Strategic plans are based on demand forecasting that needs to be coordinated with the production unit to account for the long cycle times of the products in manufacturing systems.

The massive amounts of collected data from the manufacturing systems introduce new avenues for further improvement of the system. One of the main challenges that arise in the analysis of manufacturing systems is \textit{using the collected data in identifying the most important features, evaluating the performance, and predicting the performance measures in an effective way.} Addressing this challenge is crucial in modeling and analysis of the manufacturing systems for design and control purposes. In order to address this challenge the product flow dynamics should be analyzed in different levels of detail. Higher level production planning and raw-material release decisions require basic understanding of the long term performance metrics such as the average total cycle times of the products. On the other hand, lower level decisions such as determining the amount of material in front of different equipment groups depend on the system dynamics at the machine level. 

Researchers have usually studied dynamics of the production systems under different modeling assumptions. Accordingly, the uncertainties in different levels of aggregation have been modeled with a pre-specified distribution or a process. For instance, the uncertainties in the inter-event times of the machines usually has been modeled with an exponential or phase type distribution \citep{Inman1999EmpiricalSystems, Chen1988EMPIRICALFABRICATION, Shanthikumar2007QueueingProblems}. The performance of the models build on the pre-specified assumptions has rarely been compared to that of the real production systems due to lack of the available shop-floor data.  

There is a lack of documented, comprehensive, empirical research in manufacturing systems literature that uses detailed inter-event data from the shop-floor to evaluate the main features that impact the performance of system. In this chapter, I use the detailed inter-event data of the product-movement (referred to as the lot-trace data) from the semiconductor production system of the Robert Bosch Company in Reutlingen, Germany, to evaluate the product flow dynamics by using the inter-event times in different levels of detail. In particular, I focus on the wafer fabrication of the semiconductor manufacturing system as the wafer fabrication data has a high quality. Wafer fabrication is a complex production network with a mixture of different product types. Products may spend several months in the wafer fabrication before their processing is finished. Presence of re-entrant loops, different product type mixtures, and several hundreds of production steps make wafer fabrication the most complicated manufacturing system. Analyzing the behavior of such a complex system by using the data from shop-floor is necessary in making planning, allocation, product-delivery, product-release, and product-flow decisions efficiently. 

Our lot-trace dataset consists of the inter-event data of every product processed in the 200 mm wafer fabrication in 2018 at Reutlingen plant. The raw dataset consists of 17\,223\,658 rows of inter-event data which contains the data related to 16305 unique products, categorized into 216 different parts. The products are being processed in 160 different equipment-groups which contain 500 different machines and can process 2159 different recipes. In this chapter, I use \textit{lot} and \textit{product} interchangeably. 


The aim of this study is to identify the most important features that are impacting the main performance metrics of the wafer fabrication in different levels of detail depending on the aggregation levels. First, I analyze the product flow dynamics in the wafer fabrication level. In this level of analysis, I treat the wafer fabrication as a black-box and investigate the dynamics of WIP levels, the arrival and departure processes, and the total cycle times of the products inside the wafer fabrication. It is shown that the total cycle times of the products vary substantially from several days to several months. I investigate some of the properties of the total cycle times, and different factors that possibly impact the total cycle times of the products. Second, I investigate the statistical properties of the inter-event times (inter-arrival, inter-departure and processing times) of the machines at machine aggregation level. It is shown that the inter-event times of the machines may demonstrate a significant dependency between themselves which has usually been ignored in the literature. Finally, I evaluate the waiting and processing times of the products in different layers in their product routes. The analysis shows that some layers demonstrate a significantly higher ratio of waiting to processing times in comparison to others. Each level of analysis is discussed in further detail below.

\subsection{Wafer Fabrication Level Analysis of the Production System}
Understanding the product flow dynamics in the wafer fabrication aggregation level is fundamental in taking the production planning decisions and offering delivery time to customers. Semiconductor customers value on-time delivery as high as the price of the products \citep{Batra2018QuantifyingChain}. In such a setting making the planning decisions becomes highly important. A production manager needs a thorough understanding of the total cycle times of the products in making the planning and release decisions. Releasing products into production system without considering the remaining cycle times of the current products in the system may lead to exponential increase in the waiting and consequently the total cycle times \citep{Monch2013ProductionFacilities}. 

In the wafer fabrication level of analysis, I analyze the arrival or release of products to the wafer fabrication, the departure of products from the wafer fabrication, and the total cycle times of the products inside the wafer fabrication. The analysis shows that the variation in the departure process of the system is substantially higher than that of the arrival process. The variation in the departure process is affected by variety of different factors such as the long total cycle times of the products and mixture of different product types. 

The total cycle time (TCT) of a product is defined as the difference between the time-stamp that the product starts and finishes its processing. I show that TCT of the products in the wafer fabrication varies substantially ranging from several days to several months. The analysis demonstrates that some of the variation in TCT can be explained by the type of the products. However, the variation of TCT of each product type is quite high as well. In general, half of the products spend more than 70\% of their time waiting to be processed inside the wafer fabrication. In other words, half of the products spend more than $2.34$ times their total processing time waiting to be processed. This ratio may increase or decrease based on the number of process steps of the products. The number of process-steps of a certain product consists of a pre-determined set of process-steps that the product needs to go through plus inspection and reworks steps. Figure \ref{numprocesssteps} demonstrates the number of process-steps of the products processed in the wafer fabrication. The distribution of the number of process-steps is a mixture of different distributions due to inspection and rework steps.  

\subsection{Machine Level Analysis of Inter-Event Times in a Wafer Fabrication}
In the machine level analysis, I treat each equipment in the wafer fabrication in isolation and evaluate statistical properties of thier inter-event times. The main approach in the literature to model an equipment is to model the uncertainties in the product-arrival, and service times with a given distribution \citep{Shanthikumar2007QueueingProblems}. Such an approach ignores the sequence in which the  products arrive at the system, get service, and leave the system. In other words, the dependency between the inter-event times is ignored. Employing analytical models that ignore dependency in evaluating the performance of the system may results in inefficient performance evaluation and control decisions \citep{ManafzadehDizbin2019ModellingProcesses}. The most common distributional assumption in the literature of the manufacturing systems for modeling inter-event times is using exponentially distributed inter-arrival and service times \citep{Chen1988EMPIRICALFABRICATION, Inman1999EmpiricalSystems, Shanthikumar2007QueueingProblems}. Researchers have extended these assumptions to contain more general cases by adopting models such as general or phase type distributions to account for more information from the data. However, any possible dependency between the inter-arrival and service times has been usually ignored due to challenges in modeling correlation. Almost all of the performance evaluation methods in the literature are based on the independence assumption between the inter-event times. \citet{Whitt2018UsingQueues} state that the major shortcoming of the Queuing Network Analyzer (QNA) is due to its inability in capturing the dependency between inter-arrival and service times. \citet{ManafzadehDizbin2019ModellingProcesses} show that ignoring dependence leads to errors in optimal design and control of manufacturing systems. 

I evaluate the presence of dependency between the inter-event times by analyzing the statistical properties of the processing, inter-arrival, and inter-departure times data in the wafer fabrication. In particular, I focus on the empirical distribution of the coefficient of variation and first-degree (first-lag) dependency between the inter-event times in different machines. Although there are empirical studies on the distribution of the inter-event times from automotive industry \citep{Inman1999EmpiricalSystems}, the statistical properties of the inter-event times in call centers \citep{Brown2005StatisticalCenter, Kim2014AreProcesses} and health-care systems \citep{Armony2015OnPerspective}, there are no comprehensive large-scale empirical study on the distribution and inter-dependency of the inter-event times in semiconductor wafer fabrication. Dependency in the inter-event times of the machines from the wafer fabrication exists due to complicated processes such as batch arrival, batch departure, batch processing, merging, and complicated dispatching rules. Our analysis demonstrates that different types of dependency may exist between the inter-event times. Inter-arrival and inter-departure times may demonstrate a positive or negative dependence while processing times demonstrate a positive dependence. To the best of our knowledge, \citet{Inman1999EmpiricalSystems} is the only study that evaluates the inter-event times of productions systems. \citet{Inman1999EmpiricalSystems} analyzes a few machines from automotive industry. I present a more comprehensive analysis by using the data from 500 different machines in the wafer fabrication of the semiconductor manufacturing plant of the Robert Bosch Company.  

\subsection{Layer Level Analysis of Inter-Event Times in a Wafer Fabrication}
In semiconductor wafer fabrication products go through hundreds of process steps in which different number of layers are fabricated on the surface of the wafers. Each layer consists of a set of pre-specified process steps along with random inspection steps in between. Different layers consist of the same set of equipment groups which may process different recipes on the wafers. 

\begin{figure}
	\caption{A simple production network to illustrate different levels of analysis}
	\begin{center}
		\includegraphics[scale=.5]{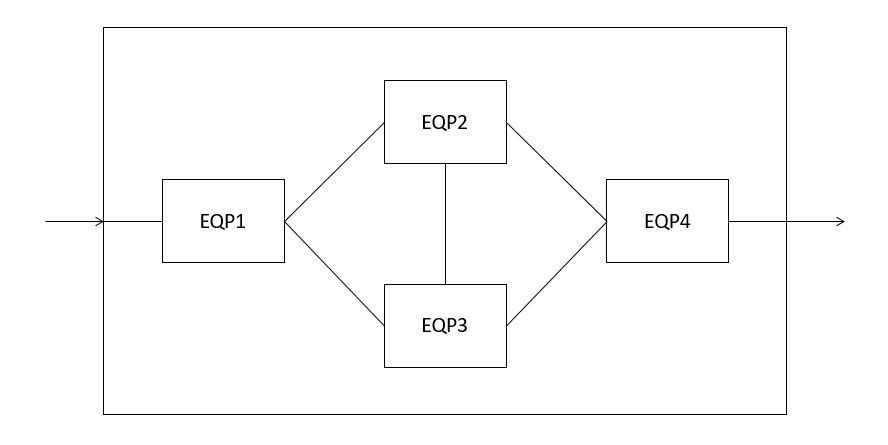}
	\end{center}
	\label{product_route_example_4}
\end{figure}

Consider the simple fabricated example given in Figure \ref{product_route_example_4} for understanding layers and recipes in the wafer fabrication. This production system consists of four machines denoted by $EQP_i$. Suppose that $EQP_1$ can process the recipes $r_1$ and $r_5$ and the rest of machines process the recipes with the same subscript (e.g., $EQP_2$ process the recipe $r_2$). Assume this simple production network manufactures two types of products, $A$ and $B$. The type $A$ products require two different layers to be fabricated on the wafers. The first ($A_1$) and second $(A_2)$ layer are fabricated on top of the wafer by sequentially processing the set of recipes $(r_{1}, r_{2}, r_{3})$, and $(r_{5}, r_{2}, r_{4})$, respectively, where $r_{i}$ is the recipe used for processing. Hence, the product route of the type $A$ products is $(r_{1}, r_{2}, r_{3}, r_{5}, r_{2}, r_{4})$. The product type $B$ has a single layer ($B_1$) in which the products go through $(r_{1}, r_{3}, r_{4})$ and then leave the system. 
 
A production manager needs to balance the amount of time that products spend in each layer to ensure a balanced production network. A thorough understanding of the processing times (total time the product spends inside the machines) and waiting times (total time the product spends waiting to be processed in front of the machines) of  the layers is necessary in doing so. For instance, assume that $EQP_4$ processes the products with the first-come-first-serve policy and has a higher processing time for products in layer $A_2$ in comparison to layer $B_1$. A higher WIP levels for products in layer $A_2$ will result in significant increase in the waiting times of the product in layer $B_1$. Consequently, failing to balance the WIP levels of different products in front of $EQP_4$ may result in inefficient system control. 

The layer level analysis aims at investigating the amount of time spend in different layers. A higher waiting time in a given layer may suggest possible bottlenecks along its production route. In the complicated semiconductor production network, the dispatching rules determine the waiting times of the products in front of different equipment group. Inefficiencies in controlling the WIP levels of different product types in different layers may substantially increase the total cycle times of the products in the system. 

The findings of this thesis has been made possible through the participation of Ko\c{c} University in the \textit{Productive 4.0}\footnote{\url{https://productive40.eu/}} project and the corresponding \textit{T\"{U}BITAK} project. My one year PhD visit at Bosch Center for Artificial Intelligence has also been another contributing factor. The visiting position in Stuttgart, Germany enabled me to visit the wafer fabrication more frequently and discuss the related issues with the experts in understanding and cleansing the data and integrating different datasets. 

\subsection{Outline of the Chapter}
The rest of this chapter is organized as follows. I introduce basics of the semiconductor manufacturing and wafer fabrication in Section \ref{semicondutorintro} and describe our dataset. In Section \ref{wfanalysis}, I analyze the arrival and departure process and the total cycle times of the products at the wafer fabrication level. In Section \ref{machine_level_analysis}, I investigate the statistical properties of the system at the machine level. In Section \ref{layer_level_analysis}, I analyze the waiting and processing times of the products in the major layers in their product routes. Finally, I summarize the main findings in Section \ref{summary_of_findings} and conclude the chapter in Section \ref{conclusions}.
 
\section{Semiconductor Manufacturing System}\label{semicondutorintro}
The structure of the semiconductor manufacturing processes can be characterized with different levels of details referred to as the top, macro and micro levels. The top level analyzes the major part of supply chain of the system, which consists of two parts, the frontend and the backend. The frontend and the backend are divided in the industry partners supply chain by a warehouse called the Diebank. The frontend consists of all the semiconductor production steps on the wafers while the backend contains all the cutting, testing and assembly steps to build a microelectronic component. The frontend can be thought as a push system where the wafers go under a network of complicated processing steps, while the backend mostly consists of production lines. The frontend contains the Wafer Fabrication (WF), and Sort or Probe production areas in the macro level. The backend on the other hand contains dicing, assembly, packaging, and final testing production areas. WF is the initial and the most important step of the semiconductor manufacturing accounting for more than 75\% of the total cycle time of the products. As defined earlier, the total cycle time of a product is defined as the difference between the time that processing of the product starts, and finishes. TCT is summation of the real processing times that machines or machines spend on a given product, the waiting times of the products before processing, and the possible transportation times. The product that flows in the semiconductor manufacturing plant differs from the frontend to backend. In the frontend, the products are wafers which are made of silicon. Raw silicon wafers enter WF and leave WF after several hundreds of  processing steps described below. Each step of the WF may add a new components to the wafers, removes some components from the wafer, clean the wafer or test it. Wafers are kept inside a container box which usually contains 25 wafers referred to as a lot or sometimes job. In this chapter, we use lot and product inter-changeably. After WF wafers are sent to Probe station to test for their basic functionality. Wafers that pass the functionality test are either sent to Assembly or Diebank. In Diebank, wafers are stocked for later usage. In Assembly, each wafer is cut into dices and then packaged for use in the electronic devices. The final step of the production is to test the ICs before delivering to customers. 
\begin{figure}
	\caption{Layout of the semiconductor manufacturing system}
	\begin{center}
		\includegraphics[scale=.4]{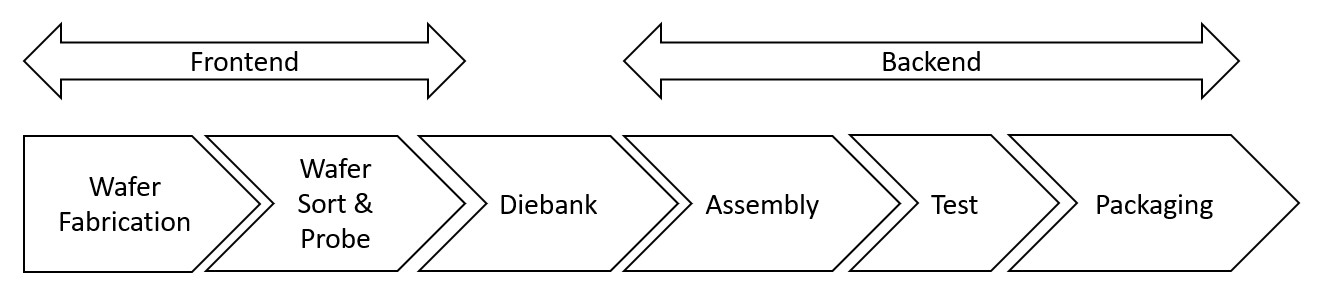}
	\end{center}
	\label{frontbackend}
\end{figure}

\subsection{Wafer Fabrication (WF)}
\begin{figure}
	\caption{Empirical distribution of the number of process-steps of the lots in the wafer fabrication}
	\begin{center}
		\includegraphics[scale=.75]{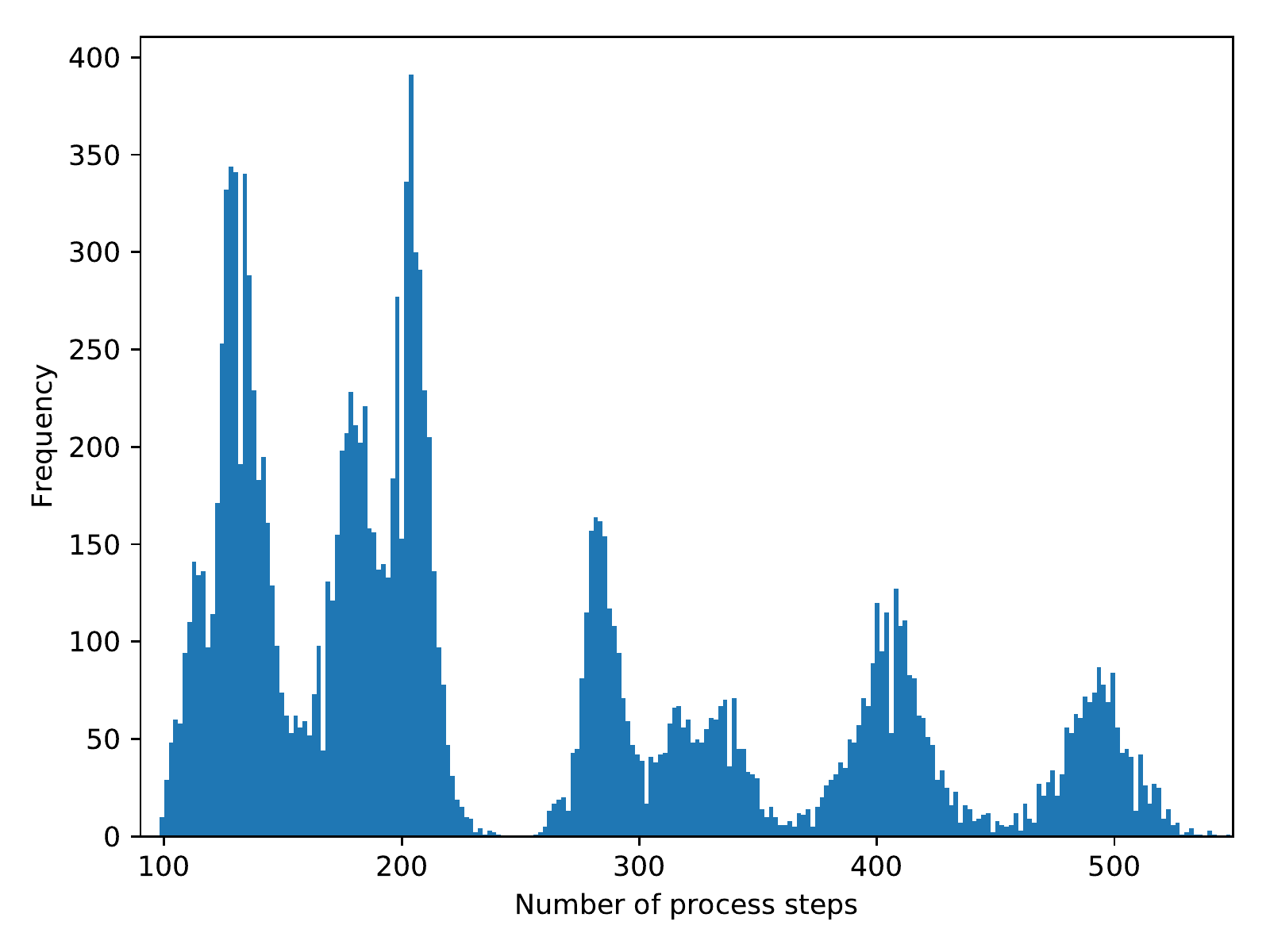}
	\end{center}
	\label{numprocesssteps}
\end{figure}
\begin{figure}
	\caption{Number of layers of the 50 most frequent products in the wafer fabrication}
	\begin{center}
		\includegraphics[scale=.75]{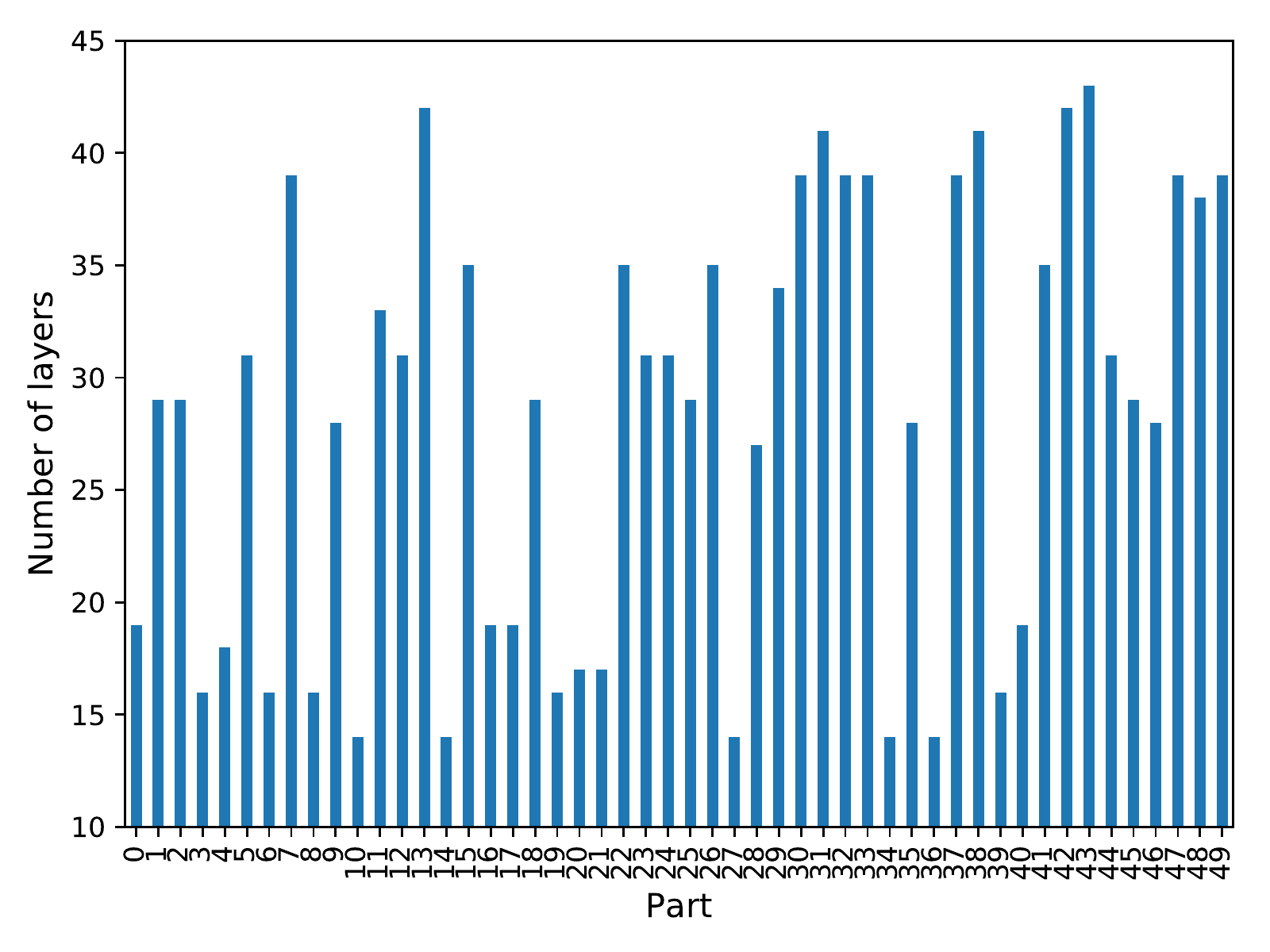}
	\end{center}
	\label{wf_parts_num_layers}
\end{figure}

WF is the most complex manufacturing system in existence. It differs from regular manufacturing systems in several ways. Products in WF go through several hundreds of productions steps. Number of process-steps of each lot in WF is the summation of main production-steps (or recipes) determined based on the technology of the product and testing steps and reworks arising from the test-failure or long waiting times. Figure \ref{numprocesssteps} demonstrates the number of process-steps of lots in the wafer fabrication of the Reutlingen plant. The number of process-steps of the lots differ significantly based on the type of the product. Each of the process-steps are either a main processing step, a cleaning step, or a test step to assess the quality of the wafer at the current production stage. A pre-defined set of main processing steps determined by the recipes that a particular product needs to go through along with cleaning and testing determine the number of production-steps of the product. The main processing steps of each lot are categorized into set of pre-defined subsets called \emph{layer}. A layer, as the name suggests adds a new physical layer on top of the wafers inside each product. The number of layers of a certain product is determined based on the technology of the product. Figure \ref{wf_parts_num_layers} shows the number of layers of the most frequent 50 products types (parts) in WF of the Reutlingen Plant of the Robert Bosch Company. 

\begin{figure}
	\caption{Work stations in the semiconductor wafer fabrication}
	\begin{center}
		\includegraphics[scale=.7]{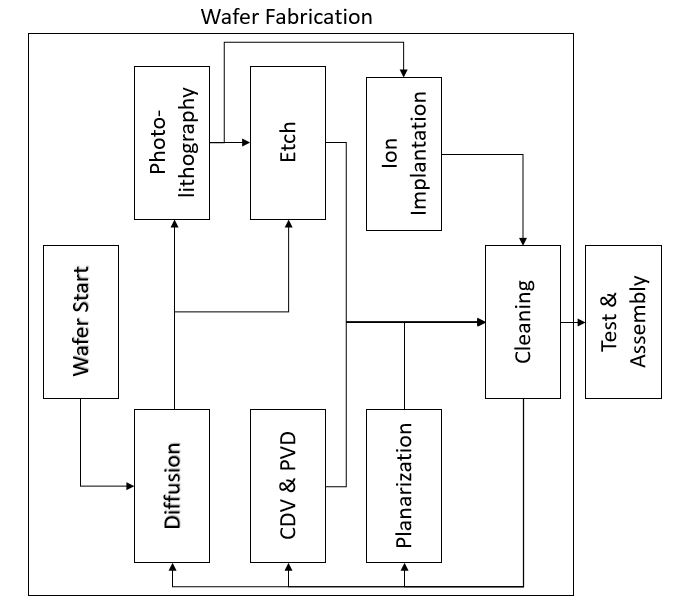}
	\end{center}
	\label{boschwflayout}
\end{figure}

Figure \ref{boschwflayout} shows the main processing steps of the wafer fabrication, and the general movement of the lots between the process-steps of a semiconductor manufacturing system. Lots usually start their processing from Diffusion and Lithography. Lithography is used to transfer a pattern from a photomask to the top of the wafer. Diffusion is a high temperature process that disperses material on the wafer surface. Oxidation is another process-step that converts silicon into silicon dioxide. Ion implantation is the process of introducing dopant impurities into the wafers. Doping is the process of adding a small percentages of foreign atoms to pure semiconductors to change their electrical properties. After ion implantation the crystal structure of the wafer is damaged which worsens the electrical properties of the wafer. Diffusion is applied to anneal the crystal defects after ion implantation or to introduce dopant atoms into silicon from a chemical vapor source. Etching is used to remove material selectively from wafer to create patterns defined by the etching mask. The etching mask protects the parts that need to remain on the wafer. There exist two types of etching in general: wet etching and dry etching. In Deposition step, multiple layers of different material are deposited on the wafer. There are two types of deposition methods in general, namely Physical Vapor Deposition (PVD) and Chemical Vapor Deposition (CVD). Chemical Mechanical Planarization (CMP) modifies the surface of the non-planar wafer surface that is created in etching, deposition, or oxidation steps. Cleaning is necessary since some steps such as Lithography require a plane surface. Cleaning is necessary before the diffusion, oxidation, and deposition steps as well. CMP planes the wafer surface with the help of a chemical slurry. In addition to the main process-steps lots goes through inspections as well. If a lot does not pass the inspection test it  may be eliminated from the production. For further information regarding each of the production steps the reader may refer to \citet{Monch2013ProductionFacilities}.  

The lots in process in WF are a mixture of different product types that requires different technologies. Even though majority  of the lots belong to 4 or 5 technology types, referred to as \emph{lottype} in our database, controlling such a mixture is not easy. Each lot type is classified to different sub-types called \emph{part} which contains the products that go through the same layers. Each lottype in WF goes through different process-steps adding to the complexity of the system. In addition, existence of factors such as reworks, and long waiting times make the number of process-steps that lots go through different. 

Although semiconductor manufacturers have been gathering intensive data from the state of machines and flow of the material in the system for a long time, there is no comprehensive study that uses the data from the production system to evaluate the performance of a semiconductor manufacturing system. The data gathered from WF are stored in different databases, making the integration of machine level, process-step level, and supply-chain level data difficult. Each group of the machines in the wafer fabrication are highly specialized machines that store different types of data. Integrating data from different sources is a challenging task in the WF since it requires close collaboration with the experts of each production area. For instance, consider the ion implantation machines. The ion implantation machines require long setup times whenever a major recipe change is required. The dataset that is gathering the activities of the lots inside the machines, which can be used for extracting the setup times, is different from the dataset that gathers the information about the machines being up or down or in maintenance, and the lot trace data. Efficient integration of these datasets is necessary for running an efficient production system and decreasing the cycle times and increasing the throughput of the system. The cycle time of the wafers is a function of variety of different factors from product release into WF, flow of the products in the system, dispatching rules of the sub-production areas and equipment groups, mixture of the product types, to down times of the machines and maintenance policies. These factors create high variability in the cycle times of the wafers. 

\subsection{Data Description} \label{datades}
I use the lot trace data of the semiconductor manufacturing plant of the Robert Bosch Company in Reutlingen, Germany. The manufacturing system produces micro electro-mechanical systems sensors to be used in the automotive and consumer electronics industry with applications such restraint systems (e.g., airbag), engine management systems, and vehicle comfort systems. The wafers are produced in two different wafer fabrication areas which process 150 mm and 200 mm wafers. I use the lot-trace data of the 200 millimeter wafers in my analysis since it is more automated than the 150 mm wafer fabrication. The lot trace data captures every event related to the movement of the lots between the production steps. The general format of the dataset is presented in \citep{Laipple2019GenericChains}. My investigations demonstrate that the inter-event data from the backend process may not be as accurate and reliable as WF data. Therefore, I focus on the wafer fabrication due to better data quality for the inter-event data in comparison to the backend processes. 
\begin{table}[]
\scalebox{0.9}{
\begin{tabular}{|l|l|}
\hline
lotid          & Unique ID given to the products upon their release to the production. \\
partid         & \begin{tabular}[c]{@{}l@{}}A unique ID given to a category of products that go through similar \\ set of process steps.\end{tabular}                                                                                                               \\
lottype        & \begin{tabular}[c]{@{}l@{}}The highest aggregation of the products. The products are \\ categorized into lottyes based on the technology used for in their \\ production. A certain lottype contain different parts as subcategories.\end{tabular} \\
prodarea       & Name of the production area (factory) \\
eqpid          & A Unique ID given to the machines \\
eqptype        & A unique ID that categories similar machines into the same category. \\
stage          & \begin{tabular}[c]{@{}l@{}}Stage determines the current state of the production of a certain lot. \\ The stage column of the data is in the $AB-CD-EF$ where $AB$ is \\ the current layer that the product in is.\end{tabular} \\
location       & Name of the area that the equipment is physically located in. \\
recpid         & Name of the recipe (machine program) used for processing the lot.  \\
queuetime      & Time stamp that the lot leaves the previous process-step.\\
trackintime    & Time stamp that the lot enters the processing step.\\                                trackouttime   & Time stamp that the lot leaves the process-step. \\                      
trackinmainqty & Number of wafers in a given lot at lot trackin \\
curmainqty     & number of wafers in a given lot at lot trackout.\\  \hline           
\end{tabular}} \label{datatemp_4}
\end{table}
Events are defined as the movement of the lots from one process step to the next. Table \ref{datatemp_4} describes each column in the data. It starts with the \emph{lotid} which is a unique id given to a certain lot. Each lot is a subset of \emph{partid} which determines the lowest level type of the product. The \emph{partid} information in the \emph{partid} column consists of two parts in a $abcd.01$ format. I extract a new column called \emph{part} from this column. The products that belong a certain part go through similar set of recipes and layers explained below. \emph{lottype} is a higher aggregation level that groups parts with similar technology into a single category. The dataset consists of a six major lot-types. A machine processes the lots by a product specific recipe given in the \emph{recid} column. \emph{stage} determines the current state of the production of the lot. The stage column is in the $AB-CD-EF$ format. The first two characters of \emph{stage} ($AB$) determines the current layer of the product. Layer refers to a physical layer fabricated on top of the wafers. \emph{lotid}, \emph{recpid}, and \emph{stage} completely current state of the production of a certain lot. \emph{eqpid} is the id of the equipment that processes the lot. Each equipment belongs to a equipment-group specified by \emph{eqptype}. The type of the machines in the same production steps may differ from each other due to different reasons such as a newer technology. For instance machines in diffusion area consists of two major sets of old and newer equipment, which differ in the number of batches of lots they process. On the  other hand, the machines in the ion implant differ from each other due to the chemical processes that they conduct on the lots. \emph{location} determines the physical place that the equipment is present in. \emph{queuetime} captures the time stamp that the lot leaves the previous equipment or equivalently enters the current equipment or process step. \emph{trackintime} and \emph{trackouttime} capture the time stamp that the lot enters and leaves the equipment, respectively. Note that the \emph{trackouttime} time of the lots is equal to the \emph{queuetime} of the next production step. Finally, \emph{prodarea} is the name of the factory that the data belongs to. Table \ref{datatempexpample} shows an example of the inter-event data. 
\begin{table}[!htb] 
	\centering
	\caption{A single row of the lot-trace data from the wafer fabrication of the semiconductor manufacturing system of the Robert Bosch Company }
	\begin{tabular}{|l|l|}
		\hline
		lotid        & C93674.1            \\
		partid       & CMP211BC\_2.01      \\
		recpid       & SOG02.03            \\
		priority     & 3                   \\
		eqpid        & NN117               \\
		eqptype      & 8\_SD           \\
		stage        & PF-DIF-OG           \\
		location     & DIFFUSION               \\
		prodarea     & WF                \\
		queuetime    & 2018-09-17 00:41:10 \\
		trackintime  & 2018-09-17 00:52:26 \\
		trackouttime & 2018-09-17 00:55:16 \\ \hline 
	\end{tabular}
\label{datatempexpample}
\end{table}

Our dataset consists of lot-trace data of every product processed in the 200 mm wafer fabrication. The 3.7 GB of raw dataset in feather format consists of 17\,223\,658 rows of inter-event data. The rows in the raw-data correspond to two different sets of lots: productive and  non-productive. The non-productive lots contain the test-lots, engineering-lots, and other types of lots that are used for different purposes such as assessing the performance of the machine or developing a new technology. On the other hand, the productive lots are the lots that move to the Diebank, or backend of the supply chain. 
The productive lots dataset consists of 10\,317\,224 rows of data. It contains the data related to 16305 unique products categorized into 216 different parts. The products are being processed in 160 different equipment-groups which contain 500 different machines and can process 2159 different recipes. 

\subsection{Computational Tools}
The analysis presented in this study has been conducted using Python programming languages and Jupyter Notebook and Spyder IDEs. The computations leading to these results has been conducted in the High Performance Computing cluster of the Ko\c{c} University. 

\section{Wafer Fabrication Level Analysis}\label{wfanalysis}
\begin{figure}
	\caption{The evolution of the mean daily WIP, arrival and departures rates in 15 days rolling horizon}
	\begin{center}
		\includegraphics[scale=1]{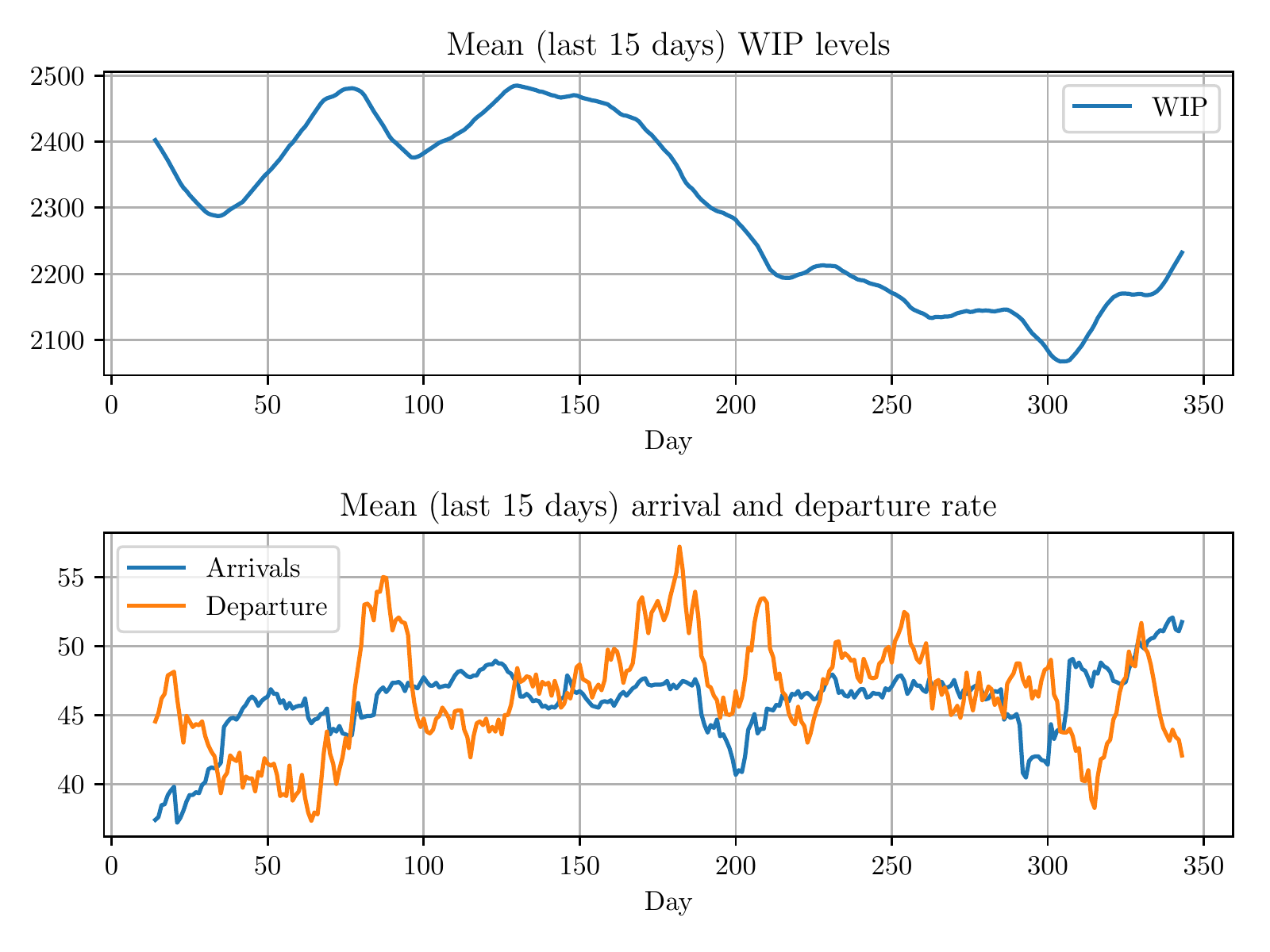}
	\end{center}
	\label{wf_wip_arriva_departure_rolling_15_evolution}
\end{figure}
In this section, I treat the wafer fabrication as a black-box and investigate the dynamics of WIP levels, inter-arrival times, inter-departure times, and total cycle times of the products in the wafer fabrication. Viewing WF as a birth-death process, I characterize the birth and death rates which correspond to the release of new products to WF and departure of products from WF, respectively. Furthermore, I investigate the WIP levels, which correspond to the state of the birth-death process, and the empirical distribution of TCT of the products in WF. Figure \ref{wf_wip_arriva_departure_rolling_15_evolution} shows the mean of WIP levels and arrival and departure process in 15 days (the first percentile of the total cycle times of all products) rolling horizon from January 10, 2018 to December 20, 2018. I eliminate the first and last 10 days of the year to eliminate the end-of-year effect on the dynamics of the system. The evolution of the departure process demonstrates significant variations in short intervals of time. Understanding these variations requires investigating the arrival process, and the time that each product spends inside the system. In the reminder of this section, I investigate the arrival and departure process, and the total cycle times of the lots to understand the dynamics of the system.  
\subsection{Arrival and Departure Processes Analysis}
\begin{figure}
	\caption{Autocorrelation function of the inter-arrival and inter-departure times of the new product released into WF and departing products from WF}
	\begin{center}
		\includegraphics[scale=.4]{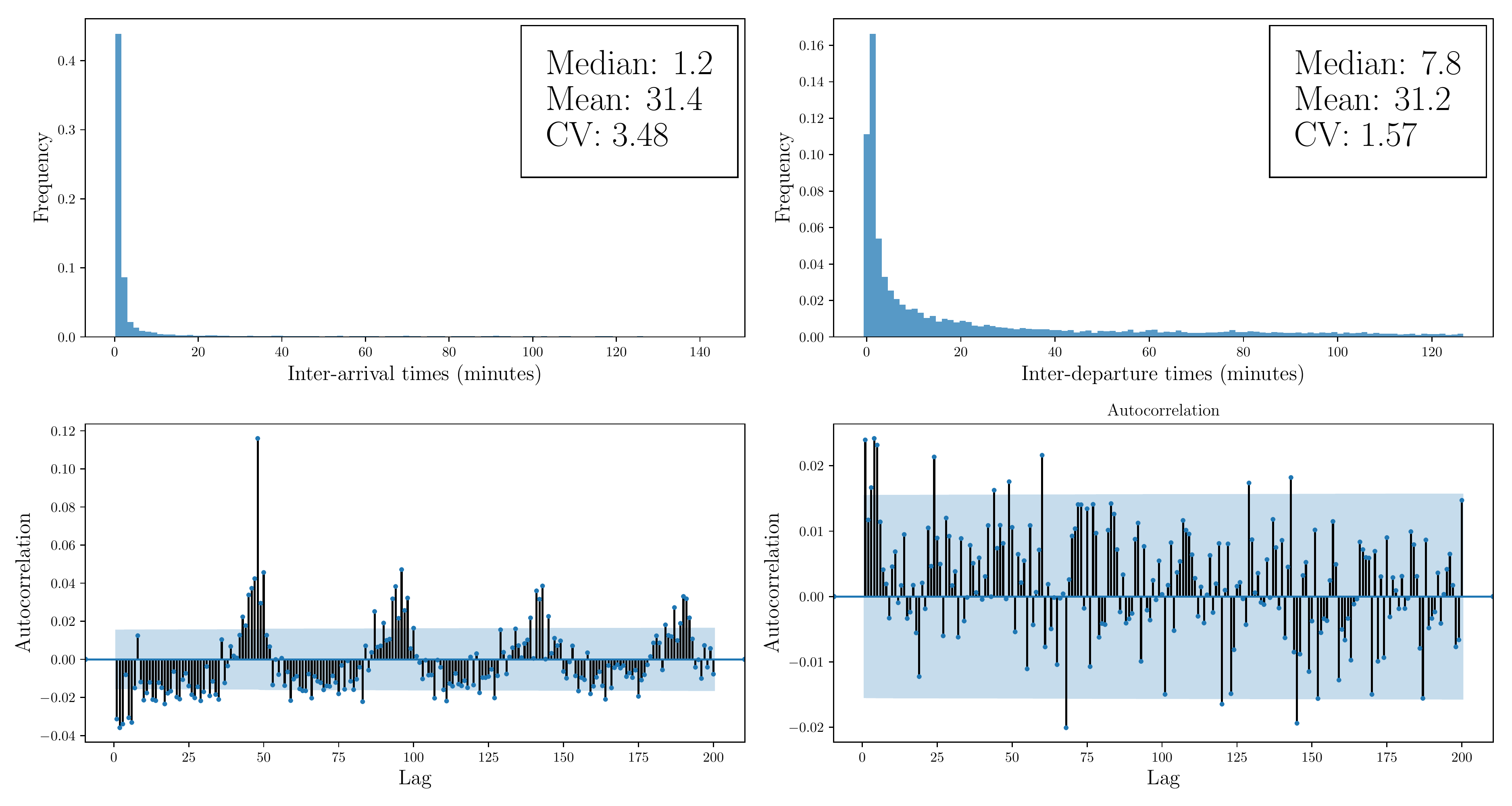}
	\end{center}
	\label{wf_inter_arrival_departure_hist_acf}
\end{figure}
\begin{figure}
	\caption{Average number of products being released into and leaving WF per hour of the day}
	\begin{center}
		\includegraphics[scale=1]{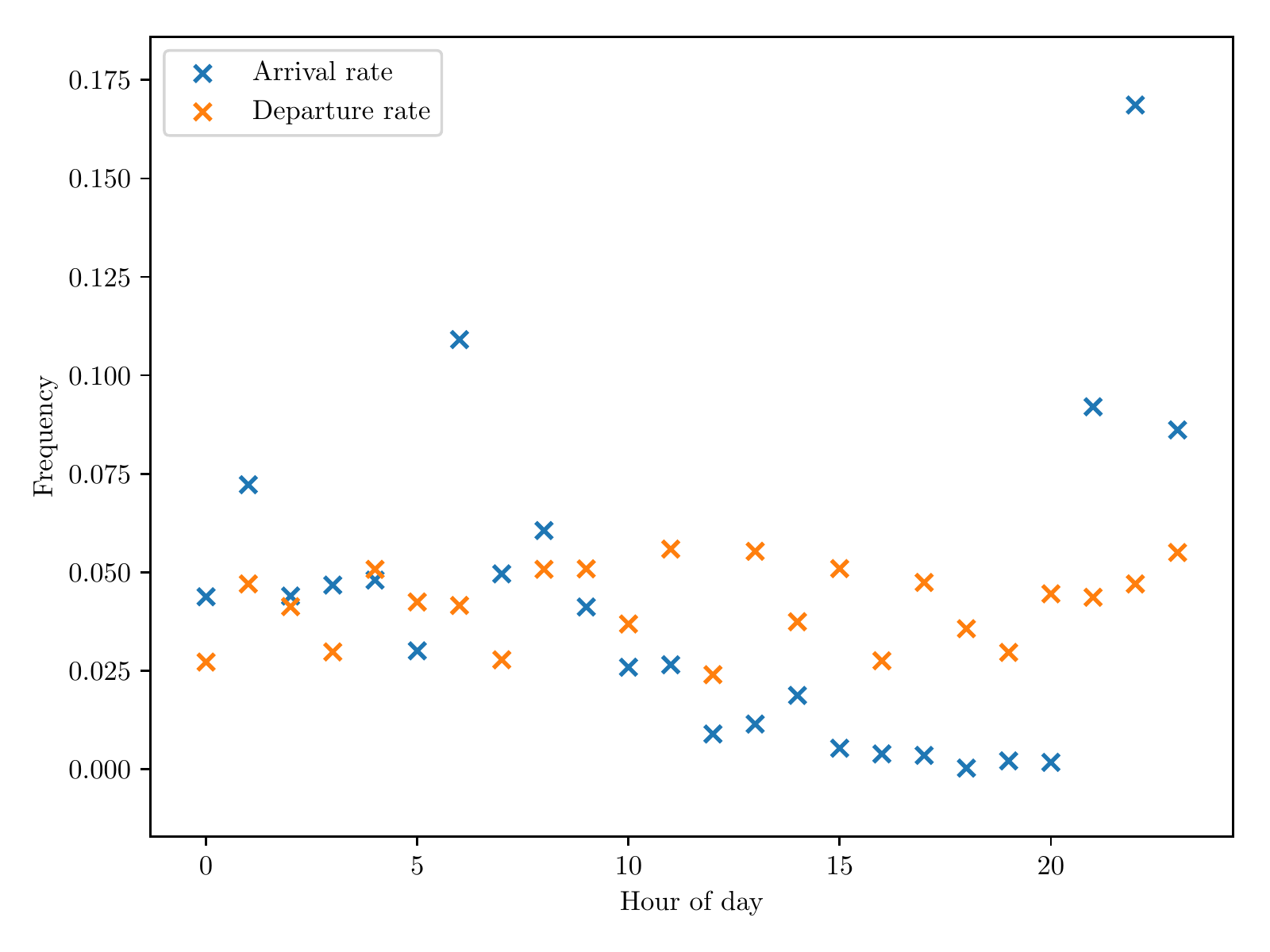}
	\end{center}
	\label{wf_arrival_departure_hourofday_freq}
\end{figure}
\begin{figure}
	\caption{Empirical distribution of the arrival (release) and departure rate of the product released into WF and leaving WF}
	\begin{center}
		\includegraphics[scale=.4]{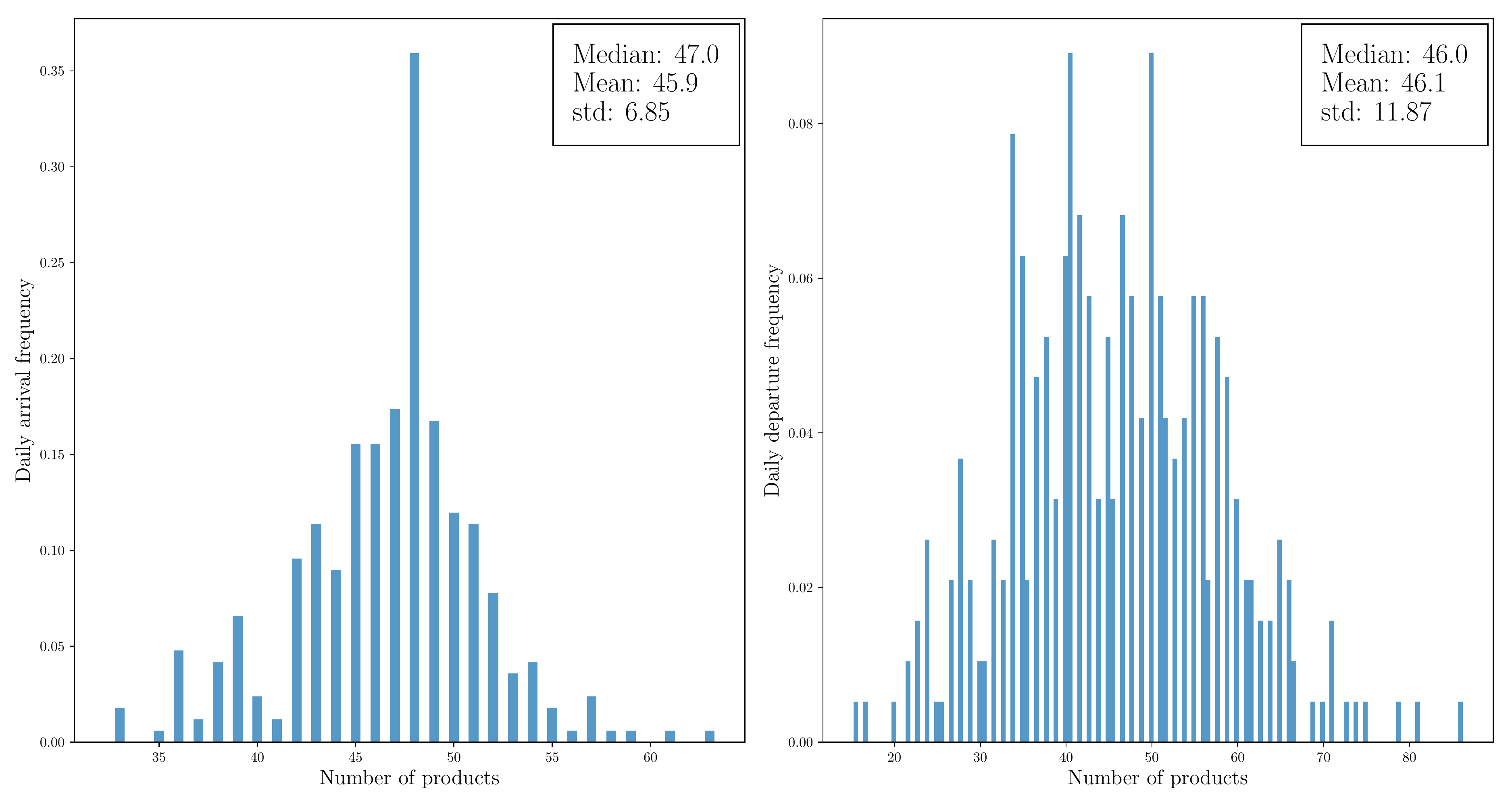}
	\end{center}
	\label{wf_daily_arrival_departure_rate_hist}
\end{figure}

In this subsection, I investigate the arrival and departure processes of the wafer fabrication. Figure \ref{wf_inter_arrival_departure_hist_acf} demonstrates the empirical distribution of the inter-arrival and inter-departure times of the products being released to and leaving WF and the inter-dependency between them. While the inter-departure times do not demonstrate significant dependencies between themselves, the inter-arrival times show a periodic inter-dependency of almost 50 lags. Further investigation of the inter-arrival times and dependency between them reveals that majority of the new products are being released to the system in short time-intervals. Figure \ref{wf_arrival_departure_hourofday_freq} shows the hourly average number of products released to WF and products departing WF. The frequency of release of new products to WF demonstrates a significant hour-of-day effect. Majority of the new products are being released to the system in a time interval between 21:00-9:00.
I investigate the daily arrival rates of the products in daily time-intervals of one hour starting at 18:00 since the arrival rates are the lowest between 18:00-19:00. Figure \ref{wf_daily_arrival_departure_rate_hist} demonstrates the empirical distribution of the number of new products released to the system, and the number of products leaving the system on a daily basis. On average 46.1 new lots are being released into the system with a standard deviation of 6.85 lots. On the other hand 45.9 products are leaving the system on daily basis. However, the standard deviation of the departure process is significantly greater than that of the arrival process. 

\begin{figure}
	\caption{Distribution of daily arrival and departure rates as a function of WIP levels}
	\begin{center}
		\includegraphics[scale=.4]{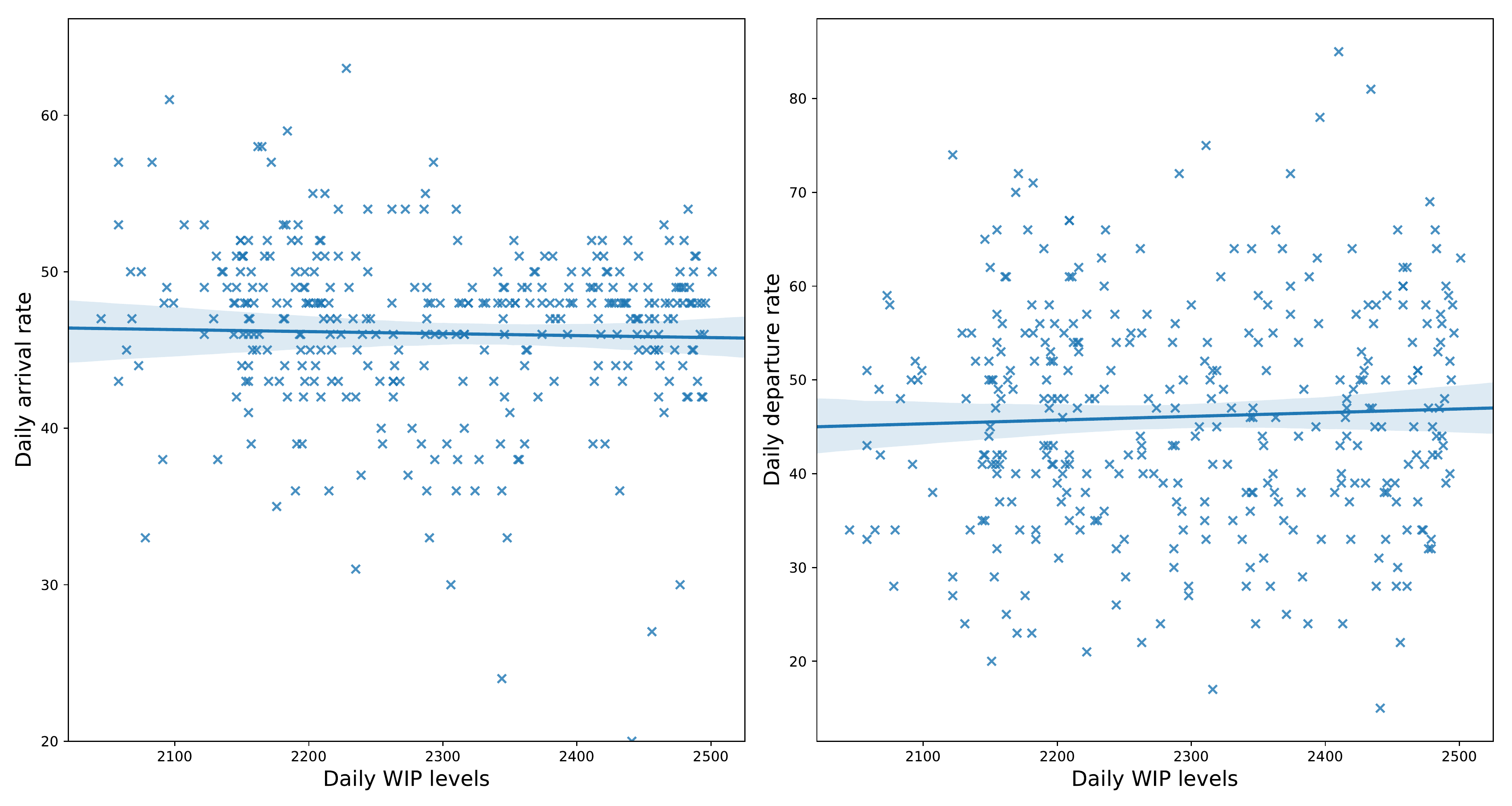}
	\end{center}
	\label{wf_daily_arrival_departure_wip}
\end{figure}
\begin{figure}
	\caption{The evolution of the mean daily WIP of different product  types and total departure rates in 15 days rolling horizon}
	\begin{center}
		\includegraphics[scale=1]{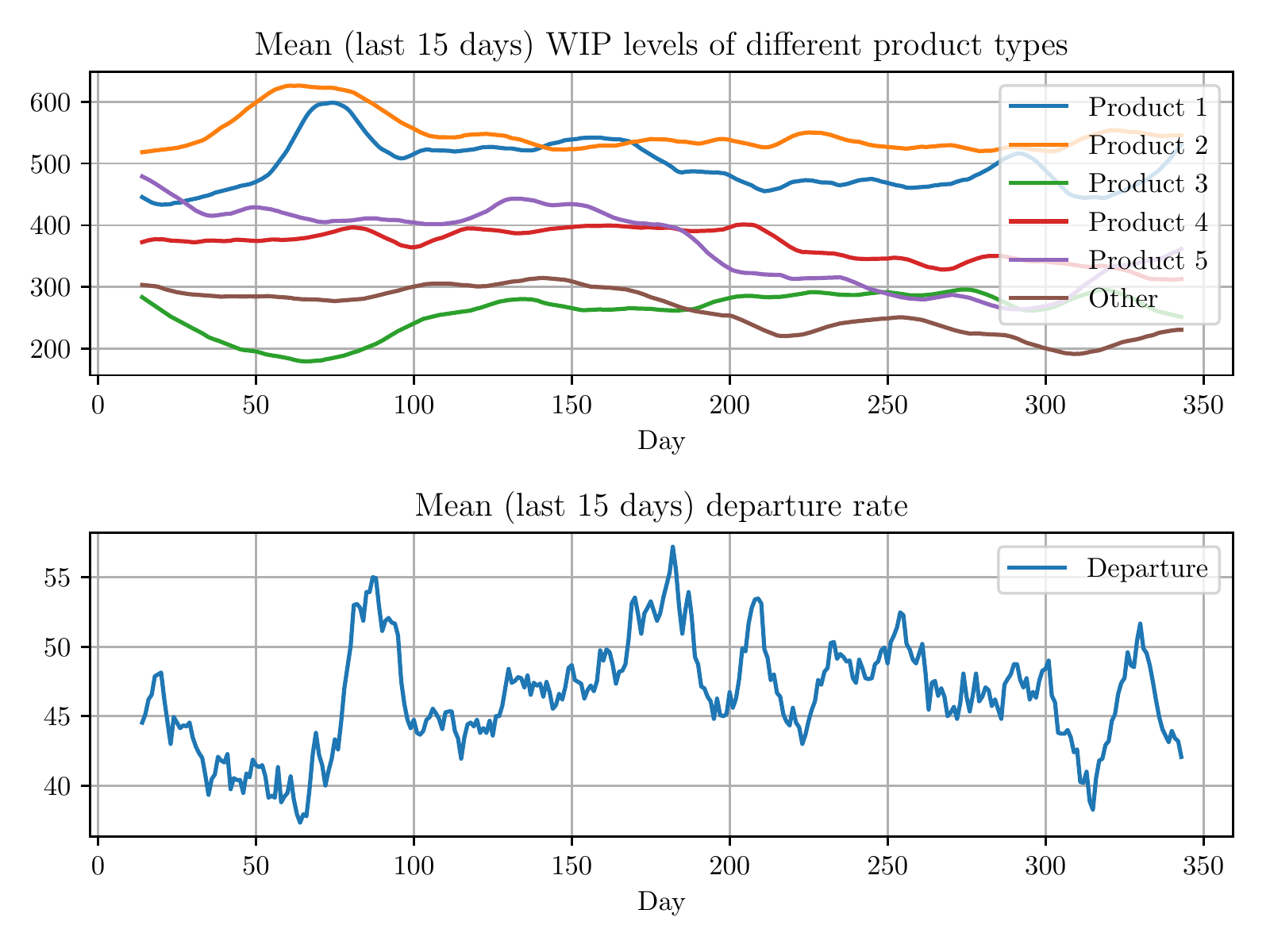}
	\end{center}
	\label{wf_product_type_wip_departure_rolling_15_evolution}
\end{figure}

In order to figure out the possible sources of variation of the arrival and departure process I investigate the impact of WIP levels on the arrival and departure process. Figure \ref{wf_wip_arriva_departure_rolling_15_evolution} shows the mean daily arrival and departure rates of the products in a 15 days rolling horizon. The arrival rates show a relatively stationary dynamics in comparison to departure process. Figure \ref{wf_daily_arrival_departure_wip} shows the scatter plot of the daily arrival and departure rates in y axis and the daily WIP levels in x axis and linear regression model fitted to them. The arrival rates show a slight decrease in the number of lots being released to the system. On the other hand, the number  of daily products leaving the system increases slightly as a function of the WIP levels. The small slope of the first-degree dependency between the arrival and departure rates and daily WIP levels suggests that the sharp increases in the departure rates may be the function of other factors such the mixture of products in the system. Figure \ref{wf_product_type_wip_departure_rolling_15_evolution} shows the mean WIP levels of different product types and the mean departure process in a 15 days rolling horizon. The evolution of the WIP levels suggests that the mean departure rates increase substantially when departures of different product types are synchronized. Note that different product types have different total cycle time distribution as discussed in the following section. 

\subsection{Empirical Analysis of the Total Cycle Times of the Products}
\begin{figure}
	\caption{Empirical distribution of the number of layers that lots go through in the wafer fabrication}
	\begin{center}
		\includegraphics[scale=.75]{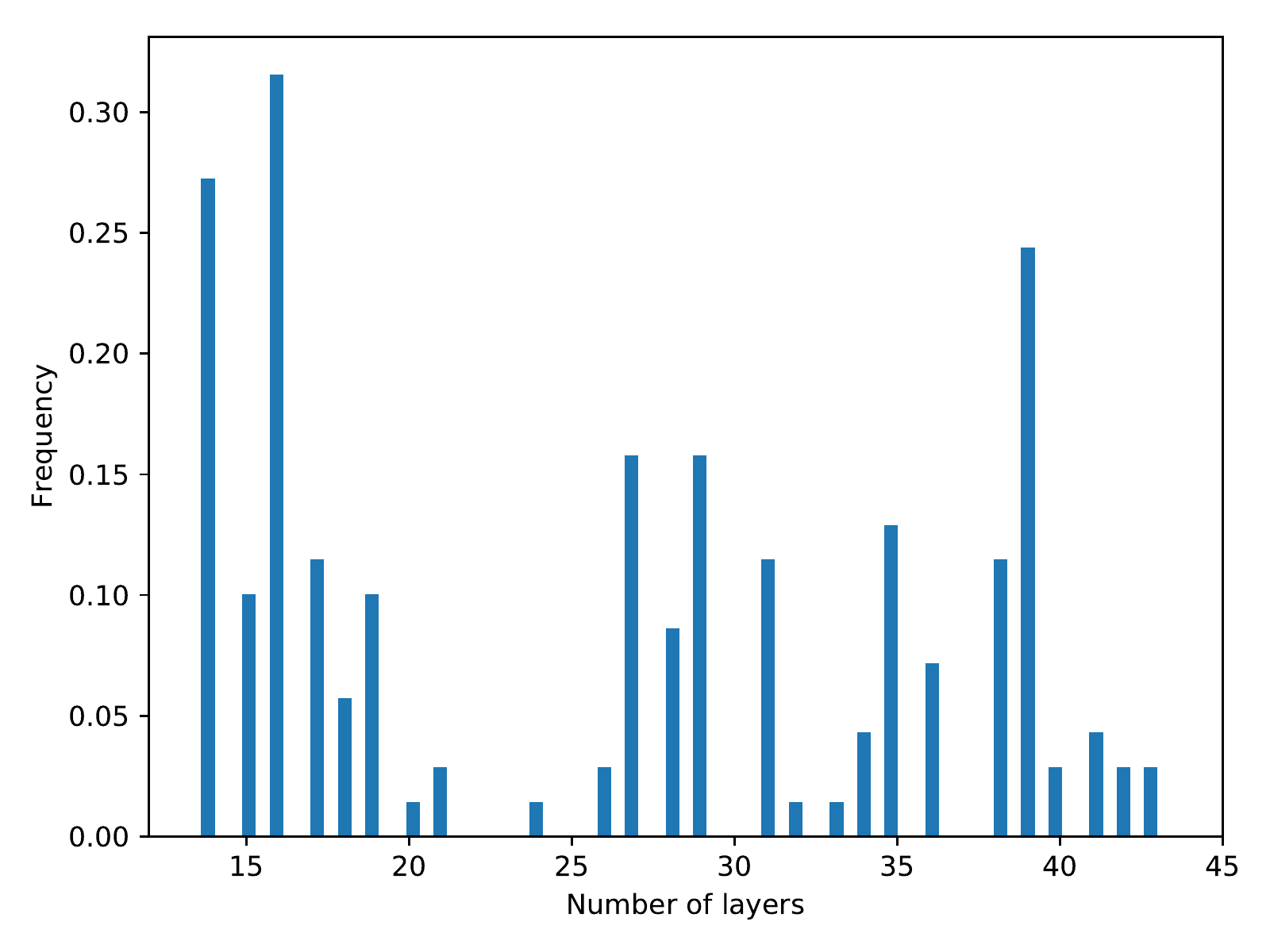}
	\end{center}
	\label{wf_parts_num_layers_hist}
\end{figure}

TCT is one of the most important performance metrics of a WF. Ideally, a lower TCT is desired as on-time delivery is valued as high as the price of products by customers \citep{Batra2018QuantifyingChain}. However, the complicated nature of WF with factors such as mixture of different products, complicated production network, machine down times, reworks, and dispatching rules create a highly uncertain TCT. Understanding this uncertainty is necessary in strategic decision making, planning, and taking release of new products decisions. Due to the capital intensive nature of the machines in WF, a high utilization of the machines is desired as well. However, releasing products into the production network without understanding its impact on the TCT may increase the TCT and its variance substantially. Different product types in WF go through different number of layers which results in different total cycle times for each of the product types. Figure \ref{wf_parts_num_layers_hist} shows the distribution of the number of layers that products go through. The number of layers  of the products differ from 13 to 43. Most of these layers are using the same set of machines.

\begin{figure}
	\caption{Empirical distribution of the normalized total cycle times of the lots and dependency between them}
	\begin{center}
		\includegraphics[scale=1]{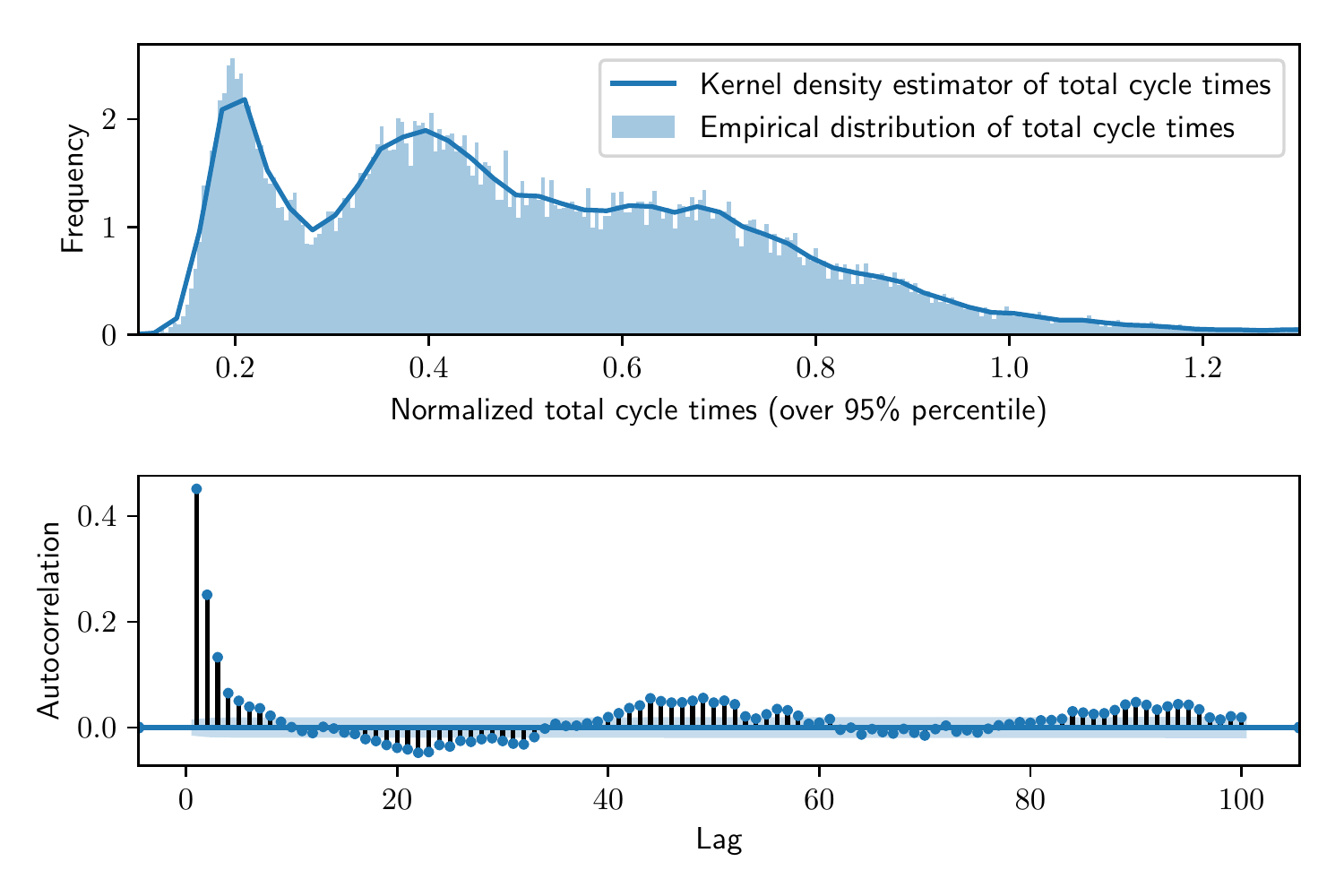}
	\end{center}
	\label{tct_dist_acf}
\end{figure}

In this section, I investigate TCT of the productive lots processed in WF. I first analyze the distribution of TCT of all product types. Then, I investigate the total cycle times of each product type in WF. I continue the analysis by evaluating distribution of the total waiting times and total processing of the lots inside WF. I then analyze the ratio of total waiting times to total cycle times for different lottypes. 
Figure \ref{tct_dist_acf} shows the empirical distribution of the total cycle times of the lots in WF (due to the confidentiality reasons, I report the  normalized TCT of the lots over their $95\%$ percentile which is approximately three months). TCT of the majority of lots ranges from almost 15 days to 100 days. The uncertainties in TCT is summation of the uncertainties in the waiting times and processing times of the lots. The uncertainties in waiting times are the result of dispatching rules or scheduling decisions in the fab, and down times of the machines. On the other hand, the uncertainties in the processing times arise from the down times, recipe-change times and possible degradation of the machines. Both the processing and recipe-change times are stochastic in nature. In addition, presence of different handlers inside machines, which are used to load and unload the wafers inside each lot, adds to the uncertainty of the processing times and correlation between them. TCT of a lot is the summation of its processing and waiting times. A high variance in waiting and processing times result in a higher variance and uncertainty in TCT. The uncertainty in TCT leads to uncertain delivery times which is highly undesirable in semiconductor manufacturing. I define TCT of the lot $j$, dented by $TCT^j$, inside WF as the difference between the time-stamp that the lot enters ($queuetime^j_{1}$) and leaves ($trackouttime^j_{N}$) the system given in Equation (\ref{tct0}), where $N$ is the number of process-steps of the lot. 
\begin{equation}\label{tct0}
TCT^j = trackouttime^j_{N} - queuetime^j_{1}.  
\end{equation}
As demonstrated in Figure \ref{numprocesssteps} the number of process-steps of the lots is uncertain. Hence, the value of $N$ is not known in advance. Alternatively, TCT can be calculated by using the summation of cycle times over all the production steps. The cycle time of the lot $j$ in its $i^{th}$ production step is defined as:
	\begin{equation}\label{loc_tct}
		ct^j_{i} = trackouttime^j_i - queuetime^j_i,  
	\end{equation}
where $queuetime^j_i$ and $trackouttime^j_i$ are the time stamp that lot $j$ in its $i^{th}$ production step enters the queue of the equipment and leaves the equipment as defined in Table \ref{datatemp}. Consequently, TCT can be calculated as a summation of the cycle times of each production step of the lot as:  
	\begin{equation}\label{tct1}
		TCT^j = \sum_{i=1}^{N} ct^j_{i}.  
	\end{equation}

Total cycle times of the lots vary substantially ranging from 0.15 (corresponding to almost 15 days) to 1.2 (120 days). Even though some of the variation can be explained by the lottype, managing such an uncertain system is difficult. In addition to variation in the distribution of TCT, products show a significant amount of correlation between themselves as well. Figure \ref{tct_dist_acf} shows the distribution and autocorrelation of the total cycle times of the products sorted by their arrival-instance to WF. The total cycle times of the products demonstrate a significant autocorrelation with the previous three lots released to the system. Furthermore, there exist a periodic structure in the autocorrelation structure of the products. The periodicity occurs almost every 50 lag. \citet{Schomig1995AutocorrelationSystems} report similar phenomenon in the total cycle times of the products. They associate the periodicity to the "closed-loop rule with a WIP of 50" that is adopted for controlling the system. The periodicity in our system may have arisen from the release of new products decisions. On average almost 46 new products are being released to the system on a daily basis. As \citet{Schomig1995AutocorrelationSystems} argue such a periodicity may not be desirable from the operations managers point of view. 

\begin{figure}
	\caption{Empirical distribution of the normalized total cycle  times of 4 major product types in WF of the Robert Bosch Company (the percentage in front of the label of each product type demonstrates its ratio in WF)}
	\begin{center}
		\includegraphics[scale=1]{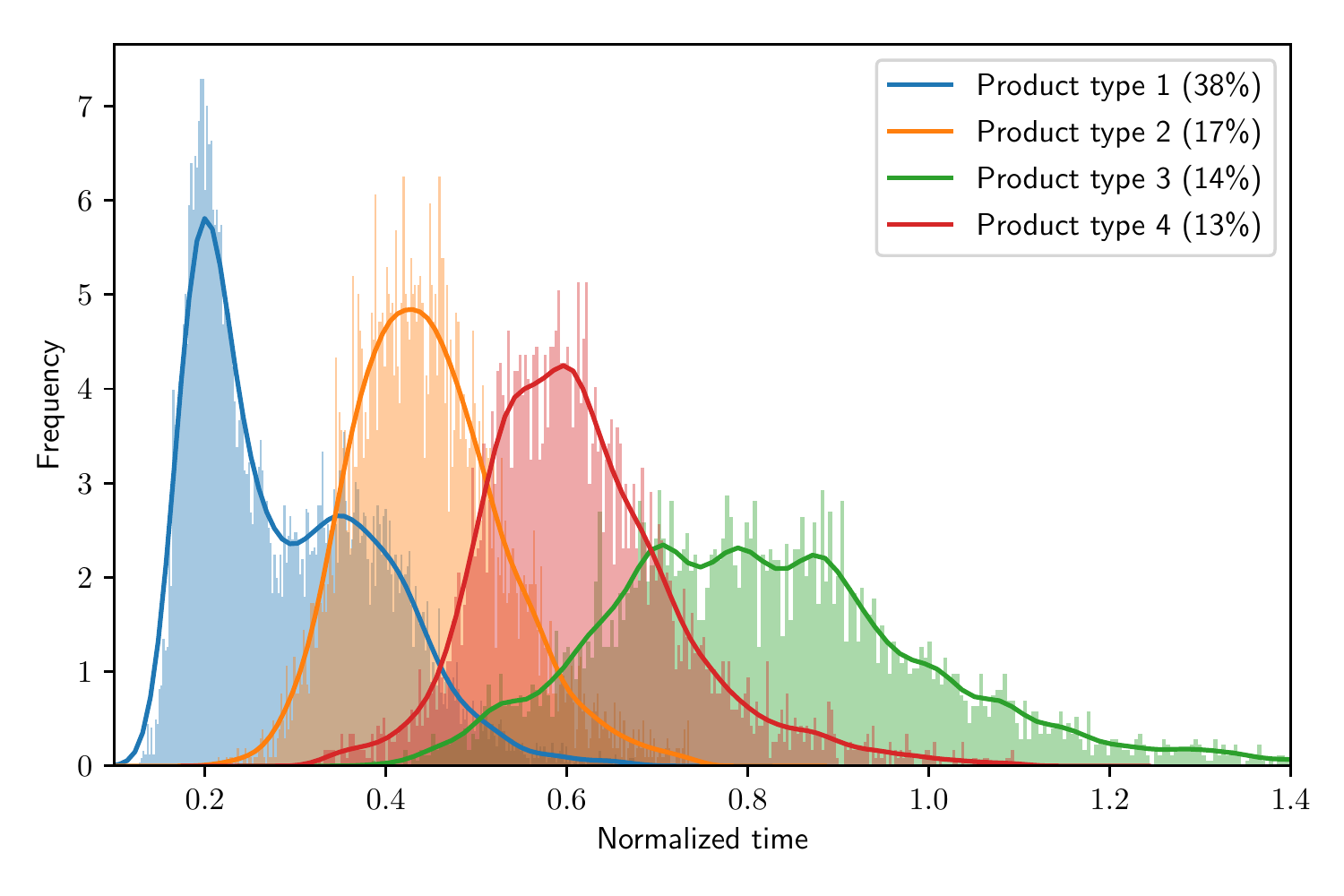}
	\end{center}
	\label{lottype_tct}
\end{figure}

Figure \ref{lottype_tct} shows TCT of the 4 major lottypes in WF. Each lottype demonstrates a different TCT distribution. The most common lottype which accounts for 38\% of the products demonstrates TCT distribution which is a mixture of two distinct distributions.  

\begin{figure}
	\caption{Empirical distribution of the normalized total processing times of the lots}
	\begin{center}
		\includegraphics[scale=1]{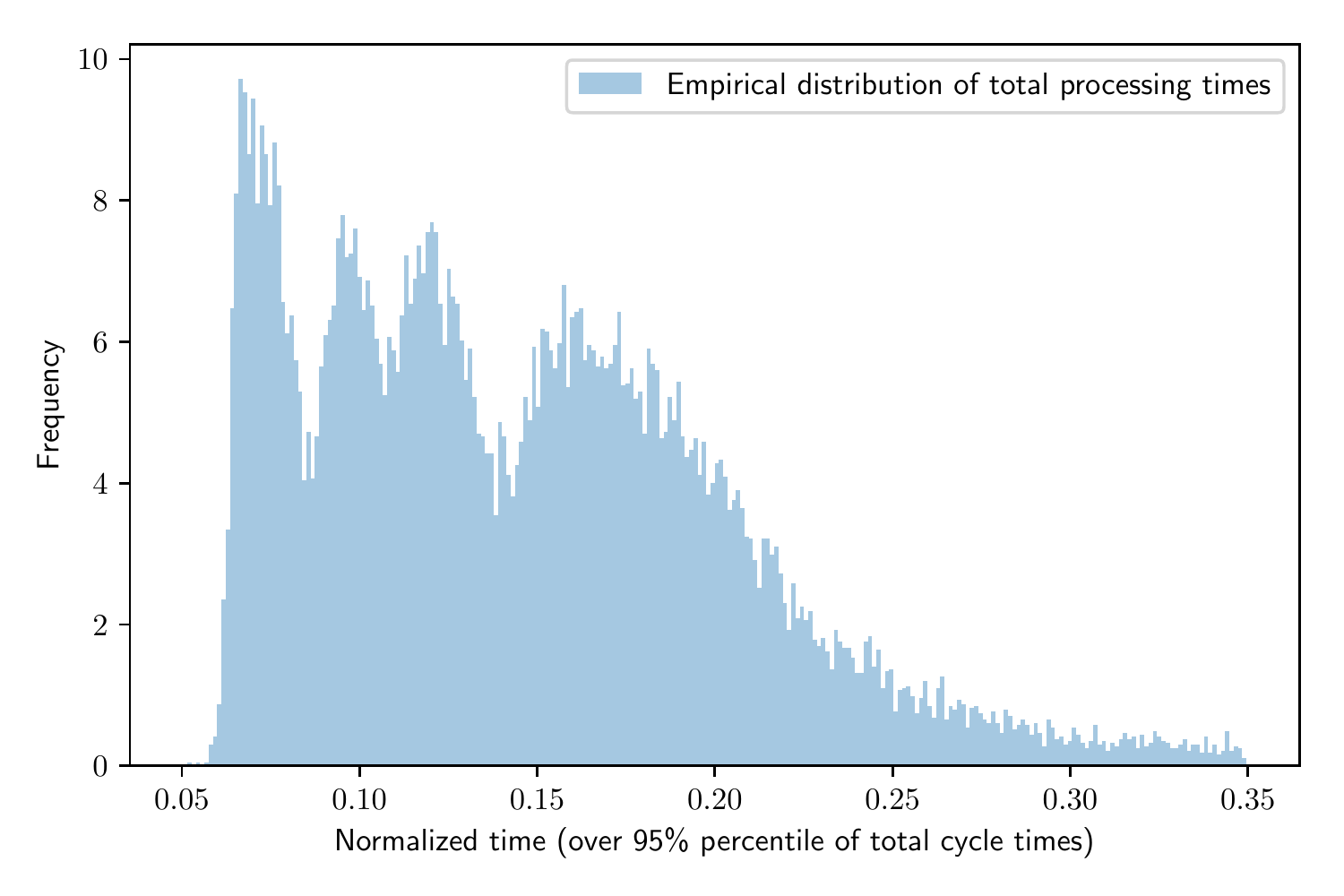}
	\end{center}
	\label{tptdist}
\end{figure}

\begin{figure}
	\caption{Empirical distribution of the normalized total waiting and cycle times of the lots}
	\begin{center}
		\includegraphics[scale=1]{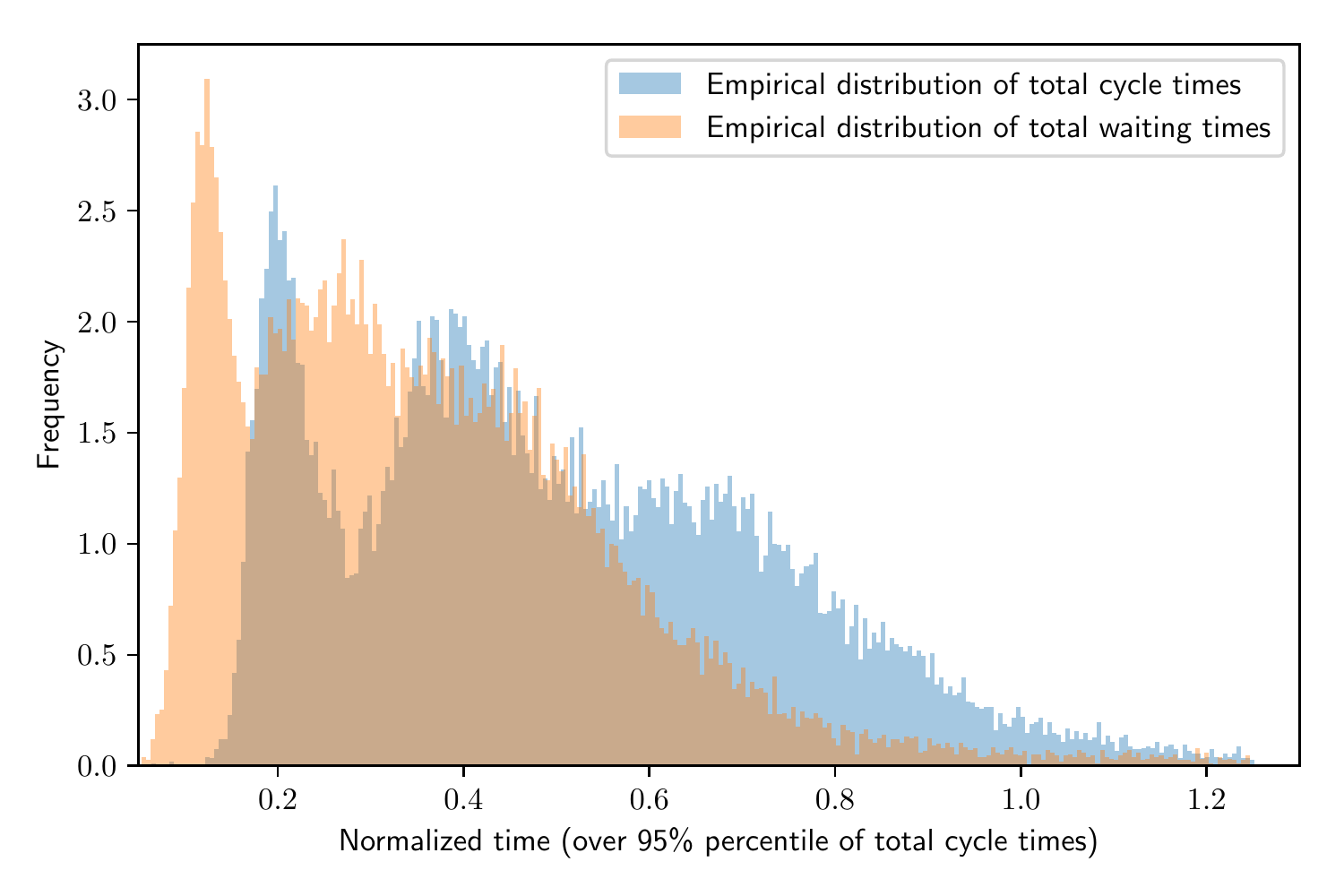}
	\end{center}
	\label{twtdist}
\end{figure}

\begin{figure}
	\caption{Empirical distribution of the ratio of total waiting and processing times of the lots to their total cycle times}
	\begin{center}
		\includegraphics[scale=1]{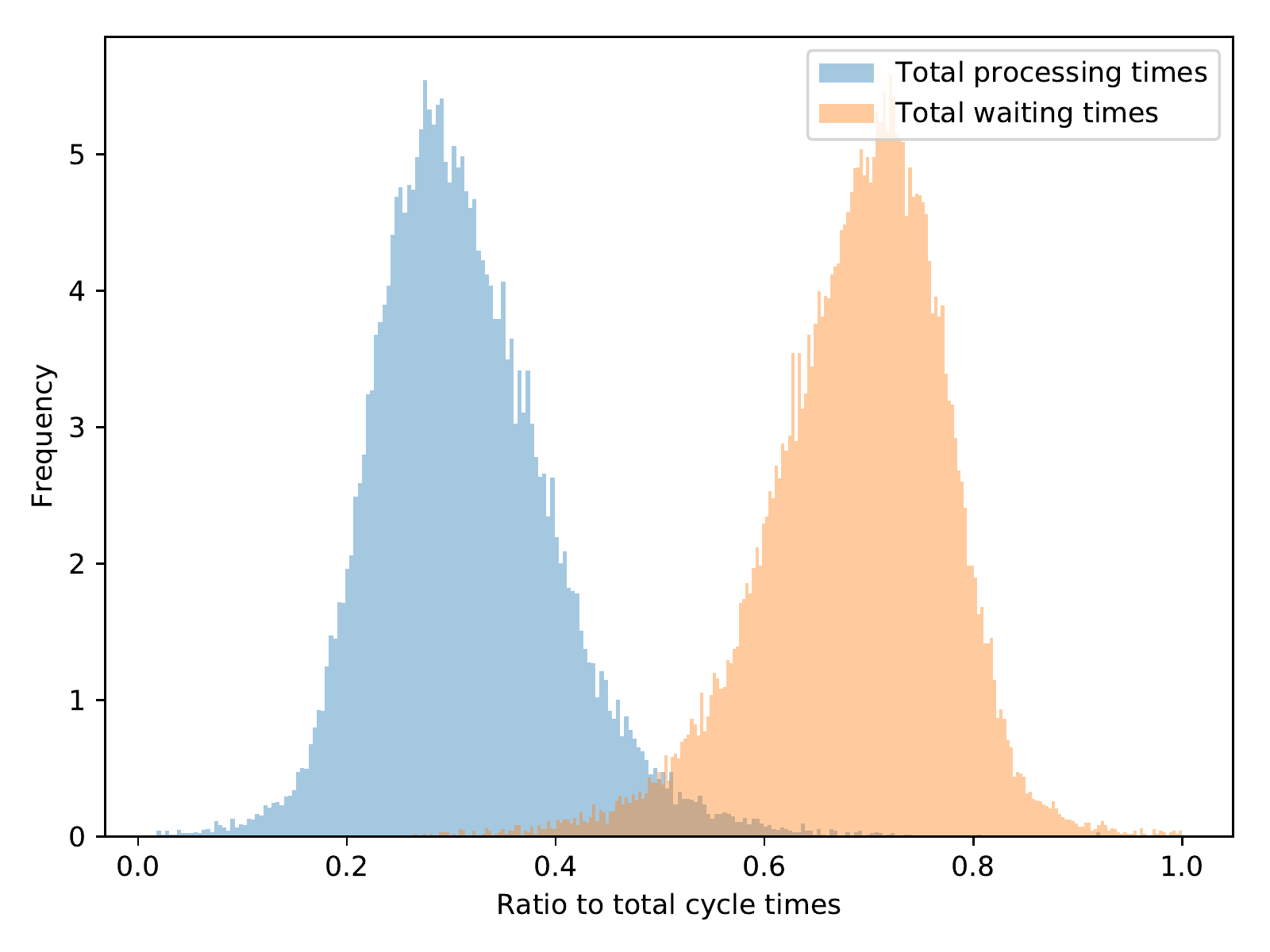}
	\end{center}
	\label{tpt_twt2tct}
\end{figure}

\begin{figure}
	\caption{Empirical distribution of the ratio of total waiting times of the lots to their total processing times}
	\begin{center}
		\includegraphics[scale=1]{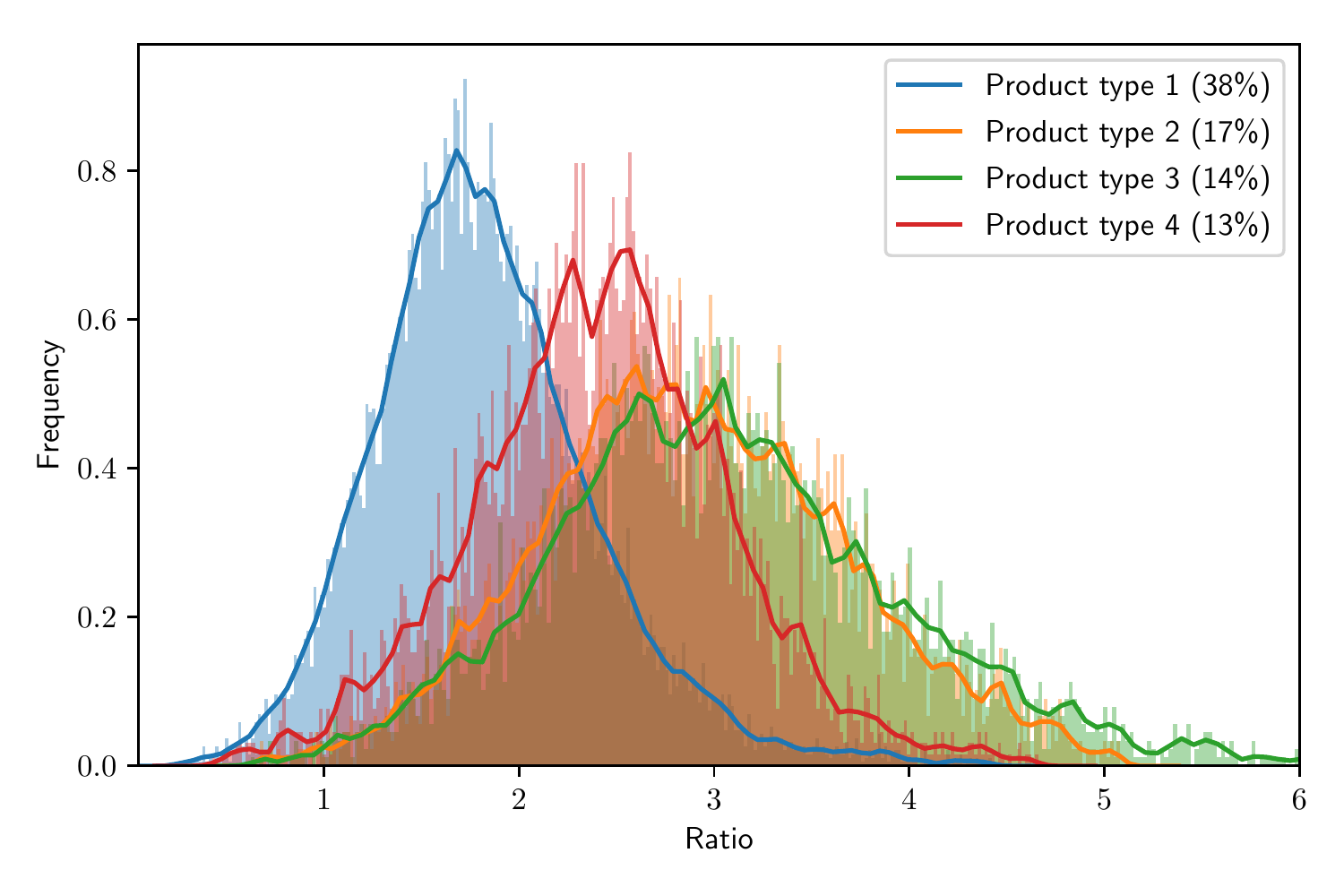}
	\end{center}
	\label{twt2tpt}
\end{figure}

As discussed earlier, TCT is the summation of total processing times (TPT), and total waiting times (TWT) of the lots. TPT is defined as the summation of the time that a lot spends inside machines in each production step. TPT of the lot $j$, denoted by ($TPT^j$) can be calculated by using Equation (\ref{tpt1}). Note that TPT may contain the recipe-change times in locations such as ion implantation, and photo-lithography and machine down-times.      
	\begin{equation}\label{tpt1}
		TPT^j = \sum_{j=1}^{N} trackouttime^j_{i} - trackintime^j_{i}. 
	\end{equation}
Figure \ref{tptdist} demonstrates the distribution of TPT of the lots inside WF normalized over the 95\% percentile of TCT distribution. The range of TPT differs from 0.05 (5 days) to 0.35 (37 days) for majority of the products. TPT is a mixture of different distributions as well. Such a distribution is seen since different product types have different number of layers and consequently process steps.   
TWT on the other hand is defined as the time that a lot waits inside WF for getting processed. There are multiple  factors that impact TWT from dispatching and planning rules to ratio of different product types inside the system. TWT of the lot $j$, denoted by $TWT^j$, can be calculates as: 
	\begin{equation}\label{twt1}
		TWT^j = \sum_{j=1}^{N} trackintime^j_{i} - queuetime^j_{i}. 
	\end{equation}
Figure \ref{twtdist} demonstrates the distribution of TWT of the lots inside the wafer fabrication. The distribution of the TWT is similar to that of the TCT. In fact, distribution of TCT could be viewed as the shifted distribution of the distribution of TWT. Such a similarity between the distributions demonstrates that TWT constitutes the main proportion of TCT.  
Figure \ref{tpt_twt2tct} demonstrates the ratio of distributions of TWT and TPT to TCT of the lots inside the wafer fabrication. This ratio varies between 0. 4 and 0.9 for TWT with a variation of 0.13. The median of the distribution is 0.70, meaning that half of the lots spend more than 70\% of their time inside the production system waiting to be processed. Equivalently, half of the lot spends less than 30\% of their time in the wafer fabrication for getting processed. The distribution of the TPT ratios shows less variability in comparison to the distribution of the TWT.  
Another performance metric for evaluating the performance of WF is the ratio of TWT to TPT. Its desired that this ratio is as low as possible. Figure \ref{twt2tpt} shows the distribution of the ratio of TWT to TPT of different lottypes. The most common lottype shows a distribution that ranges from 1 to 3. While the third lottype has a higher range of variation in TCT in comparison to other product types, it demonstrates a same ratio as the second lottype. 

\subsection{Research Opportunities}
In this section, I show that the total cycle times of the products in the wafer fabrication are highly variable, ranging from several days to several months. Some of the variability in the total cycle times is explained by the lottype of the product. However, the variation in the total cycle times are still quite significant. Hence, further research is required for explaining the main features that contribute the variability. \citet{Chen1988EMPIRICALFABRICATION} state that in order to significantly reduce the manufacturing cycle times, one must reduce the variability in the operating environment. Over the years researchers have adopted different methodologies such as queuing systems, simulation to study and model the variability in the wafer fabrication. Queuing models, despite being fast, rely on restrictive modeling assumptions which deteriorates their performance for real complex systems. Simulation models on the other hand are slow. A single replication of a simulation model may take several hours for a complex manufacturing systems. 

The wafer fabrication level of analysis can be used to reduce the variability in the total cycle times through better product release and due-date assignment decisions. Further research is required to understand the impact of release of different types of new products to the system. New stream of research in this area integrates machine learning models into the simulation models to build meta-models for estimating the cycle time–throughput (CT-TP) curves \citep{Yang2010NeuralManufacturing} that can be used for making release decisions. A possible research direction is to replace the costly simulation models with queuing systems such as infinite server queues to approximate the CT-TP curves.  

Due-date assignment is another important factor that impacts the total cycle times of products through dispatching rules and scheduling of the lots on the machines. Different set of parameters, such as the priority and due-dates of the lots, which are being used for optimizing the flow of products inside the wafer fabrication are set based on the approximate total cycle times of the products. Therefore, developing models that are able to predict delivery dates and remaining production times based on the observed data and current state of the wafer fabrication is quite important. In this line of research machines learning algorithms have captured a lot of attention for directly predicting the cycle times of the lots \citep{Wang2018ASystem}. The literature on the prediction of cycle times using learning methods has usually focused on smaller production systems with few data points. Further research is required on how to predict the cycle times of the products in a large-scale production systems where the cycle times of the products vary from several days to several months. I investigate the total cycle time prediction of the products in the wafer fabrication in the following chapter.

\section{Statistical Properties of the Inter-Event Times of the Machines in the Wafer Fabrication}\label{machine_level_analysis}
In this section, I investigate the statistical properties of the inter-event times at the machine level. The most common approach in modeling the inter-event data in manufacturing systems is to assume an i.i.d distribution for the inter-event times \citep{Shanthikumar2007QueueingProblems}. Our main observation from the data is that inter-event times demonstrate correlation between themselves. In other words, the independence assumption may not be valid for all of the inter-event times. In the following sections, I evaluate the range of coefficient of variation and dependence between processing, inter-arrival, and inter-departure  times of the products.   

\subsection{Processing Times}
\begin{figure}
	\caption{Empirical distribution of the coefficient of variation and the first-lag autocorrelation of the processing times of 500 machines}
	\begin{center}
		\includegraphics[scale=.4]{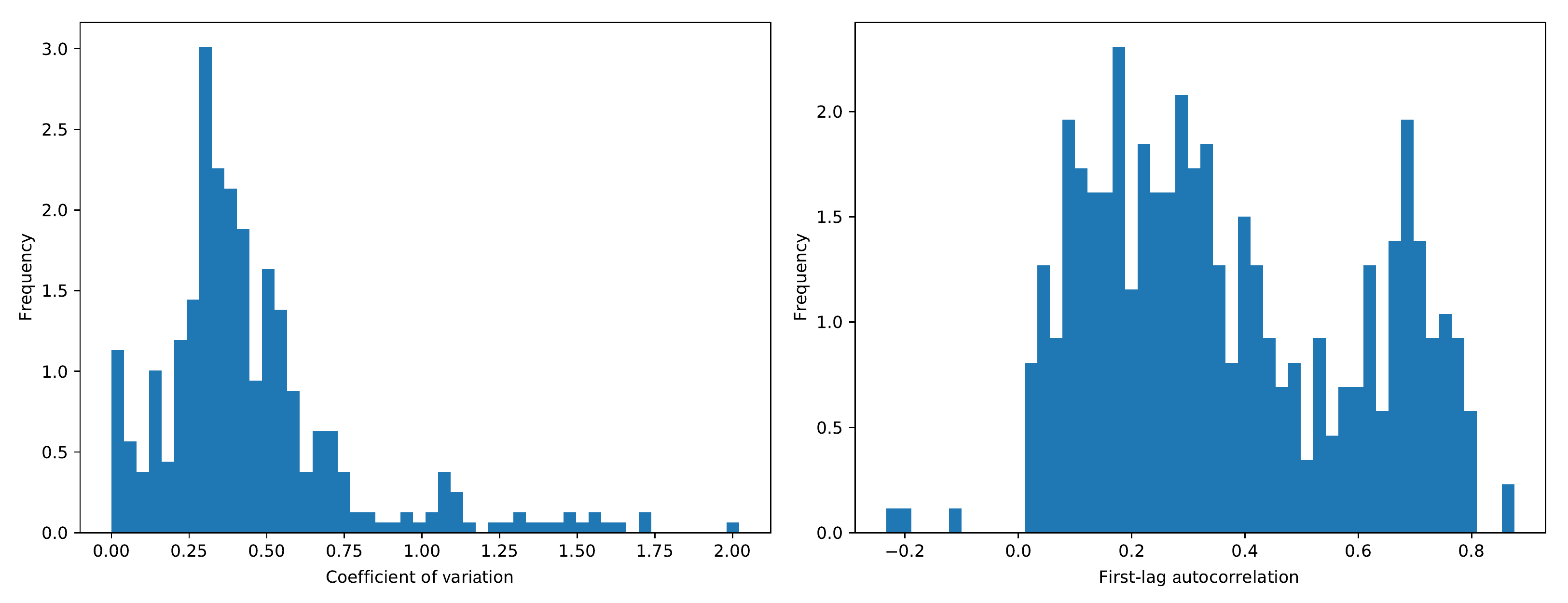}
	\end{center}
	\label{Processingdep}
\end{figure}

I define the processing times of the products as the time that the product spends inside an equipment. Our data captures the time-stamp that a given product enters and leaves the equipment. Hence, the defined processing time may contain the loading, unloading, and waiting times of the lots inside the equipment if they possess an internal buffer. A given machines in the wafer fabrication either processes a single lot or a batch of lots. Each machine can be categorized into one of four general categories: 
\begin{itemize}
	\item machines that work on one wafer or lot at a time 
	\item machines working in parallel on several wafers or lots
	\item  machines working on a batch of wafers or lots
	\item machines working on a batch of lots in parallel. 
\end{itemize}
The processing time of the lot $j$ on equipment $m$ can be calculated as  
\begin{equation}\label{pt}
pt^j_{m} = trackouttime^j_{m} -trackintime^j_{m},
\end{equation}
where $trackouttime^j_{m}$ and $trackintime^j_{m}$ are the time-stamps that the lot enters and leaves the equipment. I examine the dependency between the processing times by sorting the lots according to their entrance to the equipment. Figure \ref{Processingdep} demonstrates the empirical distribution of the coefficient of variation and first-lag autocorrelation of the processing times of the lots inside the wafer fabrication. The coefficient of variation of the lots are mainly between [0.2,  0.7]. It is a demonstration that processing times of the equipment are not as variant in compared to the exponential distribution. However, the processing times are showing a high positive correlation at the same time. \citet{ManafzadehDizbin2019ModellingProcesses} show that positive correlation affects the behavior of the production system in a similar fashion to the higher coefficient of variation. Hence, one may approach the impact of positive correlation in the processing times with caution. The positive correlation may arise from different sources depending on the equipment. Dispatching rules usually favor processing lots of the same type before changing the recipe in processing different types of lots. Such a dispatching rule will result in a positive correlation by creating clusters of processing times that are closer to each other. \citet{Inman1999EmpiricalSystems} reports a similar phenomena for three machines in automotive line where the autocorrelations are created due to having mixed recipes in two of the machines. In addition to mixed recipes, processing lots serially (the equipment is able to process different lots at the same time) creates positive correlations. Furthermore, the processing times of the lots from the same type may change over time due to the machine-degradation resulting in positive correlation. Our results are also in accordance with the findings of the \citet{Inman1999EmpiricalSystems} which finds the goodness of fit of the exponential distribution in modeling the processing times to be poor.  

\subsection{Inter-Arrival Times}
\begin{figure}
	\caption{Empirical distribution of the coefficient of variation and the first-lag autocorrelation of the inter-arrival times of 500 machines}
	\begin{center}
		\includegraphics[scale=.4]{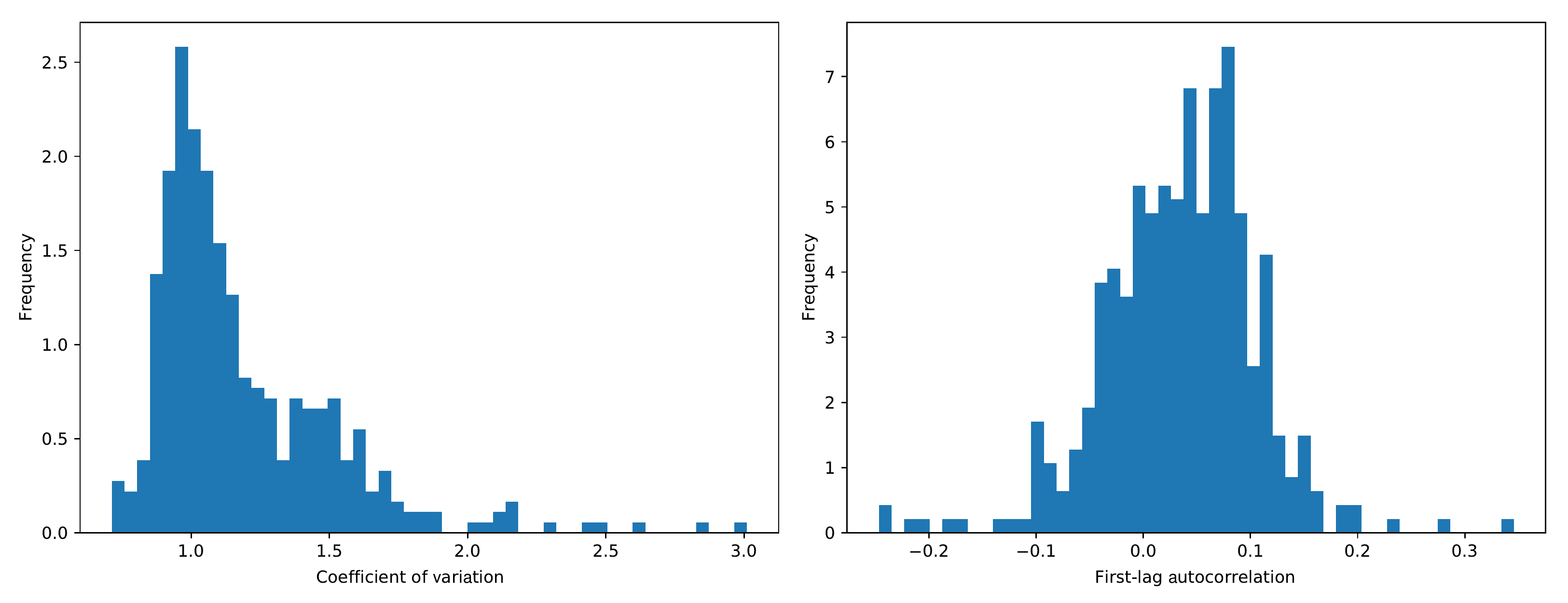}
	\end{center}
	\label{Arrivalstats}
\end{figure}

\begin{figure}[!htb]
	\label{correxparr}
	\centering
	\caption{Test machines with correlated exponential type inter-arrival times}
	\includegraphics[scale=0.45]{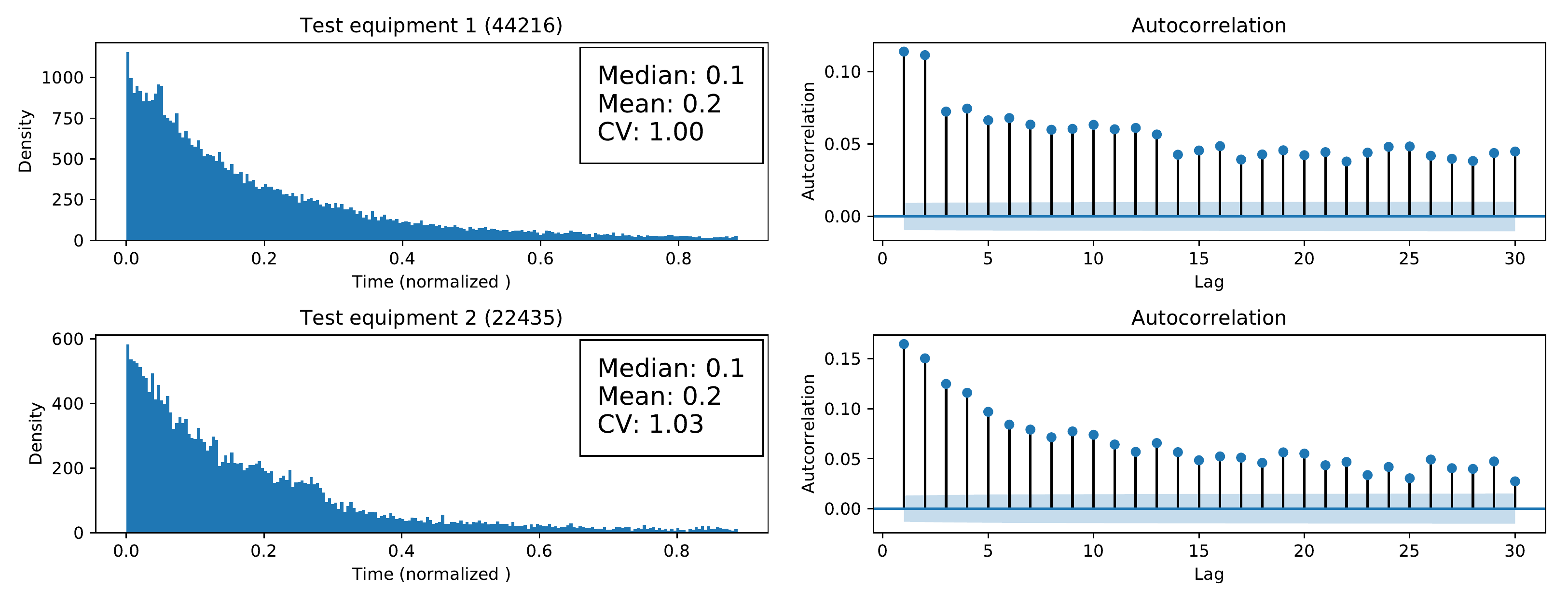}
\end{figure}

In this section, I analyze the statistical properties of the inter-arrival times of the machines. I define the inter-arrival times as the time difference between $j^{th}$ and $(j+1)^{th}$ lot arrival to the equipment $m$. It is calculated as 
\begin{equation}
ia^{j}_m = queuetime^{j+1}_m - queuetime^j_m,
\end{equation} 
where $queuetime^j_m$ is the time stamp that the lot arrives at the queue of the equipment. 
Figure \ref{Arrivalstats} demonstrate the empirical distribution of the coefficient of variation and first-lag autocorrelations of the inter-arrival times. The coefficient of variation of the majority of  the inter-arrival times is distributed between [0.8, 1.8]. While the mode of the coefficient of variation is almost one, most of the inter-arrival times do not pass the statistical test that data has an exponential distribution. 
Our analysis demonstrates that even machines that demonstrate an exponential type inter-arrival times, show dependence between themselves. Exponential type of distribution is usually observed in the inter-arrival times of the test machines. Figure \ref{correxparr} shows the normalized empirical distribution and dependency between inter-arrival times of two test machines in the wafer fabrication. Even though the empirical distribution of the inter-arrival times is close to exponential distribution, both of them demonstrate a statistically significant positive correlations (the shaded area in the autocorrelation figure demonstrate the 95\% confidence interval).  

The magnitude of the dependency between the inter-arrival times is significantly smaller than that of the processing times. The reason may lie in the fact that wafer fabrication is complex production network with re-entrant cycles, where  lots are coming from different almost independent production steps. Most of the machines demonstrate autocorrelations between [-0.1, 0.2]. Positive autocorrelation in inter-arrival means that long (short) arrivals are accompanied by long (short) arrivals in expectation. Such a behavior can be a result of the  dispatching rules, complicated processing steps, and down times of the machines in earlier stages. For instance, engineering and test times on the equipment (which are significant in the wafer fabrication environment) decreases the number of lots leaving a certain production area. As a result, the number of lots that arrive to a certain area may decrease, resulting in longer inter-arrival times, which leads to the creation of positive correlations. \citet{Inman1999EmpiricalSystems} presents the inter-arrival data from an automotive body welding closed loop production line with positive autocorrelations as well.

\begin{figure}[!htb]
	\centering
	\caption{Machines with negatively correlated inter-arrival times}
	\label{negativearrival}
	\centering
	\includegraphics[scale=0.45]{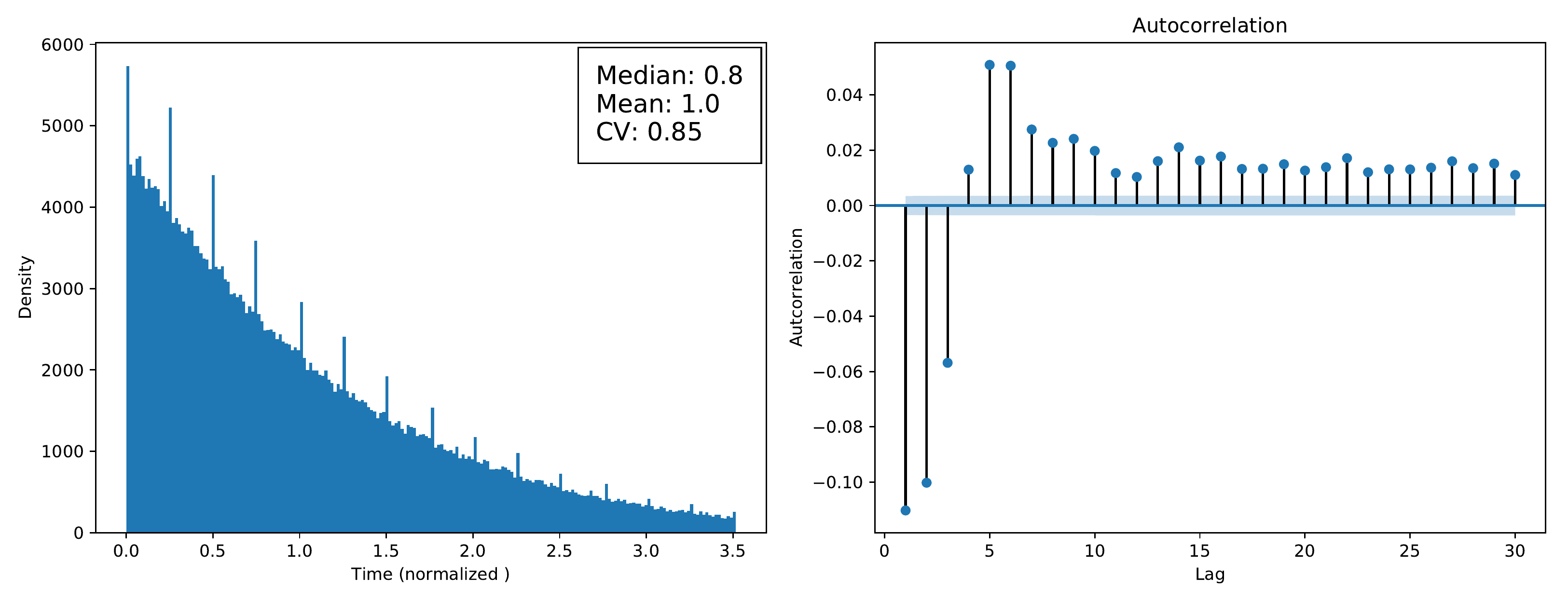}
\end{figure}

Our analysis demonstrates that negative autocorrelation may be result of departure from consecutive processing steps. For instance, consider the arrivals to machines in Figure \ref{negativearrival}. A given lot enters two other processing steps consecutively before entering these machines. Such a behavior for consecutive processing steps is reported in the literature as well. \citet{Hendricks1993TheBuffers} demonstrate by means of simulation that the output process from an open line with reliable machines is negatively correlated. \citet{ManafzadehDizbin2019ModellingProcesses} conclude the same result by calculating the autocorrelation from an open line by means of Markovian Arrival Processes. They argue that negative autocorrelation is the result of blocking in the open lines which creates shorter inter-arrival times after a longer inter-arrival times. 

\subsection{Inter-Departure Times}
\begin{figure}
	\caption{Empirical distribution of the coefficient of variation and the first-lag autocorrelation of the inter-departure times of 500 machines}
	\begin{center}
		\includegraphics[scale=.4]{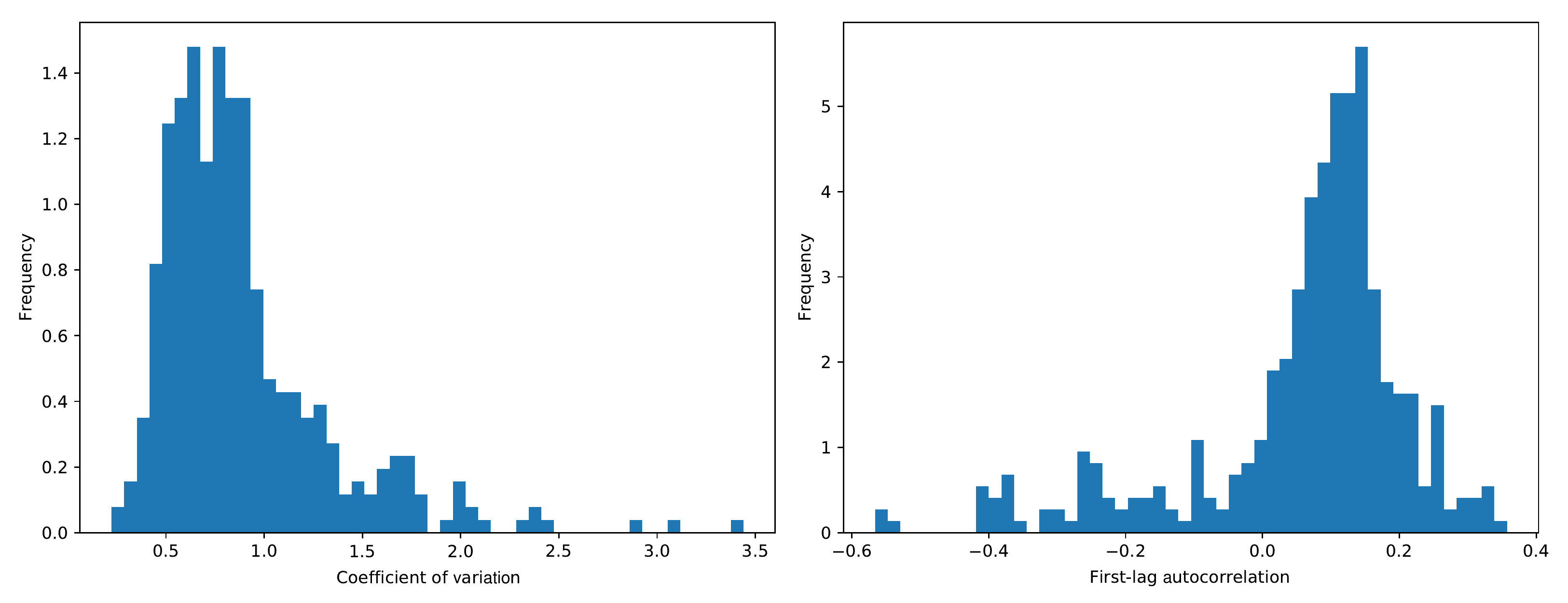}
	\end{center}
	\label{Departurestats}
\end{figure}

The inter-departure time of a given equipment is the time between departure of two consecutive lots. I calculate the inter-departure times between the $j^{th}$ and $(j+1)^{th}$ lot departure from the equipment $m$ as 
\begin{equation}
id^j_{m} = trackouttime^{j+1}_{m} - trackouttime^j_{m}.
\end{equation}
The inter-departure from one equipment constitutes the inter-arrival into other machines and production steps. Understanding the statistical characteristics of the inter-departure process is necessary to model the inter-departure times in simulation or analytical models. In this section, I analyze the empirical distribution of the coefficient of variation and the first-lag autocorrelation of the inter-departure times. Figure \ref{Departurestats} shows the empirical distribution of the coefficient of variation, and first-lag autocorrelation of the inter-departure times of the machines in the wafer fabrication. The coefficient of variations of the inter-departure times are smaller in comparison to the that of the inter-arrival times.  Majority of the machines demonstrate coefficient of variation between [0.4, 1.5]. However, the magnitude of the dependency between the inter-departure times are larger in comparison to the dependency of the inter-arrival times. This may be due to the dependency in the processing times, and down times of the machines. Note that the down times of the machines in the wafer fabrication takes significant amount of time in comparison to other manufacturing systems. The down times of the machines consists of the unscheduled failure times and scheduled engineering down times. In addition, test lots are processed in the machines  that process the lots. Test lots are usually for experimenting a new type of wafer or performance of machines. 

\subsection{Research Opportunities}
In this section, I show that the inter-event times of the products may demonstrate a significant correlation between themselves. The immediate follow up research question is: \textit{What is the impact of  ignoring correlation in the inter-vent times on analytical models and simulation-based methodologies?} \citet{ManafzadehDizbin2019ModellingProcesses, ManafzadehDizbin2020OptimalTimes} show that ignoring correlation between the inter-arrival and processing times of a simple production system with a single machine and a single product type leads to underestimation or overestimation of the performance measures of the system. Understanding the impact of correlation between inter-event times in different settings such as multi-product, multi-server, and infinite-server systems requires further investigations. In addition, the performance of simulation-based methodologies can be further improved by integrating the correlation and dependency information into the simulation model. 

\section{Layer Level Analysis of the Processing and Waiting Times of Products}\label{layer_level_analysis}
\begin{figure}
	\caption{The median of the processing and waiting times of the major layers in the wafer fabrication}
	\begin{center}
		\includegraphics[scale=1]{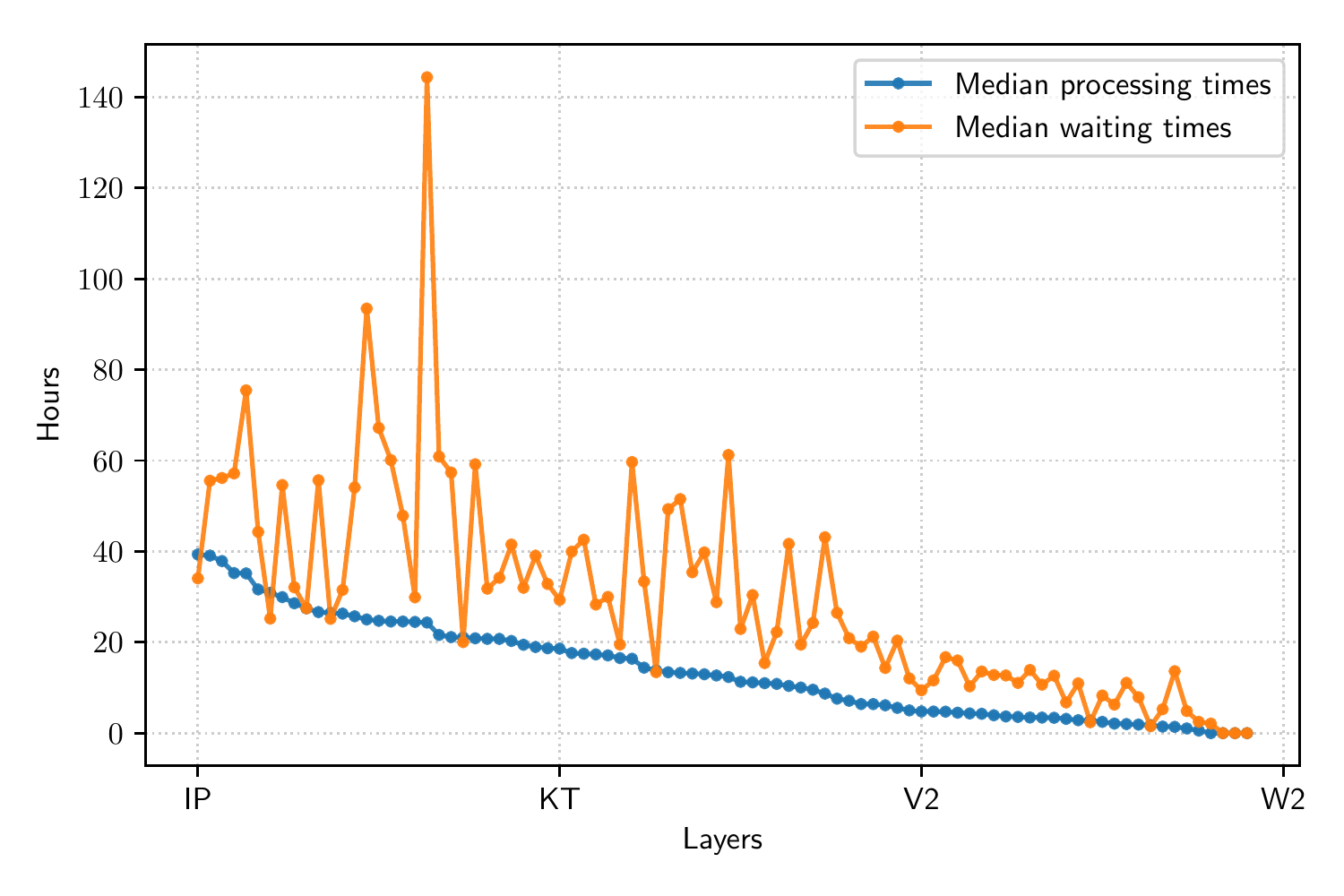}
	\end{center}
	\label{wf_layers_processing_waiting_median}
\end{figure}
In this section, I investigate the waiting and processing times of the products in different layers that products go though. I define the processing ($pt^j_{l}$) and waiting ($wt^j_{l}$) time of the lot $j$ in layer $l$ as the total amount of time that a product spends inside and in front of the equipment groups used for processing the recipes belonging to that layer, respectively. The processing and waiting time of the lot $j$ in layer $l$ can be calculated using Equation (\ref{layer_pt}) as
\begin{eqnarray}\label{layer_pt}
pt^j_{l} &=& \sum_{r\in R_l}{pt^j_{lr}}, \nonumber \\
wt^j_l &=& \sum_{r\in R_l}{wt^j_{lr}},
\end{eqnarray}
where $R_l$ is the sequence of the recipes of the layer $l$ and $pt^j_{lr}$ and $wt^j_{lr}$ are the processing and waiting time of the lot $j$ in layer $l$ and recipe $r$. For instance, consider layer $A_1$ in the fabricated example with $R_{A_1} = \{r_1, r_2, r_3\}$. The processing times of the product $j$ of type $A$ in layer $A_1$ is equal to the summation of the processing times of $\{r_1, r_2, r_3\}$ on $EQP_1$, $EQP_2$, and $EQP_3$, respectively. 

\begin{figure}
	\caption{Box-plot of the ratio of waiting times to processing times of the products for different layers with more than 40 hours of waiting times}
	\begin{center}
		\includegraphics[scale=1]{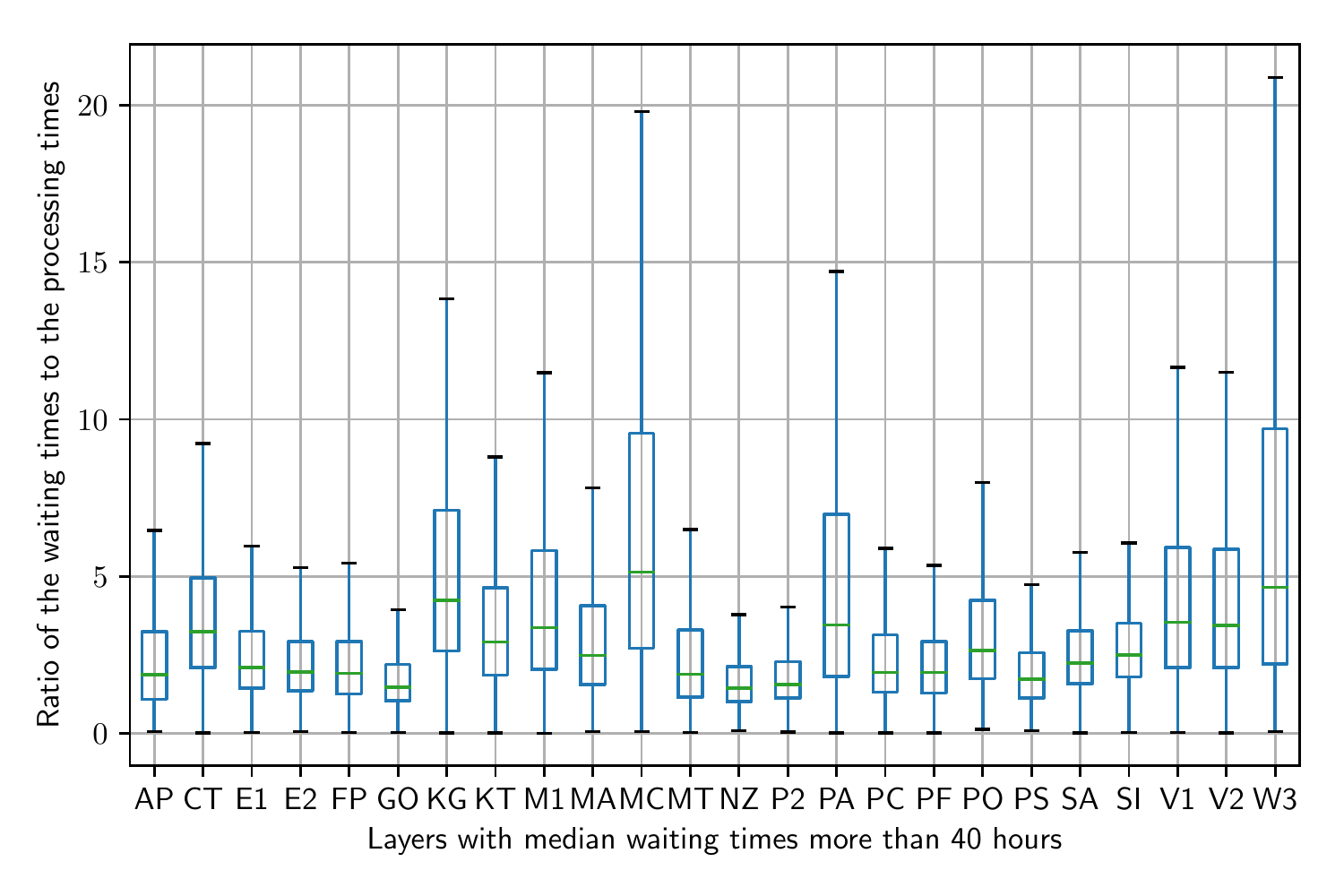}
	\end{center}
	\label{wf_layers_wt_pt_ratio_boxplot}
\end{figure}

Layer level analysis is necessary in determining the bottleneck production steps that impact the total cycle times of the products. Figure \ref{wf_layers_processing_waiting_median} shows the median of the processing, and waiting times of layers that have processed at least 1000 data-points. Some layers demonstrate a significantly higher median for the waiting times in comparison to the processing times. Even though the layers are using the same set of equipment groups, the waiting times of the same equipment groups may differ substantially from one layer to another. Figure \ref{wf_layers_wt_pt_ratio_boxplot} shows the box-plot of the ratio of the waiting times to processing times of the layers with median waiting time greater than 40 hours. Some layers demonstrate a significantly higher ratio in comparison to others. Consider $MC$ and $KG$ layers for instance. Both layers are using the etching machines. However, the waiting times of the products in the $MC$ layer for etching machines is significantly higher than that of the $KG$ layer. A production manager may need to balance the WIP levels of products in different layers for the purpose of reducing the total cycle times of the lots and its variation.  

\subsection{Research Opportunities}
A high variance of the waiting to processing times ratio of different layers suggests a possible inefficiency in balancing the WIP levels in different layers. The bottleneck recipes on the production route of different products are more likely to be part of the layers with high waiting to processing ratio than others. Further research is required on how to balance the WIP levels of the products in different layers. The higher WIP levels in certain layers lead to increase in the waiting times products in the current layer and possibly shortage of products in the consequent layers. Currently, the flow of products between the equipment groups is decided either by means of dispatching rules or scheduling algorithms. The current state of different layers needs to be integrated into the dispatching rules and optimization algorithms to balance the WIP levels and decrease the waiting times of the products in the wafer fabrication.

\section{Summary of Findings}\label{summary_of_findings}
In this section, I summarize the main findings of the exploratory data analysis reported in the preceding sections:
\begin{itemize}
	\item The departure process of the wafer fabrication is significantly more variable in comparison to the arrival process. 
	
	\item The dynamics of the arrivals process demonstrate a relatively stationary progress. The relative non-stationarity in the arrival process is the result of holidays, where no new lot had been released to the system. 
	
	\item The departure process on the other hand shows a relatively  non-stationary dynamics. The daily amount of products leaving the system may increase and decrease substantially. Such a variation in the daily amount products leaving the system is a result of mixture of different product types which have different total cycle time distribution.  
	\item Half of the products spend more than 70\% of their total cycle times waiting to be processed in the wafer fabrication.  
	\item Some of the layers in the product route demonstrate a significantly higher waiting times in comparison to the processing times inside the layer. Balancing the WIP levels of the layers may be necessary for reducing the total cycle times of the lots. 
	\item There may exist significant amount of correlation between the inter-event times of the machines. 
	\item The processing times usually demonstrate a significant positive autocorrelations between themselves. The significant amount of positive correlation between the processing times is the result of different equipment related factors, such as processing different recipes, and existence of handlers inside the equipment.  
	\item The inter-arrival and inter-departure times may demonstrate a positive, negative or no autocorrelation between themselves.
	  
	\item Even inter-arrival and inter-departure times with a distribution similar to that of exponential distribution may demonstrate a significant correlation. 
\end{itemize}

\section{Conclusion}\label{conclusions}
In this chapter, I perform an exploratory data analysis on the wafer fabrication of a semiconductor manufacturing system to investigate the main features impacting the performance metrics and present uncertainties in different levels of detail. In wafer-fabrication aggregation level, I analyze the arrival and departure process of the systems and the total cycle times of the products. I show empirically that the departure rates demonstrates significantly higher variation in comparison to daily arrival rates. The high variation in the departure process may be associated with the long total cycle times of the products. I show that total cycle times demonstrate a wide rang from several days to several months. I demonstrate that some of the uncertainty in the total cycle times can be explained by the product type. My analysis shows that half of the products spend more than 70\% of their total cycle times  waiting to be processes inside the wafer fabrication. Some of the layers in the product route contribute significantly more than the other to the waiting times of the products. 

In machine level analysis, I investigate the statistical properties of the inter-event times. The standard practice in modeling the inter-event times in manufacturing systems is to use independent and identically distributed distributions. I analyze the range of coefficient of variation of the inter-event times to provide empirical range of the coefficient of variation. Then, I investigate the independence assumption in modeling the inter-event times. The independence assumption of the inter-event times is the most common assumption in the study of the  queuing and manufacturing systems. I demonstrate that there may exist a significant correlation between the inter-event times of the machines in manufacturing systems. The processing times of the machines usually demonstrate a significant positive correlations between themselves. The positive correlation is a result of clusters of event-times that are closer to each other. For instance, machine recipe may create clusters of processing times resulting in a positive correlation. The inter-arrival and inter-departure times show a statistically significant correlations between themselves as well. However, the magnitude  of the correlation between the inter-arrival and inter-departure times is not as big as the correlation between the processing times. My analysis show that even testing machines may demonstrate correlated inter-event times. 


\chapter[Cycle Time Prediction in Semiconductor Wafer Fabrication]{Cycle Time Prediction of the Products in Semiconductor Wafer Fabrication}\label{ct_estimation_chapter}
\section*{Abstract}
In this chapter, I use the product movement data of the Reutlingen semiconductor wafer fabrication to predict the total cycle times of the products. Two sets of features, namely the product-related and system-state related features, are determined from the data to predict the total cycle times. The product-related features capture the product attributes such as the product type, information about the product route and distribution of the processing times on the product route. On the other hand, the system-state related features capture the impact of the state of system upon arrival of the lot on the total cycle times. I use these features to predict the cycle times of the products by using rolling average and LASSO methods. The rolling average method uses the product related features and recent products that have finished their processing to predict the total cycle times. On the other hand LASSO predicts the cycle times by identifying a set of parameters that minimizes its objective function. LASSO also identifies the most important features that impact the cycle times of the products. 

The dataset prepared in this study can be used by other researchers for evaluating the accuracy of the cycle time prediction methods. In addition, the identified features can be used to analyze the performance of the resources corresponding to the most important features to decrease the waiting times in a large-scale manufacturing system.

\section{Introduction}
Wafer fabrication is one of the most complicated and capital intensive manufacturing systems. Products, referred to as lots of 25 wafers go through hundreds of process steps in which different number of layers are fabricated on the surface of the wafers. Each layer consists of a set of pre-specified process steps along with random inspection steps in between. Mixture of different product types, long cycle times, reentrant loops, several hundreds of process steps, and complicated dispatching rules on one hand and highly unreliable machines, preventive maintenance, and engineering work on the other hand make wafer fabrication the most complicated manufacturing system in existence. These factors make the total cycle times of the products in the wafer fabrication highly uncertain. Furthermore, to attain a high utilization of the capital intensive machines in the wafer fabrication, production managers push as much products as possible to the system. This push strategy leads to higher waiting and consequently cycle times and increases the variation of the cycle times. 

Total cycle time is one of the most important performance metrics of the wafer fabrication. Accurate forecasting of the cycle times is the key factor for giving reliable product-delivery promises to customers, assigning internal due-dates for scheduling and product-flow purposes, and efficient production planning. Each new product is assigned a due-date upon release to the wafer fabrication. The due-date is equivalent to the release time plus a target cycle time. This due-date is used by the equipment-group leaders for dispatching and scheduling purposes to meet the specified due-dates and reduce the total cycle times of the products eventually. In addition to setting the due-dates, manufacturers use the target cycle times to evaluate the probability of order fulfillment before promising due dates to customers as order fulfillment is valued as high as the price of the products in semiconductor manufacturing \citep{Batra2018QuantifyingChain}. In practice, total cycle time reduction is possible by optimizing the product flow inside the wafer fabrication and scheduling of the products in each equipment-group. Shortening cycle time is of paramount importance in wafer fabrication setting as small improvements in the total cycle times results in huge gains to the manufactures. Long cycle times on the other hand lead to accumulation of WIP levels, high variance in the departure process, and increased risk of wafer contamination.

The aim of this chapter is two-folds. First, I use the lot trace data of the products from the wafer fabrication of the semiconductor manufacturing system of the Reutlingen plant of the Robert Bosch Company in Germany to prepare a dataset for estimating the total cycle times of the products in the wafer fabrication. In particular, I use the inter-event data of 16305 products that have been produced in 2018. Variety of prediction methodologies have been developed in the literature of the cycle time prediction. However, there are no universal test datasets for assessing the performance of the developed methodologies. Furthermore, the scale of the of the data and the wafer fabrication is small. The objective of this study is to fill this gap by preparing a dataset from the large scale wafer fabrication of the Robert Bosch Company that can be used as a testbed by other researchers for evaluating the performance of different methodologies. Second, I predict the cycle times of products using different methodologies. The complicated nature of the wafer fabrication makes the prediction of the cycle time of lots a difficult task. I adopt different statistical and machine learning methodologies for estimating the total cycle times of the products based on different attributes from the products and state of the wafer fabrication.  


\section{Related Literature}
There exist numerous types of methods in the literature to predict the cycle times of the lots in a wafer fabrication. I classify the existing methods into two major categories as analytical and discrete event simulation, and statistical and machine learning methods. The analytical and discrete event simulation methods try to predict the cycle times by creating a queuing or simulation model that represents the flow of products in wafer fabrication. These approaches are restricted in their application in practice. Queuing models are based on restrictive assumptions that limit their practical applications. On the other hand, the simulation methods are slow and computationally intensive. A single replication of the simulation model may take from several hours to several days for a complicated production system such as semiconductor wafer fabrication. 

The objective of the queuing models is to build analytical models that represent a complex production system \citep{Bitran1992ASystems, Chung2002CycleLots}. Queuing models are based on restrictive modeling assumptions, such as exponential inter-arrival and processing times \citep{Inman1999EmpiricalSystems, ManafzadehDizbin2019ModellingProcesses},  that limit their application in practice. There has been attempts to utilize the Queuing Network Analyzer (QNA) for studying manufacturing systems that can be modeled as open queuing networks with infinite buffers \citep{Segal1989AManufacturing}. However, QNA may give unrealistic performance measures when the inter-arrival and service times are dependent \citep{Whitt2018UsingQueues}. \citet{Pearn2007Due-dateEnvironment} fit a gamma distribution to the waiting time distribution of the products in the wafer fabrication. \citet{Schelasin2011UsingManufacturing} integrate the queuing methods into simulation for cycle time prediction purposes. Simulation usually has been used for estimating the cycle time-throughput curves and estimating the cycle time  percentile curves \citep{Yang2007EfficientMetamodeling} when it is not straightforward to model the system with queuing models. Simulation methods build the digital twin of the system to predict the cycle times of the lots \citep{Hsieh2014EfficientMetamodeling, Batur2018QuantileIndustry}. The input to the simulation models needs to be estimated from the inter-event data. Small errors in modeling the inter-event times of the machines may lead to a high errors in the simulated model.  The main disadvantages of the simulation methods are computational intensity and slow running time for large manufacturing systems. In addition, an enormous amount of computing resources maybe required for simulation. The execution time of the large simulation models maybe long enough that generating several data-points take long times.

The objective of the statistical and machine learning methods is to build a model that estimates the cycle times of the products by using the collected data from the wafer fabrication. The challenges in using the statistical and machine learning methods are twofolds. First, they need sufficient amount of data to be able to approximate and generalize to new products being released to the wafer fabrication. Second, they require a high quality set of features for prediction purposes. The statistical and machine learning methods adopt a variety of methods for prediction purposes. While older studies estimate the cycle times by employing the pre-specified features directly, more recent studies aim to identify the most relevant features by using methods such Principal Component Analysis (PCA) \citep{Chen2013AnFabrication, Chen2014EnhancingImplications}, Mutual Information \citep{Wang2016BigSystem} and Logistic Regression \citep{Wang2018ASystem}. I review the statistical and machine learning approaches below. 

\citet{Chen2003AFab} uses the product attributes and wafer fabrication state as inputs to a feed-forward neural networks to estimate the cycle time of the products. \citet{Sha2004Due-dateNetworks} design a three-layer Multi-Layer Perceptron for cycle time prediction of wafer lots. \citet{Chang2005EvolvingFactory} propose evolving fuzzy rule approach to classify jobs into several categories and predicting the cycle time of products. \citet{Backus2006FactoryApproach} estimate the cycle times of the lots using clustering, K-nearest neighbors and regression trees. \citet{Chen2007AnPrediction} applies K-Means clustering before estimating the cycle times with Fuzzy Back-Propagation Neural Networks (BPN). \citet{Chen2010IncorporatingPlant} integrate the product classification into the prediction approach. They incorporate the fuzzy c-means to classifying the jobs into several categories and BPN approach with a non-linear programming model to estimate the cycle times  of jobs. \citet{Chien2012ManufacturingTime} forecast the cycle times of the lots by integrating Gauss-Newton regression method and artificial neural network. \citet{Chen2013AnFabrication} transforms the initial features using PCA to generate the input feature to classification and regression tree methods in order to estimate the cycle times of the products. \citet{Tirkel2013ForecastingDatabases} uses Neural Networks and decision trees to estimate the cycle times  of the products. \citet{Chen2014EnhancingImplications} propose to classify products based on prediction purposes rather than job attributes. They adopt hybrid PCA, CART, and BPN to estimate the cycle time of the products and assess the quality of their methodology using 120 lots from a semiconductor manufacturer. \citet{Chen2014TheFabrication} proposes to separate the jobs of each branch into two parts in a BPN tree approach  for prediction. \citet{Wang2016BigSystem} use the Mutual Information methodology to recognize the most important features for estimating the cycle times. \citet{Wang2018ASystem} adopt a logistics regression based methodology to perform feature selection for prediction purposes.








\section{Data Description and Preprocessing}
\subsection{Lot Trace Data Description}
I use the lot-trace data of the 200 millimeter wafers from the wafer fabrication of the semiconductor manufacturing plant of the Robert Bosch Company in Reutlingen, Germany. The lot-trace data captures every event related to the movement of the lots between the production steps. The general format of the dataset is presented in \citet{Laipple2019GenericChains}. 
\begin{table}[]
\scalebox{0.85}{
\begin{tabular}{|l|l|}
\hline
lotid          & Unique ID given to the products upon their release to the production. \\
partid         & \begin{tabular}[c]{@{}l@{}}A unique ID given to a category of products that go through similar \\ set of process steps.\end{tabular}                                                                                                               \\
lottype        & \begin{tabular}[c]{@{}l@{}}The highest aggregation of the products. The products are categorized \\ into lottyes based on the technology used for in their production. \\ A certain lottype contain different parts as subcategories.\end{tabular} \\
prodarea       & Name of the production area (factory) \\
eqpid          & A Unique ID given to the machines \\
eqptype        & A unique ID that categories similar machines into the same category. \\
stage          & \begin{tabular}[c]{@{}l@{}}Stage determines the current state of the production of a certain lot. \\ The stage column of the data is in the $AB-CD-EF$ where $AB$ is \\ the current layer that the product in is.\end{tabular} \\
location       & Name of the area that the equipment is physically located in. \\
recpid         & Name of the recipe (machine program) used for processing the lot.  \\
queuetime      & Time stamp that the lot leaves the previous process-step.\\
trackintime    & Time stamp that the lot enters the processing step.\\                                trackouttime   & Time stamp that the lot leaves the process-step. \\                      
trackinmainqty & Number of wafers in a given lot at lot trackin \\
curmainqty     & number of wafers in a given lot at lot trackout.\\  \hline           
\end{tabular}} \label{datatemp}
\end{table}

Events are defined as the movement of the lots from one process step to the next. Table \ref{datatemp} describes each column in the data. It starts with the \emph{lotid} which is a unique id given to a certain lot. Each lot is a subset of \emph{partid} which determines the lowest level type of the product. The \emph{partid} information in the \emph{partid} column consists of two parts in a $abcd.01$ format. I extract a new column called \emph{part} from this column. The products that belong a certain part go through similar set of recipes and layers explained below. \emph{lottype} is a higher aggregation level that groups parts with similar technology into a single category. The dataset consists of a six major lot-types. A machine processes the lots by a product specific recipe given in the \emph{recid} column. \emph{stage} determines the current state of the production of the lot. The stage column is in the $AB-CD-EF$ format. The first two characters of \emph{stage} ($AB$) determines the current layer of the product. Layer refers to a physical layer fabricated on top of the wafers. \emph{lotid}, \emph{recpid}, and \emph{stage} completely current state of the production of a certain lot. \emph{eqpid} is the id of the equipment that processes the lot. Each equipment belongs to a equipment-group specified by \emph{eqptype}. The type of the machines in the same production steps may differ from each other due to different reasons such as a newer technology. For instance machines in diffusion area consists of two major sets of old and newer equipment, which differ in the number of batches of lots they process. On the  other hand, the machines in the ion implant differ from each other due to the chemical processes that they conduct on the lots. \emph{location} determines the physical place that the equipment is present in. \emph{queuetime} captures the time stamp that the lot leaves the previous equipment or equivalently enters the current equipment or process step. \emph{trackintime} and \emph{trackouttime} capture the time stamp that the lot enters and leaves the equipment, respectively. Note that the \emph{trackouttime} time of the lots is equal to the \emph{queuetime} of the next production step. Finally, \emph{prodarea} is the name of the factory that the data belongs to. Table \ref{datatempexpample} shows an example of the inter-event data.  
\begin{table}[!htb] 
	\centering
	\caption{A single row of the lot-trace data from the wafer fabrication of the semiconductor manufacturing system of the Robert Bosch Company }
	\begin{tabular}{|l|l|}
		\hline
		lotid        & C93674.1            \\
		partid       & CMP211BC\_2.01      \\
		recpid       & SOG02.03            \\
		priority     & 3                   \\
		eqpid        & NN117               \\
		eqptype      & 8\_SD           \\
		stage        & PF-DIF-OG           \\
		location     & DIFFUSION               \\
		prodarea     & Wafer Fabrication                \\
		queuetime    & 2018-09-17 00:41:10 \\
		trackintime  & 2018-09-17 00:52:26 \\
		trackouttime & 2018-09-17 00:55:16 \\ 	
		\hline 
	\end{tabular}
	\label{datatempexpample}
\end{table}

Our dataset consists of lot-trace data of every product processed in the 200 mm wafer fabrication. The raw dataset consists of 17\,223\,658 rows of inter-event data. The rows in the raw-data correspond to two different sets of lots: productive and  non-productive. The non-productive lots contain the test-lots, engineering-lots, and other types of lots that are used for different purposes such as assessing the performance of the machine or developing a new technology. On the other hand, the productive lots are the lots that move to the Diebank, or backend of the supply chain. The productive lots dataset consists of 10\,317\,224 rows of data. It contains the data related to 16305 unique products categorized into 216 different parts. The products are being processed in 160 different equipment-groups which contain 500 different machines and can process 2159 different recipes. 

\subsection{Total Cycle Time Extraction}
Total cycle time of the lot $j$ ($tct_j$) is defined as the difference between the time-stamp that the lot enters ($queuetime_j^{1}$) and leaves ($trackouttime_j^{n}$) the wafer fabrication. It can be calculated using Equation (\ref{ct0}), where $n$ is the number of process steps that the lot will go through. 
\begin{equation}\label{ct0}
 tct_j = trackouttime_j^{n} - queuetime_j^{1}.  
\end{equation}
As demonstrated in Figure \ref{numprocesssteps} the exact number of process steps of the lots, and consequently the value of $n$ is not known in advance. Alternatively, total cycle times can be calculated by using the summation of the cycle times over all the production steps. The cycle time of the lot $j$ in its $i^{th}$ production step is defined as:
\begin{equation}\label{loc_ct}
ct^i_{j} = trackouttime^i_j - queuetime^i_j,  
\end{equation}
where $queuetime$ and $trackouttime$ are defined in Table \ref{datatemp}. Consequently, the total cycle time can be calculated as a summation of the cycle times of each production step as:  
\begin{equation}\label{ct1}
 tct_j = \sum_{i=1}^{n} ct^i_{j}.  
\end{equation}

\subsection{Candidate Prediction Variables}
In this section, I describe the features (prediction variables) that I extract from the lot trace data for total cycle time prediction purposes. There are two sets of features that can be used for predicting the total cycle times of the products upon their arrival to the system, namely product related and wafer fabrication state related features. I define each feature set using a simple fabricated example described below in the following subsections.  

   
As mentioned earlier, in semiconductor wafer fabrication products go through hundreds of process steps in which a certain number of physical layers are fabricated on top of the wafers. Each layer is constructed by means of a pre-specified set of recipes which are processed on different machines. A recipe is a set of instruction that a certain equipment uses in processing the lot. For instance, consider the Ion Implantation process of the wafer fabrication. The recipe in the Ion Implantation determines the ion sources that the equipment will use as the material to be implanted on the wafers. A certain recipe can be processed by using different machines depending on the capabilities of the machines in the wafer fabrication. The processing times of a certain product on a given machine may depend on their current layer and the recipe used for processing. Let $t^{lrm}_{j}$ be the processing time of the product $j$ processed with recipe $r$ in layer $l$ on equipment $m$. I specify the layer that the product is in since a similar recipe can be used in different layers. 

Let $l-r$ represent the features related to the processing time distribution of the recipe $r$ in layer $l$. In this study, I assume the machines that process the same recipe follow a similar processing times since the machines that are able to process a certain recipe, are usually of the same type. Hence, I use $t^{lr}_{j}$ instead of $t^{lrm}_{j}$ for the analysis presented in the following sections. 

\begin{figure}
	\caption{A simple production network to define different candidate features}
	\begin{center}
		\includegraphics[scale=.5]{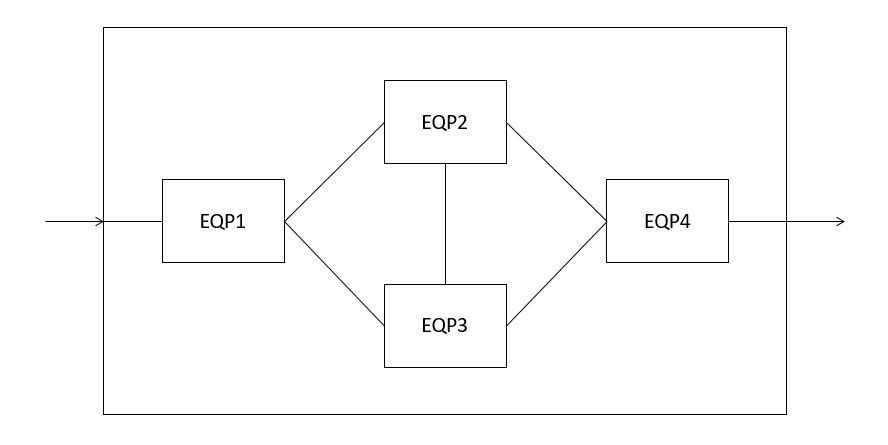}
	\end{center}
	\label{product_route_example}
\end{figure}

Consider the simple fabricated example given in Figure \ref{product_route_example} for understanding layers and recipes in the wafer fabrication. This production system consists of four machines denoted by $EQP_i$. Suppose that $EQP_1$ can process the recipes $r_1$ and $r_5$ and the rest of machines process the recipes with the same subscript (e.g., $EQP_2$ process the recipe $r_2$). Assume this simple production network manufactures two types of products, $A$ and $B$. The type $A$ products require two different layers to be fabricated on the wafers. The first ($A_1$) and second $(A_2)$ layer are fabricated on top of the wafer by sequentially processing the set of recipes $(r_{1}, r_{2}, r_{3})$, and $(r_{5}, r_{2}, r_{4})$, respectively, where $r_{i}$ is the recipe used for processing. Hence, the product route of the type $A$ products is $(r_{1}, r_{2}, r_{3}, r_{5}, r_{2}, r_{4})$. The product type $B$ has a single layer ($B_1$) in which the products go through $(r_{1}, r_{3}, r_{4})$ and then leave the system. 


\begin{figure}
	\caption{Empirical distribution of the number of process steps of the lots in the wafer fabrication}
	\begin{center}
		\includegraphics[scale=.9]{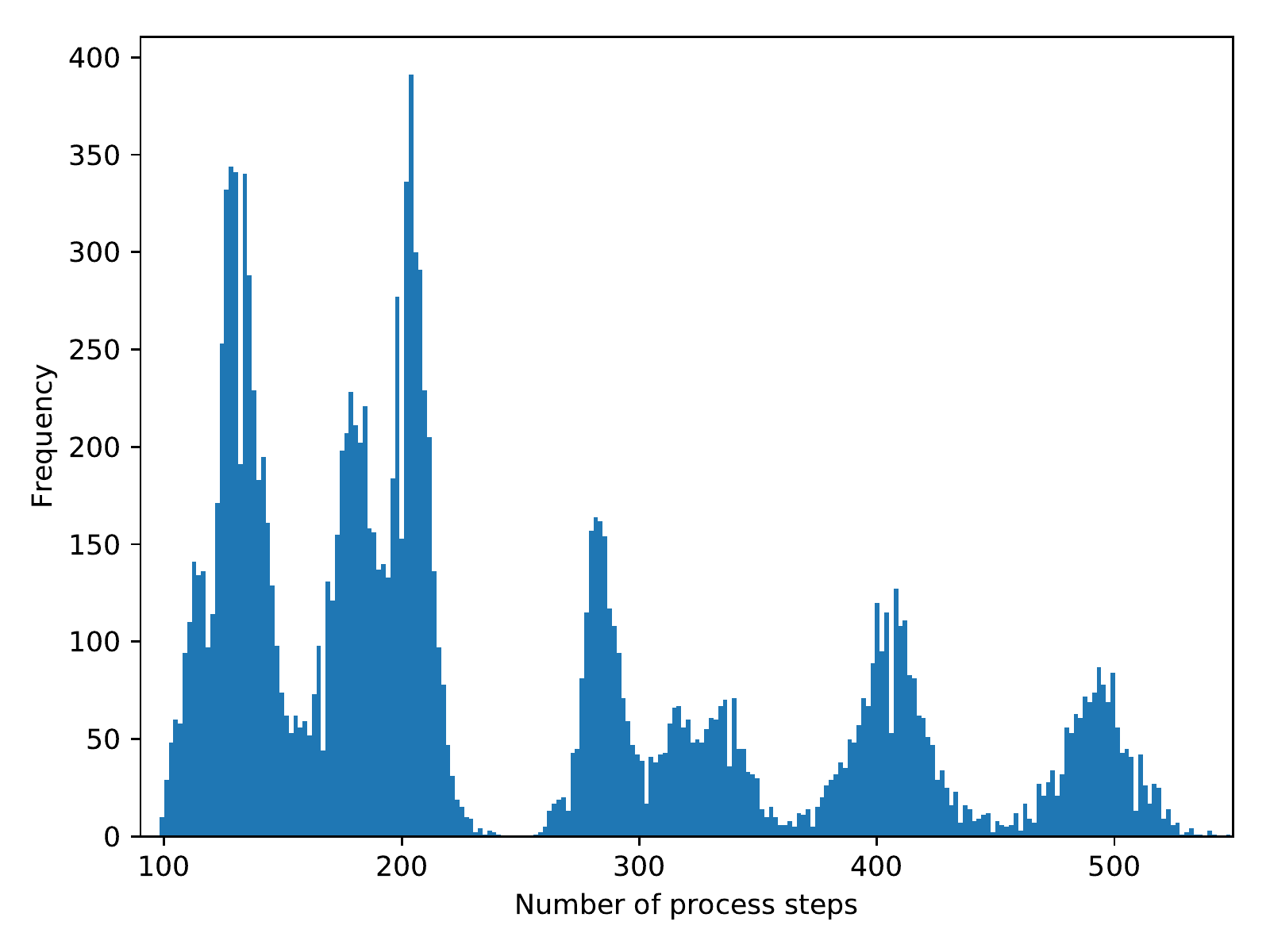}
	\end{center}
	\label{numprocesssteps}
\end{figure}

\subsubsection{Product Related Features}
Product related features are the subset of product attributes that are known upon release of the product to the wafer fabrication. Table \ref{lit_features} shows different product features that have been employed in the literature for total cycle time prediction purposes. The most common product related features are the lot size, the processing times of the machines on the process route of the lot, and the priority of the lot. I use a similar set of product features in this study. In particular, I extract two different sets of product related features, namely recipe-based and layer-based. The recipe-based product related features use the quantiles of distribution of the processing times of the training data ($\{ t^{lr}_{1}, t^{lr}_{2}, \dots, t^{lr}_{N_{train}} \}$ where $N_{train}$ is the number of products in the training data), to construct the features. I use Equation (\ref{lot_pt}) to calculate the $t^{lr}_{j}$ of the product $j$ from the lot-trace data
\begin{equation}\label{lot_pt}
t^{lr}_{j} = trackouttime^{lr}_{j} - trackintime^{lr}_{j}, 
\end{equation}
where $trackouttime^{lr}_{j}$ and $trackintime^{lr}_{j}$ are the time-stamps that the product $j$ will be processed by using recipe $r$ in layer $l$ enters and leaves a certain equipment, respectively. 

As an example, consider the fabricated example described above. Assume that we use the median ($50^{th}$ quantile) of the $\{ t^{lr}_{1}, t^{lr}_{2}, \dots, t^{lr}_{N_{train}} \}$ as the product related features for prediction purposes. Let $l-r$ represent the features related to the processing time distribution of the recipe $r$ in layer $l$.  Assume the medians of the $(A_1-r_{1}$, $A_1-r_{2}, A_1-r_{3}, A_2-r_{5}, A_2-r_{2}, A_2-r_{4})$, and $(B_1-r_{1}$, $B_1-r_{3}, B_1-r_{4})$ be (1, 2, 3, 5, 2, 4) and (1, 3, 4) for the product types $A$ and $B$, respectively. The product related feature used in estimating the cycle time of a type $A$ and $B$ products would be a 9 dimensional vector (1, 2, 3, 5, 2, 4, 0, 0, 0) and (0, 0, 0, 0, 0, 0, ,1 , 3, 4) which represents the features related to $(A_1-r_{1}$, $A_1-r_{2}, A_1-r_{3}, A_2-r_{5}, A_2-r_{2}, A_2-r_{4}, B_1-r_{1}$, $B_1-r_{3}, B_1-r_{4})$. 

Since the exact product route is not known advance, I identify the most important recipes that the product needs to go through by using two criterion. First, I identify the recipes that have processed at least 100 products. Second, I only consider recipes with $75^{th}$ quantile above 6 minutes since other recipes contribute significantly less to the cycle times of the products. These criterion decrease the number of recipes used as features from 1927 to 846 and results in a $2221\times number\_of\_quantiles$ dimensional feature vector of the layer-recipe ($l-r$) type. There are two shortcomings in adopting such product related features. First, it results in a high dimensional feature vector which could degrade the performance of the prediction methods. Second, the number of data-points for some $l-r$ combinations is small resulting in an inaccurate estimation of the processing times quantiles. My analysis show constructing the product related features by using the quantiles of $\{ t^{r}_{1}, t^{r}_{2}, \dots, t^{r}_{N_{train}}\}$ instead of $\{ t^{lr}_{1}, t^{lr}_{2}, \dots, t^{lr}_{N_{train}} \}$ results in a slightly better predictions. Hence, I use the quantiles of the processing times of the most important recipes for predicting the cycle times instead of the $l-r$ based features. Such an approach decreases the feature vector dimension to $846\times number\_of\_quantiles$. I refer to these features as a Recipe-Based Product Related Features ($RBPRF$). I use the $[0.25, 0.50, 0.75]$ quantiles from the distribution of the $\{ t^{r}_{1}, t^{r}_{2}, \dots, t^{r}_{N_{train}}\}$ which results in a 2538 recipe-based product related feature vector. In our fabricated production system the recipe-based product related features is a 5 dimensional vector which represents the medians of $(r_1, r_2, r_3, r_4, r_5)$. 

In addition to $RBPRF$, I also define a feature set based on the processing times of the products in different recipes referred to as Layer-Based Product Related Features ($LBPRF$). $LBPRF$ construct the features set based on the processing time distribution of the products in layers. The processing time of the product $j$ in the layer $l$ ($pt^{l}$) is defined as the summation of the processing times of the recipes that belong to the layer $l$. The processing times of the layer $l$ can be calculated using Equation (\ref{layer_pt}) 
\begin{equation}\label{layer_pt}
pt^{l} = \sum_{r\in R}{t^{lr}},
\end{equation}
where $R$ is the sequence of the recipes of the layer $l$. For instance, consider the layer $A_1$ in the above example with $R = \{r_1, r_2, r_3\}$. The processing times of the layer $A_1$ for the product $j$ of type $A$ is equal to the summation of the processing times of the $j$ in $r_1$, $r_2$, and $r_3$. Similar to the recipe-based product related features, I eliminate the unimportant layers before constructing the layer-base features. Layers which have processed less than 100 products are not included in the layer-based product related features. This results in a 100 dimensional layer-based product related features.   

The layer-based product related features are used in clustering the products into different categories and LASSO methods. My analysis shows that recipe-based product related features result in poor cluster means. However, they may result in a slightly better predictions using LASSO.



\begin{landscape}
	\begin{table}[]
		\caption{The product and fab related input features in some of the related works}
		\centering
		\scalebox{.7}{
			\begin{tabular}{|l|l|l|}
				\hline
				Study & Product related features & Fab related features \\ \hline
				\citet{Sha2007DevelopmentProblem}     & \begin{tabular}[c]{@{}l@{}}the processing time the lot \\ at the three bottleneck machines,\\ the sum of processing times of the\\  lot on the process rout, \\ the type of lot\end{tabular} & \begin{tabular}[c]{@{}l@{}}total WIP levels of the fab, \\ the number of jobs present in the queue\\  of the three bottleneck machines, \\ the total remaining workload of three\\  bottleneck stations for all the products in the fab,\\ the number of down machines in the \\ three bottleneck stations, the total number of \\ each job type in the fab, the average cycle time\\  of the three lots of the product-types that have\\  most recently completed their processing, sum \\ of the remaining processing times for all products in the fab\end{tabular} \\ \hline
				\citet{Chen2010IncorporatingPlant}    & the size of the lot & \begin{tabular}[c]{@{}l@{}}total WIP levels of the fab,\\ the average fab utilization, \\ total queue length on the processing route of the lot, \\ total queue length before the bottlenecks, \\ total queue length in the whole fab, \\ average waiting time of the three, \\ most recently completed jobs\end{tabular}\\ \hline
				\citet{Chen2011ApplyingProduct} & the size of the lot & \begin{tabular}[c]{@{}l@{}}total WIP levels of the fab, \\ the queue length before the bottleneck, \\ the queue length on the process route, \\ the average waiting time, \\ and the fab utilisation\end{tabular}  \\ \hline
				\citet{Chien2012ManufacturingTime}    & \begin{tabular}[c]{@{}l@{}}average layers of one wafer\\  need to be manufactured, \\ average time needed to \\ finish one layer of a wafer \\ (average cycle time of one layer)\end{tabular}   & \begin{tabular}[c]{@{}l@{}}total WIP levels of the fab,\\ the number of product output per \\ month (throughput), fab utilization \\ defined as the percentage of used \\ capacity to the maximum useful \\ capacity in a fab,  total accomplished \\ operations among all machines measured \\ by the operations per day,\end{tabular}  \\ \hline
				\citet{Chen2014EnhancingImplications} & the lot size & \begin{tabular}[c]{@{}l@{}}total WIP levels of the fab, \\ factory utilization, \\ the queue length on the product route,\\ the queue length before the bottleneck, \\ and the average waiting times of the most \\ recent three completed products\end{tabular}  \\ \hline
				\citet{Wang2016BigSystem}             & \begin{tabular}[c]{@{}l@{}}the processing time of each operation \\ on wafer lots' processing route,\\ the priority of lot\end{tabular}                                                         & \begin{tabular}[c]{@{}l@{}}total wip levels of the fab,\\ the utilization of each machine (last 24 hours),\\ the queue length of each machine\end{tabular}\\ 
				& &  \\ \hline
			\end{tabular}
		}
	\label{lit_features}
	\end{table} 
\end{landscape}

\subsubsection{Wafer Fabrication State Related Features}
\begin{figure}
	\caption{Empirical distribution of the ratio of total waiting and  processing times of the products to their total cycle times in the wafer fabrication}
	\begin{center}
		\includegraphics[scale=.9]{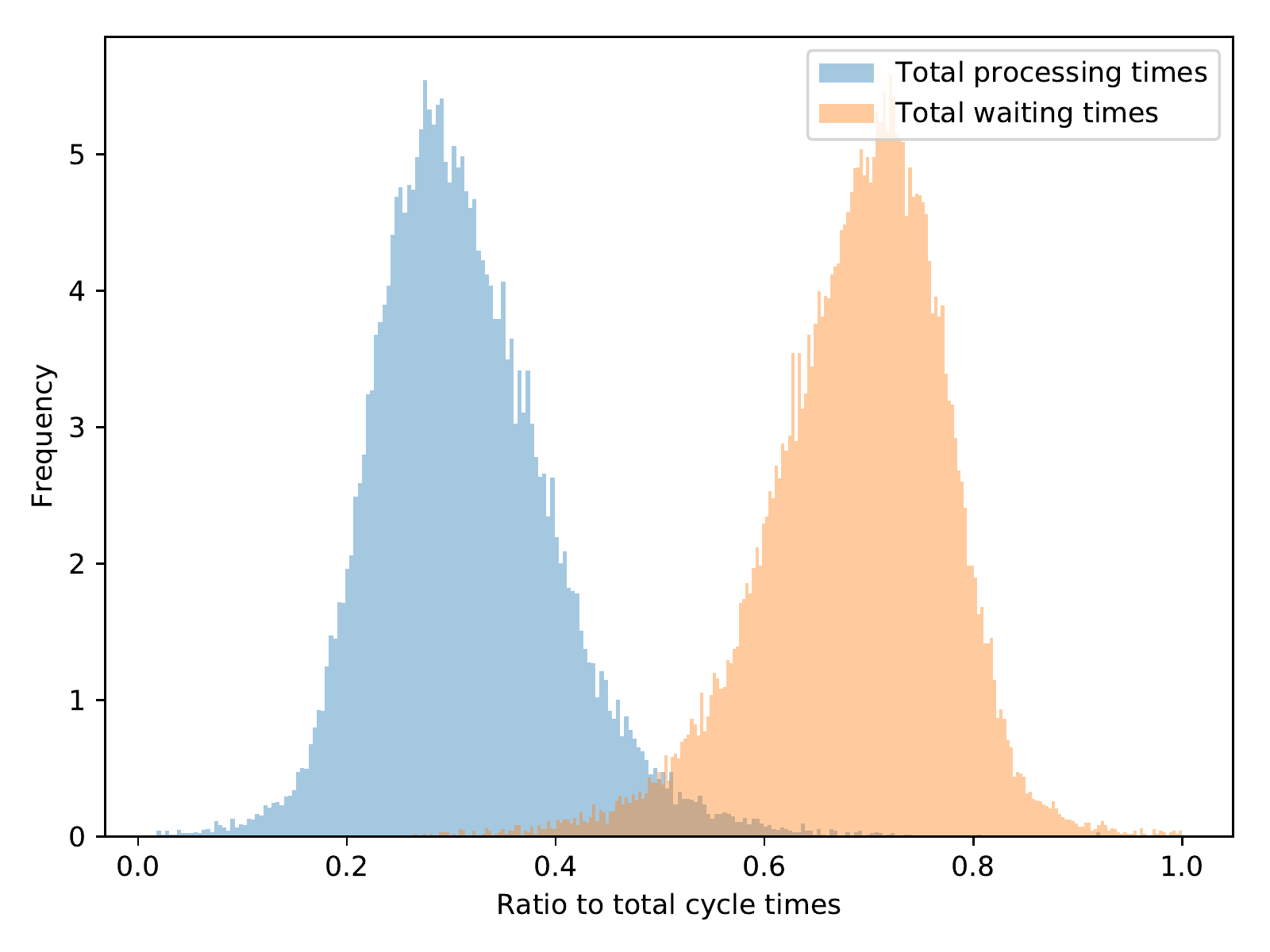}
	\end{center}
	\label{tpt_twt2ct}
\end{figure}
State of the production system in the wafer fabrication impacts the total cycle times of the products through the queue size of each equipment group on the product route. The greater the queue size, the greater the waiting times of the products in the wafer fabrication. Waiting times constitute major part of the total cycle time of the products in the production systems. Figure \ref{tpt_twt2ct} shows the ratio of total waiting and processing times of the products to their corresponding total cycle times. Half of the product in the wafer fabrication spend more than 70\% of their total cycle times waiting to be processed in front of the machines on the product route. 

\begin{figure}
	\caption{Scatter plot of the total WIP levels of the wafer fabrication upon arrival of the lot and the total cycle times of 4 different parts and corresponding fitted linear regression}
	\begin{center}
		\includegraphics[scale=.4]{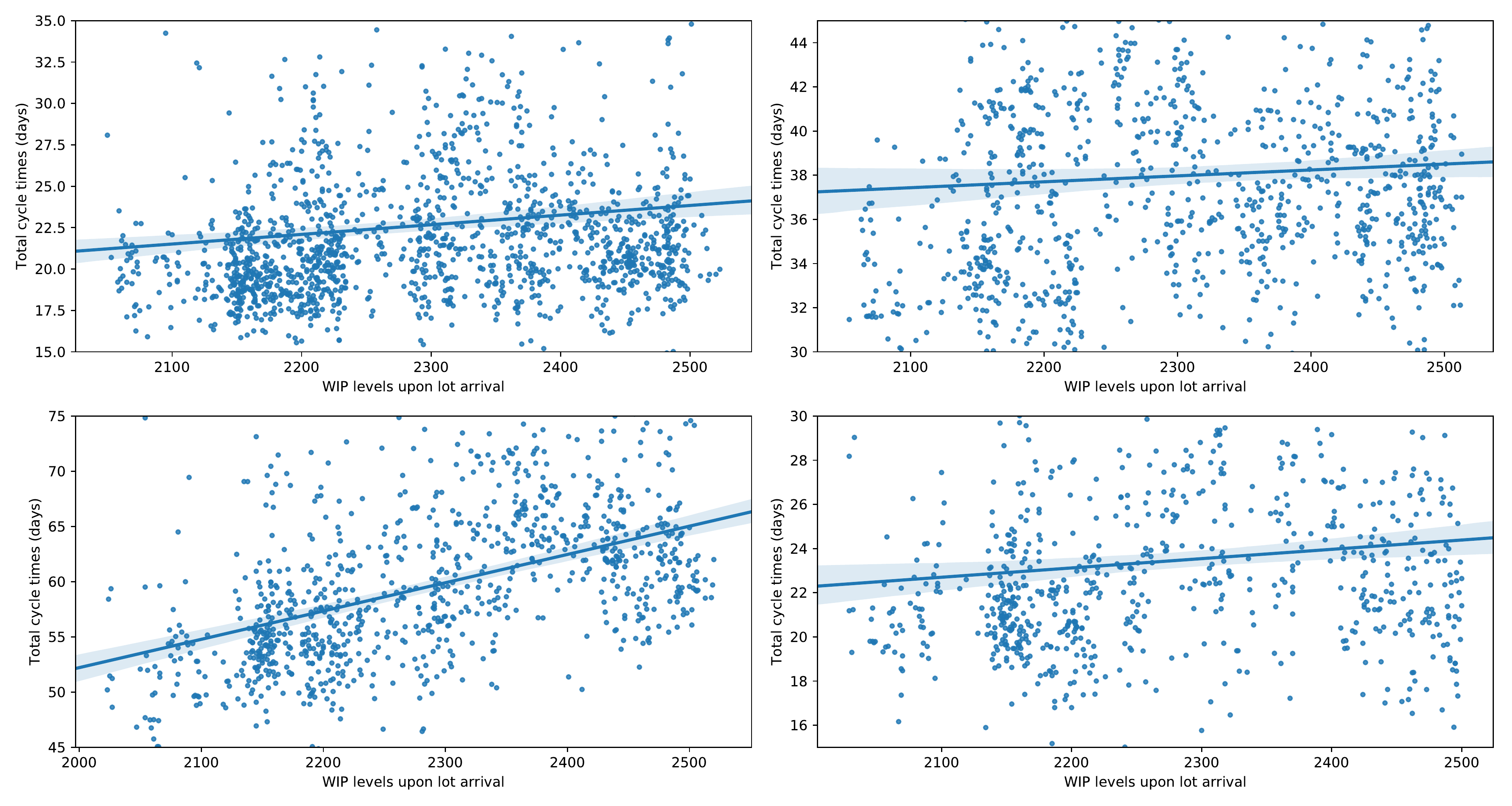}
	\end{center}
	\label{wip_ct_effect}
\end{figure}

The impact of state of the wafer fabrication can be integrated into the prediction methods through different feature sets from the state of the fab. Table \ref{lit_features} lists the features that have been used to capture the state of the wafer fabrication for prediction purposes in some of the studies in the literature. Almost all of the studies include the total WIP levels of the wafer fabrication as an input feature. The total WIP level are considered as a wafer fabrication state related feature in this study as well. Figure \ref{wip_ct_effect} demonstrates the scatter plot of the total WIP levels of the wafer fabrication upon arrival of the lots on $x$ axis and the total cycle times of the products on $y$ axis and a fitted linear regression for 4 of the majors parts in the Reutlingen wafer fabrication. The total cycle times of some parts increase substantially as a function of the total WIP levels of the wafer fabrication.

In addition to the total WIP levels, I construct another feature set based on the WIP levels of the layers. The WIP levels of the layer $l$ is defined as the summation of the WIP levels of the recipes necessary for constructing the layer by using Equation (\ref{layer_wip})
\begin{equation}\label{layer_wip}
wip^l = \sum_{r\in R}{wip^{lr}},
\end{equation}
where $wip^{lr}$ is the number of products that will be processed by using recipe $r$ in layer $l$. For instance, consider the layer $A_1$ in the simple production network above. Remember that layer $A_1$ is fabricated by being processed using  $(r_{1}, r_{2}, r_{3})$. Assume $(wip^{A_1r_1}, wip^{A_1r_2}, wip^{A_1r_3})$ is equal to (7, 8, 9). The $wip^{A_1}$ in this case would be equal to 24. The layer-based WIP levels feature set is a 100 dimensional vector where each element corresponds to the WIP levels of each of the important layers determined in the layer-based product related features.



\subsubsection{Features Sets Employed in Prediction Methods}
I use concatenation of the product related and wafer fabrication state related features as an input for the learning based methods for predicting the cycle times. The feature vector of the lot $j$ is demonstrated by $P^\alpha_j$ where the superscript $\alpha$ determines the type of features that I use in predicting the cycle times and $P^\alpha = \{P^\alpha_1, P^\alpha_2, \dots, P^\alpha_N,\}$ represents features of type $\alpha$ of the products $1, 2, \dots, N$. In this study, I use different combinations of the product related and system state related features to find a feature set that best predicts the cycle times. Table \ref{feature_sets} shows the list of different combinations of the feature sets used. 

\begin{table}[]
\caption{Different feature sets adopted in predicting the cycle times of the products}
\centering
\scalebox{0.9}{
\begin{tabular}{|l|l|l|} \hline 
Feature set & Description & Dimension \\ \hline 
$P^{L-0}$ & Layer-based product related features ($LBPRF$) & 300 \\ \hline 
$P^{R-0}$ & Recipe-based product related features ($RBPRF$) & 2538\\ \hline 
$P^{L-1}$ & Concatenation of $LBPRF$ and total WIP levels of the system & 301\\ \hline 
$P^{R-1}$ & Concatenation of $RBPRF$ and total WIP levels of the system & 2539\\ \hline 
$P^{L-2}$ & \begin{tabular}[c]{@{}l@{}}Concatenation of $LBPRF$, total WIP levels of the system, \\ and WIP levels of the layers\end{tabular} & 401\\ \hline 
$P^{R-2}$ & \begin{tabular}[c]{@{}l@{}}Concatenation of $RBPRF$, total WIP levels of the system, \\ and WIP levels of the layers\end{tabular} & 2638\\ \hline 
\end{tabular}}\label{feature_sets}
\end{table}

\section{Cycle Time Prediction} 
In this section, we introduce the methods that have been used for predicting the cycle times of the products. Two types of methods have been adopted for predicting the cycle times. The first type are rolling average based methods that use the cycle times of the other products to predict the cycle time of a new lot. The second are learning based methods that identify a set of parameters to a specific model to minimize a certain objective function such as the squared prediction errors. Each method is explained in further detail in the following subsections. 

I split the cycle times dataset into train and test dataset to find the best parameters and test the performance of different prediction methods. The training and test dataset constitute 80\%, and 20\% of the data, respectively, splatted in the chronological order in which lots have been released to the system. That is, the chronologically first 80\% of the lots comprise the training data and the remaining 20\% comprise the test data.
\subsection{Performance Metrics}
I use the following performance metrics for assessing the performance of different prediction methods
\begin{eqnarray}
RMSE & = & \sqrt{\frac{1}{N_{test}}\sum_{j=1}^{N_{test}}{\Big(tct_j - \hat{tct}_j}\Big)^2}, \nonumber \\
MAE & = & \frac{1}{N_{test}}\sum_{j=1}^{N_{test}}{\Big\lvert tct_j - \hat{tct}_j}\Big\rvert, \nonumber\\ 
MAPE & = & \frac{1}{N_{test}}\sum_{j=1}^{N_{test}}\frac{\Big\lvert tct_j - \hat{tct}_j \Big\rvert}{tct_j} \times 100,  \nonumber\\
MedAE & = & median(\{\lvert tct_1 - \hat{tct}_1\rvert, \lvert tct_2 - \hat{tct}_2\rvert, \dots, \lvert tct_{N_{test}} - \hat{tct}_{N_{test}} \rvert\}), 
\end{eqnarray}
where $N_{test}$ and $\hat{tct_j}$ are the number of test data points, and predicted cycle time of the lot $j$, respectively. In the following subsections, I introduce the prediction methodologies used in this study.

\subsection{Rolling Average Methods}
Motivated by the service operations literature \citep{Ang2016AccuratePrediction}, I investigate the performance of rolling average methods in predicting the total cycle times of the lots. The rolling average methods have been widely adopted in the service operations literature for estimating the waiting time of the arriving calls in call centers and patients in hospitals emergency rooms. In this study, I investigate the performance of $C-K$ rolling average method introduced below in predicting the cycle times of the lots in the semiconductor wafer fabrication. 
\subsubsection{$C-K$ Rolling Average}
The $C-K$ rolling average method first categorizes the products using clustering algorithms into different categories, then predicts the cycle time of an arriving lot using the last $K$ products of the same cluster that have finished their processing. The best values of the $C$ and $K$ are identified through cross-validation. I use the last 12.5\% of the training data (10\% of the whole data) in the chronological order as the validation dataset for identifying $C$ and $K$. 

\begin{algorithm2e}[H]\label{rae_algo}
	\caption{$C-K$ Rolling Average Prediction}
	\DontPrintSemicolon
	\SetAlgoLined
	\KwInput{$C$, $K$}
	\KwOutput{cycle times predictions of the test lots}
	\KwData{product related features}
	Cluster the products using a clustering algorithm and layer-based product related features into $C$ clusters
	
	\For{lot in test lots}{
		Estimate the cluster $\hat{C}$ that the lot belongs to using the clustering algorithm 
		
		Take the mean of the cycle times of the last $K$ lots of the $\hat{C}$ cluster that have finished their processing upon arrival of the new lot 
	}
\end{algorithm2e}

\begin{figure}[H]
	\caption{Kernel density estimator of the cycle time distribution of the products clustered into 5 categories denoted by $C\_i$ and 6 major lottypes}
	\begin{center}
		\includegraphics[scale=1]{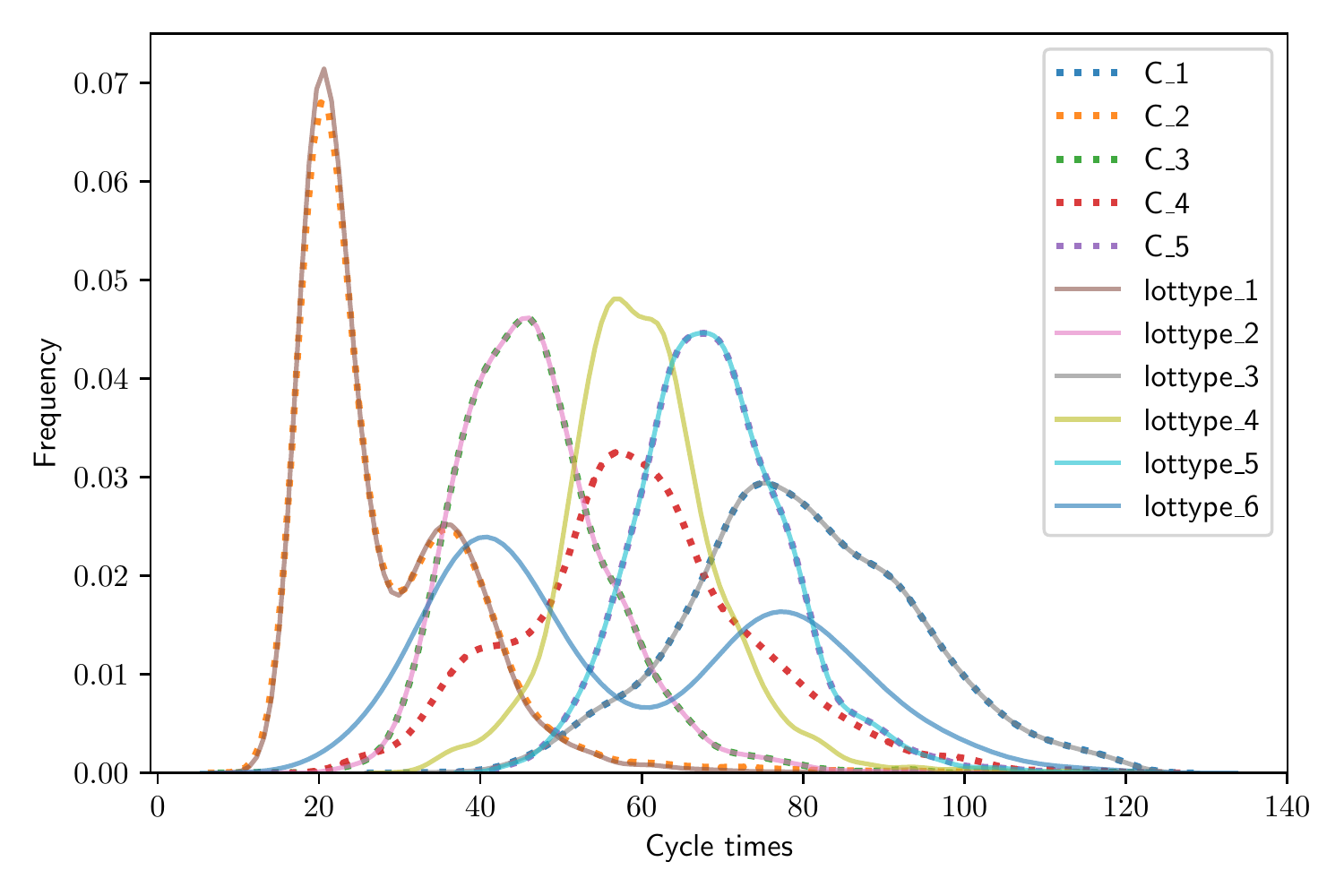}
	\end{center}
	\label{wf_product_clustrering_5}
\end{figure}

\paragraph{Clustering of the Products}
In this section, I classify the products into $C$ different categories by using the layer-based product related features. My analysis shows that the layer-based product related features results in better cluster means than the recipe-based product-related features. The \emph{lottype} column in the dataset already offers a categorization based on the technology of the products. However, the cycle time distribution of two of the lottypes ($lottype\_1$ and $lottype\_6$) is a mixture of two distributions as shown in  Figure \ref{wf_product_clustrering_5}. The variance of the cycle times distribution of those lottypes is large as well. Hence, I use clustering to categorize the products to obtain a uni-modular total cycle time distributions which can be used in the prediction methods. Figure \ref{wf_product_clustrering_5} shows the Kernel Density Estimator (KDE) of the cycle time distribution of the 6 major lottypes and 5 clusters from the clustering algorithm. KDE of the lottypes and clustering algorithms overlap except for two of the lottypes.  

\begin{figure}[H]
	\caption{Kernel density estimator of the cycle time distribution of the products clustered into 6 categories denoted by $C\_i$ and 6 major lottypes denoted by $lottype\_i$}
	\begin{center}
		\includegraphics[scale=1]{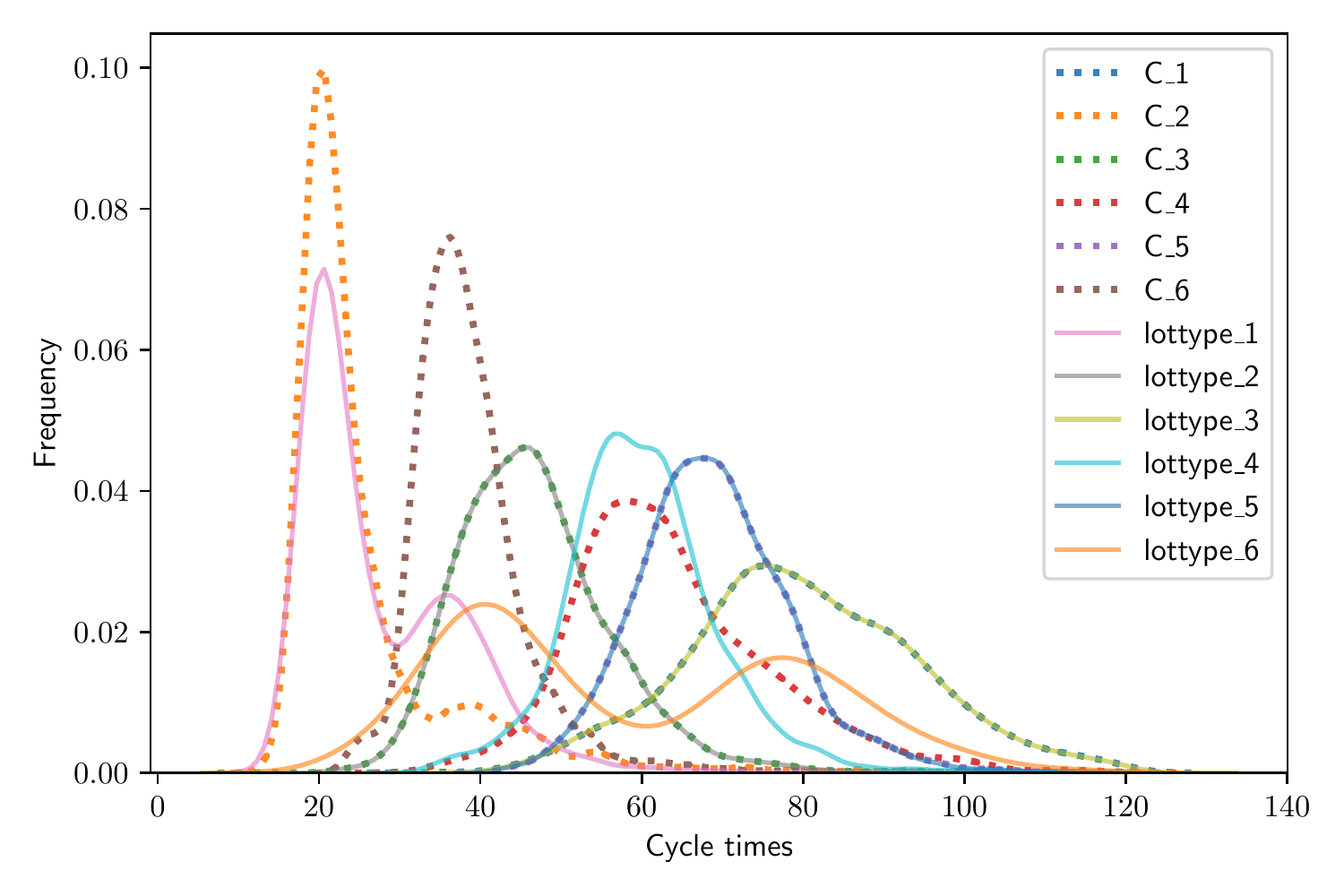}
	\end{center}
	\label{wf_product_clustrering_6}
\end{figure}

Figure \ref{wf_product_clustrering_6} shows that adopting $C = 6$ obtains better cluster means than that of 5 clusters. My analysis show that adopting more than 6 clusters result in cluster means close to each other. Table \ref{c6_stats} shows the median, mean, standard deviation and coefficient of variation of the processing, waiting and cycle times of the products in each cluster.
\begin{table}[]
\centering
\caption{Median, mean, standard deviation (std), and coefficient of variation (cv) of the processing, waiting, and cycle times of 6 different product categories clustered by the K-Means algorithm}
\scalebox{0.6}{

\begin{tabular}{|l|lll|lll|lll|lll|lll|lll|}
\hline
\multicolumn{1}{|c|}{} & \multicolumn{3}{c|}{1} & \multicolumn{3}{c|}{2} & \multicolumn{3}{c|}{3} & \multicolumn{3}{c|}{4} & \multicolumn{3}{c|}{5} & \multicolumn{3}{c|}{6} \\ \hline
& pt    & wt     & ct    & pt     & wt    & ct    & pt     & wt    & ct    & pt     & wt    & ct    & pt     & wt    & ct    & pt     & wt    & ct    \\
median                 & 7.79  & 13.20  & 21.37 & 11.20  & 33.56 & 45.78 & 13.13  & 23.44 & 37.49 & 17.35  & 40.56 & 58.81 & 20.02  & 57.62 & 79.09 & 20.65  & 46.76 & 68.45 \\ \hline
mean                   & 8.57  & 15.14  & 23.71 & 12.21  & 34.54 & 46.75 & 14.14  & 24.55 & 38.69 & 18.67  & 40.87 & 59.54 & 21.07  & 58.73 & 79.80 & 21.81  & 47.11 & 68.92 \\ \hline
std                    & 2.96  & 8.71   & 9.81  & 3.33   & 8.47  & 9.62  & 3.52   & 6.54  & 7.34  & 6.28   & 11.47 & 14.70 & 5.12   & 12.85 & 14.15 & 4.83   & 7.67  & 9.36  \\ \hline
cv                     & 0.35  & 0.57   & 0.41  & 0.27   & 0.25  & 0.21  & 0.25   & 0.27  & 0.19  & 0.34   & 0.28  & 0.25  & 0.24   & 0.22  & 0.18  & 0.22   & 0.16  & 0.14  \\ \hline
\end{tabular}}\label{c6_stats}
\end{table}

%
%

\subsection{Lasso}
Lasso is a linear prediction model that regularizes the magnitude of its coefficients to find simple, sparse models. Lasso solves the following quadratic programming equation to find the optimal values of the coefficient vector ${\beta^\alpha}$ corresponding to feature set $P^\alpha_j$ for lot j
\begin{equation}
\min _{\beta^\alpha} \frac{1}{N} \sum_{j=1}^{N}\left(tct_{j}-{\beta^\alpha}^{\top} P^\alpha_{j}\right)^{2}+\lambda\|{\beta^\alpha}\|_{1},
\end{equation}
where $\lambda\|{\beta^\alpha}\|_{1}$ is the $L_1$ regularization term. The $L_1$ regularization adds a value equal to the magnitude of the coefficients to find a sparse model and prevent overfitting. The value of $\lambda$ is set through cross validation.



\section{Performance of the Prediction Methods in Test Dataset}
In this section, I investigate the performance of the prediction methods in predicting the cycle times of the  products in the test dataset by using different feature sets.

\subsection{$C-K$ Rolling Average}
\begin{figure}
	\caption{$RMSE$ of different combinations of $C$ and $K$ in predicting the total cycle times of the validation dataset}
	\begin{center}
		\includegraphics[scale=1]{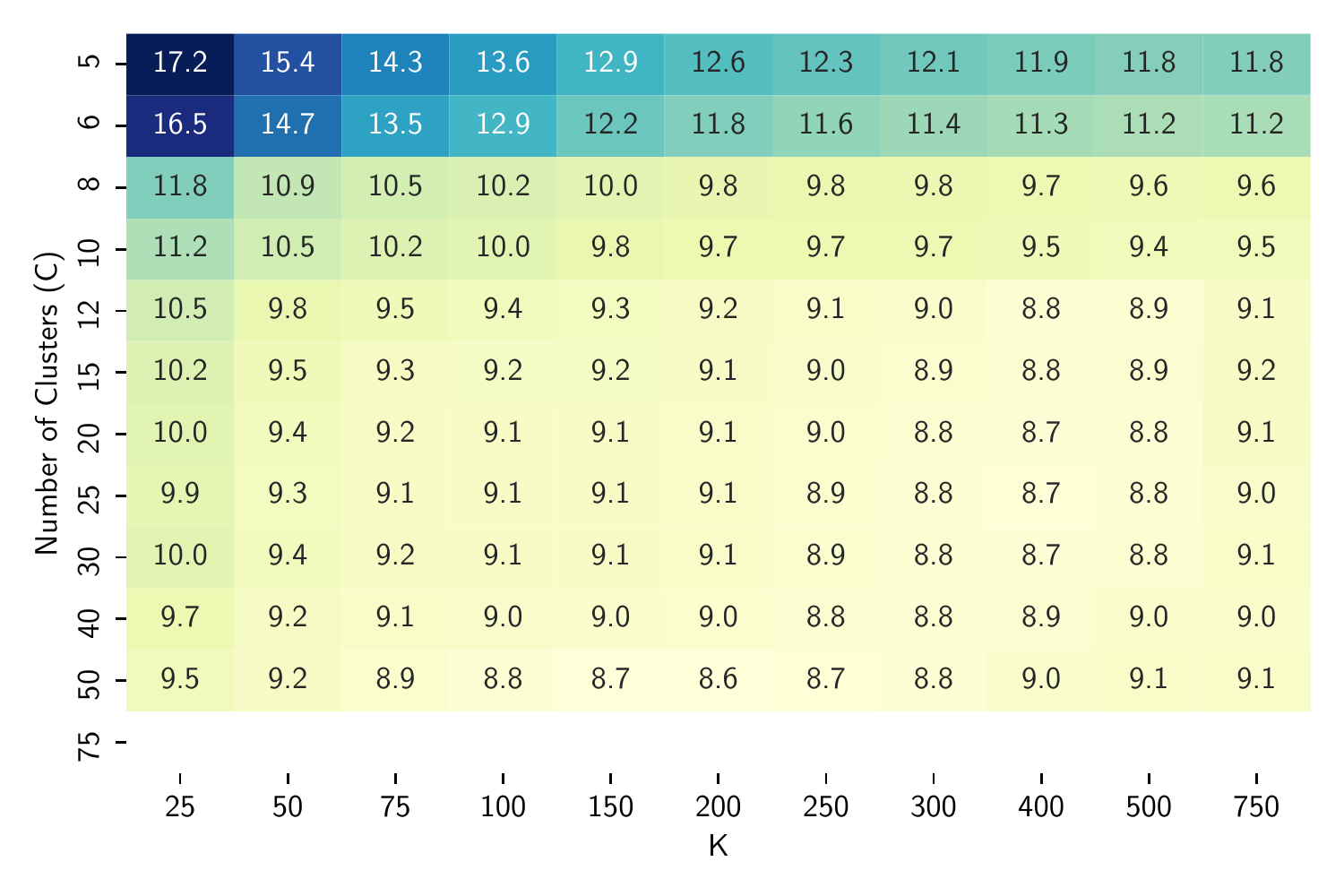}
	\end{center}
	\label{rap_rmse_val}
\end{figure}
\begin{figure}
	\caption{$RMSE$ of different combinations of $C$ and $K$ in predicting the total cycle times of the test dataset}
	\begin{center}
		\includegraphics[scale=1]{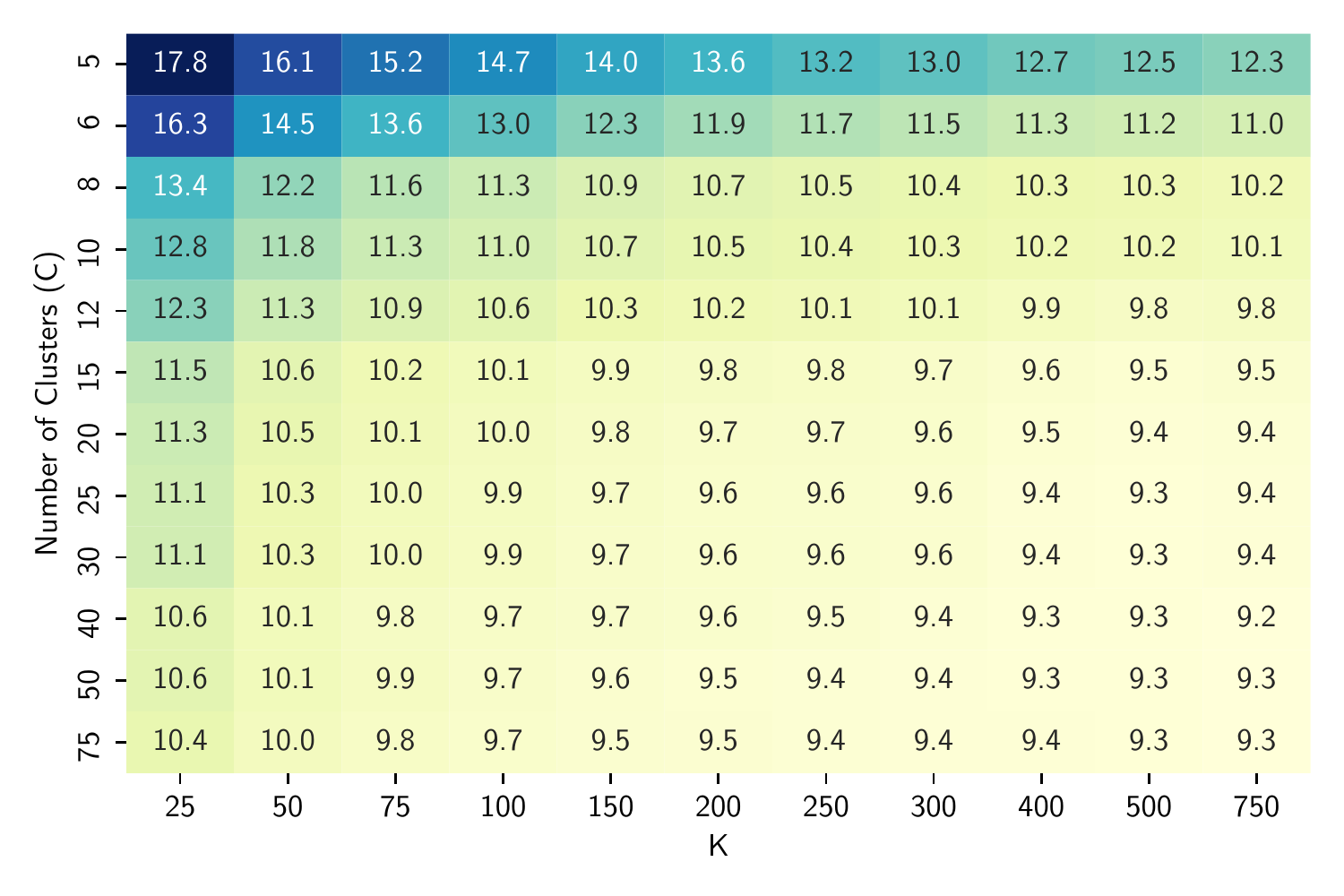}
	\end{center}
	\label{rap_rmse_test}
\end{figure}
\begin{figure}
	\caption{$MAE$ of different combinations of $C$ and $K$ in predicting the total cycle times of the validation dataset}
	\begin{center}
		\includegraphics[scale=1]{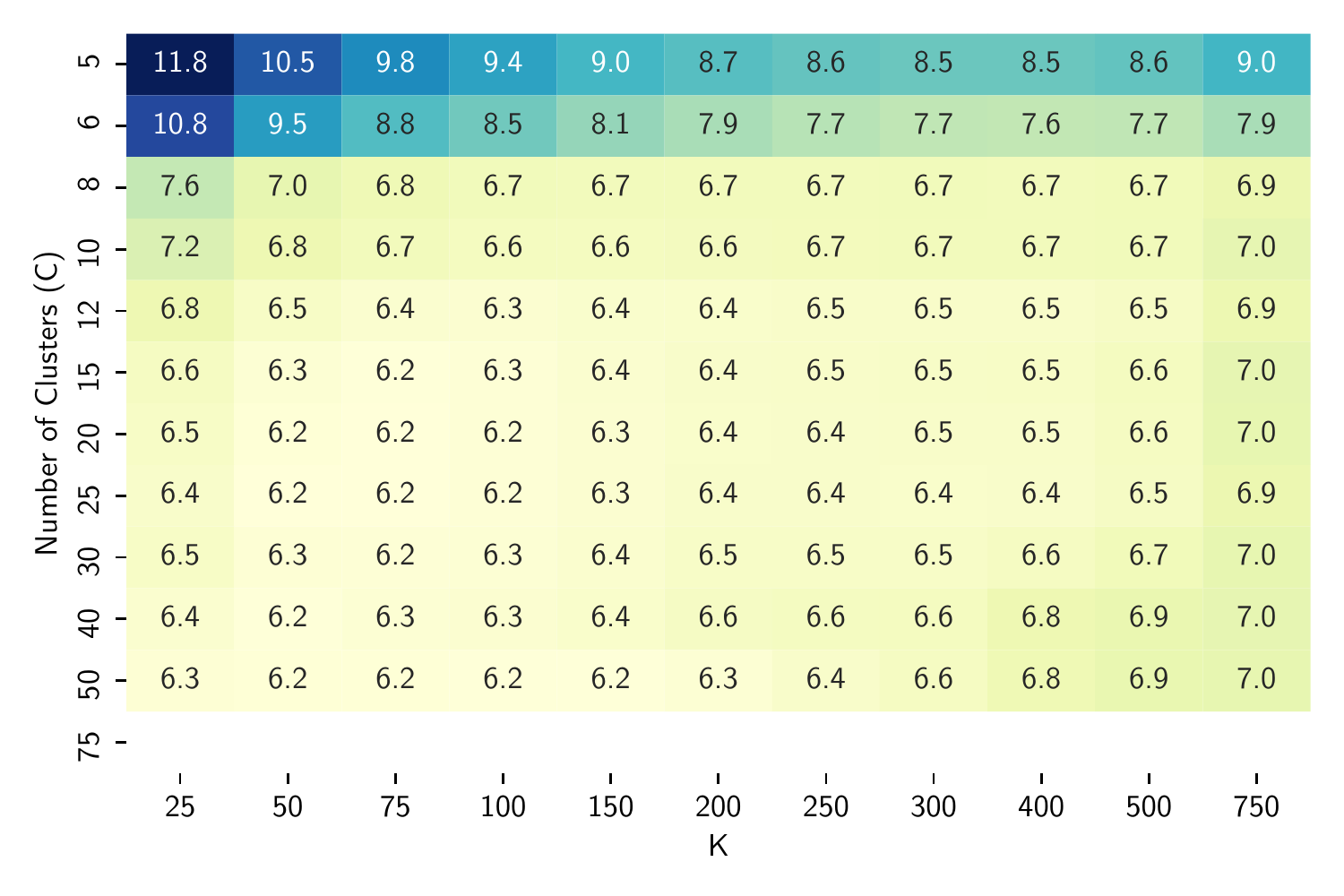}
	\end{center}
    \label{rap_mae_val}
\end{figure}
\begin{figure}
	\caption{$MAE$ of different combinations of $C$ and $K$ in predicting the total cycle times of the test dataset}
	\begin{center}
		\includegraphics[scale=1]{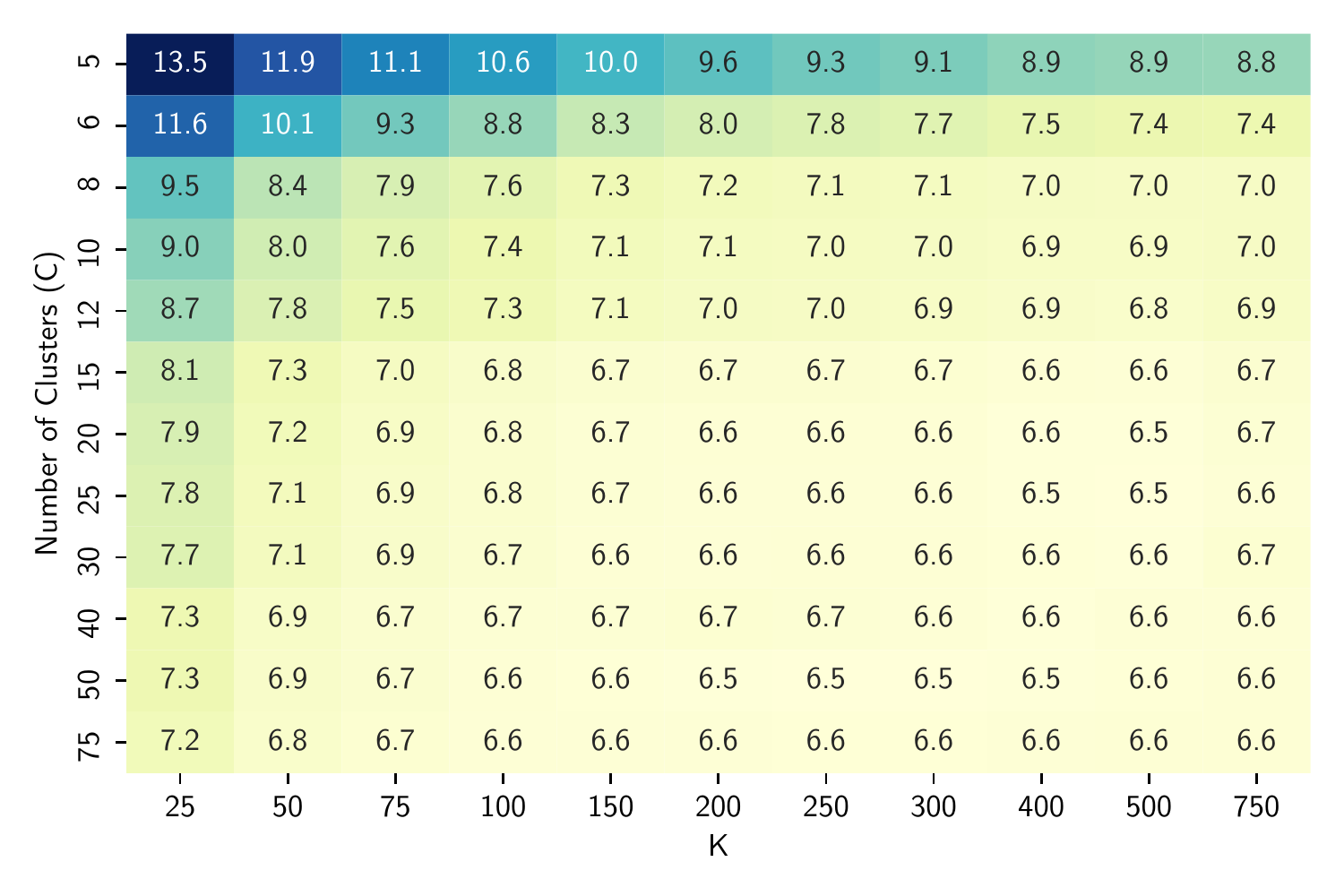}
	\end{center}
	\label{rap_mae_test}
\end{figure}

Figure \ref{rap_rmse_val} shows the $RMSE$ of the $C-K$ rolling average method in predicting the cycle times of the products in the validation dataset. Clustering the product into 20 or 25 categories and using the last 400 products of the same category in predicting the cycle times results in the lowest $RMSE$ in the validation dataset. A similar pattern is seen in the test dataset with a slight increase in the $RMSE$. Figures \ref{rap_rmse_test} shows the performance of different combinations of the $C-K$ in predicting the cycle times of the products in the test dataset. Using $C=25$ and $K=400$ to predict the cycle times of the products in the test dataset results in a $RMSE=9.4$. Even though setting $K=500$ and $K=750$ increases the $RMSE$ in the validation dataset, they show a slightly better performance in the test dataset than setting $K=400$. Using $C=25$ and $K=400$ results in a lower $MAE$ in the test dataset as well as shown in Figure \ref{rap_mae_test}. This is in contrast to the $MAE$ of the validation dataset where using smaller values of $K$ result in a better $MAE$ as shown in Figure \ref{rap_mae_val}. Table  \ref{c-k-rap_performance} shows the performance metrics of the $C-K$ rolling average method for $C \in \{20, 25\}$ and $K \in \{75, 150, 200, 250, 400\}$. $MAPE$ of the validation dataset increases as $K$ increases. Setting $K=75$ results in $MAPE=13.5\%$ which is almost 2\% less than setting $K=400$. 

The main shortcoming of the $C-K$ rolling average in predicting the cycle times is that it uses the products that have started their processing at least 20 days to three months ago to predict the cycle time of a new product. The state of the wafer fabrication may change significantly until the processing of the new lot is finished, increasing or decreasing the waiting times of the products in the system.
\begin{table}[]
\caption{Performance metrics of the $C-K$ rolling average method for $C \in \{20, 25\}$ and $K \in \{75, 150, 200, 250, 400\}$ in validation dataset}
\centering
\begin{tabular}{|l|cccc|cccc|}
\hline
C   & \multicolumn{4}{c|}{20}           & \multicolumn{4}{c|}{25}           \\ \hline
K   & $RMSE$ & $MAE$ & $MAPE$ & $MedAE$ & $RMSE$ & $MAE$ & $MAPE$ & $MedAE$ \\ \hline
75  & 9.2    & 6.2   & 13.5   & 3.8     & 9.1    & 6.2   & 13.5   & 3.8     \\
150 & 9.1    & 6.3   & 14.2   & 4.1     & 9.1    & 6.3   & 14.3   & 4.1     \\
200 & 9.1    & 6.4   & 14.6   & 4.4     & 9.1    & 6.4   & 14.7   & 4.4     \\
250 & 9.0    & 6.4   & 14.9   & 4.6     & 8.9    & 6.4   & 14.9   & 4.6     \\
300 & 8.8    & 6.5   & 15.1   & 4.8     & 8.8    & 6.4   & 15.1   & 4.7     \\
400 & 8.7    & 6.5   & 15.4   & 4.9     & 8.7    & 6.4   & 15.4   & 4.9     \\ \hline
\end{tabular}\label{c-k-rap_performance}
\end{table}

 
\subsection{LASSO}
\begin{figure}
	\caption{Scatter plot of the real and predicted cycle times by LASSO using the layer-based product related features ($P^{L-1}$) and layer-based product related features along with the total WIP level ($P^{L-0}$) }
	\begin{center}
		\includegraphics[scale=1]{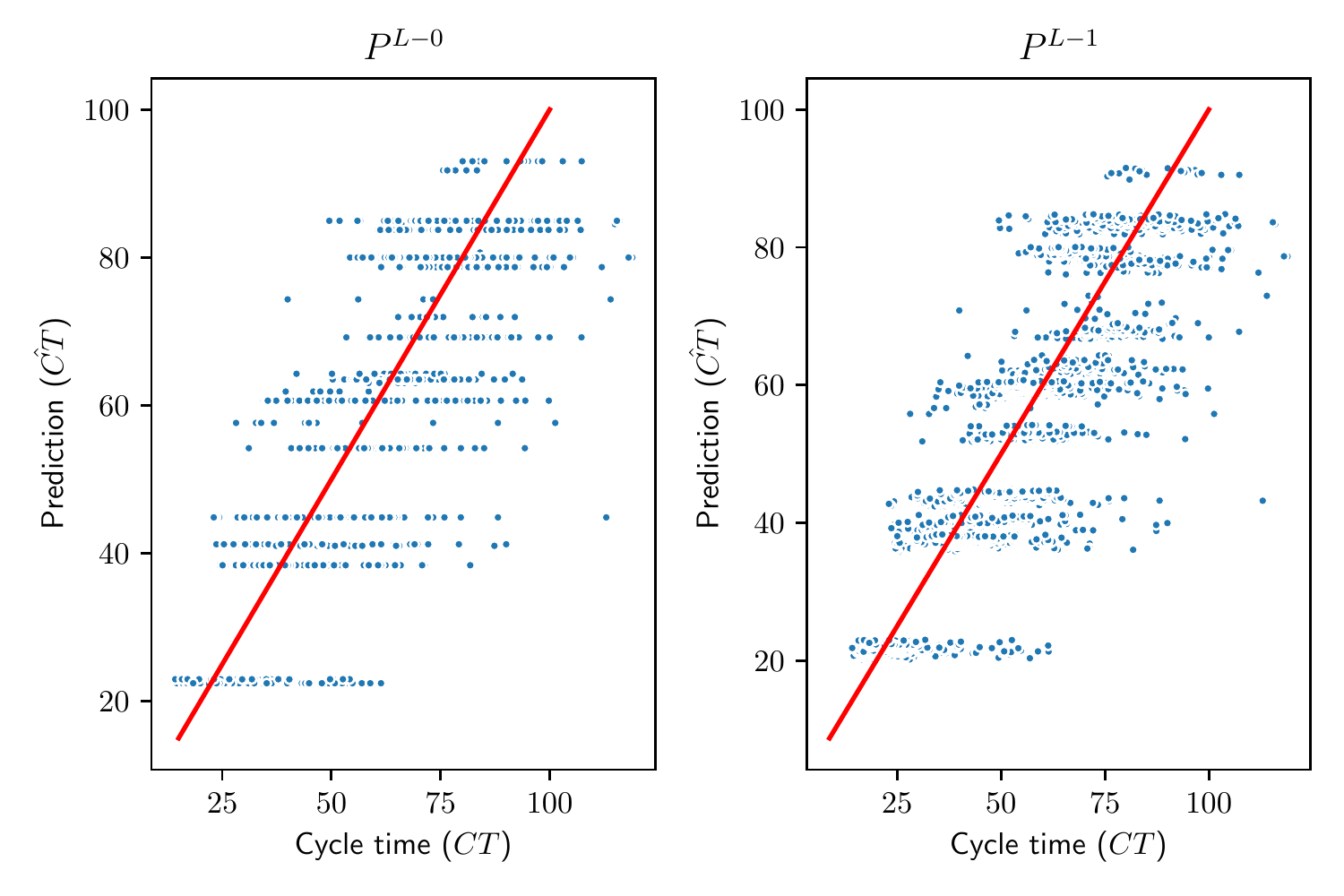}
	\end{center}
	\label{lasso_estimation_p01}
\end{figure}

\begin{table}[]
\caption{Performance metrics of the LASSO method in predicting the cycle times using different feature sets}
\centering
\begin{tabular}{|l|llllll|}\hline
 & $P^{L-0}$ & $P^{R-0}$ & $P^{L-1}$ & $P^{R-1}$ & $P^{L-2}$ & $P^{R-2}$ \\ \hline 
$RMSE$  & 9.4       & 9.4       & 9.4       & 9.8       & 9.5       & 9.5       \\
$MAE$   & 6.6       & 6.6       & 6.5       & 6.7       & 6.5       & 6.5       \\
$MAPE$  & 14.5      & 14.3      & 13.4      & 13.8      & 13.3      & 13.3      \\
$MedAE$ & 4.6       & 4.5       & 4.3       & 4.5       & 4.3       & 4.3     \\ \hline 
\end{tabular}\label{lasso_performance_metrics}
\end{table}
In  this section, I use different sets of features to predict the total cycle times of the products and the most important features impacting the cycle times using LASSO. Table \ref{lasso_performance_metrics} shows the performance metrics of the LASSO in predicting the cycle times using different feature sets. Using layer-based ($P^{L-0}$) and recipe-based ($P^{R-0}$ ) product related feature result in  14.5\% and 14.3\% $MAPE$. Adding the total WIP levels to $P^{L-0}$ decreases the $MAPE$ by 1.1\% while keeping the $RMSE$ almost the same. Figure \ref{lasso_estimation_p01} shows the scatter plots of the LASSO predictions using $P^{L-0}$ and $P^{L-1}$ feature sets. Adding the total WIP levels to the layer-based related features improves the prediction quality by accounting for some of the variability associated with the waiting times. 

\subsubsection{The Most Important Candidate Features}
\begin{figure}[]
	\caption{The most important features selected by LASSO using the layer-based product related features (${P}^{1-L}$) and layer-based product related features along with the total WIP level (${P}^{2-L}$). $AB-0.00$ denotes the name of the chosen layer ($AB$) and its quantile ($0.00$).}  
	\begin{center}
		\includegraphics[scale=1]{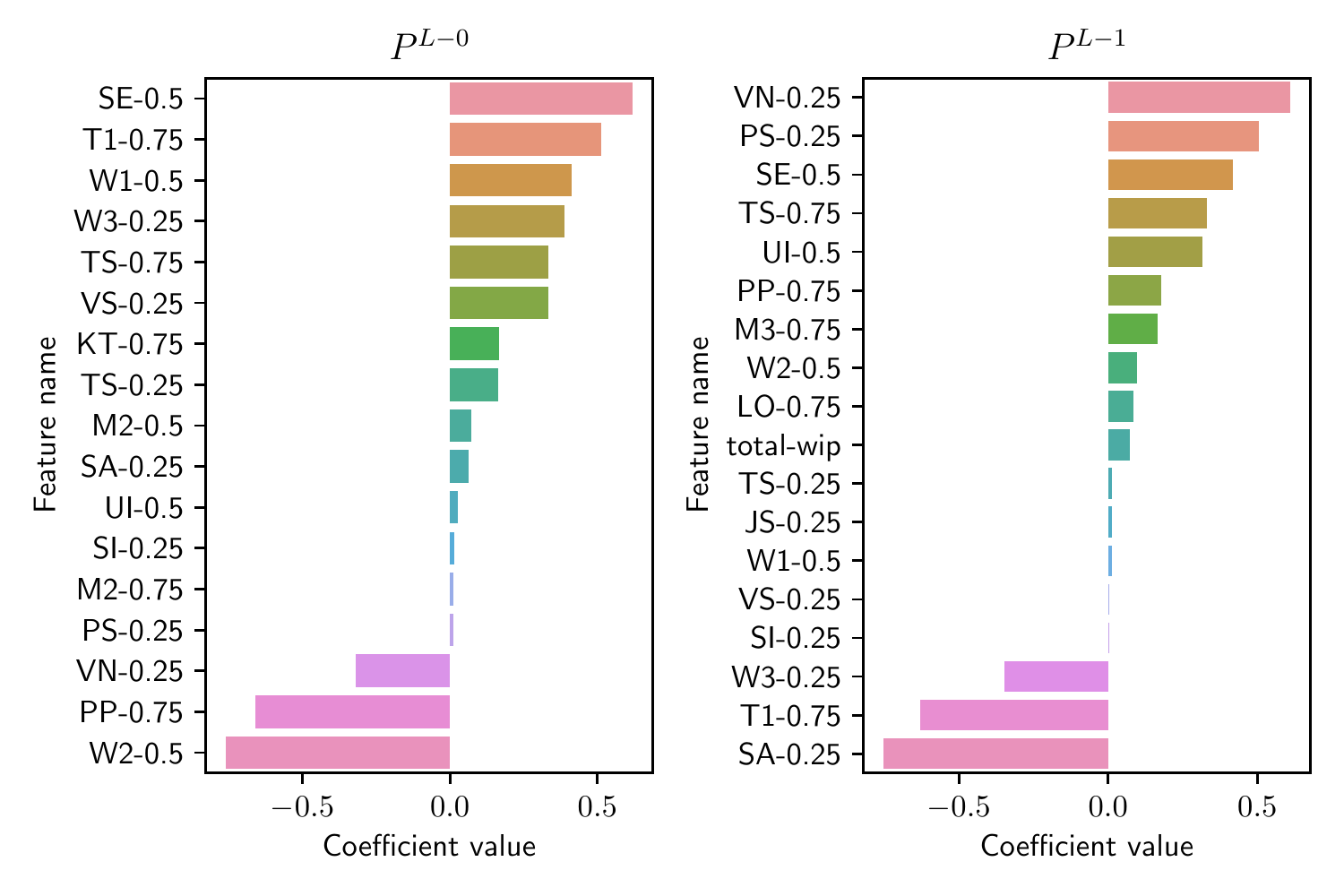}
	\end{center}
	\label{lasso_feature_importance}
\end{figure}

\begin{figure}[H]
	\caption{The most important features selected by LASSO using the layer-based product related features (${P}^{1-L}$) and layer-based product related features along with the total WIP level (${P}^{2-L}$) }  
	\begin{center}
		\includegraphics[scale=1]{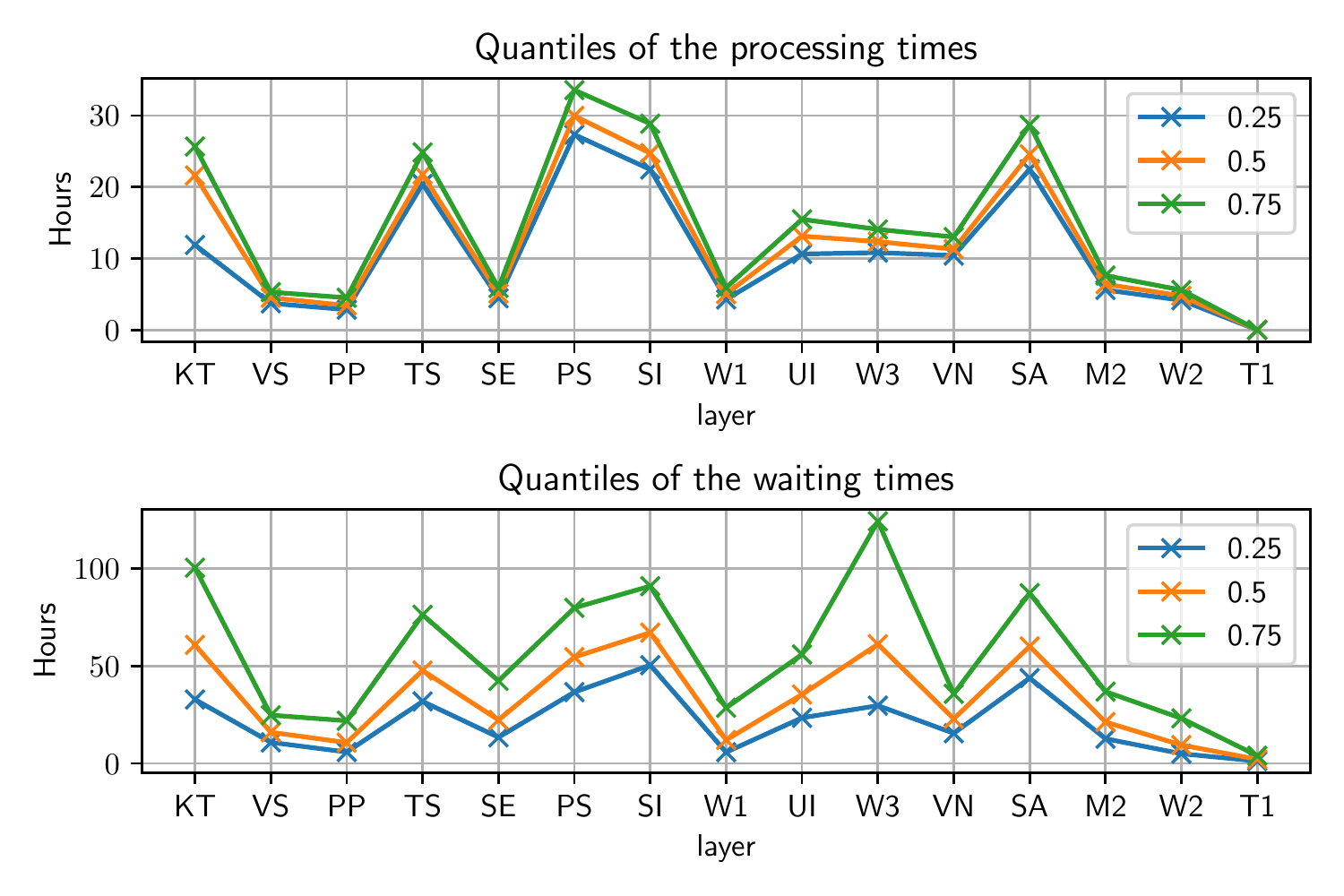}
	\end{center}
	\label{lasso_features_wt_pt_quantiles}
\end{figure}

Figure \ref{lasso_feature_importance} shows the coefficient values of the most important features selected by LASSO. LASSO assigns positive and negative coefficients to different quantiles of the layers. Even though LASSO chooses a similar set of features using $P^{L-0}$ and $P^{L-1}$, adding the total WIP levels into the layer-based product related features changes the coefficient of some layers significantly.    
Figure \ref{lasso_features_wt_pt_quantiles} shows quantiles of the processing and waiting times distribution of the chosen layers. Some layers demonstrate a significant waiting and processing times quantiles.  

\section{Conclusions and Future Research Directions}\label{conclusion}
In this chapter, I use the lot trace data from the Reutlingen semiconductor wafer fabrication to predict the total cycle times of the products. Two sets of features, namely the product-related and system-state related features, are extracted from the lot trace data to predict the total cycle times. The product-related features capture the product attributes such as the product type, information about the product route and distribution of the processing times on the product route. On the other hand, the system-state related features capture the impact of the state of system upon arrival of the lot on the total cycle times. I use these features to predict the cycle times of the products by using rolling average and LASSO methods. The rolling average method uses the product related features and recent products that have finished their processing to predict the total cycle times. On the other hand LASSO predict the cycle times by identifying a set of parameters that minimizes its objective function. In addition to predicting the cycle times LASSO also identifies the most important features that impact the cycle times of the products.

My objective is to prepare a dataset that can be used by other researchers for developing cycle time prediction methods. The prediction methods used in this study can be used as a benchmarks by other researchers to assess the performance of the developed methods. In addition, the identified features can be used by the practitioners to analyze the performance of the resources corresponding to the most important features to decrease the waiting times. 

This research can be extended in several directions. The first direction is to further investigate and integrate new sets of features that are important in explaining the variability in the cycle times. For instance, one of the most important features that impact the waiting and consequently cycle times of the products is the priority of the products. The priority of the lot changes in its route in the Reutlingen semiconductor wafer fabrication. Further discussion with the experts is needed to integrate the priority related features into the prediction methods. In this study, we only consider predicting the total cycle times of the products. A possible future research direction is to predict the remaining cycle times of the products in every step of their process route. Predicting the remaining cycle times of the product is important in setting the internal due-dates and priorities of the lots.

\chapter{Conclusions and Future Research Directions}\label{conclusion_chapter}

Technological advances allow manufacturers to collect and access data from a production system more easily and effectively. The objective of data collection is deploying the collected data in developing decision support systems for performance evaluation, problem identification, and production control. The collected data is used for different purposes from predictive maintenance to performance evaluation, production control, to supply chain optimization. Despite collection of huge amounts of data, there is a lack of documented, comprehensive, empirical research on manufacturing systems that uses detailed data from shop-floor to evaluate performance and optimize the manufacturing systems efficiency. The huge amounts of collected data introduces new challenges in the study of manufacturing systems that needs to be addressed. In this thesis, I address some of these challenges in performance evaluation and efficient control of manufacturing systems through analytical and empirical studies. I use the product flow or inter-event data from the Reutlingen semiconductor wafer fabrication in addressing these challenges.

In  the first part of the thesis, I investigate the question \textit{How can the collected data from the shop-floor be used in efficient control and design of manufacturing systems?} I analyze the statistical properties of the inter-event times in machine level to investigate the answers to this question. My empirical analysis of the statistical properties of the inter-event times dataset analyzed in this thesis shows that the inter-event times of a production system such as the inter-arrival and processing times may demonstrate a significant dependency between themselves. Such a dependency has been ignored in most of the analytical studies in control and design of production systems. In order to investigate the impact of possible dependency in the inter-event times on the optimal control and performance measures of the system, I analyze a manufacturing system that is controlled by using a single-threshold policy. I show that ignoring autocorrelation in inter-arrival or service times can lead to overestimation of the optimal threshold level for negatively correlated processes, and underestimation of the optimal threshold level for the positively correlated processes. In other words, I show that ignoring autocorrelation of a correlated inter-event times results in setting the base-stock level at a higher or lower level in comparison to the optimal threshold level. I conclude that Markovian Arrival Processes can be used to develop data-driven models and control manufacturing systems more effectively by using shop-floor inter-event data.   

I then identify the optimal control policy of a production system with correlated inter-arrival and service times. I consider a production/inventory control problem with correlated demand arrival and service process modeled as Markovian Arrival Processes. The objective of the control problem is minimizing the expected average cost of the system in steady-state by controlling when to produce a new product. I show that the optimal control policy of a fully observable system is a state-dependent threshold policy. I compare the performance of the optimal policy with that of the optimal single-threshold or base-stock policy where the threshold level is set independent of the state of the system by using Matrix Geometric methods. My analysis demonstrate that the state-independent policy performs near-optimal for the negatively correlated processes. However, when the inter-event times are positively correlated, using a state-dependent threshold policy improves the performance. 

I conclude the first part of thesis by stating that researchers and production managers should approach the correlation between the inter-event times with caution. Modeling or simulating a production system without taking the correlation into consideration may result in misleading results and increased variability of the performance metrics. Recently there has been some attempts \citep{Whitt2018UsingQueues} to integrate the correlation in the inter-event data into the analysis of queuing and manufacturing systems. However, there is a need for further research in modelling and analysis of multi-product, multi-server systems with a correlated inter-event data and impact of correlation on the optimal control of the system.  

In the second part of thesis, I investigate the answer to the question \emph{How can the collected data be used directly in identifying the most important features, evaluating the performance, and predicting the performance measures?} by performing an exploratory data analysis on the major factors that are impacting the main performance metrics of the semiconductor wafer fabrication in different levels of detail. My analysis shows that half of the products spend more than 70\% of their time inside wafer fabrication waiting to be processed.  The analysis reveals that some subsets of production steps in the product route contribute significantly more than others to the waiting and eventually the total cycle times of the products. I show that the total cycle times of the products in the wafer fabrication are highly variable, ranging from several days to several months. Some of the variability in the total cycle times is explained by the lottype of the product. However, the variation in the total cycle times are still quite significant. Hence, further research is required for explaining the main features that contribute the variability. According to  \citet{Chen1988EMPIRICALFABRICATION} to significantly reduce the manufacturing cycle times, one must reduce the variability in the operating environment. Over the years researchers have adopted different methodologies such as queuing systems and simulation to study and reduce the variability in the wafer fabrication. Queuing models, despite being fast, rely on restrictive modeling assumptions such as independent inter-event times which deteriorate their performance for real complex systems. Simulation models on the other hand are slow. A single replication of a simulation model may take several hours for a complex manufacturing systems. 

There are two main factors that impact the variability of the total cycle times through better due-date assignment and product release decisions. Due-date assignment impacts the total cycle times of the products through dispatching rules and scheduling of the lots on the machines. Different set of parameters, such as the priority and due-dates of the lots, which are being used for optimizing the flow of products inside the wafer fabrication are set based on the predictions of the total cycle times of products. Therefore, developing models that are able to predict delivery dates and remaining production times based on the observed data and current state of the wafer fabrication with acceptable accuracy is quite important. I use the findings of the empirical investigation of the total cycle times to extract two sets of features, namely product related and system state related features, that can be used to predict the total cycle times of the products. I predict the total cycle times of the products by using rolling average and learning methods. In addition to predicting the cycle times, I also identify the most important features that are impacting the total cycle times of the products. These features can be used by practitioners to improve the performance of the corresponding production areas. My objective is to prepare a dataset based on the mentioned feature sets that can be used by other researchers to develop cycle time prediction methods. The mentioned prediction methods provide a benchmark for assessing the performance of different prediction methods. 

Other important factors that impacts the variability of total waiting and cycle times are the product-release decisions and production system's WIP level balance in different production areas. Further research is required to understand the impact of release of different types of new products to the system. New stream of research in this area integrates machine learning models into the simulation models to build meta-models for predicting the cycle time–throughput (CT-TP) curves \citep{Yang2010NeuralManufacturing} that can be used for making release decisions. A possible research direction is to replace the costly simulation models with queuing systems such as infinite server queues to approximate the CT-TP curves.  

The imbalanced ratio of the waiting to processing times in different areas of the wafer fabrication suggests a possible inefficiency in controlling the wafer fabrication. The new products are usually released into the wafer fabrication based on a push strategy to maximize the utilization of the capital intensive machines. Releasing the products into wafer fabrication without taking into account its impact on the waiting times in front different sets of resources may lead to exponential increase in the waiting times and their variance. In wafer fabrication products are assigned to machines by using local optimizers such as dispatching rules. There is a need for further research that balances the WIP levels and consequently the waiting times of the products in front of different set of resources. A global optimizer that is integrated to the local optimizers and takes into account the current state of different set of resources is needed to optimize the WIP levels.

The analytical and empirical results presented in this dissertation show that the effective use of the collected data from a manufacturing system enables controlling the manufacturing system effectively and predicting its main performance measures accurately. 

\printbibliography

@inproceedings{Reference71,
    title = {{A Fast EM Algorithm for Fitting Marked Markovian Arrival Processes with a New Special Structure}},
    year = {2013},
    booktitle = {Computer Performance Engineering},
    author = {Horv{\'{a}}th, Gábor and Okamura, Hiroyuki},
    editor = {Balsamo, Maria Simonetta and Knottenbelt, William J and Marin, Andrea},
    pages = {119--133},
    publisher = {Springer Berlin Heidelberg},
    address = {Berlin, Heidelberg},
    isbn = {978-3-642-40725-3}
}

@inproceedings{Reference3,
    title = {{A MAP fitting approach with independent approximation of the inter-arrival time distribution and the lag correlation}},
    year = {2005},
    booktitle = {Second International Conference on the Quantitative Evaluation of Systems (QEST'05)},
    author = {Horvath, G and Buchholz, P and Telek, M},
    month = {9},
    pages = {124--133}
}

@article{Reference35,
    title = {{An example illustrating the possibilities of renewal theory and waiting-time theory for Markov-dependent arrival intervals}},
    year = {1961},
    journal = {Proc. Ser. A. Kon. Neder. Akad.Weten},
    author = {Runnenburg, J.Th},
    pages = {560--576},
    volume = {64}
}

@inproceedings{Reference22,
    title = {{Autocorrelation effects in manufacturing systems performance: A simulation analysis}},
    year = {2012},
    booktitle = {Proceedings of the 2012 Winter Simulation Conference (WSC)},
    author = {Pereira, D C and del Rio Vilas, D and Monteil, N R and Prado, R R and del Valle, A G},
    month = {12},
    pages = {1--12},
    issn = {0891-7736}
}

@inproceedings{Reference6,
    title = {{Autocorrelation of cycle times in semiconductor manufacturing systems}},
    year = {1995},
    booktitle = {Winter Simulation Conference Proceedings, 1995.},
    author = {Schomig, A K and Mittler, M},
    month = {12},
    pages = {865--872}
}

@inproceedings{butools,
    title = {{BuTools 2: A Rich Toolbox for Markovian Performance Evaluation}},
    year = {2017},
    booktitle = {Proceedings of the 10th EAI International Conference on Performance Evaluation Methodologies and Tools on 10th EAI International Conference on Performance Evaluation Methodologies and Tools},
    author = {Horvath, Gabor and Telek, Miklos},
    pages = {137--142},
    series = {VALUETOOLS{\&}{\#}39;16},
    publisher = {ICST (Institute for Computer Sciences, Social-Informatics and Telecommunications Engineering)},
    address = {Brussels, Belgium},
    isbn = {978-1-63190-141-6}
}

@article{IEC,
    title = {{Factory of the Future}},
    year = {2015},
    journal = {IEC-International Electrotechnical Commission White Paper},
    author = {{IEC}}
}

@book{Reference1,
    title = {{Fundamentals of Matrix-Analytic Methods}},
    year = {2013},
    author = {He, Qi-Ming},
    publisher = {Springer Publishing Company, Incorporated}
}

@article{Reference38,
    title = {{Further Results on Queues with Partial Correlation}},
    year = {1985},
    journal = {Operations Research},
    author = {Hadidi, Nasser},
    number = {1},
    pages = {203--209},
    volume = {33}
}

@book{Reference2,
    title = {{Input Modeling with Phase-Type Distributions and Markov Models: Theory and Applications}},
    year = {2014},
    author = {Buchholz, P and Kriege, J and Felko, I},
    publisher = {Springer Briefs in Mathematics}
}

@book{Reference14,
    title = {{Introduction to Queuing System with Telecommunication Applications}},
    year = {2013},
    author = {Lakatos, L and Szeidl, L and Telek, M},
    publisher = {Springer}
}

@article{Reference12,
    title = {{Marked Point Processes as Limits of Markovian Arrival Streams}},
    year = {1993},
    journal = {Journal of Applied Probability},
    author = {Asmussen, Søren and Koole, Ger},
    number = {2},
    pages = {365--372},
    volume = {30},
    publisher = {Applied Probability Trust}
}

@mastersthesis{Reference98,
    title = {{Modeling and Analysis of Autocorrelation in Manufacturing Systems}},
    year = {2016},
    author = {Manafzadeh Dizbin, Nima},
    school = {Ko{\c{c}} University},
    address = {Istanbul, Turkey}
}

@article{Reference64,
    title = {{Modeling and Generating Multivariate Time-series Input Processes Using a Vector Autoregressive Technique}},
    year = {2003},
    journal = {ACM Trans. Model. Comput. Simul.},
    author = {Biller, Bahar and Nelson, Barry L},
    number = {3},
    month = {7},
    pages = {211--237},
    volume = {13},
    publisher = {ACM},
    address = {New York, NY, USA},
    issn = {1049-3301}
}

@article{Reference27,
    title = {{MR/GI/1 Queues with Positively Correlated Arrival Stream}},
    year = {1994},
    journal = {Journal of Applied Probability},
    author = {Szekli, R and Disney, R L and Hur, S},
    number = {2},
    pages = {497--514},
    volume = {31},
    publisher = {Applied Probability Trust}
}

@article{Reference77,
    title = {{Multi-class Markovian arrival processes and their parameter fitting}},
    year = {2010},
    journal = {Performance Evaluation},
    author = {Buchholz, P and Kemper, P and Kriege, J},
    number = {11},
    pages = {1092--1106},
    volume = {67}
}

@article{Reference62,
    title = {{Non-Markovian State-Space Models in Dependability Evaluation}},
    year = {2013},
    journal = {Quality and Reliability Engineering International},
    author = {Distefano, Salvatore and Trivedi, Kishor S},
    number = {2},
    pages = {225--239},
    volume = {29}
}

@article{Reference74,
    title = {{On performance comparison of MR/GI/1 queues}},
    year = {1994},
    journal = {Queueing Systems},
    author = {Szekli, R and Disney, R L and Hur, S},
    number = {3},
    month = {9},
    pages = {451--470},
    volume = {17}
}

@article{Reference41,
    title = {{On the Exact Analysis of a Discrete-Time Queueing System with Autoregressive Inputs}},
    year = {2003},
    journal = {Queueing Systems},
    author = {Hwang, Gang Uk and Sohraby, Khosrow},
    number = {1},
    month = {1},
    pages = {29--41},
    volume = {43},
    issn = {1572-9443}
}

@article{Reference32,
    title = {{Performance Decay in a Single Server Exponential Queueing Model with Long Range Dependence}},
    year = {1997},
    journal = {Operations Research},
    author = {Resnick, Sidney and Samorodnitsky, Gennady},
    number = {2},
    pages = {235--243},
    volume = {45}
}

@inproceedings{Reference51,
    title = {{Simulation of a stochastic model for a service system}},
    year = {2010},
    booktitle = {Proceedings of the 2010 Winter Simulation Conference},
    author = {Brickner, C and Indrawan, D and Williams, D and Chakravarthy, S R},
    month = {12},
    pages = {1636--1647},
    issn = {0891-7736}
}

@article{Reference40,
    title = {{Single-Server Queue with Markov-Dependent Inter-Arrival and Service Times}},
    year = {2003},
    journal = {Queueing Systems},
    author = {Adan, I.J.B.F. and Kulkarni, V G},
    number = {2},
    month = {10},
    pages = {113--134},
    volume = {45}
}

@article{Reference36,
    title = {{Some numerical results on waiting-time distributions for dependent arrival-intervals}},
    year = {1962},
    journal = {Statistica Neerlandica},
    author = {Runnenburg, J Th.},
    number = {4},
    pages = {337--347},
    volume = {16},
    publisher = {Blackwell Publishing Ltd},
    issn = {1467-9574}
}

@inproceedings{Reference57,
    title = {{Structured Markov Chains Solver: Algorithms}},
    year = {2006},
    booktitle = {Proceeding from the 2006 Workshop on Tools for Solving Structured Markov Chains},
    author = {Bini, D A and Meini, B and Steff{\'{e}}, S and Van Houdt, B},
    series = {SMCtools '06},
    publisher = {ACM},
    address = {New York, NY, USA},
    isbn = {1-59593-506-1}
}

@article{Reference19,
    title = {{The Case for Modeling Correlation in Manufacturing Systems}},
    year = {2001},
    journal = {IIE Transactions},
    author = {Altiok, Tayfur and Melamed, Benjamin},
    number = {9},
    month = {9},
    pages = {779--791},
    volume = {33},
    issn = {1573-9724}
}

@article{Reference25,
    title = {{The effect of autocorrelated demand in JIT production systems}},
    year = {1998},
    journal = {International Journal of Production Research},
    author = {Takahashi, K and Nakamura, N},
    number = {5},
    pages = {1159--1176},
    volume = {36},
    publisher = {Taylor {\&} Francis}
}

@article{Reference65,
    title = {{The Effect of Correlated Arrivals on Queues}},
    year = {1993},
    journal = {IIE Transactions},
    author = {Patuwo, B Eddy and Disney, Ralph L and McNickle, Donald C},
    number = {3},
    pages = {105--110},
    volume = {25},
    publisher = {Taylor {\&} Francis}
}

@article{Reference33,
    title = {{The effect of long-memory arrivals on queue performance}},
    year = {2001},
    journal = {Operations Research Letters},
    author = {Dahl, Travis A and Willemain, Thomas R},
    number = {3},
    pages = {123--127},
    volume = {29},
    issn = {0167-6377}
}

@article{Reference76,
    title = {{The effects of the dependency in the Markov renewal arrival process on the various performance measures of exponential server queues}},
    year = {1989},
    journal = {PhD Dissertation, Virginia Polytechnic Institute and State University, Blacksburg, VA},
    author = {Patuwo, B E}
}

@article{Reference20,
    title = {{The Impact of Autocorrelation on Queuing Systems}},
    year = {1993},
    journal = {Management Science},
    author = {Livny, Miron and Melamed, Benjamin and Tsiolis, Athanassios K},
    number = {3},
    pages = {322--339},
    volume = {39}
}

@article{Reference21,
    title = {{The Impact of Dependence on Queuing Systems}},
    year = {2009},
    journal = {Working paper, Carnegie Mellon University, Pittsburgh},
    author = {Civelek, I and Biller, B and SchellerWolf, A}
}

@article{JEMAI,
    title = {{The influence of demand variability on the performance of a make-to-stock queue}},
    year = {2005},
    journal = {European Journal of Operational Research},
    author = {Jemai, Zied and Karaesmen, Fikri},
    number = {1},
    pages = {195--205},
    volume = {164},
    issn = {0377-2217}
}

@article{Sethi1997Demand,
    title = {{ Optimality of (s,S) policies in inventory models with Markovian demand}},
    year = {1997},
    journal = {Operations Research},
    author = {Sethi, S P and Cheng, F},
    number = {},
    month = {8},
    pages = {931--939},
    volume = {45}
}

@article{Hu2016sDemands,
    title = {{(s, S) Inventory Systems with Correlated Demands}},
    year = {2016},
    journal = {INFORMS Journal on Computing},
    author = {Hu, J and Zhang, C and Zhu, C},
    number = {4},
    month = {8},
    pages = {603--611},
    volume = {28}
}

@article{Wang2018ASystem,
    title = {{A Data Driven Cycle Time Prediction With Feature Selection in a Semiconductor Wafer Fabrication System}},
    year = {2018},
    journal = {IEEE Transactions on Semiconductor Manufacturing},
    author = {Wang, Junliang and Zhang, Jie and Wang, Xiaoxi},
    number = {1},
    month = {2},
    pages = {173--182},
    volume = {31},
    url = {http://ieeexplore.ieee.org/document/8242675/},
    doi = {10.1109/TSM.2017.2788501},
    issn = {0894-6507}
}

@article{Janakiraman2009ASystems,
    title = {{A decomposition approach for a class of capacitated serial systems}},
    year = {2009},
    journal = {Operations Research},
    author = {Janakiraman, G and Muckstadt, J},
    number = {},
    month = {8},
    pages = {1384--1393},
    volume = {57}
}

@article{Chen2003AFab,
    title = {{A fuzzy back propagation network for output time prediction in a wafer fab}},
    year = {2003},
    journal = {Applied Soft Computing Journal},
    author = {Chen, Toly},
    number = {3},
    month = {1},
    pages = {211--222},
    volume = {2},
    publisher = {Elsevier BV},
    doi = {10.1016/S1568-4946(02)00066-2},
    issn = {15684946}
}

@article{Manuel2008ACustomers,
    title = {{A perishable inventory system with service facilities and retrial customers}},
    year = {2008},
    journal = {Computers {\&} Industrial Engineering},
    author = {Manuel, Paul and Sivakumar, B. and Arivarignan, G.},
    number = {3},
    month = {4},
    pages = {484--501},
    volume = {54},
    publisher = {Pergamon},
    url = {https://www.sciencedirect.com/science/article/abs/pii/S0360835207002021},
    doi = {10.1016/J.CIE.2007.08.010},
    issn = {0360-8352}
}

@article{Manuel2007ATimes,
    title = {{A perishable inventory system with service facilities, MAP arrivals and PH — Service times}},
    year = {2007},
    journal = {Journal of Systems Science and Systems Engineering},
    author = {Manuel, Paul and Sivakumar, B. and Arivarignan, G.},
    number = {1},
    month = {3},
    pages = {62--73},
    volume = {16},
    publisher = {Systems Engineering Society of China},
    url = {http://link.springer.com/10.1007/s11518-006-5025-3},
    doi = {10.1007/s11518-006-5025-3},
    issn = {1004-3756}
}

@inproceedings{Segal1989AManufacturing,
    title = {{A Queueing Network Analyzer for Manufacturing}},
    year = {1989},
    booktitle = {Teletraffic Science for New Cost-Effective Systems, Networks and Services, ITC 12 },
    author = {Segal, Moshe and Whitt, Ward},
    pages = {1146--1152},
    url = {https://www.semanticscholar.org/paper/A-Queueing-Network-Analyzer-for-Manufacturing-Segal-Whitt/b326371efbf2a1aa14a1a4d29de48da9aacfede1},
    address = {North-Holland, Amsterdam}
}

@article{Zhao2011ACustomers,
    title = {{A queueing-inventory system with two classes of customers}},
    year = {2011},
    journal = {International Journal of Production Economics},
    author = {Zhao, Ning and Lian, Zhaotong},
    number = {1},
    month = {1},
    pages = {225--231},
    volume = {129},
    publisher = {Elsevier},
    url = {https://www.sciencedirect.com/science/article/pii/S0925527310004044},
    doi = {10.1016/J.IJPE.2010.10.011},
    issn = {0925-5273}
}

@article{Bitran1992ASystems,
    title = {{A review of open queueing network models of manufacturing systems}},
    year = {1992},
    journal = {Queueing Systems},
    author = {Bitran, Gabriel R. and Dasu, Sriram},
    number = {1-2},
    month = {3},
    pages = {95--133},
    volume = {12},
    publisher = {Kluwer Academic Publishers},
    url = {http://link.springer.com/10.1007/BF01158637},
    doi = {10.1007/BF01158637},
    issn = {0257-0130}
}

@article{Muharremoglu2008ASystems,
    title = {{A Single-Unit Decomposition Approach to Multiechelon Inventory Systems}},
    year = {2008},
    journal = {Operations Research},
    author = {Muharremoglu, Alp and Tsitsiklis, John N.},
    number = {5},
    month = {10},
    pages = {1089--1103},
    volume = {56},
    publisher = { INFORMS },
    url = {http://pubsonline.informs.org/doi/abs/10.1287/opre.1080.0620},
    doi = {10.1287/opre.1080.0620},
    issn = {0030-364X}
}

@article{Liberopoulos2000ASystems,
    title = {{A unified framework for pull control mechanisms in multi‐stage manufacturing systems}},
    year = {2000},
    journal = {Annals of Operations Research},
    author = {Liberopoulos, George and Dallery, Yves},
    number = {1/4},
    pages = {325--355},
    volume = {93},
    publisher = {Kluwer Academic Publishers},
    url = {http://link.springer.com/10.1023/A:1018980024795},
    doi = {10.1023/A:1018980024795},
    issn = {02545330}
}

@article{Neuts1979AProcess,
    title = {{A versatile Markovian point process}},
    year = {1979},
    journal = {Journal of Applied Probability},
    author = {Neuts, Marcel F.},
    number = {04},
    month = {12},
    pages = {764--779},
    volume = {16},
    publisher = {Applied Probability Trust},
    url = {https://www.cambridge.org/core/product/identifier/S0021900200033465/type/journal_article},
    doi = {10.2307/3213143},
    issn = {0021-9002}
}

@article{Ang2016AccuratePrediction,
    title = {{Accurate emergency department wait time prediction}},
    year = {2016},
    journal = {Manufacturing and Service Operations Management},
    author = {Ang, Erjie and Kwasnick, Sara and Bayati, Mohsen and Plambeck, Erica L. and Aratow, Michael},
    number = {1},
    month = {12},
    pages = {141--156},
    volume = {18},
    publisher = {INFORMS Inst.for Operations Res.and the Management Sciences},
    url = {https://pubsonline.informs.org/doi/abs/10.1287/msom.2015.0560},
    doi = {10.1287/msom.2015.0560},
    issn = {15265498}
}

@article{Chen2013AnFabrication,
    title = {{An effective fuzzy collaborative forecasting approach for predicting the job cycle time in wafer fabrication}},
    year = {2013},
    journal = {Computers and Industrial Engineering},
    author = {Chen, Toly},
    number = {4},
    month = {12},
    pages = {834--848},
    volume = {66},
    publisher = {Pergamon},
    doi = {10.1016/j.cie.2013.09.010},
    issn = {03608352}
}

@article{Chen2007AnPrediction,
    title = {{An intelligent hybrid system for wafer lot output time prediction}},
    year = {2007},
    journal = {Advanced Engineering Informatics},
    author = {Chen, Toly},
    number = {1},
    month = {1},
    pages = {55--65},
    volume = {21},
    publisher = {Elsevier},
    doi = {10.1016/j.aei.2006.10.002},
    issn = {14740346}
}

@article{Jiang2015AnBalance,
    title = {{An Optimization Model for Inventory System and the Algorithm for the Optimal Inventory Costs Based on Supply-Demand Balance}},
    year = {2015},
    journal = {Mathematical Problems in Engineering},
    author = {Jiang, Qingsong and Xing, Wei and Hou, Ruihuan and Zhou, Baoping},
    month = {12},
    pages = {1--11},
    volume = {2015},
    publisher = {Hindawi},
    url = {http://www.hindawi.com/journals/mpe/2015/508074/},
    doi = {10.1155/2015/508074},
    issn = {1024-123X}
}

@article{Chen2011ApplyingProduct,
    title = {{Applying the hybrid fuzzy c-means-back propagation network approach to forecast the effective cost per die of a semiconductor product}},
    year = {2011},
    journal = {Computers and Industrial Engineering},
    author = {Chen, Toly},
    number = {3},
    month = {10},
    pages = {752--759},
    volume = {61},
    publisher = {Pergamon},
    doi = {10.1016/j.cie.2011.05.007},
    issn = {03608352}
}

@article{Kim2014AreProcesses,
    title = {{Are call center and hospital arrivals well modeled by nonhomogeneous poisson processes?}},
    year = {2014},
    journal = {Manufacturing and Service Operations Management},
    author = {Kim, Song Hee and Whitt, Ward},
    number = {3},
    month = {6},
    pages = {464--480},
    volume = {16},
    publisher = {INFORMS Inst.for Operations Res.and the Management Sciences},
    doi = {10.1287/msom.2014.0490},
    issn = {15265498}
}

@article{Schomig1995AutocorrelationSystems,
    title = {{Autocorrelation of cycle times in semiconductor manufacturing systems}},
    year = {1995},
    journal = {Proceeding WSC '95 Proceedings of the 27th conference on Winter simulation},
    author = {Schomig, A K and Mittler, M},
    number = {},
    month = {8},
    pages = {865--872},
    volume = {}
}

@article{Beyer1997AverageDemands,
    title = {{Average cost optimality in inventory models with markovian demands}},
    year = {1997},
    journal = {Journal of Optimization Theory Applications},
    author = {Beyer, D and Sethi, S P},
    number = {},
    month = {8},
    pages = {497--526},
    volume = {92}
}

@article{Wang2016BigSystem,
    title = {{Big data analytics for forecasting cycle time in semiconductor wafer fabrication system}},
    year = {2016},
    journal = {International Journal of Production Research},
    author = {Wang, Junliang and Zhang, Jie},
    number = {23},
    month = {12},
    pages = {7231--7244},
    volume = {54},
    publisher = {Taylor {\&} Francis},
    url = {https://www.tandfonline.com/doi/full/10.1080/00207543.2016.1174789},
    doi = {10.1080/00207543.2016.1174789},
    issn = {0020-7543}
}

@article{Bright1995CalculatingProcesses,
    title = {{Calculating the equilibrium distribution in level dependent quasi-birth-and-death processes}},
    year = {1995},
    journal = {Communications in Statistics. Stochastic Models},
    author = {Bright, L. and Taylor, P.G.},
    number = {3},
    month = {1},
    pages = {497--525},
    volume = {11},
    publisher = { Marcel Dekker, Inc. },
    url = {http://www.tandfonline.com/doi/abs/10.1080/15326349508807357},
    doi = {10.1080/15326349508807357},
    issn = {0882-0287}
}

@article{Duri2000ComparisonKanbanb,
    title = {{Comparison among three pull control policies: kanban, base stock, and generalized kanban}},
    year = {2000},
    journal = {Annals of Operations Research},
    author = {Duri, C. and Frein, Y. and Di Mascolo, M.},
    number = {1/4},
    pages = {41--69},
    volume = {93},
    publisher = {Kluwer Academic Publishers},
    url = {http://link.springer.com/10.1023/A:1018919806139},
    doi = {10.1023/A:1018919806139},
    issn = {02545330}
}

@article{Nasr2015ContinuousDemand,
    title = {{Continuous (s, S) policy with MMPP correlated demand}},
    year = {2015},
    journal = {European Journal of Operational Research},
    author = {Nasr, Walid W. and Maddah, Bacel},
    number = {3},
    month = {11},
    pages = {874--885},
    volume = {246},
    publisher = {North-Holland},
    url = {https://www.sciencedirect.com/science/article/abs/pii/S0377221715004191},
    doi = {10.1016/J.EJOR.2015.05.029},
    issn = {0377-2217}
}

@article{Chung2002CycleLots,
    title = {{Cycle time estimation for wafer fab with engineering lots}},
    year = {2002},
    journal = {IIE Transactions},
    author = {Chung, Shu Hsing and Huang, Hung Wen},
    number = {2},
    pages = {105--118},
    volume = {34},
    publisher = { Taylor {\&} Francis Group },
    url = {https://www.tandfonline.com/doi/abs/10.1080/07408170208928854},
    doi = {10.1080/07408170208928854},
    issn = {15458830}
}

@inproceedings{Susto2016DealingProblems,
    title = {{Dealing with time-series data in Predictive Maintenance problems}},
    year = {2016},
    booktitle = {2016 IEEE 21st International Conference on Emerging Technologies and Factory Automation (ETFA)},
    author = {Susto, G A and Beghi, A},
    doi = {10.1109/ETFA.2016.7733659}
}

@article{Gershwin2000DesignPolicy,
    title = {{Design and operation of manufacturing systems: The control-point policy}},
    year = {2000},
    journal = {IIE Transactions},
    author = {Gershwin, Stanley B},
    number = {},
    month = {8},
    pages = {891--906},
    volume = {32}
}

@article{Sha2007DevelopmentProblem,
    title = {{Development of a regression-based method with case-based tuning to solve the due date assignment problem}},
    year = {2007},
    journal = {International Journal of Production Research},
    author = {Sha, D. Y. and Storch, R. L. and Liu, C. H.},
    number = {1},
    pages = {65--82},
    volume = {45},
    publisher = { Taylor {\&} Francis Group },
    url = {https://www.tandfonline.com/doi/abs/10.1080/00207540500507435},
    doi = {10.1080/00207540500507435},
    issn = {00207543}
}

@inproceedings{Pearn2007Due-dateEnvironment,
    title = {{Due-date assignment for wafer fabrication under demand variate environment}},
    year = {2007},
    booktitle = {IEEE Transactions on Semiconductor Manufacturing},
    author = {Pearn, W. L. and Chung, S. H. and Lai, C. M.},
    number = {2},
    month = {5},
    pages = {165--175},
    volume = {20},
    doi = {10.1109/TSM.2007.895215},
    issn = {08946507}
}

@article{Sha2004Due-dateNetworks,
    title = {{Due-date assignment in wafer fabrication using artificial neural networks}},
    year = {2004},
    journal = {The International Journal of Advanced Manufacturing Technology},
    author = {Sha, D. Y. and Hsu, S. Y.},
    number = {9-10},
    month = {5},
    pages = {768--775},
    volume = {23},
    publisher = {Springer-Verlag},
    url = {http://link.springer.com/10.1007/s00170-003-1644-8},
    doi = {10.1007/s00170-003-1644-8},
    issn = {0268-3768}
}

@article{Berman2001DynamicChains,
    title = {{Dynamic order replenishment policy in internet-based supply chains}},
    year = {2001},
    journal = {Mathematical Methods of Operations Research (ZOR)},
    author = {Berman, Oded and Kim, Eungab},
    number = {3},
    month = {7},
    pages = {371--390},
    volume = {53},
    publisher = {Springer-Verlag Berlin Heidelberg},
    url = {http://link.springer.com/10.1007/s001860100116},
    doi = {10.1007/s001860100116},
    issn = {1432-2994}
}

@book{Bertsekas2005DynamicControl,
    title = {{Dynamic programming and optimal control}},
    year = {2005},
    author = {Bertsekas, Dimitri P.},
    publisher = {Athena Scientific},
    isbn = {9781886529441}
}

@article{Hsieh2014EfficientMetamodeling,
    title = {{Efficient development of cycle time response surfaces using progressive simulation metamodeling}},
    year = {2014},
    journal = {International Journal of Production Research},
    author = {Hsieh, Liam Y. and Chang, Kuo-Hao and Chien, Chen-Fu},
    number = {10},
    month = {5},
    pages = {3097--3109},
    volume = {52},
    publisher = {Taylor {\&} Francis},
    url = {http://www.tandfonline.com/doi/abs/10.1080/00207543.2013.864055},
    doi = {10.1080/00207543.2013.864055},
    issn = {0020-7543}
}

@article{Yang2007EfficientMetamodeling,
    title = {{Efficient generation of cycle time-throughput curves through simulation and metamodeling}},
    year = {2007},
    journal = {Naval Research Logistics},
    author = {Yang, Feng and Ankenman, Bruce and Nelson, Barry L.},
    number = {1},
    month = {2},
    pages = {78--93},
    volume = {54},
    publisher = {John Wiley {\&} Sons, Ltd},
    url = {http://doi.wiley.com/10.1002/nav.20188},
    doi = {10.1002/nav.20188},
    issn = {0894069X}
}

@article{Chen1988EMPIRICALFABRICATION,
    title = {{Empirical Evaluation of a Queuing Network Model for Semiconductor Wafer Fabrication}},
    year = {1988},
    journal = {Operations Research},
    author = {Chen, Hong and Harrison, J. Michael and Mandelbaum, Avi and Ackere, Ann Van and Wein, Lawrence M.},
    number = {2},
    month = {4},
    pages = {202--215},
    volume = {36},
    publisher = {INFORMS},
    url = {https://pubsonline.informs.org/doi/abs/10.1287/opre.36.2.202},
    doi = {10.1287/opre.36.2.202}
}

@article{Inman1999EmpiricalSystems,
    title = {{Empirical evaluation of exponential and independence assumptions in queueing models of manufacturing systems}},
    year = {1999},
    journal = {Production and Operations Management},
    author = {Inman, Robert R.},
    number = {4},
    month = {1},
    pages = {409--432},
    volume = {8},
    publisher = {Wiley/Blackwell (10.1111)},
    url = {http://doi.wiley.com/10.1111/j.1937-5956.1999.tb00316.x},
    doi = {10.1111/j.1937-5956.1999.tb00316.x}
}

@article{Chen2014EnhancingImplications,
    title = {{Enhancing the Effectiveness of Cycle Time Estimation in Wafer Fabrication-Efficient Methodology and Managerial Implications}},
    year = {2014},
    journal = {Sustainability},
    author = {Chen, Toly and Wang, Yu-Cheng},
    number = {8},
    month = {8},
    pages = {5107--5128},
    volume = {6},
    publisher = {MDPI AG},
    url = {http://www.mdpi.com/2071-1050/6/8/5107},
    doi = {10.3390/su6085107},
    issn = {2071-1050}
}

@inproceedings{Heindl2004ETAQAProcess,
    title = {{ETAQA truncation models for the MAP/MAP/1 departure process}},
    year = {2004},
    booktitle = {First International Conference on the Quantitative Evaluation of Systems, 2004. QEST 2004. Proceedings.},
    author = {Heindl, A. and Zhang, Q. and Smirni, E.},
    pages = {100--109},
    publisher = {IEEE},
    url = {http://ieeexplore.ieee.org/document/1348024/},
    isbn = {0-7695-2185-1},
    doi = {10.1109/QEST.2004.1348024}
}

@article{Song1996EvaluationPolicies,
    title = {{Evaluation of base-stock policies in multiechelon inventory systems with state-dependent demands: Part I: State-independent policies}},
    year = {1996},
    journal = {Naval Research Logistics},
    author = {Song, J S and Zipkin, P},
    number = {},
    month = {8},
    pages = {715--728},
    volume = {39}
}

@article{Song1996EvaluationPoliciesb,
    title = {{Evaluation of base-stock policies in multiechelon inventory systems with state-dependent demands:Part II: State-dependent depot policies}},
    year = {1996},
    journal = {Naval Research Logistics},
    author = {Song, J S and Zipkin, P},
    number = {},
    month = {8},
    pages = {381--396},
    volume = {43}
}

@article{Chang2005EvolvingFactory,
    title = {{Evolving fuzzy rules for due-date assignment problem in semiconductor manufacturing factory}},
    year = {2005},
    journal = {Journal of Intelligent Manufacturing},
    author = {Chang, Pei Chann and Hieh, Jih Chang and Liao, T. Warren},
    number = {4-5},
    month = {10},
    pages = {549--557},
    volume = {16},
    publisher = {Springer},
    url = {https://link.springer.com/article/10.1007/s10845-005-1663-4},
    doi = {10.1007/s10845-005-1663-4},
    issn = {09565515}
}

@article{Backus2006FactoryApproach,
    title = {{Factory cycle-time prediction with a data-mining approach}},
    year = {2006},
    journal = {IEEE Transactions on Semiconductor Manufacturing},
    author = {Backus, Phillip and Janakiram, Mani and Mowzoon, Shahin and Runger, George C. and Bhargava, Amit},
    number = {2},
    month = {5},
    pages = {252--258},
    volume = {19},
    doi = {10.1109/TSM.2006.873400},
    issn = {08946507}
}

@article{Liu2014FlexibleClasses,
    title = {{Flexible service policies for a Markov inventory system with two demand classes}},
    year = {2014},
    journal = {International Journal of Production Economics},
    author = {Liu, Mingwu and Feng, Mengying and Wong, Chee Yew},
    month = {5},
    pages = {180--185},
    volume = {151},
    publisher = {Elsevier},
    url = {https://www.sciencedirect.com/science/article/pii/S0925527313004519},
    doi = {10.1016/J.IJPE.2013.10.010},
    issn = {0925-5273}
}

@article{Tirkel2013ForecastingDatabases,
    title = {{Forecasting flow time in semiconductor manufacturing using knowledge discovery in databases}},
    year = {2013},
    journal = {International Journal of Production Research},
    author = {Tirkel, Israel},
    number = {18},
    month = {9},
    pages = {5536--5548},
    volume = {51},
    publisher = { Routledge },
    url = {http://www.tandfonline.com/doi/abs/10.1080/00207543.2013.787168},
    doi = {10.1080/00207543.2013.787168},
    issn = {0020-7543}
}

@inproceedings{Laipple2019GenericChains,
  author={G. {Laipple} and S. {Dauzère-Pérès} and T. {Ponsignon} and P. {Vialletelle}},
  booktitle={2018 Winter Simulation Conference (WSC)}, 
  title={Generic data model for semiconductor manufacturing supply chains}, 
  year={2018},
  volume={},
  number={},
  pages={3615-3626},
  }

@article{Chen2010IncorporatingPlant,
    title = {{Incorporating the FCM-BPN approach with nonlinear programming for internal due date assignment in a wafer fabrication plant}},
    year = {2010},
    journal = {Robotics and Computer-Integrated Manufacturing},
    author = {Chen, Toly and Wang, Yi Chi},
    number = {1},
    month = {2},
    pages = {83--91},
    volume = {26},
    publisher = {Pergamon},
    doi = {10.1016/j.rcim.2009.04.001},
    issn = {07365845}
}

@article{Song1993InventoryEnvironment,
    title = {{Inventory control in a fluctuating demand environment}},
    year = {1993},
    journal = {Operations Research},
    author = {Song, J S and Zipkin, P},
    number = {},
    month = {8},
    pages = {351--370},
    volume = {41}
}

@article{Bayraktar2010InventoryDemand,
    title = {{Inventory management with partially observed nonstationary demand}},
    year = {2010},
    journal = {Annals of Operations Research},
    author = {Bayraktar, E and Ludkovski, M},
    number = {1},
    month = {8},
    pages = {7--39},
    volume = {176}
}

@article{Beyer1998InventoryGrowth,
    title = {{Inventory models with markovian demands and cost functions of polynomial growth}},
    year = {1998},
    journal = {Journal of Optimization Theory and Applications},
    author = {Beyer, D and Sethi, S P and Taksar, M},
    number = {2},
    month = {8},
    pages = {281--323},
    volume = {98}
}

@article{Ozekici1999InventoryEnvironment,
    title = {{Inventory models with unreliable suppliers in a random environment}},
    year = {1999},
    journal = {Annals of Operations Research},
    author = {{\"{O}}zekici, S and Parlar, M},
    number = {},
    month = {8},
    pages = {123--136},
    volume = {91}
}

@article{Chien2012ManufacturingTime,
    title = {{Manufacturing intelligence to forecast and reduce semiconductor cycle time}},
    year = {2012},
    journal = {Journal of Intelligent Manufacturing},
    author = {Chien, Chen-Fu and Hsu, Chia-Yu and Hsiao, Chih-Wei},
    number = {6},
    month = {12},
    pages = {2281--2294},
    volume = {23},
    publisher = {Springer US},
    url = {http://link.springer.com/10.1007/s10845-011-0572-y},
    doi = {10.1007/s10845-011-0572-y},
    issn = {0956-5515}
}

@article{Asmussen1993MarkedStreams,
    title = {{Marked point processes as limits of Markovian arrival streams}},
    year = {1993},
    journal = {Journal of Applied Probability},
    author = {Asmussen, Søren and Koole, Ger},
    number = {02},
    month = {6},
    pages = {365--372},
    volume = {30},
    publisher = {Applied Probability Trust},
    url = {https://www.cambridge.org/core/product/identifier/S0021900200117371/type/journal_article},
    doi = {10.2307/3214845},
    issn = {0021-9002}
}

@article{ManafzadehDizbin2019ModellingProcesses,
    title = {{Modelling and analysis of the impact of correlated inter-event data on production control using Markovian arrival processes}},
    year = {2019},
    journal = {Flexible Services and Manufacturing Journal},
    author = {Manafzadeh Dizbin, N. and Tan, B.},
    number = {4},
    pages = {1042--1076},
    volume = {31},
    doi = {10.1007/s10696-018-9329-7},
    issn = {19366590}
}

@article{Koole2006MonotonicityApplications,
    title = {{Monotonicity in Markov Reward and Decision Chains: Theory and Applications}},
    year = {2006},
    journal = {Foundations and Trends in Stochastic Systems},
    author = {Koole, Ger},
    number = {1},
    pages = {1--76},
    volume = {1},
    publisher = {Now Publishers, Inc.},
    url = {http://www.nowpublishers.com/article/Details/STO-002},
    doi = {10.1561/0900000002}
}

@article{Bauerle1997MonotonicityQueues,
    title = {{Monotonicity results for MR/GI/1 queues}},
    year = {1997},
    journal = {Journal of Applied Probability},
    author = {B{\"{a}}uerle, Nicole},
    number = {02},
    month = {6},
    pages = {514--524},
    volume = {34},
    publisher = {Applied Probability Trust},
    url = {https://www.cambridge.org/core/product/identifier/S0021900200101147/type/journal_article},
    doi = {10.2307/3215390},
    issn = {0021-9002}
}

@article{Yang2010NeuralManufacturing,
    title = {{Neural network metamodeling for cycle time-throughput profiles in manufacturing}},
    year = {2010},
    journal = {European Journal of Operational Research},
    author = {Yang, Feng},
    number = {1},
    month = {8},
    pages = {172--185},
    volume = {205},
    publisher = {North-Holland},
    doi = {10.1016/j.ejor.2009.12.026},
    issn = {03772217}
}

@inproceedings{Baumann2010NumericalProcesses,
    title = {{Numerical solution of level dependent quasi-birth-and-death processes}},
    year = {2010},
    booktitle = {Procedia Computer Science},
    author = {Baumann, Hendrik and Sandmann, Werner},
    number = {1},
    pages = {1561--1569},
    volume = {1},
    publisher = {Elsevier B.V.},
    doi = {10.1016/j.procs.2010.04.175},
    issn = {18770509}
}

@article{Armony2015OnPerspective,
    title = {{On Patient Flow in Hospitals: A Data-Based Queueing-Science Perspective}},
    year = {2015},
    journal = {Stochastic Systems},
    author = {Armony, Mor and Israelit, Shlomo and Mandelbaum, Avishai and Marmor, Yariv N. and Tseytlin, Yulia and Yom-Tov, Galit B.},
    number = {1},
    month = {6},
    pages = {146--194},
    volume = {5},
    publisher = { INFORMS },
    url = {http://pubsonline.informs.org/doi/10.1287/14-SSY153},
    doi = {10.1287/14-SSY153},
    issn = {1946-5238}
}

@article{Tan2017OnBlocking,
    title = {{On the output dynamics of production systems subject to blocking}},
    year = {2017},
    journal = {IISE Transactions},
    author = {Tan, Barış and Lagershausen, Svenja},
    number = {3},
    month = {3},
    pages = {268--284},
    volume = {49},
    publisher = {Taylor {\&} Francis},
    url = {https://www.tandfonline.com/doi/full/10.1080/0740817X.2016.1222470},
    doi = {10.1080/0740817X.2016.1222470},
    issn = {2472-5854}
}

@article{He2002OptimalSystem,
    title = {{Optimal and near-optimal inventory control policies for a make-to-order inventory–production system}},
    year = {2002},
    journal = {European Journal of Operational Research},
    author = {He, Q.-M. and Jewkes, E.M. and Buzacott, J.},
    number = {1},
    month = {8},
    pages = {113--132},
    volume = {141},
    publisher = {North-Holland},
    url = {https://www.sciencedirect.com/science/article/abs/pii/S0377221701002570},
    doi = {10.1016/S0377-2217(01)00257-0},
    issn = {0377-2217}
}

@article{Veatch1994OptimalSystem,
    title = {{Optimal Control of a Two-Station Tandem Production/Inventory System}},
    year = {1994},
    journal = {Operations Research},
    author = {Veatch, Michael H. and Wein, Lawrence M.},
    number = {2},
    month = {4},
    pages = {337--350},
    volume = {42},
    publisher = { INFORMS },
    url = {http://pubsonline.informs.org/doi/abs/10.1287/opre.42.2.337},
    doi = {10.1287/opre.42.2.337},
    issn = {0030-364X}
}

@article{ManafzadehDizbin2020OptimalTimes,
    title = {{Optimal control of production-inventory systems with correlated demand inter-arrival and processing times}},
    year = {2020},
    journal = {International Journal of Production Economics},
    author = {Manafzadeh Dizbin, N. and Tan, B.},
    number = {107692},
    pages = {},
    volume = {228},
    doi = {10.1016/j.ijpe.2020.107692},
    issn = {09255273}
}

@article{Xia2017OptimalQueue,
    title = {{Optimal Control of State-Dependent Service Rates in a MAP/M/1 Queue}},
    year = {2017},
    journal = {IEEE Transactions on Automatic Control},
    author = {Xia, Li and He, Qi-Ming and Alfa, Attahiru Sule},
    number = {10},
    month = {10},
    pages = {4965--4979},
    volume = {62},
    url = {http://ieeexplore.ieee.org/document/7873328/},
    doi = {10.1109/TAC.2017.2679139},
    issn = {0018-9286}
}

@article{Chen2001OptimalDemand,
    title = {{Optimal policies for multi-echelon inventory problems with Markov modulated demand}},
    year = {2001},
    journal = {Operations Research},
    author = {Chen, F and Song, J S},
    number = {},
    month = {8},
    pages = {226--234},
    volume = {49}
}

@article{deVericourt2002OptimalSystem,
    title = {{Optimal Stock Allocation for a Capacitated Supply System}},
    year = {2002},
    journal = {Management Science},
    author = {de V{\'{e}}ricourt, Francis and Karaesmen, Fikri and Dallery, Yves},
    number = {11},
    month = {11},
    pages = {1486--1501},
    volume = {48},
    publisher = {INFORMS},
    url = {http://pubsonline.informs.org/doi/abs/10.1287/mnsc.48.11.1486.263},
    doi = {10.1287/mnsc.48.11.1486.263},
    issn = {0025-1909}
}

@article{Gurkan2007OptimalOptimization,
    title = {{Optimal threshold levels in stochastic fluid models via simulation-based optimization}},
    year = {2007},
    journal = {Discrete Event Dynamic Systems: Theory and Applications},
    author = {Gurkan, Gul and Karaesmen, Fikri and Ozdemir, Ozge},
    number = {1},
    month = {2},
    pages = {53--97},
    volume = {17},
    publisher = {Kluwer Academic Publishers-Plenum Publishers},
    url = {http://link.springer.com/10.1007/s10626-006-0002-z},
    doi = {10.1007/s10626-006-0002-z}
}

@article{Cheng1999OptimalitySales,
    title = {{Optimality of state-dependent (s,S) policies in inventory models with Markov-modulated demand and lost sales}},
    year = {1999},
    journal = {Production and Operations Management},
    author = {Cheng, F and Sethi, S},
    number = {2},
    month = {8},
    pages = {183--192},
    volume = {8}
}

@article{He2000PerformanceSystem,
    title = {{Performance measures of a make-to-order inventory-production system}},
    year = {2000},
    journal = {IIE Transactions},
    author = {He, QI-MING and Jewkes, E.M.},
    number = {5},
    month = {5},
    pages = {409--419},
    volume = {32},
    publisher = {Taylor {\&} Francis Group},
    url = {http://www.tandfonline.com/doi/abs/10.1080/07408170008963917},
    doi = {10.1080/07408170008963917}
}

@article{Neuts1975ProbabilityType,
    title = {{Probability distributions of phase type}},
    year = {1975},
    journal = {Liber Amicorum Prof. Emeritus H. Florin},
    author = {Neuts, Marcel F.},
    publisher = {Department of Mathematics, University of Louvain},
    url = {https://ci.nii.ac.jp/naid/10020985825/}
}

@article{Sharifnia1988ProductionStates,
    title = {{Production control of a manufacturing system with multiple machine states}},
    year = {1988},
    journal = {IEEE Transactions on Automatic Control},
    author = {Sharifnia, A.},
    number = {7},
    month = {7},
    pages = {620--625},
    volume = {33},
    url = {http://ieeexplore.ieee.org/document/1270/},
    doi = {10.1109/9.1270},
    issn = {00189286}
}

@article{Tan2002ProductionUncertainty,
    title = {{Production control of a pull system with production and demand uncertainty}},
    year = {2002},
    journal = {IEEE Transactions on Automatic Control},
    author = {Tan, B.},
    number = {5},
    month = {5},
    pages = {779--783},
    volume = {47},
    url = {http://ieeexplore.ieee.org/document/1000272/},
    doi = {10.1109/TAC.2002.1000272},
    issn = {0018-9286}
}

@article{Gershwin2009ProductionDemand,
    title = {{Production control with backlog-dependent demand}},
    year = {2009},
    journal = {IIE Transactions},
    author = {Gershwin, Stanley B. and Tan, Bariş and Veatch, Michael H.},
    number = {6},
    month = {4},
    pages = {511--523},
    volume = {41},
    publisher = { Taylor {\&} Francis Group },
    url = {http://www.tandfonline.com/doi/abs/10.1080/07408170801975040},
    doi = {10.1080/07408170801975040},
    issn = {0740-817X}
}

@article{Tan2018ProductionUncertainty,
    title = {{Production Control with Price, Cost, and Demand Uncertainty}},
    year = {2018},
    journal = {OR Spectrum},
    author = {Tan, Barış},
    month = {12},
    pages = {1--29},
    publisher = {Springer Berlin Heidelberg},
    url = {http://link.springer.com/10.1007/s00291-018-0542-2},
    doi = {10.1007/s00291-018-0542-2},
    issn = {0171-6468}
}

@book{Monch2013ProductionFacilities,
    title = {{Production Planning and Control for Semiconductor Wafer Fabrication Facilities}},
    year = {2013},
    author = {M{\"{o}}nch, Lars and Fowler, John W. and Mason, Scott J.},
    series = {Operations Research/Computer Science Interfaces Series},
    volume = {52},
    publisher = {Springer New York},
    url = {http://link.springer.com/10.1007/978-1-4614-4472-5},
    address = {New York, NY},
    isbn = {978-1-4614-4471-8},
    doi = {10.1007/978-1-4614-4472-5}
}

@article{Karabag2019PurchasingPrices,
    title = {{Purchasing, production, and sales strategies for a production system with limited capacity, fluctuating sales and purchasing prices}},
    year = {2019},
    journal = {IISE Transactions},
    author = {Karaba{\u{g}}, Oktay and Tan, Bariş},
    number = {9},
    month = {9},
    pages = {921--942},
    volume = {51},
    publisher = {Taylor {\&} Francis},
    url = {https://www.tandfonline.com/doi/full/10.1080/24725854.2018.1535217},
    doi = {10.1080/24725854.2018.1535217},
    issn = {2472-5854}
}

@article{Batra2018QuantifyingChain,
    title = {{Quantifying the semiconductor supply chain}},
    year = {2018},
    journal = {McKinsey},
    author = {Batra, Gaurav and Nolde, Kristian and Santhanam, Nick and Vrijen, Rutger},
    url = {https://www.mckinsey.com/industries/semiconductors/our-insights/right-product-right-time-right-location-quantifying-the-semiconductor-supply-chain}
}

@article{Batur2018QuantileIndustry,
    title = {{Quantile regression metamodeling: Toward improved responsiveness in the high-tech electronics manufacturing industry}},
    year = {2018},
    journal = {European Journal of Operational Research},
    author = {Batur, Demet and Bekki, Jennifer M. and Chen, Xi},
    number = {1},
    month = {1},
    pages = {212--224},
    volume = {264},
    publisher = {Elsevier B.V.},
    doi = {10.1016/j.ejor.2017.06.020},
    issn = {03772217}
}

@article{Shanthikumar2007QueueingProblems,
    title = {{Queueing Theory for Semiconductor Manufacturing Systems: A Survey and Open Problems}},
    year = {2007},
    journal = {IEEE Transactions on Automation Science and Engineering},
    author = {Shanthikumar, J. George and Ding, Shengwei and Zhang, Mike Tao},
    number = {4},
    month = {10},
    pages = {513--522},
    volume = {4},
    url = {http://ieeexplore.ieee.org/document/4312834/},
    doi = {10.1109/TASE.2007.906348}
}

@article{Brown2005StatisticalCenter,
    title = {{Statistical Analysis of a Telephone Call Center}},
    year = {2005},
    journal = {Journal of the American Statistical Association},
    author = {Brown, Lawrence and Gans, Noah and Mandelbaum, Avishai and Sakov, Anat and Shen, Haipeng and Zeltyn, Sergey and Zhao, Linda},
    number = {469},
    month = {3},
    pages = {36--50},
    volume = {100},
    publisher = {Taylor {\&} Francis},
    url = {http://www.tandfonline.com/doi/abs/10.1198/016214504000001808},
    doi = {10.1198/016214504000001808},
    issn = {0162-1459}
}

@article{Hendricks1993TheBuffers,
    title = {{The Output Process of Serial Production Lines of General Machines with Finite Buffers}},
    year = {1993},
    journal = {Management Science},
    author = {Hendricks, Kevin B. and McClain, John O.},
    number = {10},
    month = {10},
    pages = {1194--1201},
    volume = {39},
    publisher = { INFORMS },
    url = {http://pubsonline.informs.org/doi/abs/10.1287/mnsc.39.10.1194},
    doi = {10.1287/mnsc.39.10.1194},
    issn = {0025-1909}
}

@article{Chen2014TheFabrication,
    title = {{The symmetric-partitioning and incremental-relearning classification and back-propagation-network tree approach for cycle time estimation in wafer fabrication}},
    year = {2014},
    journal = {Symmetry},
    author = {Chen, Toly},
    number = {2},
    pages = {409--426},
    volume = {6},
    publisher = {MDPI AG},
    doi = {10.3390/sym6020409},
    issn = {20738994}
}

@article{Whitt2018UsingQueues,
    title = {{Using robust queueing to expose the impact of dependence in single-server queues}},
    year = {2018},
    journal = {Operations Research},
    author = {Whitt, Ward and You, Wei},
    number = {1},
    month = {1},
    pages = {184--199},
    volume = {66},
    publisher = {INFORMS Inst.for Operations Res.and the Management Sciences},
    doi = {10.1287/opre.2017.1649},
    issn = {15265463}
}

@inproceedings{Schelasin2011UsingManufacturing,
    title = {{Using static capacity modeling and queuing theory equations to predict factory cycle time performance in semiconductor manufacturing}},
    year = {2011},
    booktitle = {Proceedings of the 2011 Winter Simulation Conference (WSC)},
    author = {Schelasin, Roland},
    month = {12},
    pages = {2040--2049},
    publisher = {IEEE},
    url = {http://ieeexplore.ieee.org/document/6147917/},
    isbn = {978-1-4577-2109-0},
    doi = {10.1109/WSC.2011.6147917}
}

\end{document}